\documentclass[twocolumn]{aastex631}
\usepackage{epsfig,amsmath,footmisc,longtable}

\shorttitle{Infrared galaxies detected by the ACT}
\shortauthors{Kilerci et al.}
\graphicspath{{./}{figures/}}

\begin{document}

\title{Infrared galaxies detected by the Atacama Cosmology Telescope}

\correspondingauthor{Ece Kilerci} \email{ece.kilerci@sabanciuniv.edu}

\author[0000-0002-9119-2313]{Ece Kilerci}
\affiliation{Sabanc{\i} University, Faculty of Engineering and Natural Sciences, 34956, Istanbul, Turkey}

\author[0000-0001-7228-1428]{Tetsuya Hashimoto}
\affiliation{Department of Physics, National Chung Hsing University, No. 145, Xingda Rd., South Dist., Taichung, 40227, Taiwan (R.O.C.)}

\author[0000-0002-6821-8669]{Tomotsugu Goto}
\affiliation{Institute of Astronomy, National Tsing Hua University, 101, Section 2. Kuang-Fu Road, Hsinchu, 30013, Taiwan (R.O.C.)}

\author[0000-0002-5274-6790]{Ersin G\"{o}\u{g}\"{u}\c{s}}
\affiliation{Sabanc{\i} University, Faculty of Engineering and Natural Sciences, 34956, Istanbul, Turkey}

\author[0000-0001-9970-8145]{Seong Jin Kim}
\affiliation{Institute of Astronomy, National Tsing Hua University, 101, Section 2. Kuang-Fu Road, Hsinchu, 30013, Taiwan (R.O.C.)}

\author[0000-0002-8560-3497]{Simon C.-C. Ho}
\affiliation{Research School of Astronomy and Astrophysics, The Australian National University, Canberra, ACT 2611, Australia}

\author[0000-0001-6236-6882]{Yi Hang Valerie Wong}
\affiliation{Department of Astrophysical and Planetary Sciences, University of Colorado Boulder, CO 80309, USA}



\begin{abstract}

We report on 167 infrared (IR) galaxies selected by AKARI and IRAS and detected in the Atacama Cosmology Telescope (ACT) Data Release 5 (DR5) sky maps at the 98, 150 and 220 GHz frequency bands. Of these detections, 134 (80\%) of the millimeter counterparts are first-time identifications with ACT. 
We expand the previous ACT extragalactic source catalogs, by including new 98 GHz detections and measurements from ACT DR5. 
We also report flux density measurements at the 98, 150, and 220 GHz frequency bands. We compute  $\alpha_{98-150}$, $\alpha_{98-220}$, and $\alpha_{150-220}$ millimeter-wave spectral indices and far-IR to millimeter-wave spectral indices between 90 $\mu$m and 98, 150, and 220 GHz. We specify the galaxy type, based on $\alpha_{150-220}$. We combine publicly available multiwavelength data-including ultraviolet, optical, near-IR, mid-IR, far-IR, and the millimeter measurements obtained in this work-and perform spectral energy distribution (SED) fitting with CIGALE.  With the radio emission decomposition advantage of CIGALE V2022.0, we identify the origins of the millimeter emissions for 69 galaxies in our sample. 
Our analysis also shows that millimeter data alone indicates the need for a radio synchrotron component in the SEDs that are produced by active galactic nuclei (AGNs) and/or star formation.
We present SEDs and measured physical properties of these galaxies, such as the  dust luminosity, AGN luminosity, the total IR luminosity, and the ratio of the IR and radio luminosity. We quantify the relationships between the total IR luminosity and the millimeter-band luminosities, which can be used in the absence of SED analysis. 
\end{abstract}


\section{Introduction} \label{sec:intro}
Infrared (IR) galaxies \citep[e.g.,][]{Sanders1996,Lonsdale2006} emit most of 
their radiation at mid-IR to far-IR (FIR) wavelengths, due to their dust content. 
Dust absorbs  the optical and ultraviolet (UV) radiation produced by star formation (SF) and active galactic nuclei (AGNs), then re-emits it in the IR, submillimeter, and millimeter wavelengths. 
IR galaxies are classified according to their total IR luminosity ($L_{\rm{IR}}$, integrated over -$3 - $1000 $\mu$m) as luminous IR galaxies (LIRGs; 10$^{11}L_{\sun} < L_{\rm{IR}} <$ 10$^{12}$ $L_{\sun}$) and ultraluminous IR galaxies (ULIRGs; 10$^{12}L_{\sun} < L_{\rm{IR}} < $ 10$^{13} L_{\sun}$). 
It is noteworthy that LIRGs and ULIRGs, which are relatively rare in the local universe  \citep[e.g.,][]{LeFloch2005,Goto2010,Rodighiero2010,KilerciEser2018}, are major systems of SF at the peak ($z \sim  1 - 2$) of cosmic SF activity  \citep[e.g.,][]{Rowan-Robinson2001,Goto2010}. Therefore, it is highly important to have a complete understanding of IR galaxies and the evolution of their SF.  The main processes controlling the evolutions of galaxies are SF and AGN activity (i.e. black hole accretion). After their peak, between $z \sim  1 - 3$, these two phenomena declined toward the present epoch. At the present epoch, however, the dusty star-forming IR galaxies do not follow this decline, representing the higher-redshift population of IR galaxies. For normal galaxies at the present epoch especially, AGN feedback is expected to reduce the star formation rate (SFR). On the other hand, both processes are working simultaneously in the IR galaxies at this epoch. These separate processes can be quantified by accurate measurements of the physical parameters of galaxies, such as the stellar mass, SFR, dust mass, and the AGN contribution to the total luminosity. All of these parameters can be directly obtained from spectral energy distributions (SEDs). 

The first attempts to model the SEDs of IR galaxies only included the parameterization of the IR luminosity \citep[e.g.,][]{Sanders1996,Chary2001,Rieke2009}. 
In addition, models obtained from observed galaxy SEDs with one or two parameters have been widely used for IR SEDs \citep[e.g.,][]{Dale2002,Dale2014}. 
Recently, we have attained a better understanding of the physical properties of IR galaxies, by using more sophisticated SED tools, such as the Code for Investigating Galaxy Emission \citep[CIGALE;][]{Noll2009,Serra2011,Boquien2019}, which requires the 
energy balance between the absorbed UV-to-optical emission and the re-emitted IR emission, with the presence of AGNs as well. 
 Due to the limited wavelength coverage of current space telescopes--- such as AKARI \citep{Murakami2007}, the Spitzer Space Telescope \citep[][]{Werner2004}, and the Herschel Space Observatory \citep[Herschel;][]{Pilbratt2010}--- the longest FIR wavelength to be included in SEDs is $\sim 600 \mu$m.  
Beyond the FIR region, the IR---radio correlation \citep{Helou1985} between the $L_{\rm{IR}}$ and the radio luminosity at 1.4 GHz is well known for radio galaxies, and can be used to 
trace SF \citep[e.g.,][]{Murphy2011,Murphy2012}. On the other hand, the millimeter region of the SED---that is, between the peak of the SED and the radio region---is not a deeply 
explored domain. In particular, the information that could be gained from the millimeter data has not been carefully examined for SEDs of IR galaxies. 
The millimeter data have the potential to constrain the SED beyond the peak ($\sim$90--100 $\mu$m), especially in the absence of FIR measurements.  
Additionally, the millimeter bands also provide an effective tool for selecting higher-redshift dusty star-forming galaxies \citep[DSFGs; see, e.g.,][]{Casey2021,Zavala2021}, due to the negative \textit{K}-correction (especially in the 2 mm band), and for measuring the contributions of DSFGs to the cosmic SF beyond redshift $z = 3$.
The scope of this work is to systematically explore the millimeter regions in the SEDs of a large sample of IR galaxies, to better describe the SEDs up to radio, to investigate the origin of the millimeter emission, and to obtain better insights into the physical properties of IR galaxies, such as their stellar masses, dust masses, SFRs and AGN contributions to the total IR luminosity. 

The main purpose of millimeter-wave surveys is to measure primordial fluctuations in the cosmic microwave background  \citep[CMB; e.g.;][]{Das2011,Story2013,Planck2014a,Planck2014b,Planck2020}. Foreground extragalactic sources also emit at millimeter wavelengths, between 90 and 220 GHz (1.3$-$3.0 mm). 
These sources are mainly AGNs, DSFGs, gravitationally lensed DSFGs, and 
galaxy clusters. AGN accretion disks produce jets composed of ionized relativistic matter traveling around magnetic field lines. 
When the direction of such a jet is toward us, the system is called a blazar \citep{Urry_Padovani1995}. The blazar jets generate synchrotron emission \citep{Ulrich1997}. DSFGs are also referred as IR galaxies, due to the significant IR radiation that is produced by the dust that is heated by the radiation of the SF 
and/or the AGN continuum radiation. This thermal dust emission extends to the millimeter wavelength region. AGNs at the centers of such galaxies also have contributions to the millimeter-wave emission as nonthermal radiation. Galaxy clusters can be mapped due to the thermal Sunyaev--Zel`dovich effect \citep[e.g.,][]{SunyaevZeldovich1972}, as a distortion of the CMB spectrum at millimeter wavelengths \citep[e.g.,][]{Hilton2021}. 

Due to the technical capacities of telescopes, such as the large beam sizes and the limited small sky surveys (mostly up to 10 deg$^{2}$), the 
study of the millimeter-wave properties of extragalactic sources has recently become a developing area 
 \citep[e.g.,][]{Vieira2010,Marriage2011a,Mocanu2013,Marsden2014,Planck2014a,Gralla2020,Everett2020,Reuter2020}. The Planck space telescope \citep{Planck2011} has observed the entire sky in the 30$-$875 GHz (9993--343 $\mu$m), frequency regime 
and cataloged several extragalactic sources \citep{Planck2011b,PlanckCollaboration2016b,Lianou2019}. 
With its large beam size of $\sim 5'-7'$, Planck can detect bright sources above $\gtrsim 400$\,mJy at 143 and 217 GHz \citep{PlanckCollaboration2016b}. 
 Additionally, the South Pole Telescope \citep[SPT;][]{Vieira2010, Carlstrom2011}, which is a 10 m millimeter telescope with arcminute resolution that is located at the National Science Foundation Amundsen--Scott South Pole Station in Antarctica, has scanned the southern hemisphere. The SPT had three millimeter bands, at 95, 150, and 220 GHz with 9.8, 5.8, and 20.5 mJy detection limits, respectively \citep[][]{Everett2020}. A large sample of millimeter-wave extragalactic  sources (more than 4000) have been identified and cataloged \citep[][]{Vieira2010,Mocanu2013,Everett2020,Reuter2020} in the SPT Sunyaev--Zel'dovich survey \citep[][]{Vieira2010}, which was completed between 2008 and 2011, over an area of $\sim 2530$ square degrees. 

The Atacama Cosmology Telescope \citep[ACT;][]{Swetz2011,Thornton2016} is a 6 meter ground-based telescope that is designed to survey the CMB. 
ACT started observing in 2007, with its first-generation Millimeter Bolometric Array Camera \citep[][]{Fowler2007,Swetz2011}, 
then between 2013 and 2015 it used ACTPol  camera \citep{Thornton2016}, which improved the sensitivity of polarization measurements. Since 2016, ACT has been observing with the Advanced ACTPol camera \citep{Henderson2016}, which has  
mapped almost half of the sky in five frequency bands, at 30, 40, 98, 150, and 220 GHz, at an angular resolution of a few arcminutes, with increased sensitivity. 
Recently, the latest ACT Data Release 5 \citep[DR5;][]{Naess2020} included merged maps at 90, 150, and 220 GHz obtained between the 2008 and 2018 observing seasons. Note that the Planck measurements at 143 and 217 GHz are  analogous to those at the 150 and 220 GHz frequencies of ACT. However, with a finer angular resolution ($\sim 1.'5$) and a lower detection threshold \citep[$\sim 20$ mJy at 150 GHz; ][]{Datta2019}, ACT has a great advantage in detecting sources with lower brightness.

The first catalog of extragalactic millimeter sources detected by ACT was presented by  \citet{Marriage2011a}. 
Their report 157 extragalactic sources, including clusters and radio sources, which were detected at 148 GHz during the 2008 observations. 
\citet{Marriage2011a} noted that most of these sources were detected at lower radio frequencies. 
They also determined the dominant radiation processes and spectral indices for these systems by cross-identifications with other catalogs, including the NASA/IPAC Extragalactic Database (NED), the REFLEX cluster catalog \citep{Bohringer2004}, the Infrared Astronomical Satellite (IRAS) Point Source Catalog \citep[][]{HelouWalker1988}, the Sydney University Molonglo Sky Survey (SUMSS) 0.84 GHz catalog \citep{Mauch2003}, the SPT 2.0 mm catalog \citep{Vieira2010}, and the Australia Telescope 20 GHz (AT20G) 5, 8, and 20 GHz catalogs \citep{Murphy2010}. Additionally, they examined the color-color comparisons at 5--20 GHz and 20--148 GHz for 109 ACT--AT20G sources, finding median spectral indices of 
$\alpha_{5-20} = -0.07 \pm 0.37$ and $\alpha_{20-148} = -0.39 \pm 0.24$, indicating spectral steeping between these frequencies \citep{Marriage2011a}. 

\citet{Marsden2014} presented an extended catalog, containing 191 extragalactic sources detected by ACT from 148 and 218 GHz maps that were obtained between 2007 and 2010. 
Their sample included 167 radio galaxies that were powered by AGNs and 24 DSFGs. \citet{Marsden2014} identified counterparts of their sample at 0.84 GHz
\citep[from SUMSS; ][]{Mauch2003}, 4.86 GHz \citep[from the Parkes-MIT-NRAO survey; ][]{Wright1994}, and 
20 GHz \citep[from the AT20G survey; ][]{Murphy2010}, as well as in the 1.4/2.0 mm \citep[from the SPT survey;][]{Vieira2010} and 12--100 $\mu$m 
bands \citep[from IRAS;][]{HelouWalker1988}. 
\citet{Marsden2014} included 20 GHz measurements from follow-up ATCA observations for 41 sources, reporting a 
steepening slope of the SED from 20 to 218 GHz for synchrotron-dominated galaxies. 
The nature of the millimeter sources  could be identified from the spectral index, $\alpha$, of the power-law spectra $S_{\nu} \sim \nu^{\alpha}$ \citep{Marsden2014}. The dust-dominated galaxies in their sample exhibited a median spectral index of $\alpha_{148-218}=3.7^{+0.62} _{-0.86}$. 
They reported a spectral index of $\alpha_{148-218} = -0.6$ for synchrotron-dominated sources \citep{Marsden2014}. 

\citet{Datta2019} presented three extragalactic point-source catalogs derived from ACT Data Release 3 (DR3), covering three sky regions. The ACT DR3 observations were obtained  with ACTPol between 2013 and 2014. The ACT DR3 point-source catalogs that mostly included synchrotron-dominated blazars provided
flux densities at 150 GHz and polarization properties. Based on the crossmatched counterpart flux densities from the National Radio Astronomy Observatory Very Large Array Sky Survey \citep[][]{Condon1998} and the Very Large Array$-$Faint Images of the Radio Sky at Twenty-Centimeters \citep[VLA-FIRST;][]{Becker1995}, AT20G, ACT Equatorial, and Herschel Spectral and Photometric Receiver (SPIRE) surveys, they derived mean spectral indices of $\alpha_{20-148} = -0.36$, $\alpha_{1.4-148} = -0.21$. 
\citet{Datta2019} listed 19 potential DSFG candidates based on their spectral indices. Recently, \citet{Gralla2020} have prepared a catalog of ACT  Equatorial survey extragalactic sources, including 510 AGNs and 287 DSFGs. Their catalog reports the flux density measurements at the 148 GHz, 218 GHz, and 277 GHz frequency bands. They also present the spectral properties of DSFGs, including both lensed sources and nearby unlensed sources \citep{Gralla2020}.

FIR counterparts observed with IRAS, AKARI, Herschel, and Spitzer of some of the local DSFGs identified at ACT wavelengths have been reported by \citet{Marsden2014} and \citet{Gralla2020}. The catalog of \citet{Marsden2014} has a few nearby IRAS galaxies,  with 
synchrotron- or dust-dominated spectra. AKARI completed the most recent all-sky survey in the mid-IR and FIR,  detecting thousands of IR galaxies in the low-redshift ($z\leq 0.3$) universe \citep{KilerciEser2018}. Therefore, we base our investigation on the AKARI- detected galaxies and select IR galaxies that overlap with the area mapped by the ACT. We identify and select IR galaxies that have been detected by IRAS or AKARI  in the publicly released ACT DR5 sky maps at 98, 150, and 220 GHz. Throughout the paper, we refer to the ACT frequency band centered at 224 GHz as 220 GHz band. 
In order to avoid DSFGs that have possibly been lensed by foreground galaxies, we restrict our investigation to  spectroscopically confirmed low-redshift ($z < 0.35$) IR galaxies with IRAS$-$AKARI detections at FIR bands. 

 Our goal is to investigate the origins of the measured millimeter emission from the SEDs and to measure physical properties of the local DSFGs that have been detected  by ACT.
In order to investigate the origins of the millimeter emission, for the first time, we take advantage of the latest version of CIGALE, which can decompose the radio emissions from SF and AGNs separately.
We also aim to examine the relationships between the total IR luminosity and the millimeter-band luminosities that have not been reported for  low-redshift ($z<0.35$) IR galaxies with millimeter counterparts.

The structure of this paper is organized as follows.
In \S \ref{S:data}, we present the IRAS--AKARI--ACT sample selection, the ACT source detection, and the multiwavelength data of  our ACT-detected IRAS--AKARI IR galaxy sample.  
In \S \ref{S:analysis}, we present our analysis of the spectral indices and the SEDs.  In \S \ref{S:dis}, we discuss our results. 
Our conclusions are summarized in \S \ref{S:conc}.
Throughout this paper, we adopt a flat $\Lambda$ cold dark matter cosmology, with 
$H_0=72$\,km s$^{-1}$ Mpc$^{-1}$, $\Omega_\Lambda = 0.7$, and $\Omega_{\rm m} = 0.3$ \citep{Spergel2003}. 

\section{Sample Selection and Data} \label{S:data}

\subsection{IRAS--AKARI--ACT Sample Selection}\label{S:samples}

The goal of this work is to study the UV-to-millimeter wave emission properties of IR galaxies that have been detected    with IRAS or AKARI and ACT. 
 IRAS and AKARI satellite are the two major telescopes that have scanned the whole sky and observed IR galaxies. IRAS
scanned all the sky in four IR bands, centered at 12 $\mu$m, 25 $\mu$m, 60 $\mu$m, and 100 $\mu$m. The IRAS Point Source Catalog Redshift (PSCz) survey \citep{Saunders2000} includes 15,411 IRAS galaxies with secure spectroscopic redshifts. Among those, 4954 galaxies are  within the ACT DR5 sky region.  

The AKARI satellite \citep{Murakami2007} conducted mid-IR and FIR all-sky surveys at 9 and 18 $\mu$m and at 65, 90, 140, and 160 $\mu$m, respectively. 
The AKARI/FIS all-sky survey bright source catalog version  2\footnote[1]{\url{http://www.ir.isas.jaxa.jp/AKARI/Archive/Catalogues/FISBSCv2/}} \citep[][]{Yamamura2018}  \citep[][]{Yamamura2018} 
includes  FIR flux measurements of 918,056 IR sources. These sources are detected due to their dust emissions, with the SEDs peaking around 100\,$\mu$m. 
From these, we select 166,625 IR sources within the ACT DR5 sky coverage ($-180^{\circ} < $ RA $ < 180^{\circ}$ and $-63^{\circ} < $ decl. $ < 23^{\circ}$). 
In order to identify the galaxies among these IR sources, we crossmatch with the optical counterparts within 20\arcsec\ from the
Sloan Digital Sky Survey (SDSS) Data Release 17 \citep[DR17;][]{Abdurrouf2022}, the Six-degree Field Galaxy Survey \citep[6dFGS;][]{Jones2004,Jones2005,Jones2009}, the 
Two Micron All Sky Survey \citep[2MASS;][]{Huchra2005,Huchra2012}, and the IRAS PSCz \citep[][]{Saunders2000} spectroscopic redshift catalogs. 
As a result, we identify 12,559 spectroscopically confirmed IR galaxies as detected by AKARI that overlap with the ACT DR5 sky coverage. 
 1145 of these 12,559 galaxies are among the 4954 IRAS galaxies. The IRAS galaxy sample includes 3809 galaxies that are not in the AKARI galaxy sample. The AKARI sample has 11,414 galaxies that are not in the IRAS galaxy sample. Therefore, our initial IRAS--AKARI infrared galaxy sample includes 16,368 galaxies.

\subsection{ACT Source Detection}\label{S:actdetection}
In the following, we explain the ACT source detection procedure in two steps: (i) the applied empirical flux calibration; and (ii) the source detection with Source- Extractor.

\subsubsection{Empirical Flux Calibration}\label{S:actfluxcal}
The latest ACT DR5 includes the deepest data obtained 
between 2008 and 2018 \citep{Naess2020}. We use the 
coadded 98 GHz, 150 GHz, and 220 GHz day--night maps, combined 
with Planck data. 
The point-source FWHMs (${\rm FWHM}_{\rm PSF}$) of the coadded maps are 
2.$^{\prime}$1, 1.$^{\prime}$3, and 1.$^{\prime}$0 at 
98 GHz, 150 GHz, and 220 GHz, respectively. 
These  ${\rm FWHM}_{\rm PSF}$ are larger than the individual galaxy sizes, except 
for very nearby sources. Therefore, individual galaxies are 
expected to appear as point sources in most cases. 
However, the DR5 point-source catalog has not yet been released, as of 2022 November. The available point-source catalog is for DR3 at 150 GHz \citep{Datta2019}.
Therefore, we use the DR3 map and its point-source 
catalog to empirically calibrate our source detection in
DR5, as described in the following paragraphs.

\begin{figure}
    \includegraphics[width=\columnwidth]{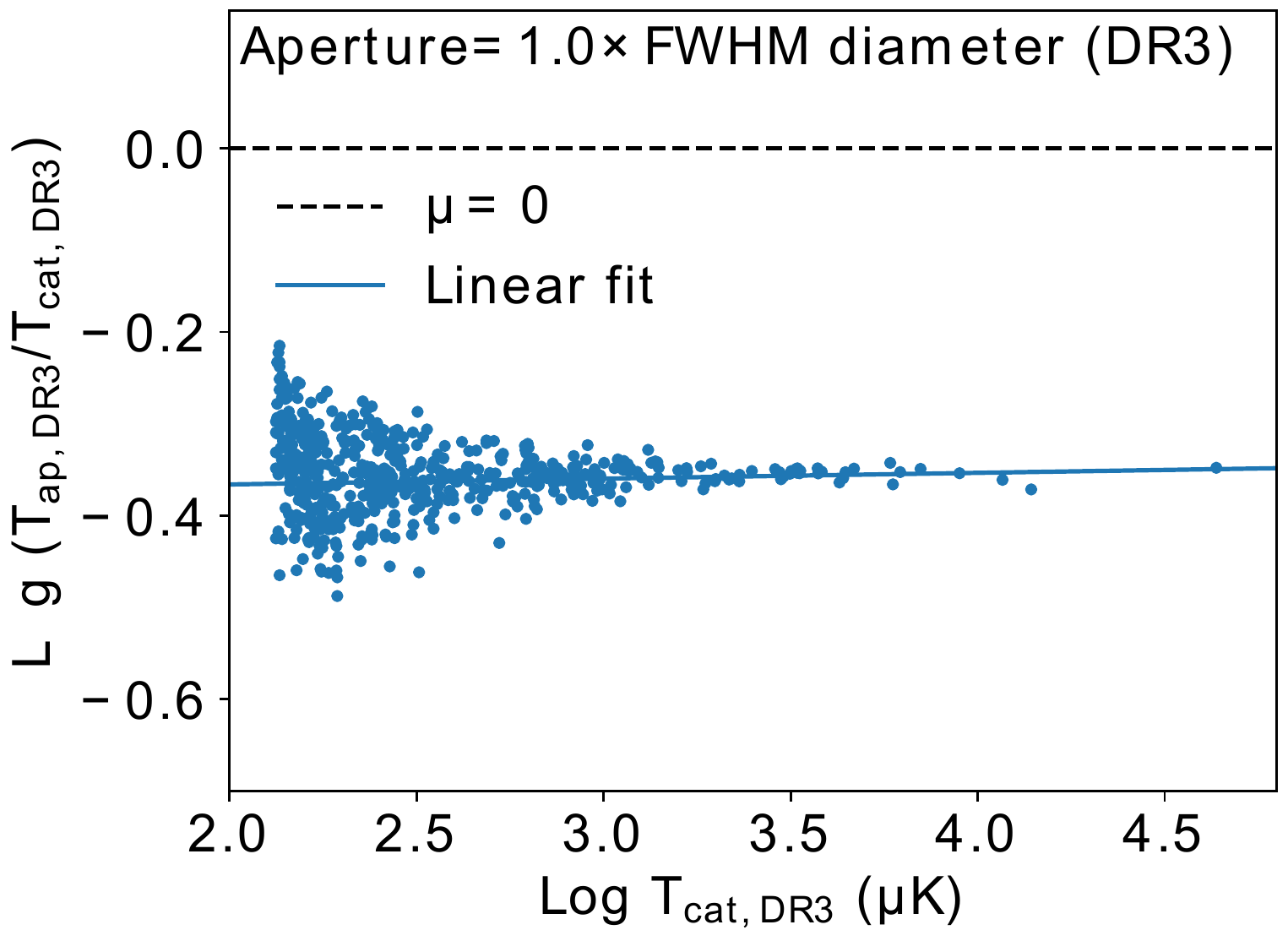}
    \caption{
    Log $(T_{\rm ap,DR3}/T_{\rm cat,DR3})$ as a function of $\log T_{\rm cat,DR3}$ for sources in the DR3 point-source catalog \citep{Datta2019}.
    Here, $T_{\rm ap,DR3}$ is the DR3  differential CMB temperature measured within an aperture with a 1.0 $\times$  ${\rm FWHM}_{\rm PSF}$ diameter.
    $T_{\rm cat,DR3}$ is the  differential CMB temperature in the DR3 point-source catalog.
    The observed frequency is 150 GHz.
    The horizontal dashed black line indicates $y = 0$, while the solid blue line is the best-fit linear function ($y = ax + b$) to the data points. In this linear fit, $a = 0.006 \pm 0.001$ and $b = -0.379 \pm 0.003$ and $\sigma_{{\rm log}T}=0.041$.   
    $\sigma_{{\rm log}T}$ indicates the vertical data dispersion around the best-fit linear function in logarithm scale.
    $\sigma_{{\rm log}T}=0.041$ corresponds to $\sim 10$\% dispersion in linear scale.
    }
    \label{fig:fig1}
\end{figure}

We subtract background radiation from the DR3 map using Source-Extractor V2.19.5 \citep{Bertin1996}.
The adopted background mesh size is 4 pixels, which is larger than the DR3 FWHM size of 2.8 pixels.
The GLOBAL background type is selected in this procedure. 
Aperture photometry is performed for the background-subtracted DR3 map, with a 1.0 $\times$  ${\rm FWHM}_{\rm PSF}$ diameter.
In Fig. \ref{fig:fig1}, $\log (T_{\rm ap,DR3}/T_{\rm cat,DR3})$ as a function of $\log T_{\rm cat,DR3}$ in units of $\mu$K are shown for sources in the DR3 point-source catalog \citep{Datta2019}.
Here, $T_{\rm ap,DR3}$ and $T_{\rm cat,DR3}$ are the DR3  differential CMB temperature measured within the aperture and the  differential CMB temperature in the DR3 point-source catalog, respectively.
The observed frequency is 150 GHz. 
If $T_{\rm ap,DR3} = T_{\rm cat,DR3}$, the data should be distributed around $y = 0$ (dashed black line).
The slope of the best-fit linear function to the data (solid blue line) is $\sim 0$, whereas $\log (T_{\rm ap,DR3}/T_{\rm cat,DR3})$ shows a systematic offset of $-0.379$ dex from $y = 0$.
This systematic offset is due to the aperture effect, which misses the flux density at the outer part of the point spread function  (PSF). 
We use this $-0.379$ dex for the aperture correction in the following analyses.

\begin{figure}
    \includegraphics[width=\columnwidth]{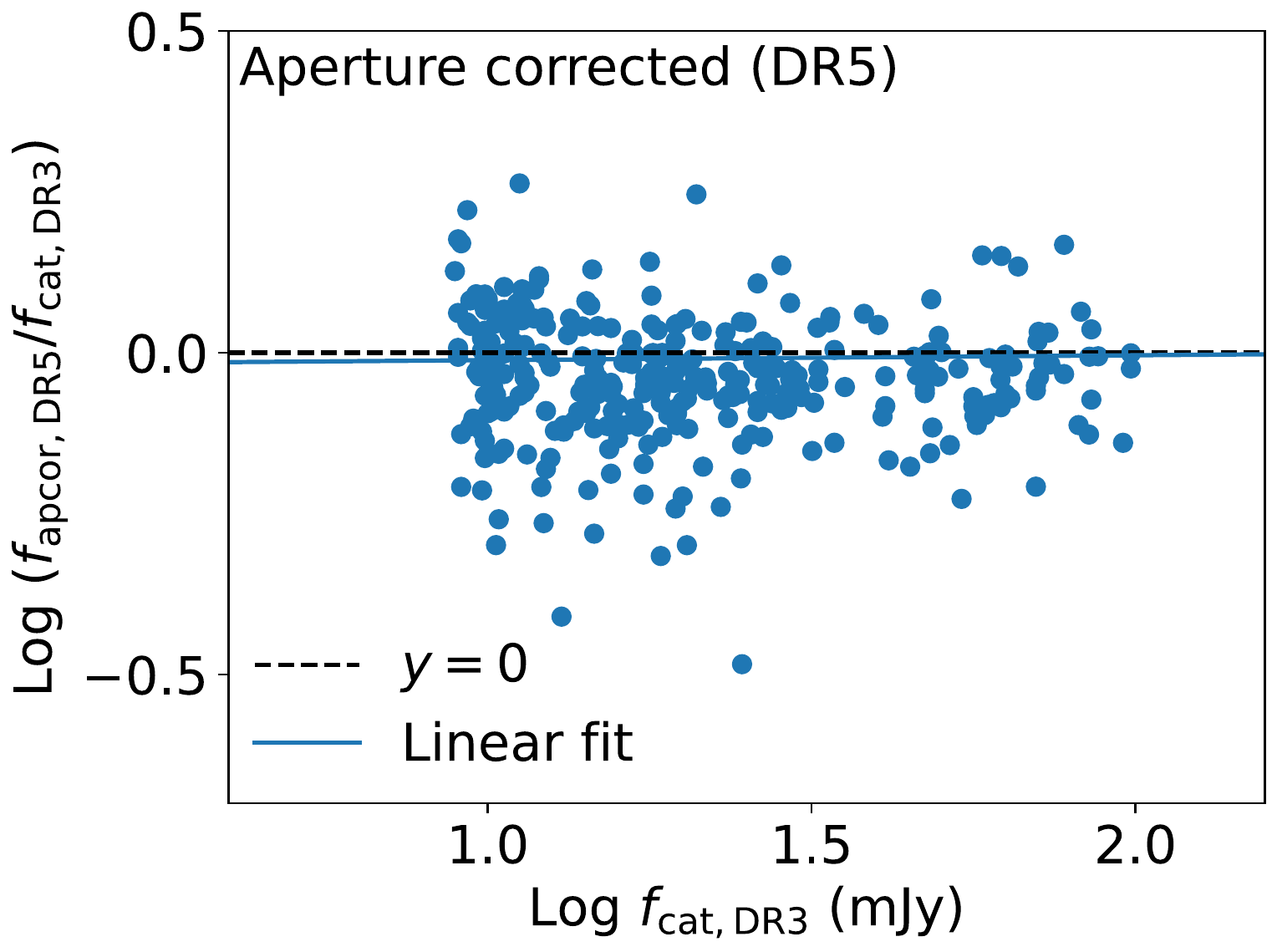}
    \caption{
    Log $(f_{\rm apcor,DR5}/f_{\rm cat,DR3})$ as a function of $\log f_{\rm cat,DR3}$ for sources in the DR3 point-source catalog \citep{Datta2019}.
    Here, $f_{\rm apcor,DR5}$ is the measured DR5 flux density corrected for aperture effect.
    $f_{\rm cat,DR3}$ is the flux density in the DR3 point-source catalog. 
    The observed frequency is 150 GHz for both DR3 and DR5.
    The horizontal dashed black line indicates $y = 0$, while the solid blue line is the best-fit linear function to the data points. Here, $a = 0.008 \pm 0.020$ and $b = -0.019 \pm 0.030$ and $\sigma_{{\rm log}f}=0.096$.
    $\sigma_{{\rm log}f}$ indicates the vertical data dispersion around the best-fit linear function in logarithm scale.
    $\sigma_{{\rm log}f}=0.096$ corresponds to $\sim 20$\% dispersion in linear scale.
    }
    \label{fig:fig2}
\end{figure}

The aperture photometry for the DR5 map at 150 GHz is conducted by following the same procedure as for the DR3 map. 
The DR5 map is provided in units of $\mu$K. 
The measured DR5  differential CMB temperatures are corrected for aperture effect and are empirically converted from $\mu$K to mJy ($f_{\rm apcor,DR5}$), using the median value of the conversion factors in the DR3 point-source catalog.
In Fig. \ref{fig:fig2}, $\log (f_{\rm apcor,DR5}/f_{\rm cat,DR3})$ as a function of $\log f_{\rm cat,DR3}$ in units of mJy are shown for sources in the DR3 point-source catalog.
$f_{\rm cat,DR3}$ is the flux density in the DR3 point-source catalog.
The slope and intercept of the best-fit linear function to the data points (solid blue line) are both $\sim0$, indicating that the systematic uncertainty of our flux calibration is negligible.
However, the data dispersion around the best-fit linear function is $\sigma_{{\rm log}f}=0.096$, which corresponds to a $\sim20$\% statistical uncertainty of the flux density in linear scale.
Therefore, the measured flux densities in this work include uncertainties arising from this flux calibration and from background noises.
We take both into account in the SED fitting analysis described in \S \ref{S:analysis}.

The flux calibration mentioned above is for ACT maps at 
150\,GHz. The  ${\rm FWHM}_{\rm PSF}$ of DR5 maps at 98 and 220\,GHz are 
different from that at 150\,GHz. Therefore, the conversion factors between $\mu$K and mJy are also different at 98 and 220\,GHz. 
 Following the literature \citep[e.g.,][]{Planck2014a}\footnote[2]{We also follow the formula presented at \url{https://science.nrao.edu/facilities/vla/proposing/TBconv}}, we scale the conversion factor at 150\,GHz to derive those at 98 and 220\,GHz, assuming 
 
\begin{equation}\label{Eq:1}
T\propto\frac{S}{{\rm FWHM}_{\rm PSF}^{2}(dB_{\nu}/dT)_{T_{0}}},
\end{equation}

where $T$ is the  differential CMB temperature, $S$ is the flux density, and  $(dB_{\nu}/dT)_{T_{0}}$ is the Planck function at the CMB temperature of $T_{0} = 2.7255$K \citep{Fixsen2009} and the observed frequency of $\nu_{\rm obs}$.

\subsubsection{Source detection}\label{S:sourcedet}
The background-subtracted DR5 maps at 98, 150, and 220 GHz are utilized to detect new sources with Source-Extractor.
 The adopted aperture diameters are $1.0 \times {\rm FWHM}_{\rm PSF}$, corresponding to 2.$^{\prime}$1, 1.$^{\prime}$3, and 1.$^{\prime}$0 at 98 GHz, 150 GHz, and 220 GHz, respectively.
These apertures take into account the different beam sizes for the different ACT bands.
Therefore, the point-source aperture correction should be almost identical among the ACT bands, unless the beam shapes were to be significantly different.
We adopt the same aperture correction ($-0.379$ dex) for the three ACT bands in this work.

The coordinates of the detected ACT DR5 sources are matched with those of our AKARI  and IRAS samples, as described in Section \ref{S:samples}.
The matching radius  between the AKARI and ACT DR5 sources is 1.0 $ \times$  ${\rm FWHM}_{\rm PSF}$ of the DR5 data, for each DR5 band.
 We use a 4$^{\prime}$ matching radius between the IRAS and ACT DR5 sources for all the DR5 bands, because the typical IRAS PSF size is $\sim4^{\prime}$ at 100 $\mu$m which is larger than those for ACT DR5.
The background noises are estimated by random aperture photometries around matched sources.
For each matched source and each band, we iterate 10,000 random aperture photometries around the source, to derive the standard deviation of the flux density, i.e., 1 $\sigma$ noise. The random aperture photometry provides 3 $\sigma$ upper limits of the source flux densities, if they are not detected in one or two ACT bands.

\begin{figure*}
 \begin{center}$
 \begin{array}{c}
 \includegraphics[scale=0.7]{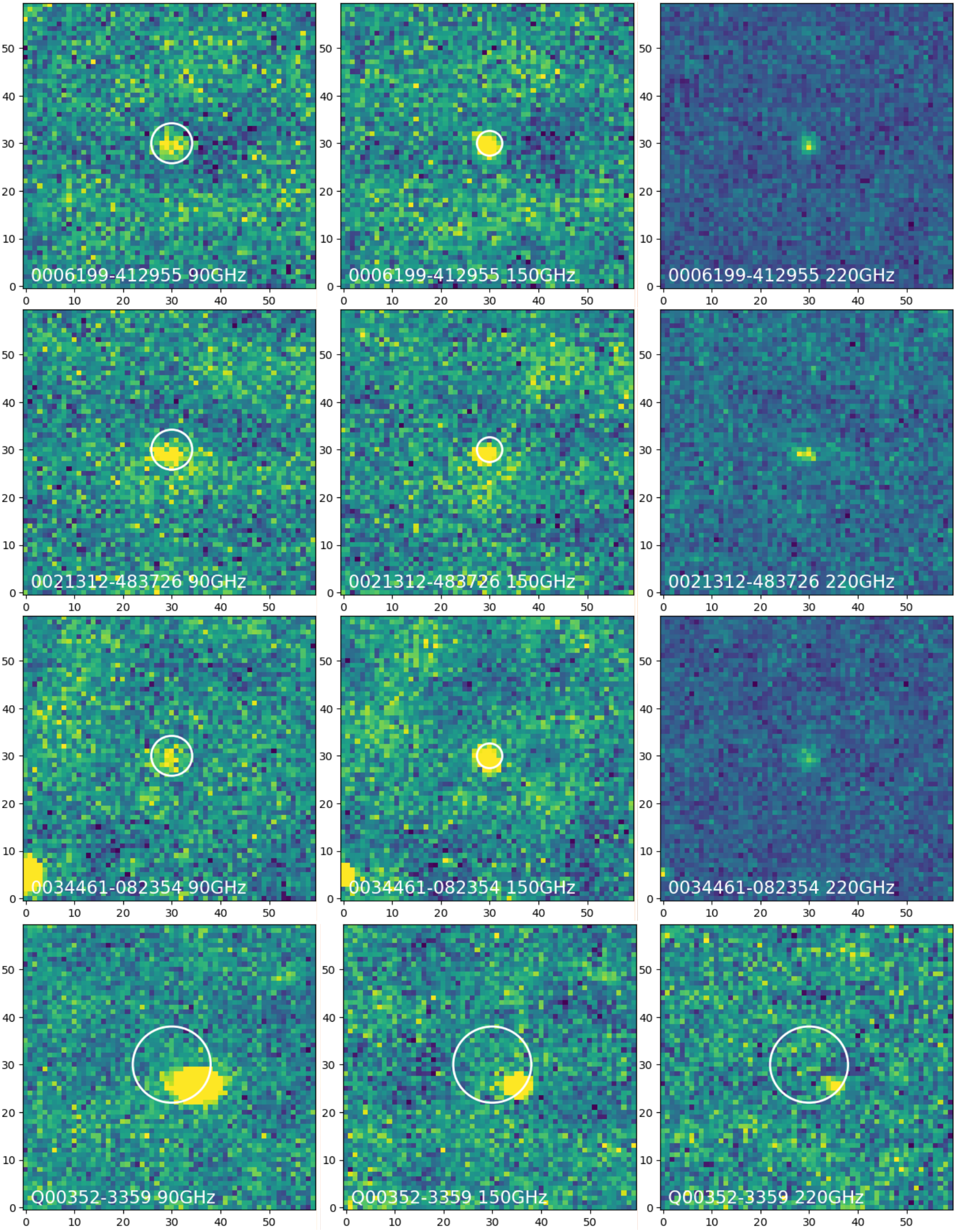}
 \end{array}$
 \end{center}
 \caption{Example of ACT images of  four of the sources (with S/N $\ge 5$) in our sample, detected at 98 GHz (left column), 150 GHz (middle column), and 220\,GHz (right column), respectively. The source detections are shown inside the adopted searching radii (white circles) for matching. The image coordinates are provided in pixel units (1 pixel corresponds to 0.$^{\prime}$5). The first three rows indicate AKARI-matched ACT sources, while the bottom row shows an IRAS-matched ACT source. The center of each panel corresponds to the position of the AKARI or IRAS source.}
 \label{fig:fig3}
 \end{figure*}

After the visual inspection, 101, 135, and 61 sources are found to match with the IRAS--AKARI sample with the signal-to-noise ratio (S/N) $ > 5$ at 98, 150, and 220 GHz, respectively. 
In total, 167 galaxies are detected at any ACT frequency with S/N $ > 5$, with 39 galaxies being detected with S/N $ > 5$ at all three DR5 frequencies. Moreover, 76 galaxies are detected at 98\,GHz and 150\,GHz. 39 galaxies are detected at 98\,GHz and 220\,GHz. 54 galaxies are detected at 150\,GHz and 220\,GHz. As an example, ACT detection images of four of the galaxies that are detected in the three ACT bands are shown in Figure \ref{fig:fig3}.

\subsubsection{Aperture correction for spatially extended sources}\label{S:sourceext}
The aperture correction derived in Section \ref{S:actfluxcal} ($-0.379$ dex) is for ACT point sources. 
This correction could underestimate the flux densities of the spatially extended sources in the ACT images.
To select such extended sources, we utilize the \texttt{FWHM\_IMAGE} parameter (the FWHM of each source) measured by Source-Extractor.
For the S/N $ > 5$ sample in each ACT band, we estimate the standard deviation of the \texttt{FWHM\_IMAGE} distribution ($\sigma_{\rm FWHM}$).
Spatially extended sources are selected if the \texttt{FWHM\_IMAGE} are higher than $3 \sigma_{\rm FWHM} + {\rm FWHM}_{\rm PSF}$ in each ACT band.
The thresholds are $4.^{\prime}2, 4.^{\prime}0$, and $4.^{\prime}3$ at 98 GHz, 150 GHz, and 220 GHz, respectively.
By applying these thresholds to the S/N $ > 5$ samples, we identify 4, 7, and 3 extended sources at 98 GHz, 150 GHz, and 220 GHz, respectively.

We adopt optimized apertures for the spatially extended sources using ``growth curves'', i.e., the total flux density within the aperture as a function of the aperture size.
The flux densities are measured by using the optimized apertures for the extended sources, instead of applying the aperture correction of $-0.379$ dex (see \S \ref{S:appendix} for details).

\subsubsection{Source matching with previous ACT extragalactic source catalogs}\label{S:sourcemat}

We crossmatch our IRAS--AKARI--ACT sample with the previous ACT extragalactic source catalogs of \citet{Marriage2011a}, \citet{Marsden2014}, \citet{Datta2019}, and \citet{Gralla2020}. 
We use a matching radius based on the beam size of the relevant frequency, namely 1.$^{\prime}$3 for 150\,GHz and  1.$^{\prime}$0 for 220\,GHz. 
We have 5, 23, and 9 overlapping sources with the catalogs of \citet{Marsden2014}, \citet{Datta2019}, and \citet{Gralla2020}, respectively. Since  four sources among these are listed in both the 
catalogs of \citet{Datta2019} and \citet{Gralla2020}, our sample includes 33 galaxies from these ACT catalogs. Therefore, compared to previous ACT extragalactic source catalogs, we have 134 new ACT-detected extragalactic sources in our sample. 

We list the properties of the ACT-detected IRAS--AKARI IR galaxy sample in Table \ref{tab:table1}. Namely, we provide the IRAS/AKARI ID (column 1); the IRAS/AKARI 
coordinates (columns 2 and 3); the optical counterpart name (column 4); the redshift (column 5); the S/N at 98 GHz and $F_{\rm{98 GHz}}$ (columns 6 and 7); the 
S/N at 150 GHz and $F_{\rm{150 GHz}}$ (columns 8 and 9); the S/N at 220 GHz and $F_{\rm{220 GHz}}$ (columns 10 and 11); 
a note of a new ACT detection of the source in this work and a flag for extended sources (column 12); and the radio detection information (column 13). Note that 12 ACT sources listed in Table \ref{tab:table1} (AKARI/IRAS IDs 0215076-034302, 0529020-140507, 0727048-381415, Q00088-1449, L00125-3928, L00446-2101, Q05238-4602, 0610082+091503, 0626161+200221, 0854487+200629, 1058299+013401, and 1140184+174600) either do not have a reliable redshift measurement or have more than one counterpart candidate. Therefore, their optical counterpart and redshift entries are left empty.

As noted by \citet{Gralla2020}, galaxies with $z \ge 0.05$ would be unresolved for the ACT beam size. Our sample includes a few galaxies at $z \sim 0.07$. A visual inspection of them indicates that they are local IR galaxies. Therefore, they are included in our sample for further analysis. Our sample also includes a couple of higher-redshift galaxies, at $z \sim 0.1$, which could be lensed systems. Since their number is low compared to the local galaxies in our sample, we keep them in our sample for the SED analysis in \S \ref{S:seds}.  

\begin{longrotatetable}
\begin{deluxetable*}{lrrcccccccccc}
\tablecolumns{13}
\tabletypesize{\scriptsize}
\setlength{\tabcolsep}{0.02in}
\tablewidth{0pt}
\tablecaption{IRAS--AKARI IR galaxy sample detected by ACT.}
\tablehead{
\colhead{AKARI/\tablenotemark{a}} &
\colhead{R.A.\tablenotemark{b}} &
\colhead{Decl.\tablenotemark{b}}&
\colhead{Optical\tablenotemark{c}} &
\colhead{$z$\tablenotemark{d}}&
\colhead{S/N\tablenotemark{e}} &
\colhead{$F_{\rm{98\,GHz}}$\tablenotemark{f}} &
\colhead{S/N\tablenotemark{e}} &
\colhead{$F_{\rm{150\,GHz}}$\tablenotemark{f}} &
\colhead{S/N\tablenotemark{e}} &
\colhead{$F_{\rm{220\,GHz}}$\tablenotemark{f}}&
\colhead{Note\tablenotemark{g}} & 
\colhead{Radio\tablenotemark{h}}\\
\colhead{IRAS} &
\colhead{(J2000)} &
\colhead{(J2000)} &
\colhead{Counterpart} &
\colhead{} &
\colhead{98 } &
\colhead{(mJy)} &
\colhead{150 } &
\colhead{(mJy)} &
\colhead{220 } &
\colhead{(mJy)}  &
\colhead{} &
\colhead{Reference}\\
\colhead{ID} &
\colhead{} &
\colhead{} &
\colhead{} &
\colhead{} &
\colhead{GHz} &
\colhead{} &
\colhead{GHz} &
\colhead{} &
\colhead{GHz} &
\colhead{} &
\colhead{} &
\colhead{} \\
\colhead{(1)} &
\colhead{(2)} &
\colhead{(3)} &
\colhead{(4)} &
\colhead{(5)} &
\colhead{(6)} &
\colhead{(7)} &
\colhead{(8)} &
\colhead{(9)} &
\colhead{(10)} &
\colhead{(11)} &
\colhead{(12)} &
\colhead{(13)}}
\startdata
 \hline
0006199-412955  &    1.58380  &  -41.49875  &  ESO 293-IG 034           &     0.0050  &       6  &      13.00 $\pm$ 2.01  &      15  &      26.88 $\pm$ 1.79  &       7  &    52.34 $\pm$ 6.91  &  *   &        \\
0009096-003659  &    2.29051  &   -0.61673  &  SDSS J000909.89-003705.3 &     0.0717  &         &         &       7  &      11.89 $\pm$ 1.58  &         &       &  1  &        \\
0010203-462510  &    2.58531  &  -46.41978  &  ESO 241-G 021            &     0.0203  &         &            &       5  &       9.85 $\pm$ 1.75  &       4  &       &  *   &        \\
Q00088-1449  &    2.85957  &  -14.55523  &   &       &      36  &     140.18 $\pm$ 3.84  &      26  &     109.06 $\pm$ 4.09  &       5  &    55.21 $\pm$ 10.53  &  *   & i      \\
L00125-3928  &    3.75411  &  -39.20403  &  &       &       5  &     105.65 $\pm$ 19.32  &       5  &     337.39 $\pm$ 58.42  &       5  &  1467.80 $\pm$ 254.74  & *, E(98\,GHz, 150\,GHz)    &        \\
0021312-483726  &    5.38062  &  -48.62416  &  NGC 0092                 &     0.0112  &       9  &      18.56 $\pm$ 1.94  &      12  &      21.38 $\pm$ 1.77  &       5  &    37.81 $\pm$ 6.58  &   *  &        \\
0023303+010053  &    5.87690  &    1.01450  &  WISEAJ002329.14+010054.9 &     0.8286  &      10  &      25.52 $\pm$ 2.39  &      10  &      15.94 $\pm$ 1.53  &         &       &  1  &        \\
0029300-044140  &    7.37550  &   -4.69461  &  2MASX J00293024-0441350  &     0.0202  &      13  &      34.05 $\pm$ 2.47  &         &            &         &       &  *   & i      \\
0030227-331430  &    7.59520  &  -33.24195  &  NGC 0134                 &     0.0053  &       5  &      25.95 $\pm$ 5.09  &       5  &      90.72 $\pm$ 17.80  &      13  &    96.37 $\pm$ 7.12  &  *, E(98\,GHz , 150\,GHz)   & i      \\
0034461-082354  &    8.69275  &   -8.39862  &  NGC 0157                 &     0.0056  &       5  &      15.01 $\pm$ 2.64  &      12  &      24.24 $\pm$ 1.88  &       5  &    54.13 $\pm$ 10.25  &  1  &        \\
\enddata
\tablecomments{A machine-readable version of the full Table \ref{tab:table1} is available. A portion is shown here for guidance regarding its form and content.}
\tablenotetext{a}{ AKARI ID as listed by the AKARI/FIS all-sky survey bright source catalog version 2.; IRAS ID from the IRAS PSCz \citep{Saunders2000}. }
\tablenotetext{b}{AKARI coordinates from the AKARI/FIS all-sky survey bright source catalog version 2.; IRAS coordinates from IRAS PSCz \citep{Saunders2000}.}
\tablenotetext{c}{ Optical counterpart name from NED; empty if there is more than one counterpart candidate within the matching radius.}
\tablenotetext{d}{Redshift of the optical counterpart from the SDSS Data Release 17, 6dFGS, 2MASS, or IRAS PSCz spectroscopic redshift catalogs; empty if the redshift measurement is not reliable.}
\tablenotetext{e}{Measured S/N of the source in the map of the given frequency.}
\tablenotetext{f}{Measured flux density and its uncertainty from the map of the given frequency.}
\tablenotetext{g}{The ``*'' symbol indicates the galaxies that have new ACT identifications, as reported in this work. References for the previous ACT extragalactic source catalogs: (1) \citet{Datta2019}; (2) \citet{Marsden2014}; and (3) \citet{Gralla2020}. ``E'' is a flag indicating that a source is extended, with a more than 3$\sigma$ FWHM$\_$IMAGE size than the nominal PSF size at the noted band.}
\tablenotetext{h}{The radio catalog references are: (i) \citet{Condon1998}; (ii) \citet{vanvelzen2012}; (iii) \citet{Helfand2015}; and (iv) \citet{Murphy2010}. }
 \label{tab:table1}
\end{deluxetable*}
\end{longrotatetable}

\subsection{Multiwavelength Data of  ACT-detected IRAS--AKARI IR Galaxy Sample } \label{S:multidata}

For the SED analysis in \S\ref{S:seds} we collect the UV, optical, near-IR, mid-IR, FIR and submillimeter archival flux densities of the counterparts, when available. 
The Galaxy Evolution Explorer \citep[\textit{GALEX;}][]{Martin2005} scanned all the sky in the near-UV (NUV; $\lambda_\mathrm{eff} = 2267 $\AA) and far-UV (FUV; $\lambda_\mathrm{eff} = 1516$\AA) bands. 
We use the Barbara A. Mikulski Archive for Space Telescopes Portal\footnote[3]{\url{http://mast.stsci.edu}} to match the optical counterpart positions of our sample within  5\arcsec\  with the GALEX source catalog \citep{Martin2005}, 
Data Release 6/7, and find 77 matches. We eliminate any unreliable measurements ( \textsc{nuv/fuv\_artifact}$ = 4, 64 $) and select the longest duration, if multiple measurements are listed for any source. 
GALEX photometry is corrected  for Galactic foreground extinction using the listed $E_{B-V}$ values \citep[based on the extinction maps of][]{Schlegel1998} in the GALEX source catalog. 

We get the Galactic extinction-corrected $u$-, $g$-, $r$-, $i$-, and $z$-band magnitudes from the SDSS \citep[][]{York2000} Data 
DR17 \citep[][]{Abdurrouf2022} \textit{PhotoObjAll} catalog\footnote[4]{\url{https://skyserver.sdss.org/dr17/}}. 
After avoiding unreliable photometry, with a \textsc{clean} = 0 flag, we use SDSS photometry only for 13 sources in SDSS DR17. 
Since the available SDSS photometry is very limited, in order to include more optical counterpart photometry, our sample is crossmatched (within a search radius of 5\arcsec) 
with the second data release (DR2) of the SkyMapper Southern Survey \citep[SMSS;][]{Onken2019}, to include more optical counterpart photometry. 
SMSS \citep{Wolf2018}, which covers the entire southern sky, provides optical photometry in six bands: $u$, $v$, $g$, $r$, $i$, and $z$. 
We use 5\arcsec\ diameter aperture photometry in six bands from the DR2 photometry table\footnote[5]{\label{note5}\url{http://skymapper.anu.edu.au/table-browser/}} 
for 86 matched galaxies in our sample. 
We consider only the reliable measurements,  with \textsc{FLAGS} $ < 4$ and \textsc{NIMAFLAGS} $ < 5$. We use the given formalism in \citet{Casagrande2019} to convert the SMSS magnitudes 
to the band fluxes. We  deredden the SMSS photometry in six bands, with the $E_{B-V}$  values from \citet[][]{Schlegel1998}, as listed in the DR2  master table\footref{note5}, 
based on the \citet{Cardelli1989} extinction law.

For the 161 sources that have cross-identifications (within 3\arcsec\ of the optical counterpart position) in the 2MASS Point Source Catalog \citep{Cutri2003,2massallskypoint} 
and the 2MASS Extended Source Catalog \citep[2MASS XSC;][]{Jarrett2000,2massextended}, we use the $J$-, $H$-, and $K_{s}$-band magnitudes. From 2MASS XSC, we use isophotal fiducial elliptical aperture magnitudes (`20mag/sq.') for 102 sources. For the remaining 59 sources, we use the default magnitudes from the 
 2MASS Point Source Catalog. We use reliable photometry, without any artifact contamination and/or confusion flags (cc\_flags). 
We apply Galactic reddening correction to the
2MASS magnitudes, based on the \citet{Schlegel1998} maps and the extinction law of \citet{Cardelli1989}. 

The AllWISE Source Catalog\footnote[6]{\url{http://wise2.ipac.caltech.edu/docs/release/allwise/}} \citep[][]{allwise} lists the photometric measurements that have been obtained by 
the WISE all-sky survey in the $w1$ (centered at 3.4$\mu$m), $w2$ (centered at 4.6$\mu$m), $w3$( centered at 12$\mu$m), and $w4$ (centered at 23$\mu$m) mid-IR bands \citep{Cutri2013}. The optical counterpart coordinates from our sample are crossmatched with the AllWISE Source Catalog within a 5\arcsec\ search radius.  
Of the available WISE photometry for 128 sources, we only use reliable measurements---with zero $cc\_flags$ values 
($cc\_flags$ = '0000'), to avoid contaminated measurements; without high scattered moonlight contamination (i.e., `{\ttfamily moon$\_$lev}' $< 
 5$ ); with an S/N level greater than 2, not to include upper limits; with quality flag ($w1,2,3,4 flg$) values between 1 and 32, not to use upper limits; and without \textit{null} uncertainties, to avoid upper limits or no measurements. The elliptical aperture magnitudes (``$wngmag$'', where $n$ is the band number) are used for the extended  sources (with $ext\_flg > 0$).

We obtain the AKARI mid-IR photometry in the  9 and 18 $\mu$m bands from the AKARI/IRC all-sky survey point-source catalog, version 1\footnote[7]{\url{http://www.ir.isas.jaxa.jp/AKARI/Observation/PSC/Public/RN/AKARI-IRC\_PSC\_V1\_RN.pdf}} \citep[][]{AKARIIRC}, by crossmatching the AKARI/FIS coordinates within 20\arcsec.  
For the AKARI FIR photometry in the  65, 90, 140, and 160 $\mu$m bands, we use the  AKARI/FIS all-sky survey bright source catalog, version 2 \citep[][]{Yamamura2018}. In order to use reliable AKARI data, we only include the photometric measurements of high quality ($FQUAL= 3$). 

For the 31 galaxies that are only detected by IRAS, we use the good-quality flux densities (with the flux quality flag 2 or 3) at 12, 25, 60, 100\,$\mu$m, from the PSCz  survey catalog \citep{Saunders2000}. We also use the upper limit on the flux density, given by a quality flag of 1.

The Herschel \citep[][]{Pilbratt2010} Photodetector Array Camera and Spectrometer  \citep[PACS;][]{Poglitsch2010} 
performed FIR measurements at 70, 100, and 160 $\mu$m, while SPIRE \citep[][]{Griffin2010} performed 
submillimeter measurements at 250, 350 and 500 $\mu$m. In order to include Herschel  photometry in our analysis, we crossmatch (within 5\arcsec) the optical counterpart coordinates of our sample with the galaxy samples of \citet{Viero2014} and \citet{Clark2018}. 
In total, we find Herschel counterparts for 23 sources. When available, we include the  Herschel SPIRE measurements of these sources in our analysis. Since the AKARI FIR bands already cover  the Herschel PACS bands, we only use Herschel PACS photometry for one source, for which the AKARI 90 $\mu$m photometry is not of good quality. 

To account for the IR emission measurement discrepancies between the galaxies, due to the spatial resolution differences among IRAS, AKARI, WISE, 
and Herschel, we follow \citet{Clements2019} and apply their beam corrections that depend on the extendedness parameter  
\citep[see equation 2 of][]{Clements2019} to AKARI 65 $\mu$m and 90 $\mu$m measurements. To be conservative about the small beam size differences between AKARI/IRC \citep[9\arcsec,][]{Ishihara2010} and WISE  \citep[6\arcsec--12 \arcsec;][]{Wright2010} we measure the 
mean difference between the two closest pairs of bands--- namely, between the AKARI/IRC 9 $\mu$m and WISE 12 $\mu$m and 
the AKARI/IRC 18 $\mu$m and WISE 22 $\mu$m fluxes. We add the typical flux difference of $\sim$25\% in quadrature to the photometric uncertainties of the 9 $\mu$m, 12 $\mu$m,  18 $\mu$m, and 22 $\mu$m fluxes. 

\subsection{Radio Identification of ACT-detected IRAS--AKARI IR Galaxy Sample } \label{S:radiodata}

We make use of the available radio galaxy catalogs of \citet{Condon1998}, \citet{vanvelzen2012}, \citet{Helfand2015}, \citet{Kuzmicz2018}, and  \citet{Murphy2010} to check the radio identifications of the galaxies in our sample. 
To identify the radio counterparts, we match the optical positions with the optical (or radio) counterpart coordinates listed in the investigated catalogs, within 20\arcsec. 
We find radio identifications at $\nu = 1.4$ GHz for 106, 23, and 48 galaxies from the catalogs of \citet{Condon1998}, \citet{vanvelzen2012} and \citet{Helfand2015}, respectively. Once the common sources in the different catalogs have been taken into account, a total of 113 galaxies in our sample have radio identifications at $\nu = 1.4$\,GHz. 
Nine galaxies have radio identifications in the AT20G source catalog  \citep{Murphy2010}, at $\nu = 20$ GHz. Eight of these galaxies had already been identified within the $\nu = 1.4$ GHz sample.  
We mark the galaxies with radio identifications (now 114 in total)  in Table \ref{tab:table1} (column 13), with their references. 

\section{Analysis and Results} \label{S:analysis}

\subsection{Spectral indices}\label{S:spectralindices}

\begin{figure*}
\begin{center}$
\begin{array}{lll}
\includegraphics[scale=0.66]{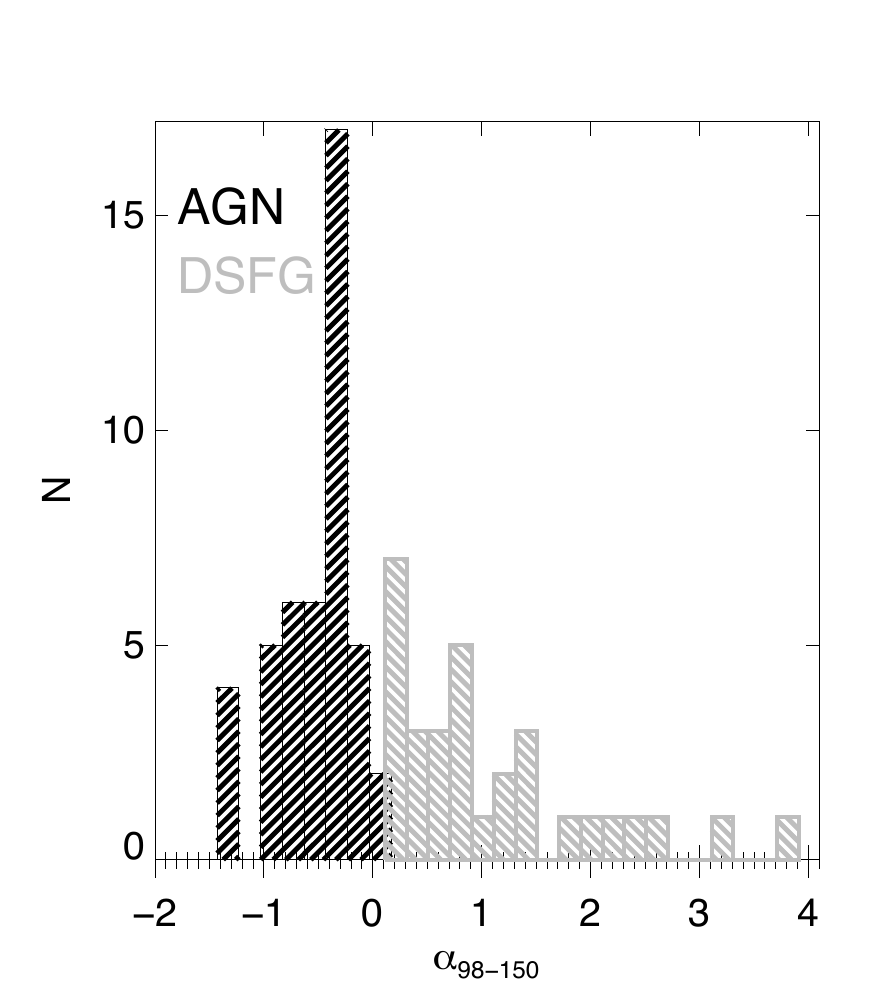}&
\includegraphics[scale=0.66]{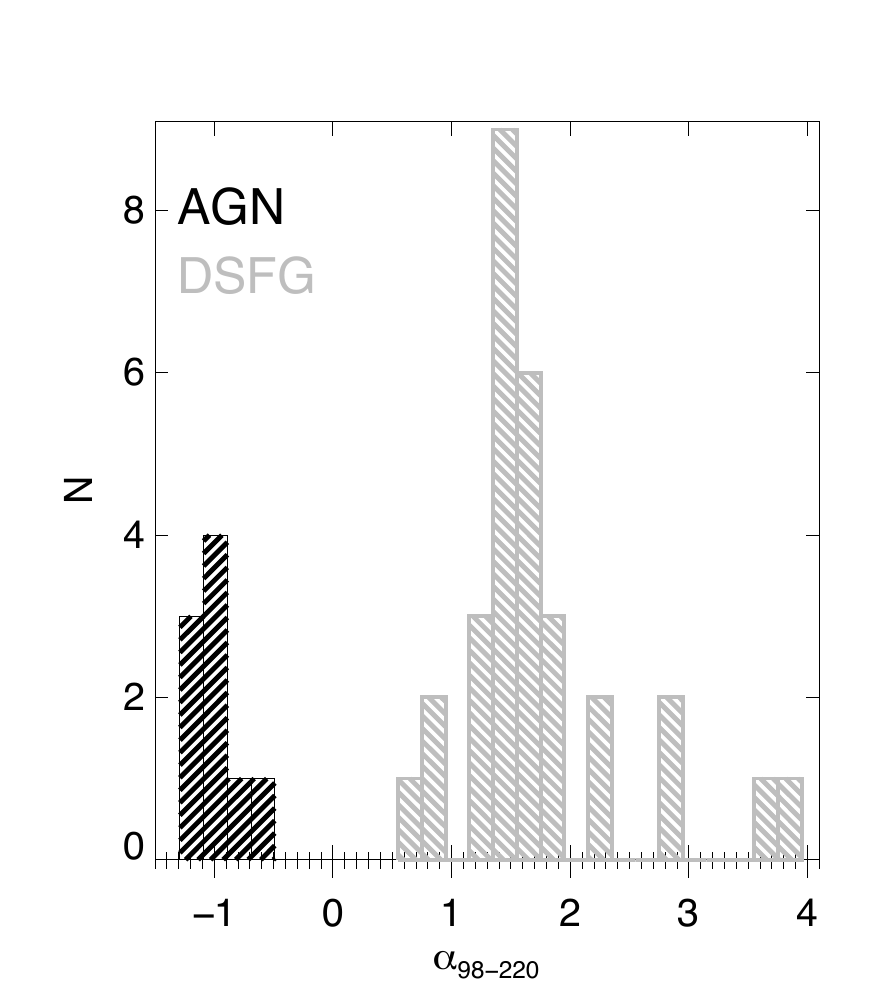}&
\includegraphics[scale=0.66]{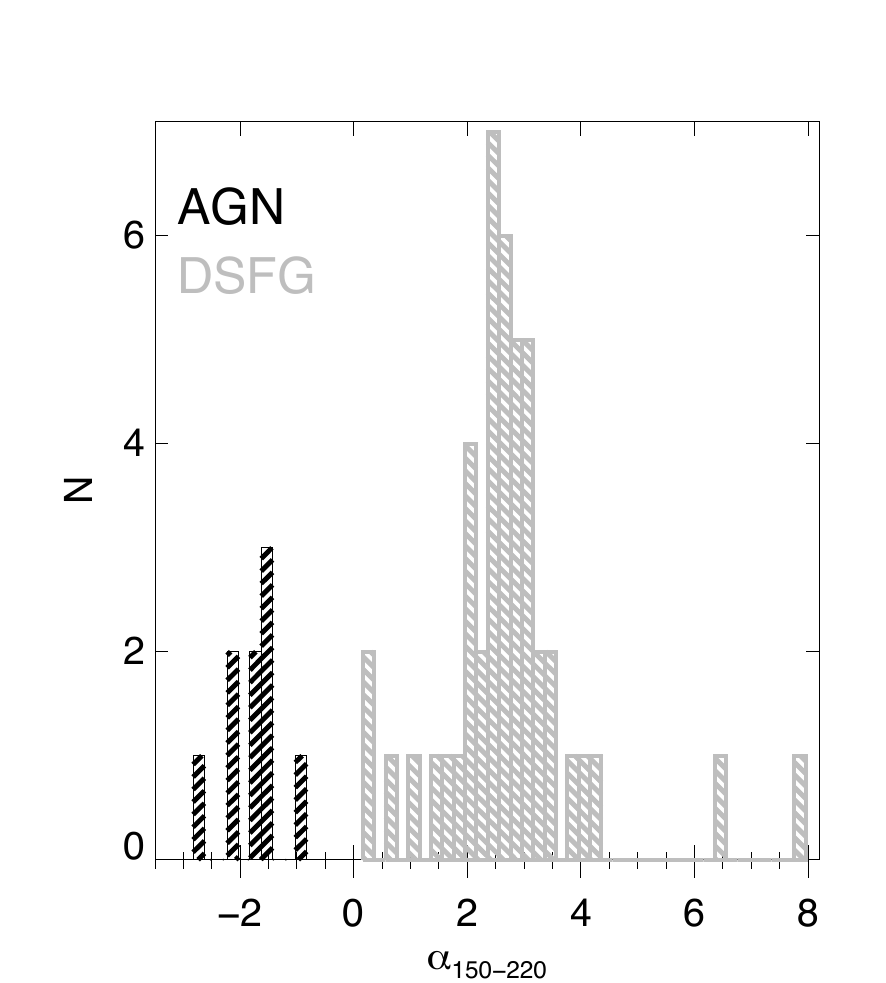}\\
\end{array}$
\end{center}
\caption{The distributions of the $\alpha_{98-150}$ (left), $\alpha_{98-220}$ (middle), and $\alpha_{150-220}$ (right) spectral indices. The black distributions represent galaxies that are classified as AGNs, according to the spectral indices, while the gray ones represent the galaxies that are classified as DSFGs.} 
\label{fig:fig4}
\end{figure*}

Millimeter-wave spectral indices can be used to determine the emission mechanism of the galaxy. AGN or dust-dominated sources can be designated from the 150$-$220 GHz spectral index. While synchrotron radiation powered by AGNs generates a negative $\alpha$ (less than unity), thermal dust-dominated radiation has $\alpha$ values greater than 2 \citep{Hall2010,Mocanu2013,Marsden2014,Gralla2020}. We compute millimeter-wave spectral indices of our IR galaxy sample over available frequency intervals---namely $\alpha_{98-150}$, $\alpha_{98-220}$, and $\alpha_{150-220}$---from the listed flux densities in Table \ref{tab:table1}, as $\alpha_{\nu_{2} - \nu_{1}}= \rm{log}(F_{2}/F_{1})/log(\nu_{2}/\nu_{1})$. We list the millimeter-wave spectral indices in Table \ref{tab:table2}.

When available, we classify each source, based on the $\alpha_{150-220}$ and $\alpha_{98-150}$ spectral indices, as a DSFG or synchrotron-dominated AGN. Our sample includes seven AGNs and 27 DSFGs among the 34 sources with available $\alpha_{150-220}$ and $\alpha_{98-150}$. This is consistent with our initial selection of the galaxies in our sample as dusty IR sources. The distributions of the spectral indices of our sample are shown in Fig. \ref{fig:fig4}. As seen in this figure, the DSFGs in our sample have spectral indices between: (i) $0.11 \le \alpha_{98-150} \le 3.71 $; (ii) $0.55 \le \alpha_{98-220} \le 3.91 $; and (iii) $0.16 \le \alpha_{150-220} \le 7.83 $. The AGNs in our sample have spectral indices between: (i) $-1.43 \le \alpha_{98-150} \le -0.01 $; (ii) $-1.29 \le \alpha_{98-220} \le -0.65 $; and (iii) $-2.83 \le \alpha_{150-220} \le -0.99 $.
The median spectral indices of the DSFGs in our sample are:  $\alpha_{98-150}$ = 0.75 $\pm$ 0.65; $\alpha_{98-220}$ = 1.56 $\pm$ 0.73; and $\alpha_{150-220}$ = 2.59 $\pm$ 1.29. The median spectral indices of the AGNs in our sample are: 
$\alpha_{98-150}$ = $-$0.40 $\pm$ 0.35; $\alpha_{98-220}$ = $-$1.04 $\pm$ 0.20; and $\alpha_{150-220}$ = $-$1.67 $\pm$ 0.51. The median spectral indices of  $\alpha_{150-220}$ = 2.59 $\pm$ 1.29 obtained for the DSFGs in this work do agree (within uncertainties) with the reported values of 3.7 $\pm$ 0.6 by \citet{Marsden2014} and 3.7 $\pm$ 1.8 reported by \citet{Gralla2020}.

\begin{deluxetable}{lccccc}
\tablecolumns{6}
\tabletypesize{\scriptsize}
\setlength{\tabcolsep}{0.01in}
\tablewidth{0pt}
\tablecaption{Spectral Indices of the IRAS--AKARI IR Galaxy Sample Detected by ACT. Only the sources with at least two ACT band detections and at least one spectral index available are listed. }
\tablehead{
\colhead{Source\tablenotemark{a}} &
\colhead{$\alpha_{98-150}$\tablenotemark{b}} &
\colhead{$\alpha_{98-220}$\tablenotemark{b}} &
\colhead{$\alpha_{150-220}$\tablenotemark{b}}&
\colhead{Type\tablenotemark{c}} &
\colhead{Type\tablenotemark{c}} \\
\colhead{Name} &
\colhead{ } &
\colhead{ } &
\colhead{} &
\colhead{$\alpha_{98-150}$ } &
\colhead{$\alpha_{150-220}$ } \\
\colhead{(1)} &
\colhead{(2)} &
\colhead{(3)} &
\colhead{(4)} &
\colhead{(5)} &
\colhead{(6)}}
\startdata
 \hline
ESO\,293-IG\,034           &   1.4 $\pm$ 0.1  &   1.6 $\pm$ 0.0  &   1.7 $\pm$ 0.1  &  DSFG  &  DSFG  \\
WISEAJ001126.18-143319.9 &  -0.5 $\pm$ 0.0  &  -1.0 $\pm$ 0.1  &  -1.8 $\pm$ 0.2  &  AGN   &  AGN   \\
SSTSL2J001501.00-391216.9&   2.3 $\pm$ 0.0  &   2.9 $\pm$ 0.0  &   3.8 $\pm$ 0.0  &  DSFG  &  DSFG  \\
NGC\,0092                 &   0.3 $\pm$ 0.0  &   0.8 $\pm$ 0.0  &   1.5 $\pm$ 0.1  &  DSFG  &  DSFG  \\
WISEAJ002329.14+010054.9 &  -0.9 $\pm$ 0.0  &          &          &        &  AGN   \\
NGC\,0134                 &   2.5 $\pm$ 0.0  &   1.5 $\pm$ 0.1  &   0.2 $\pm$ 0.1  &  DSFG  &  DSFG  \\
NGC\,0157                 &   0.9$\pm$ 0.1  &   1.4 $\pm$ 0.0  &   2.1 $\pm$ 0.1  &  DSFG  &  DSFG  \\
ESO\,350-G040              &  -0.1 $\pm$ 0.0  &  -1.3 $\pm$ 0.0  &  -2.8 $\pm$ 0.1  &  AGN   &  AGN   \\
UGC\,00439                &  -1.2 $\pm$ 0.0  &          &          &        &  AGN   \\
SSTSL2J004705.98-204459.4&          &          &   7.8 $\pm$ 0.1  &  DSFG  &        \\
 \enddata
 \tablecomments{Only the sources with at least two ACT band detections and at least one availablr spectral index are listed. A machine-readable version of the full Table \ref{tab:table2} is available. Only a portion of this table is shown here for guidance regarding its form and content.}
\tablenotetext{a}{The optical counterpart name (when available) or the IRAS/AKARI ID, as listed in Table \ref{tab:table1}.}
\tablenotetext{b}{Spectral indices derived from the listed flux densities in Table \ref{tab:table1} as $\alpha_{\nu_{2} - \nu_{1}}= \rm{log}(F_{2}/F_{1})/log(\nu_{2}/\nu_{1})$.}
\tablenotetext{c}{Galaxy type, assigned from the specified spectral indices; AGNs are chosen for  negative spectral indices, while DSFGs are chosen for positive values. }
 \label{tab:table2}
\end{deluxetable}

\subsection{SED Analysis}\label{S:seds}

To model the SEDs of the IR galaxies from UV to millimeter wavelengths, it is necessary to combine the emissions from the different major components of these galaxies. The major emission components that contribute to the total luminosity of the IR galaxies are the SF, stellar population, dust, and AGN. The energy balance method \citep[e.g.,][]{ Efstathiou2003,daCunha2008,Noll2009,Boquien2019} is one of the main tools for modeling SEDs from UV to submillimeter wavelengths. Therefore, we use CIGALE \citep[][]{Noll2009,Serra2011,Boquien2019,Yang2022},  version 2022.1 \citep{Yang2022}, for the SED fitting, based on the multiwavelength data
described in \S\ \ref{S:multidata}. The broadband filters of the collected multiwavelength data are listed in  Table \ref{tab:table3}. 
CIGALE\footnote[8]{\url{http://cigale.lam.fr/}} is an ideal galaxy SED modeling tool for IR galaxies and AGNs \citep[e.g.,][]{Malek2017,Vika2017,Jarvis2020}, due to the energy conservation that is applied between the UV/optical emission 
that is then re-emitted by the dust in the IR. CIGALE models the unattenuated stellar component from the stellar population model and the star formation history (SFH). The attenuation is obtained for the given attenuation law, and the attenuated emission is then re-emitted in the IR. The the emission components of the AGN \citep[e.g.,][]{Fritz2006} and the synchrotron radiation in radio are then added. CIGALE creates a grid of precomputed model SEDs for each possible combination of the discrete parameters of each emission component (our user-defined parameters for each emission component model are listed in Table \ref{tab:table4}). The dimension of the grid depends on the user-defined parameters. CIGALE obtains the model fluxes after convolving the model SED with the filters. It computes the likelihood (corresponding to $\chi^{2}$) of each model, by comparing the model fluxes with the observed fluxes. During the likelihood computing process, the models are scaled by a free scaling factor, to ensure extensive physical properties with acceptable values for the end product physical parameters, like the IR luminosity and the stellar mass,  etc. In this Bayesian approach, the best-fitting SED is selected from the precomputed set of models, as the one with the lowest $\chi^{2}$, which is used as a maximum likelihood estimator. The Bayesian analysis of CIGALE employs the so-called  pdf$\_$analysis \citep{Boquien2019}, which evaluates the probability from the likelihoods (the lowest $\chi^{2}$) of all the combined models of the stellar, dust, AGN, and radio synchrotron emission components and generates the probability distribution function (PDF) for each physical property (like the IR luminosity, dust mass, stellar mass, etc.). The finally obtained physical property and its uncertainty are the probability-weighted mean and standard deviation of its PDF. We note that in this SED fitting procedure, the obtained physical parameters are not always expected to be the unique possible values, because of the possible degeneracy between the input parameters. However, PDF distributions with a single peak ensure that the obtained physical parameters are reliable. 

\begin{deluxetable}{llc}
 \label{tab:table3}
\tabletypesize{\scriptsize}
\tablewidth{0pt}
\tablecolumns{3}
\tablecaption{  Photometric filters used in the SED analysis with CIGALE} 
\tablehead{       
\colhead{Survey/Telescope} & \colhead{Filter} & \colhead{ $\lambda_\mathrm{eff}$($\mu$m)}} 
\startdata
 \textit{GALEX}  & FUV   		 & 0.15     \\
                          & NUV  		 & 0.23    \\
            SDSS          & $u$       		 & 0.36    \\
            SMSS          & $u$        		 & 0.35    \\
            SMSS          & $v$        		 & 0.39    \\
            SDSS              & $g$        		 & 0.46   \\
            SMSS     & $g$        		 & 0.49    \\
            SMSS     & $r$        		 & 0.60    \\
             SDSS             & $r$         		 & 0.61   \\
             SDSS             & $i$         		& 0.74   \\
            SMSS     & $i$        		 & 0.77    \\
             SDSS             & $z$        		& 0.89   \\
            SMSS     & $z$        		 & 0.90    \\
            \hline              
           2MASS  & $J$        		& 1.24    \\
                          & $H$       		& 1.66    \\
                          &$Ks$      		& 2.16   \\
           \hline               
  WISE    & W1      & 2.25  \\
                          & W2      & 4.60    \\
                          & W3      & 11.56    \\
                          & W4      & 22.09    \\
   AKARI  & S9W    & 9   \\
                          & L18W  & 18   \\
    IRAS & IRAS1   &  12    \\
                 & IRAS2   &  25    \\
          \hline          
    IRAS & IRAS3   &  60    \\
                 &  IRAS4   & 100  \\  
    AKARI  & N60     & 65  \\
                          & WIDE-S   & 90.0   \\
                          & WIDE-L   & 140.0   \\
                          & N160        & 160.0   \\
 Herschel  & PACS blue  &   68.927   \\
                           & PACS red  & 153.95   \\
         \hline                  
 Herschel  &PSW  & 242.82   \\
                           &PMW  & 340.89   \\
                           &PLW   & 482.25     \\   
         \hline                  
 ACT &220\,GHz  & 1362.69   \\
                      &150\,GHz  & 1998.61   \\
                       &98\,GHz  & 3059.11     \\ 
            \hline
    VLA & 1.4\,GHz & 214137.47 \\
    \hline
\enddata
\end{deluxetable}

We use six modules to account for the different emission
components in the SEDs. The stellar component of the SED is produced by the stellar population models, which are convolved with the adopted SFH model. In our analysis, we adopt the standard stellar population synthesis models of \citet{Maraston2005}, since they represent the near-IR stellar populations well \citep{Maraston2009}. These stellar population synthesis models are built from the Salpeter initial mass function \citep{Salpeter1955} with solar metallicity. 
Our sample of nearby galaxies galaxies ($z < 0.05$) have old stellar ages that cannot be well described with a constant SFH model. Therefore, we adopt the double exponentially
declining SFH model \textsc{SFH2EXP}, in the form of a simple analytic function that is provided by CIGALE \citep{Ciesla2015, Ciesla2016}, which implements an SFH composed of two decreasing exponentials. We consider main stellar age values between 500 and 13,000 Myr. We use a large range of values for the e-folding time of the main stellar population model ($\tau_{\rm{main}}$), between 1 and 10,500 Myr. For the e-folding time of the late starburst population ($\tau_{\rm{burst}}$), we use values between 1 and 5000 Myr. We set the main stellar population age to be between 500 and 13,000 Myr. We use the age of the late burst between 1 and 1000 Myr.
We note that most of the listed parameter values in Table \ref{tab:table4} are chosen after several test runs of CIGALE.

\begin{deluxetable*}{lc}
 \label{tab:table4}
\tabletypesize{\scriptsize}
\tablewidth{0pt}
\tablecolumns{3}
\tablecaption{Parameters Used in the SED fitting by CIGALE. }  
\tablehead{       
\colhead{Parameter} & \colhead{Value}} 
\startdata
            \hline
            & Double Exponential SFH (sfh2exp) \\
            \hline
            Main stellar age [Myr]     & 500, 1000, 1050, 1500, 2000, 4000, 5000, 6000, 7200, 10,000, 10,400, 11,000, 11,300, 11,500, 12,000, 12,500, 13,000 \\
            Burst age [Myr]          & 1, 5, 10, 20, 50, 90, 100, 200, 800, 1000 \\
            $\tau_{\rm{main}}$ [Myr] &  1, 5, 10, 50, 100, 150, 200, 300, 350, 400, 700, 800, 1000, 2000, 10,500\\
            $\tau_{\rm{burst}}$ [Myr]       &  1, 5, 10, 20, 40, 45, 50, 100, 200, 300, 400, 700, 800, 5000 \\
            \hline
                              & Dust Emission \\
           \hline
            q$_{\rm{PAH}}$  &  0.47, 1.12, 1.77, 2.50, 3.19, 4.58, 5.26, 5.95, 6.63 \\
            $U_{\rm{min}}$  & 0.12, 0.15, 0.20, 0.35, 0.50, 0.60, 0.70, 0.80, 1.00, 1.20,  1.50, 1.70, 2.00, 2.50, 3.00, 3.50, 4.00, 5.00,  \\
                       & 6.00, 7.00, 8.00, 10.00, 12.00, 17.00, 20.00 \\
            $\alpha$      & 1.9, 2.0, 2.1, 2.2, 2.3, 2.4, 2.5, 2.8, 2.9, 3.0 \\
            $\gamma$   &  0.02, 0.1, 0.2, 0.7 \\
		\hline
		               & Dust Attenuation Law \\
		\hline
		Slope ISM   &  --0.7, --0.5, --0.4, --0.3, --0.1 \\
		Slope birth clouds & --1.3, --0.7, --0.5 \\
		\hline
		               & AGN \\
		\hline
		$R_{max}/R_{min}$    & 10, 30, 60, 100, 150 \\
		$\tau$                          & 0.1, 0.6, 1.0, 2.0, 3.0, 6.0, 10.0 \\
		$\beta$            & --1.00, --0.75, --0.50, --0.25, 0.00 \\
		$\gamma$           & 0.0, 2.0, 4.0, 6.0 \\
		$\theta$               & 60., 100., 140 \\
		$\psi$        &0.001, 10.100, 20.100,  30.100, 40.100, 50.100, 60.100, 80.100, 89.990 \\
		frac$_{\rm{AGN}} $             & 0.01, 0.02, 0.05, 0.09, 0.10, 0.15, 0.20, 0.25, 0.30, 0.35, 0.40, 0.50, 0.60 \\
		\hline
		               & Synchrotron Radio Emission \\
                 \hline
     	$q_{\rm{IR}}$    & 2.4,2.5,2.58,2.6,2.7\\
		$\alpha_{\rm{SF}}$  & 0.8 \\
		$R_{\rm{AGN}}$   & 0,0.01,0.02,0.1,0.2,1,2,3,10,...,50,100,1000 \\
		$\alpha_{\rm{AGN}}$  & 0.3,0.4,0.5,0.6,0.7,0.8,0.9,1.0 \\
		\hline
\enddata
\end{deluxetable*}

We adopt the \citet{Charlot&Fall2000} attenuation law for the dust attenuation, since it can account for different attenuation for young and old stars. In this attenuation law, there are two main attenuation sources: (i) the interstellar medium (ISM), which would affect both young and old stars (older than 10$^7$ yr); and (ii) the dust in the birth clouds, which would affect the young stars. The attenuations of these two sources are modeled by power laws. We use a combination of the allowed slope values for the ISM and birth clouds, as listed in Table \ref{tab:table4}. The dust model of \citet{Draine2014} is adopted for the dust component in the mid-IR and FIR regions. This model includes the mass fraction of polycyclic aromatic hydrocarbons (PAHs; ${q}_{\mathrm{PAH}}$), with possible values between 0.47 and 7.32, and the minimum radiation field ($U_{\mathrm{min}}$, with possible values between 0.1 and 50.0. The dust emission is in the power-law form of $dM_{\rm d}/dU \propto U^{-\alpha}$ where $M$ is the total dust mass and $\alpha$ is the slope within the 1$-$3 range. The $\gamma$ represents the fraction of the dust mass (between 0 and 1) that is heated by a greater radiation field than $U_{\mathrm{min}}$. As suggested by \citet{Toba2020}, we use different $\gamma$ values, to obtain a better fit of the FIR emission.
The modified \citet{Fritz2006} models are used for the AGN component. 
These include the isotropic point-like central source power-law radiation (in the 0.001--20 $\mu$m wavelength range) and the dust emission in a toroidal structure. 
In this model, dust (mainly silicate and graphite grains) can scatter the continuum radiation from the center or it can absorb and re-emit it at 1--1000 $\mu$m. 
The model includes three components: the central AGN, the scattered emission, and the thermal dust emission from the torodial structure. 
The size of the torodial structure is given by the ratio of the maximum outer radius to the minimum innermost radius ($R_{max}/R_{min}$), with allowed values between 10 and 150. The absorption feature of the silicate grains at 9.7 $\mu$m is taken into account by the optical depth parameter $\tau$, which can have values of  0.1, 0.3, 0.6, 1.0, 2.0, 3.0, 6.0, or 10.0.  
The radiation of the torodial structure depends on the density distribution $r^{\beta} e^{-\gamma |cos\theta |}$, where $\beta$ represents the radial dust distribution, with allowed values between  --1 and 0, and $\gamma$ represents the angular dust distribution, with allowed values between 0.0 and 6.0. Here,  $\theta$ is the opening angle of the torodial structure, with allowed values of 60$^{\circ}$, 100$^{\circ}$, or 140$^{\circ}$. 
The $\psi$ parameter is the angle between the equatorial axis and the line of sight, with allowed values between 0.001 and 89.990. 
While Type 1 AGNs have $\psi = 90^{\circ}$, a Type 2 AGN can have $\psi = 0^{\circ}$. 
We include most of the allowed parameter values of the  \citet{Fritz2006} models, to include the models of different AGN types. 
The contribution of the fractional AGN emission  total IR emission is shown by the frac$_\mathrm{AGN}$ parameter. In order to be able to model AGN components with different strengths (both low and high), we consider frac$_\mathrm{AGN}$ values starting from 0.0 to 0.6. The synchrotron radio emission includes two different model components: (i) the radio emission from the SF; and (ii) the AGN radio emission. The radio synchrotron emission produced by the SF is modeled as a cutoff power-law component, with a slope ($\alpha_{\rm{SF}}$), whose value of 0.8 is adopted as default parameter in our analysis. 
The radio$-$IR correlation coefficient, $q_{\rm{IR}}$, which is the normalization of the SF radio emission \citep[][]{Helou1985} at 1.4 GHz, has a default value of 2.58 in CIGALE.
Following \citet{Yang2022}, we allow $q_{\rm{IR}}$ to vary between 2.4 and 2.7 in our analysis, to be consistent with the observations conducted by \citet{Delvecchio2021}. 
The AGN radio emission is modeled with a simple power-law component, between 0.1 and 1000 mm, with a slope $\alpha_{\rm{AGN}} = 0.7$ \citep{Randall2012}. 
In our analysis, we allow $\alpha_{\rm{AGN}}$ to be between 0.3 and 1.0, to have statistically reliable fits. 
The AGN radio emission model also includes the radio-loudness parameter $R_{\rm{AGN}}$, which is the ratio of the monochromatic luminosities at 5 GHz and 2500 $\AA$ \citep[e.g., Equation (9) of][]{Yang2022}. 
We allow $R_{\rm{AGN}}$ to have a large range, between 0 and 1000.

With the  parameter settings listed in Table \ref{tab:table4} and the available photometric data listed in Table \ref{tab:table3}, we model the stellar, AGN, and dust emission components, by using at most 39 photometric points from UV to radio wavelengths.
In order to apply CIGALE, at least one detection at each of the UV--optical, near-IR, mid-IR, and FIR regions is required, to obtain an accurate model. 
To have a reliable dust SED, we also need  detections at both the shorter and longer  wavelengths  of the SED peak near 100 $\mu$m. We also use the 3$\sigma$ noise levels in each ACT band as the nondetection upper limits in the SED analysis.
These criteria allow us to perform the SED analyses for 87 galaxies in our sample. 

In the first run of CIGALE, we use only the stellar, AGN, and dust components, without the radio data at 1.4 GHz. After the first run, all the obtained SEDs are visually checked. In most of the SEDs, the millimeter-band fluxes are higher compared to the dust component. 
Based on the millimeter measurements, we see that more than half of our sample requires a radio component. Therefore, after the first run, the radio synchrotron emission component is added, to model the SEDs between the 100 $\mu m$ and  millimeter regions. We also quantify the FIR--millimeter wave color range that would require a radio component (see \S \ref{S:infraredACTcol}), based on our SED analysis.

\begin{figure*}
\begin{center}$
\begin{array}{llll}
\includegraphics[scale=0.29]{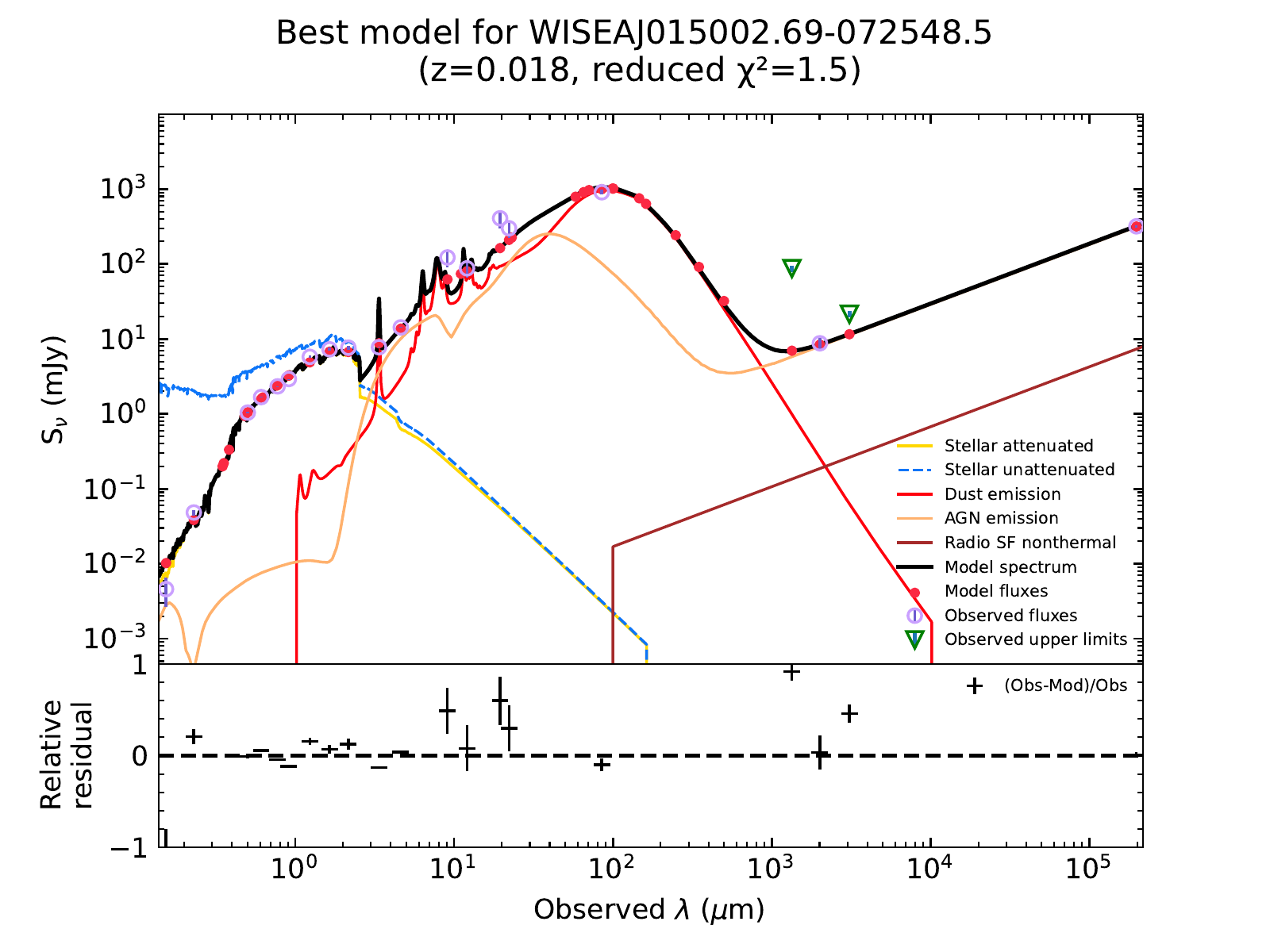}&
\includegraphics[scale=0.29]{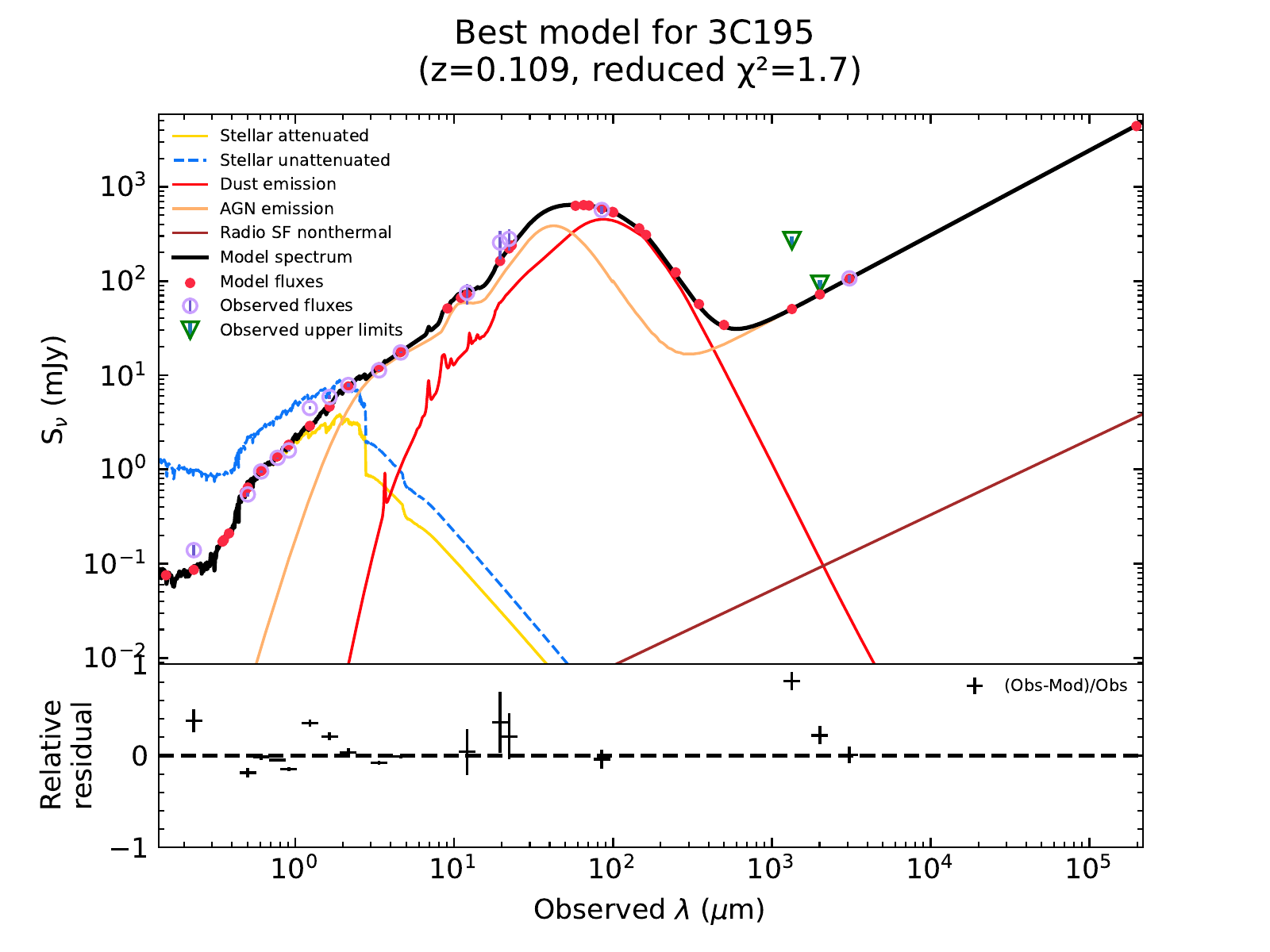}&
\includegraphics[scale=0.29]{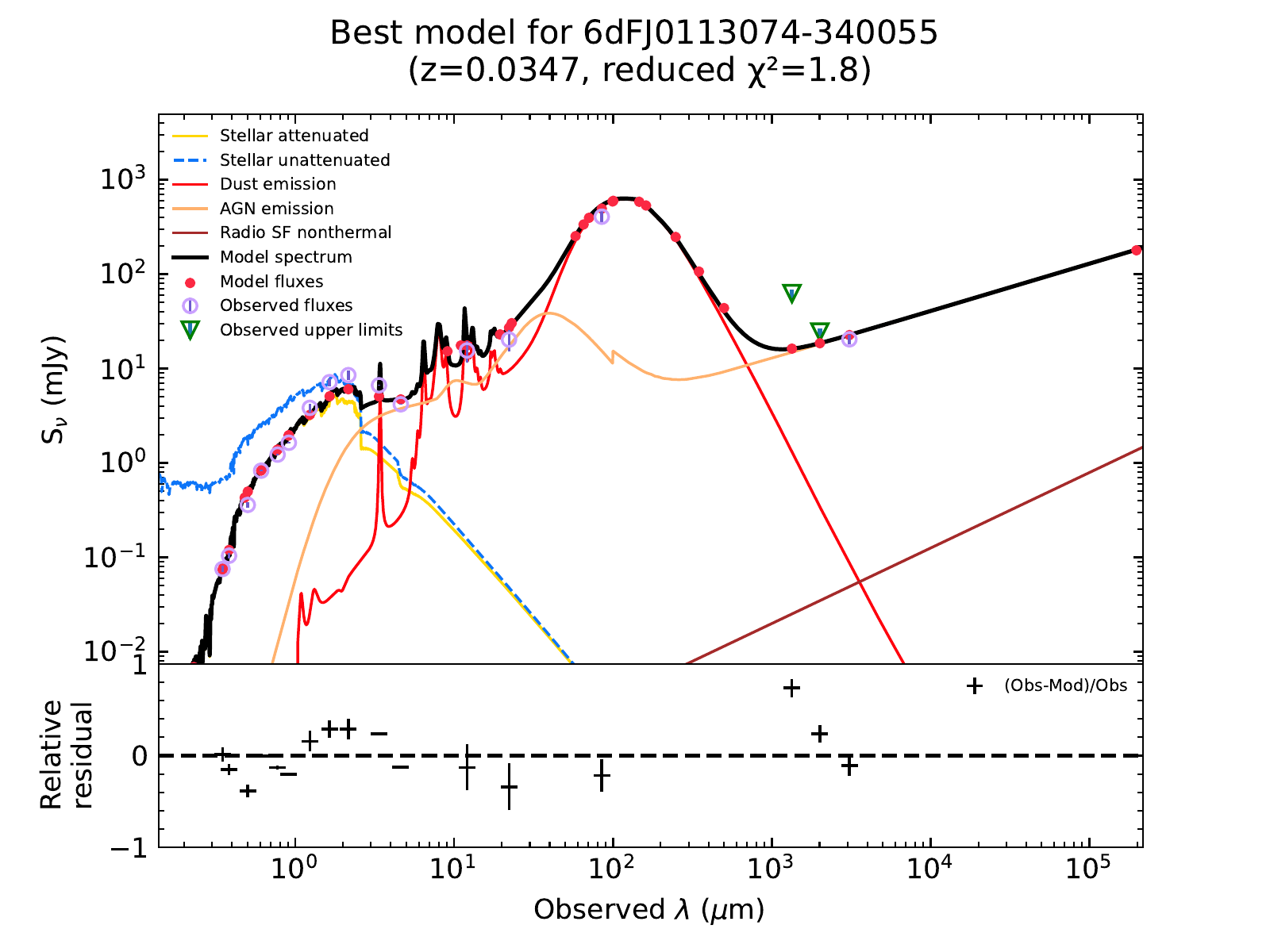}&
\includegraphics[scale=0.29]{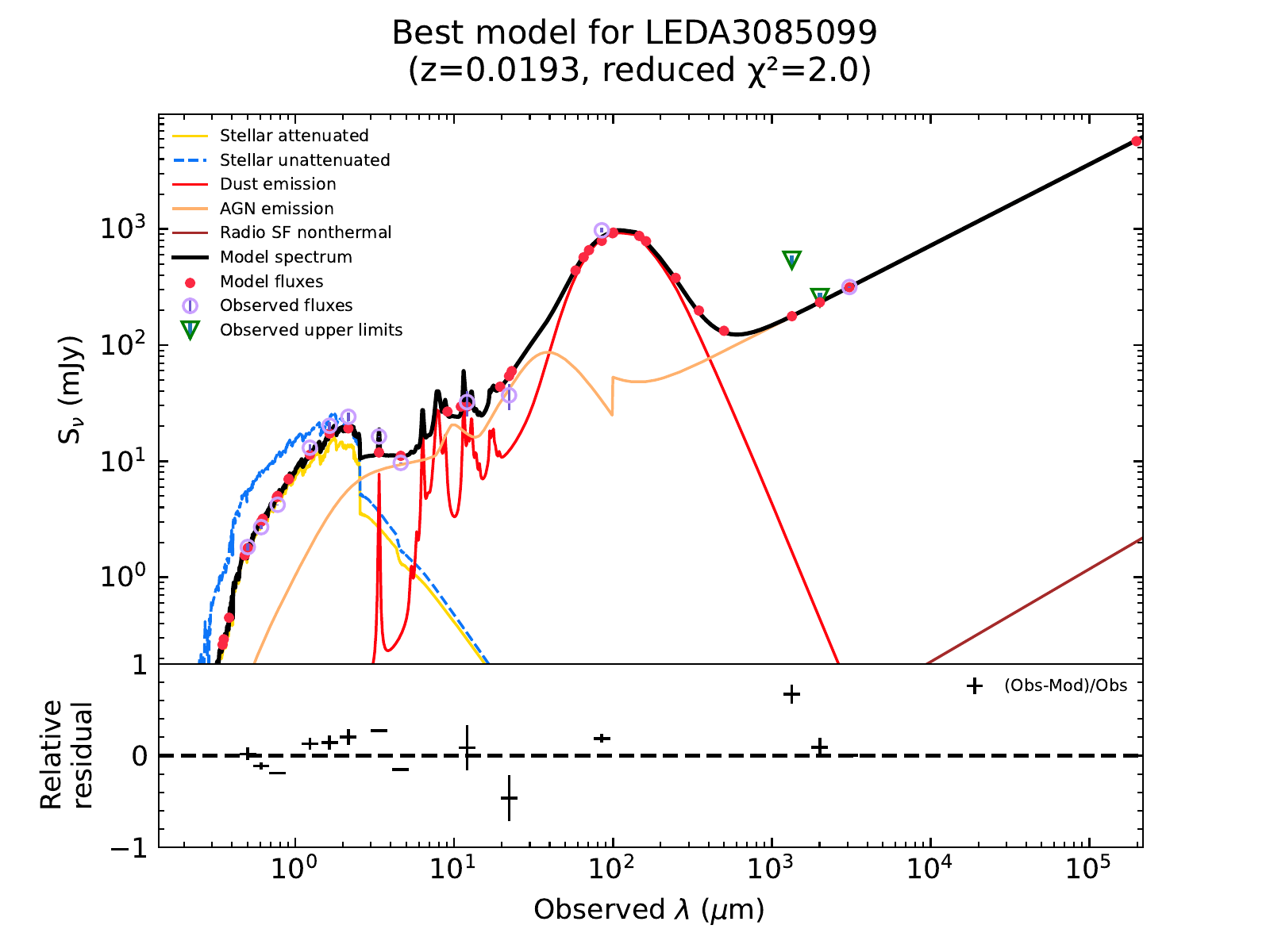}\\
\includegraphics[scale=0.29]{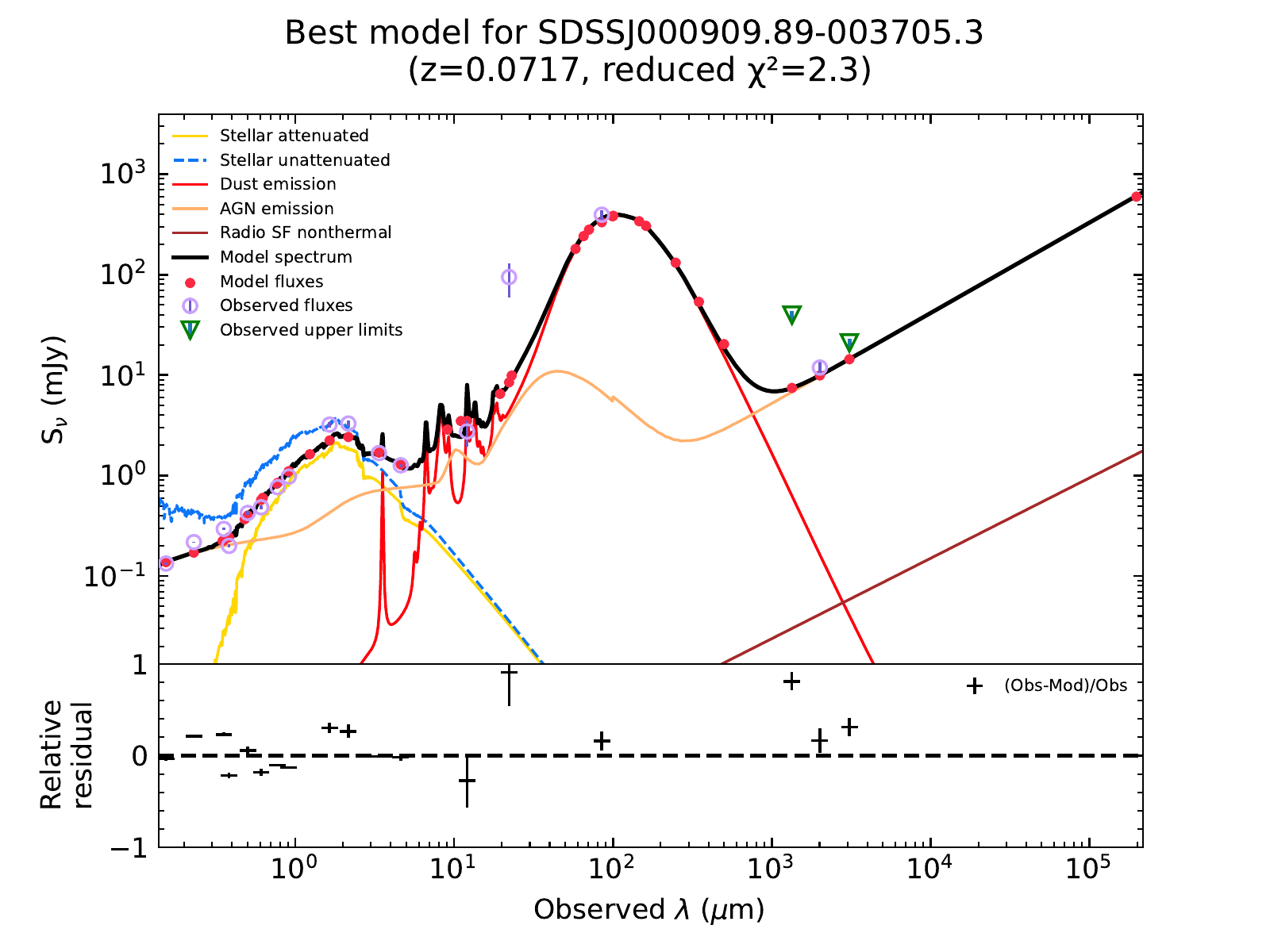}&
\includegraphics[scale=0.29]{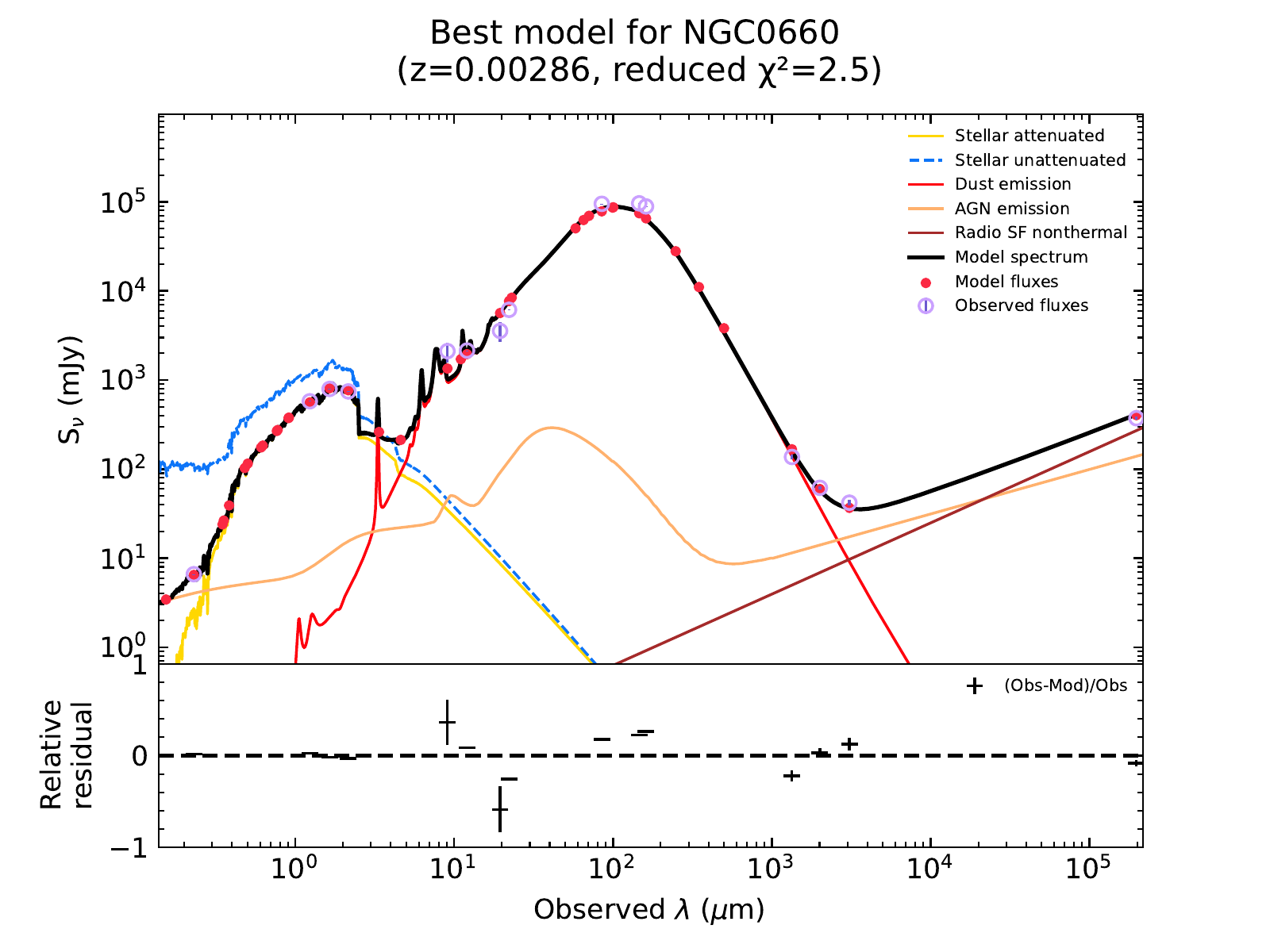}&
\includegraphics[scale=0.29]{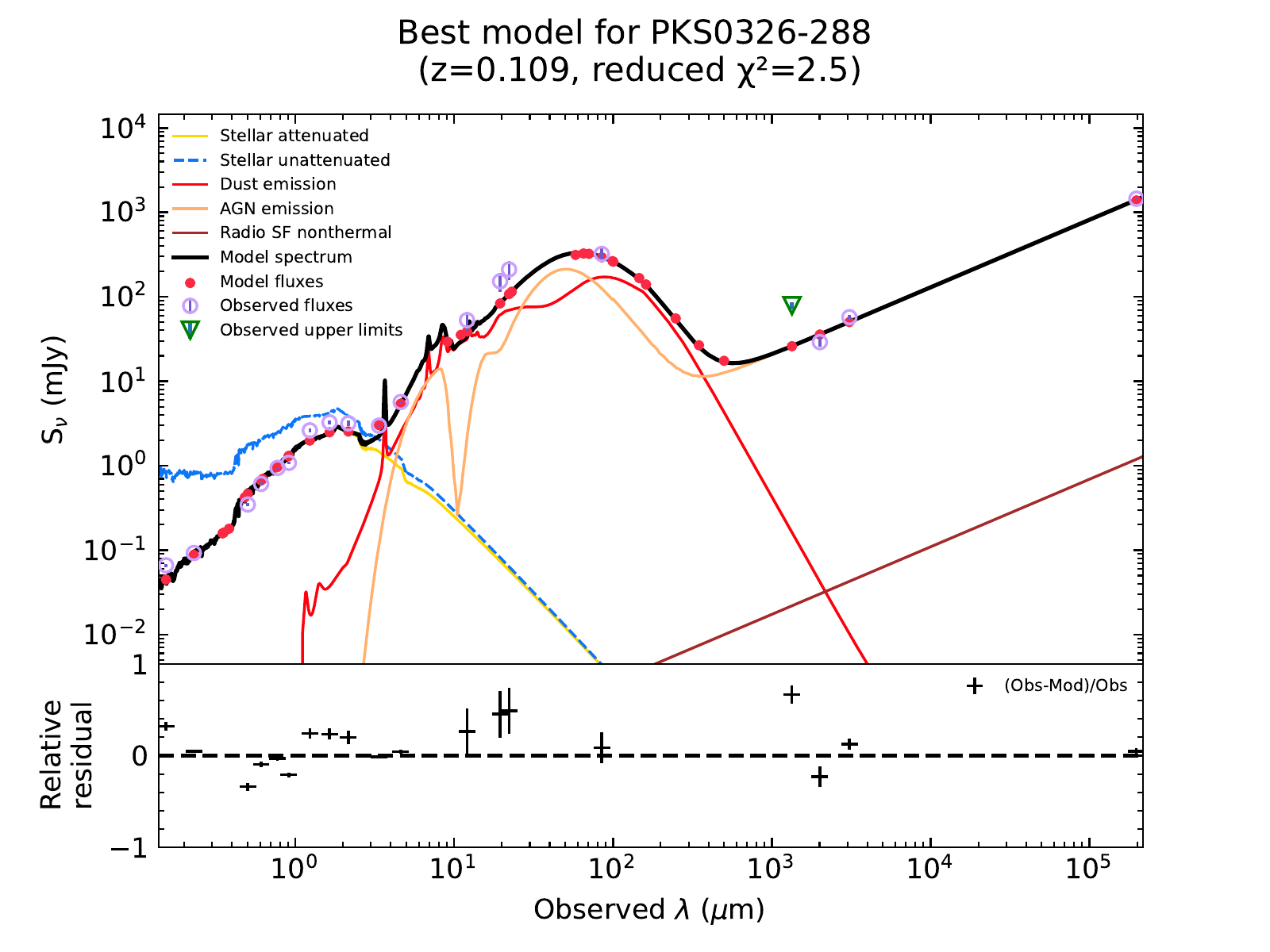}&
\includegraphics[scale=0.29]{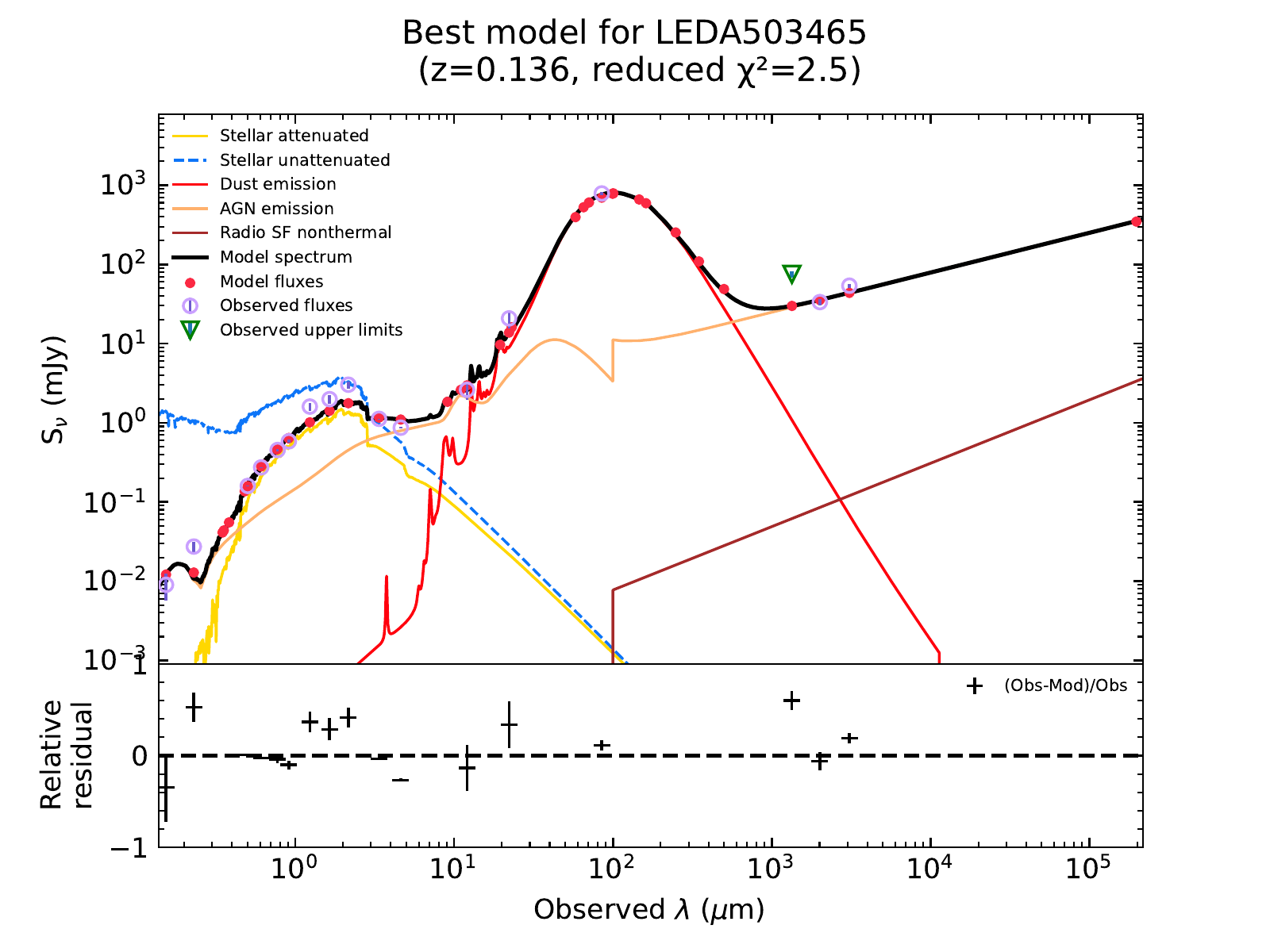}\\
\includegraphics[scale=0.29]{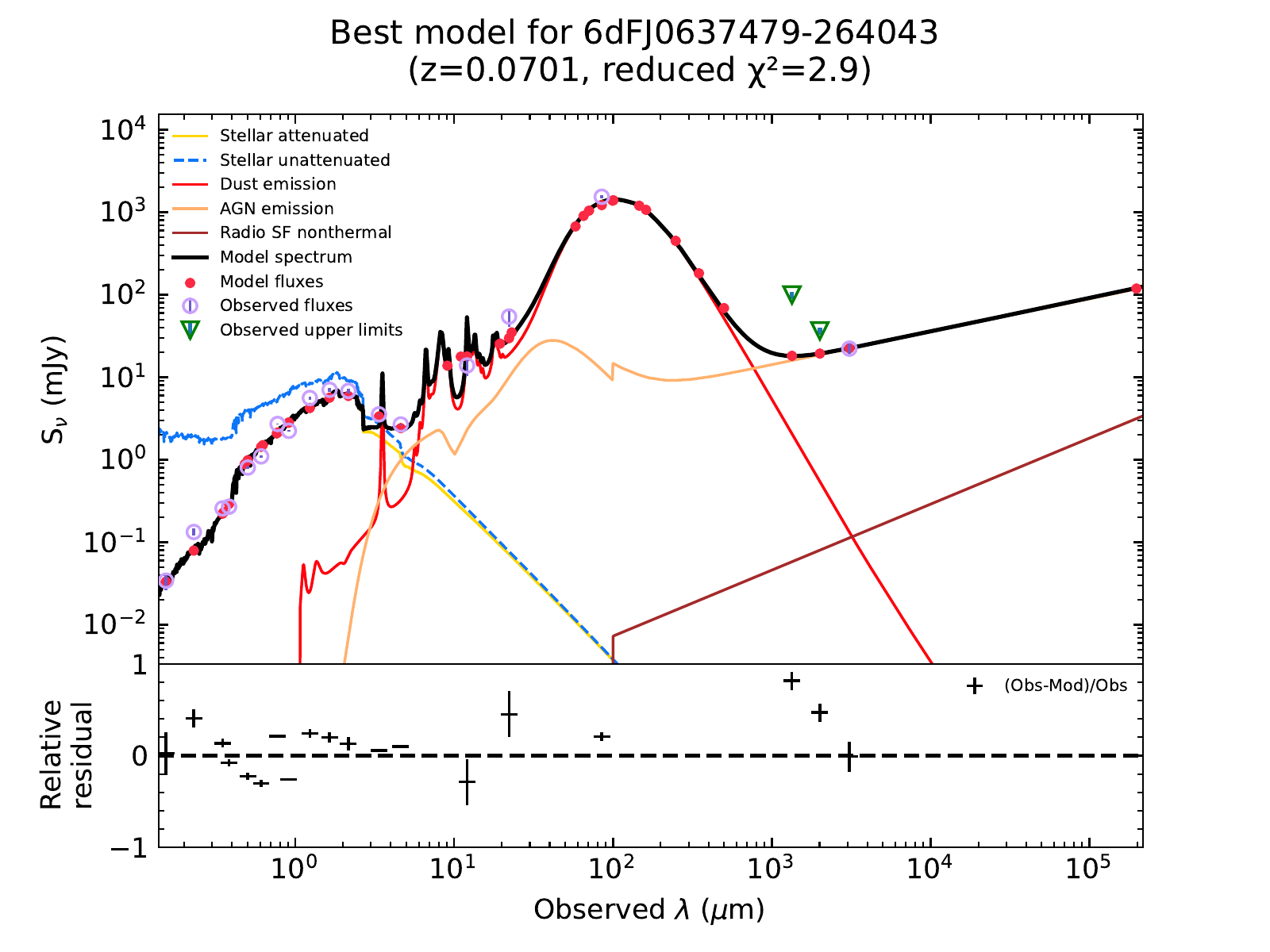}&
\includegraphics[scale=0.29]{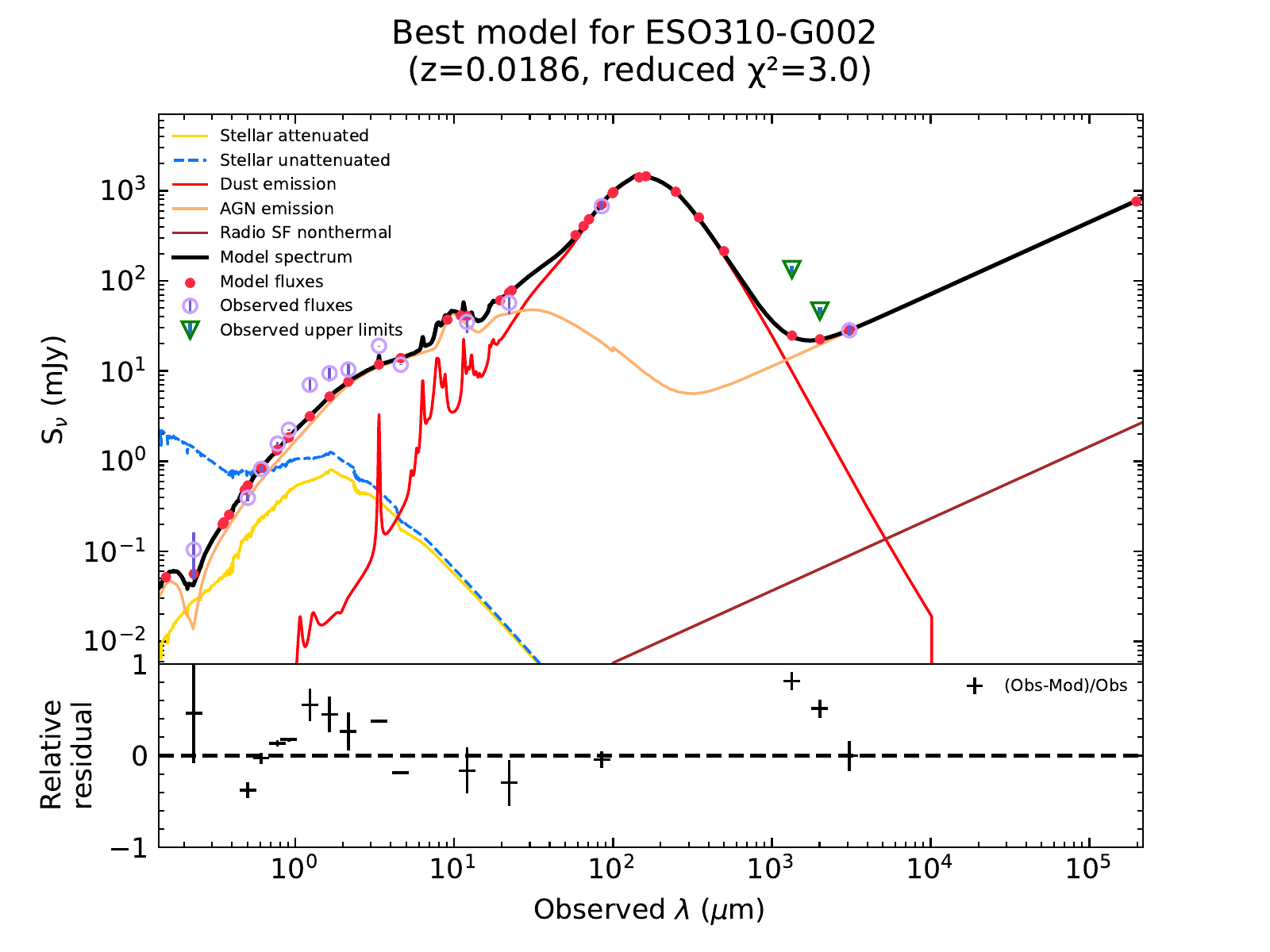}&
\includegraphics[scale=0.29]{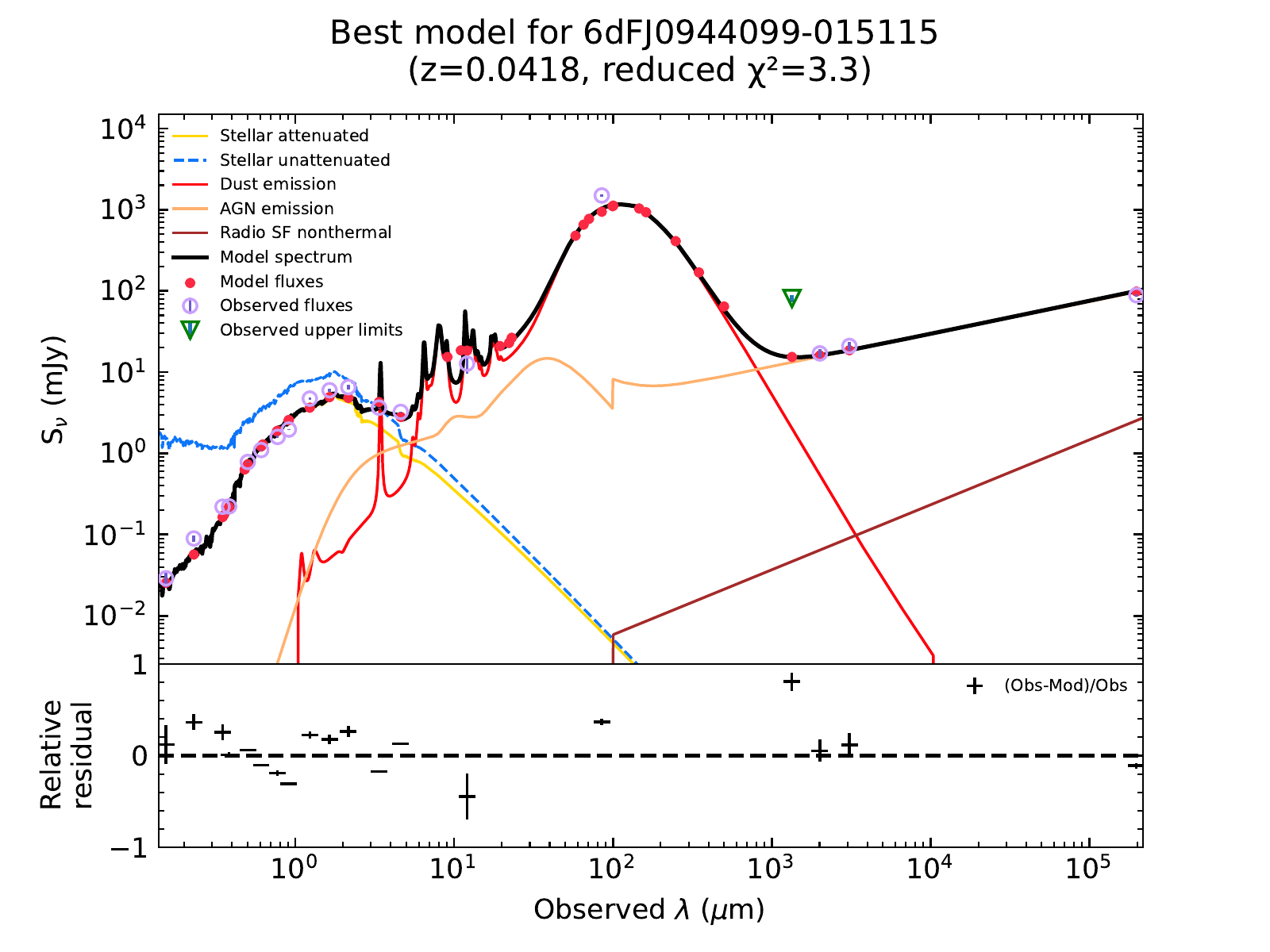}&
\includegraphics[scale=0.29]{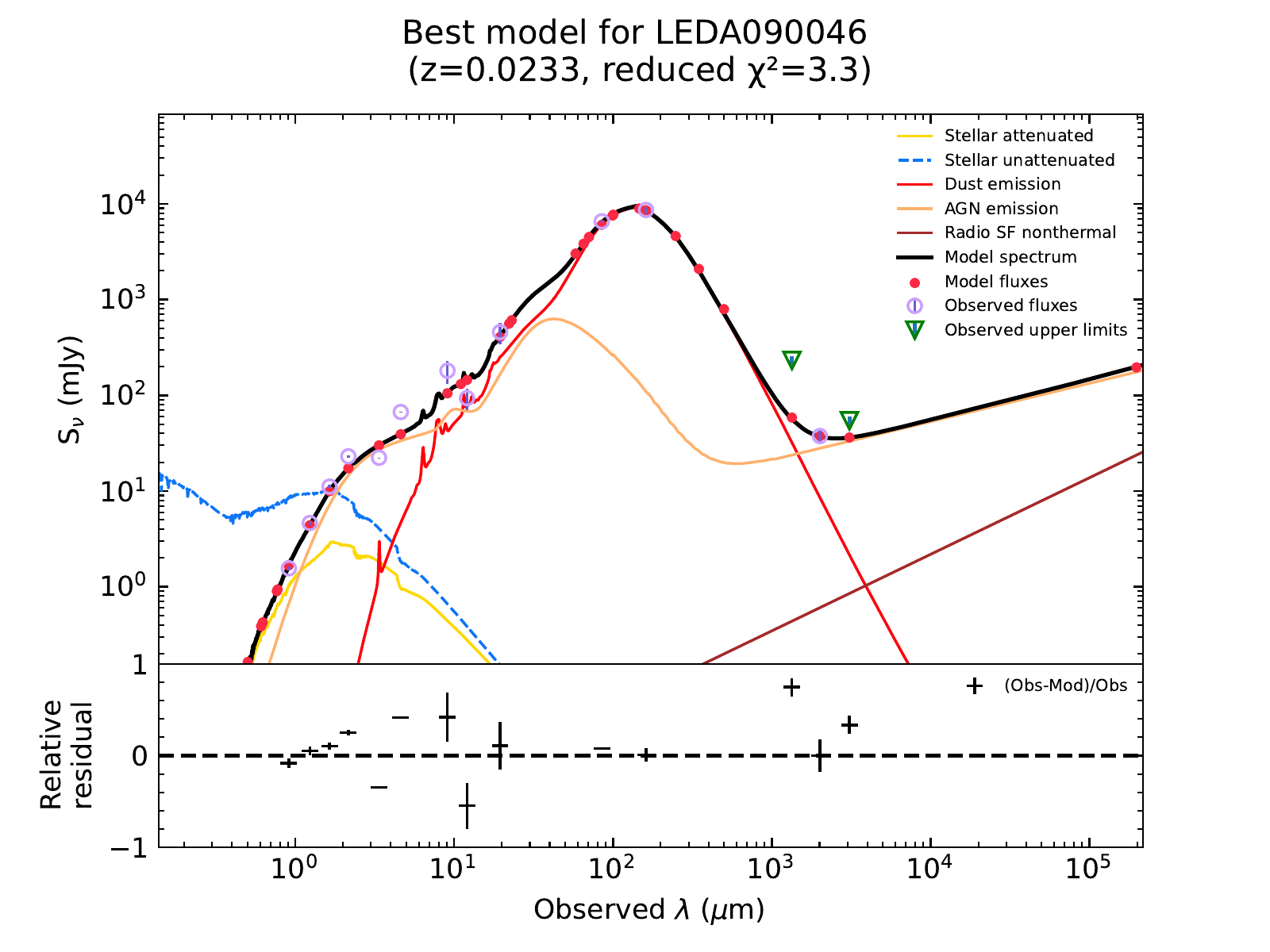}\\
\includegraphics[scale=0.29]{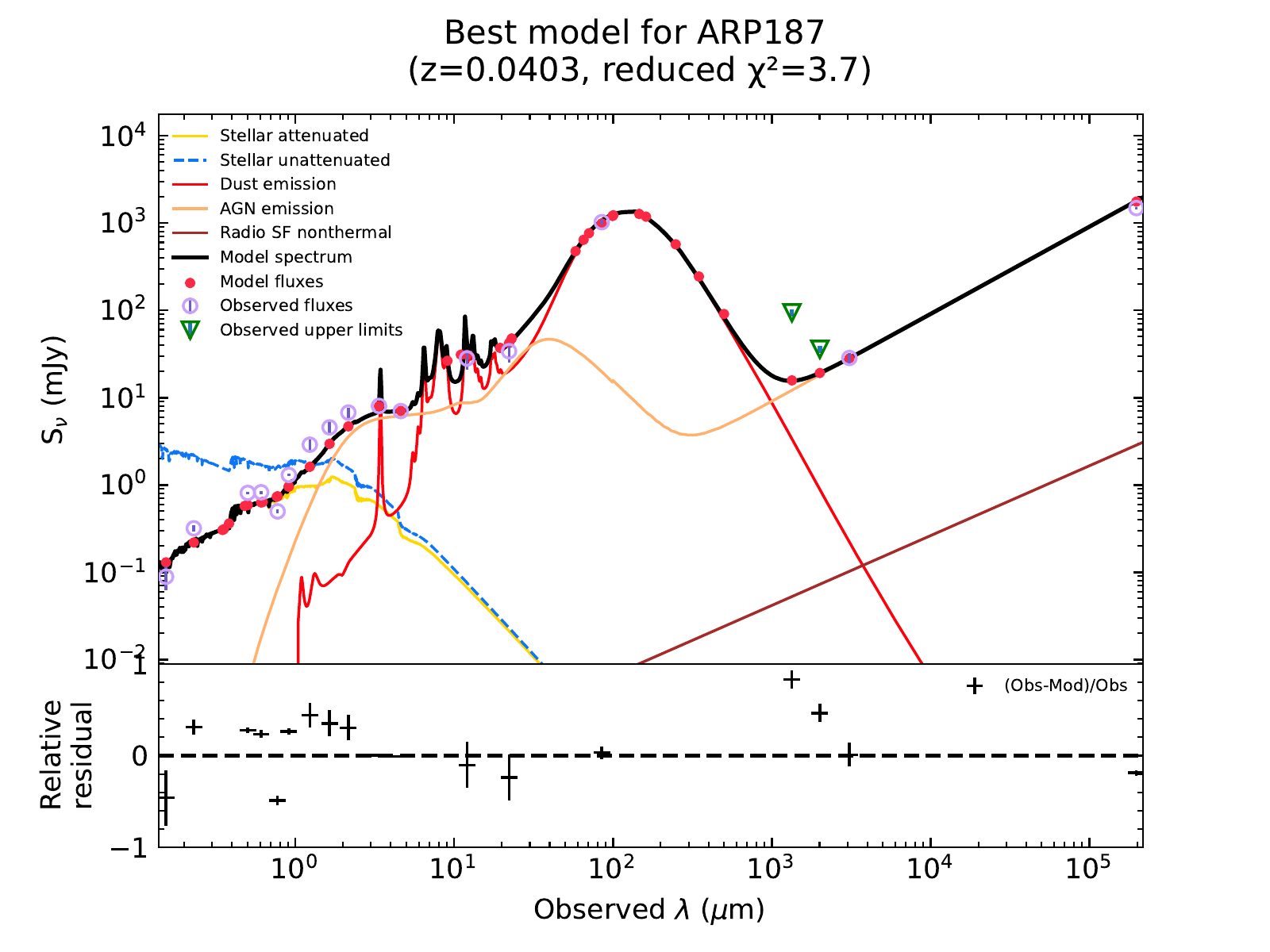}&
\includegraphics[scale=0.29]{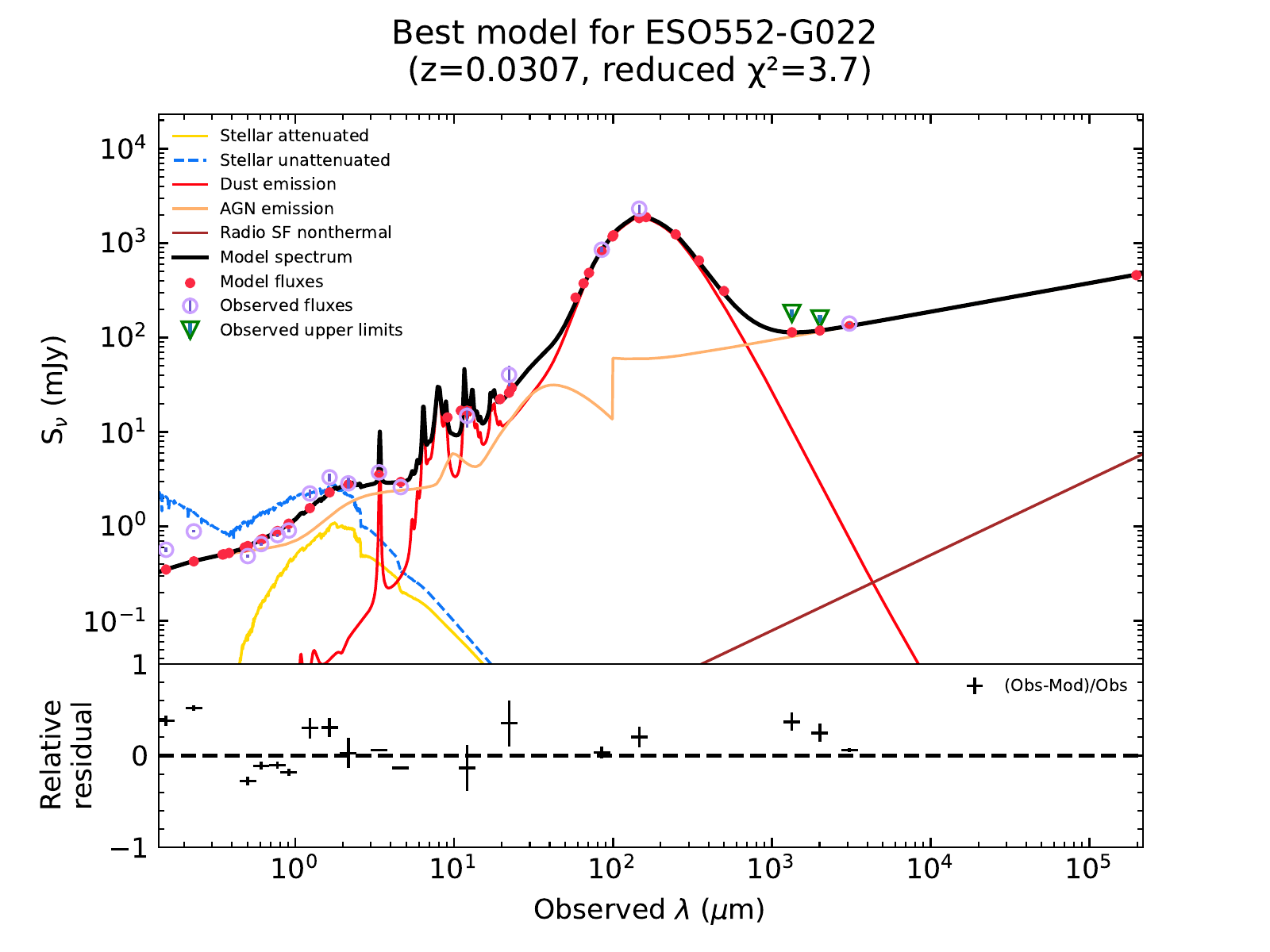}&
\includegraphics[scale=0.29]{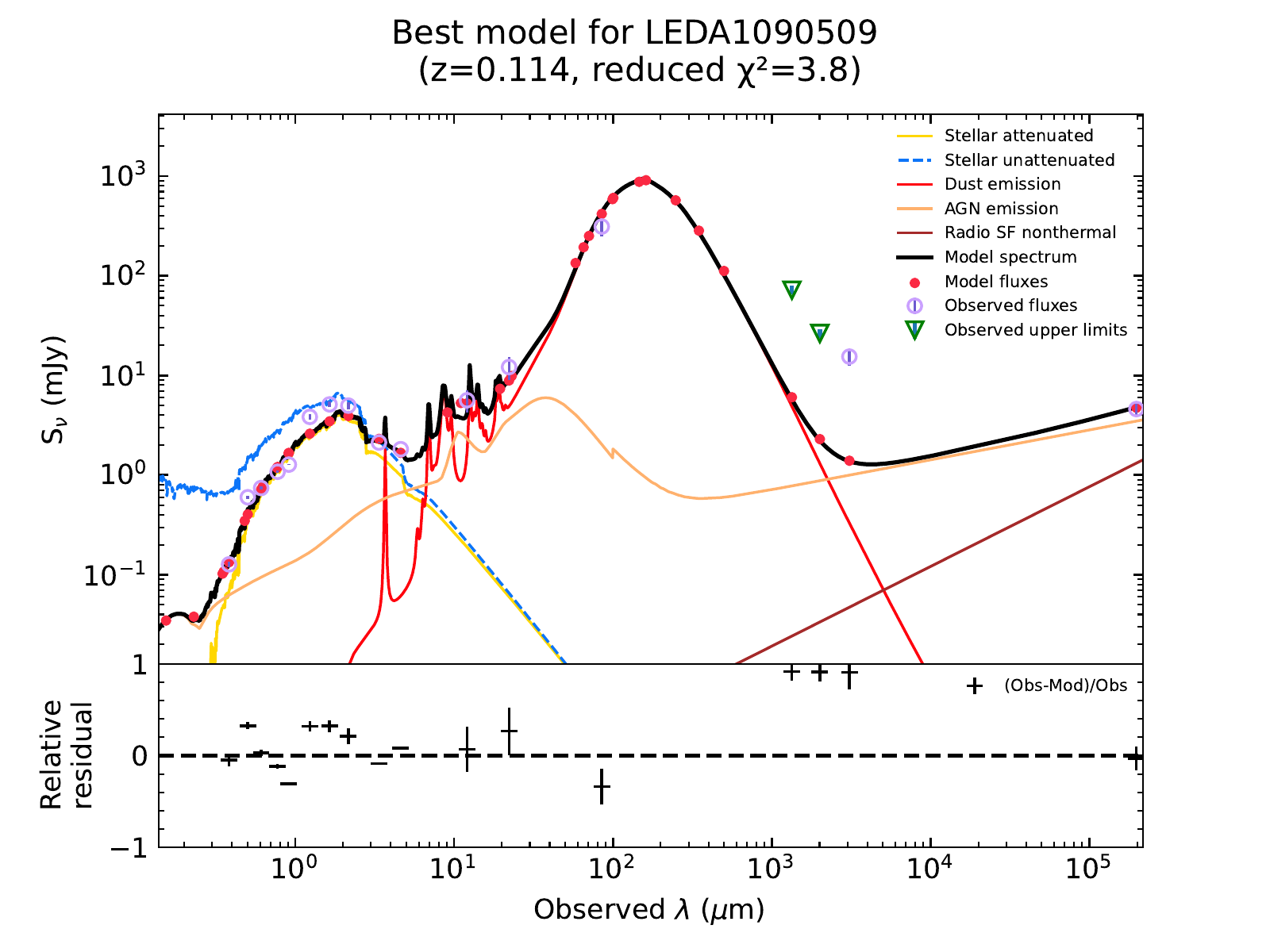}&
\includegraphics[scale=0.29]{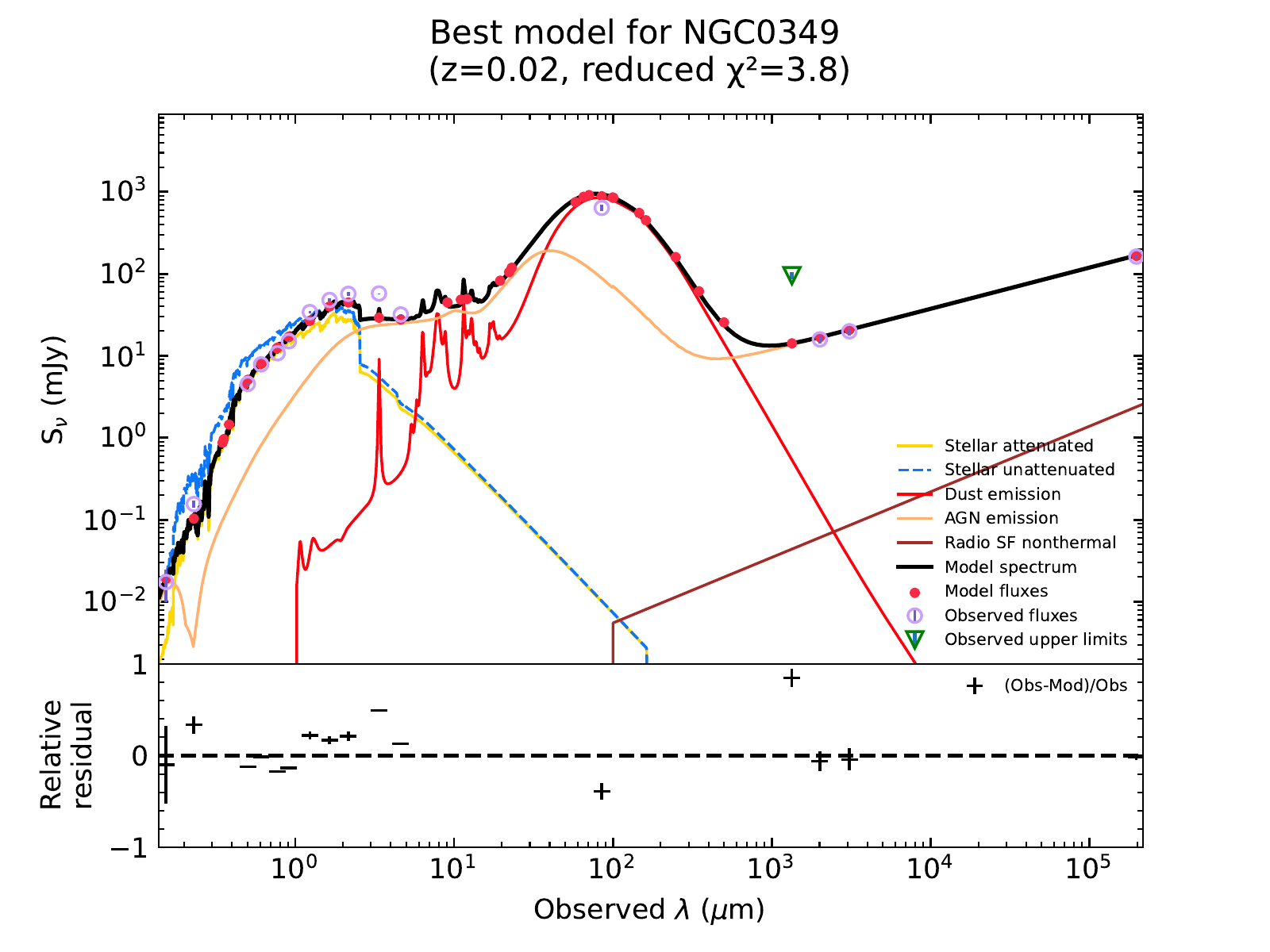}\\
\includegraphics[scale=0.29]{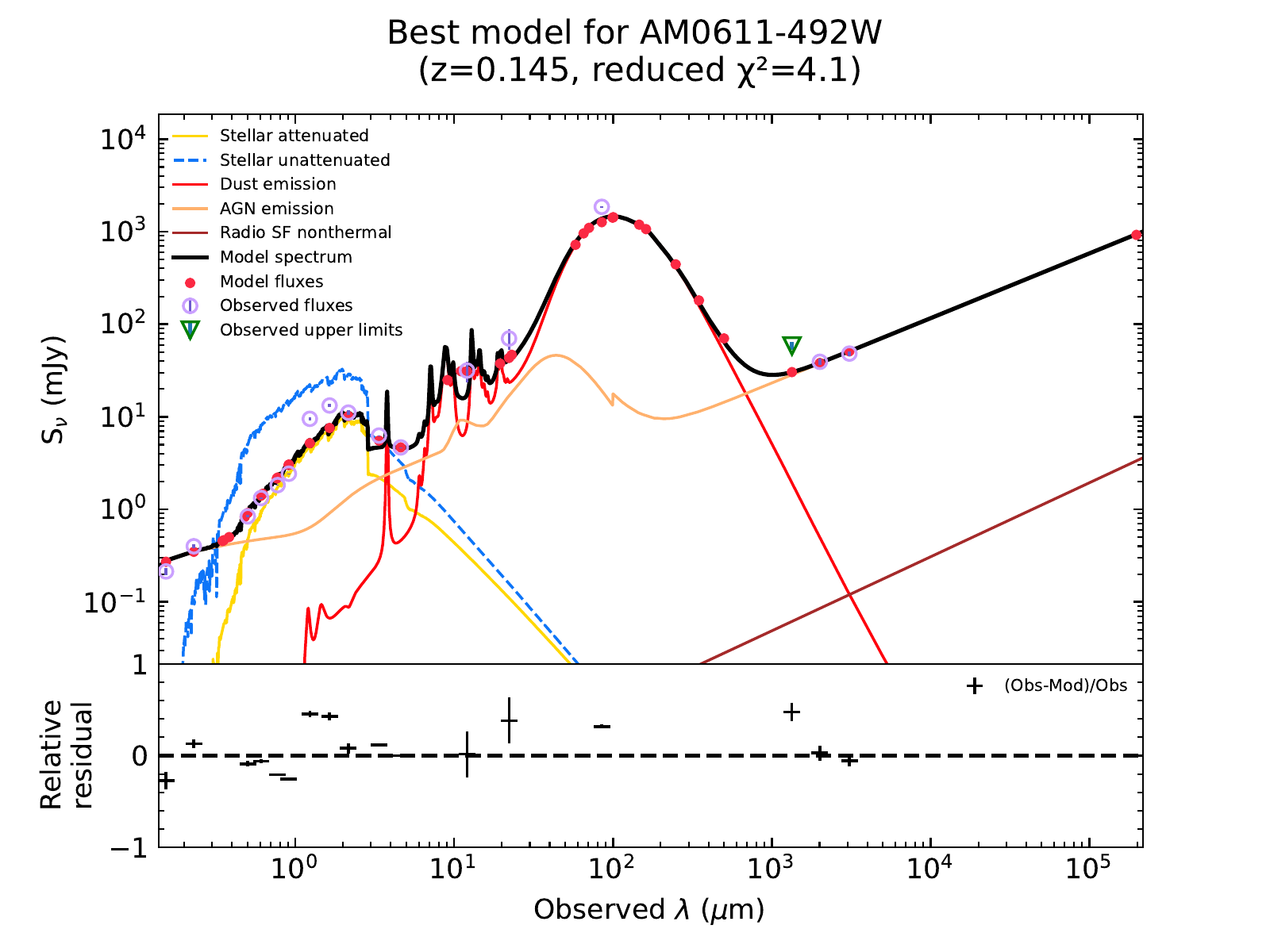}&
\includegraphics[scale=0.29]{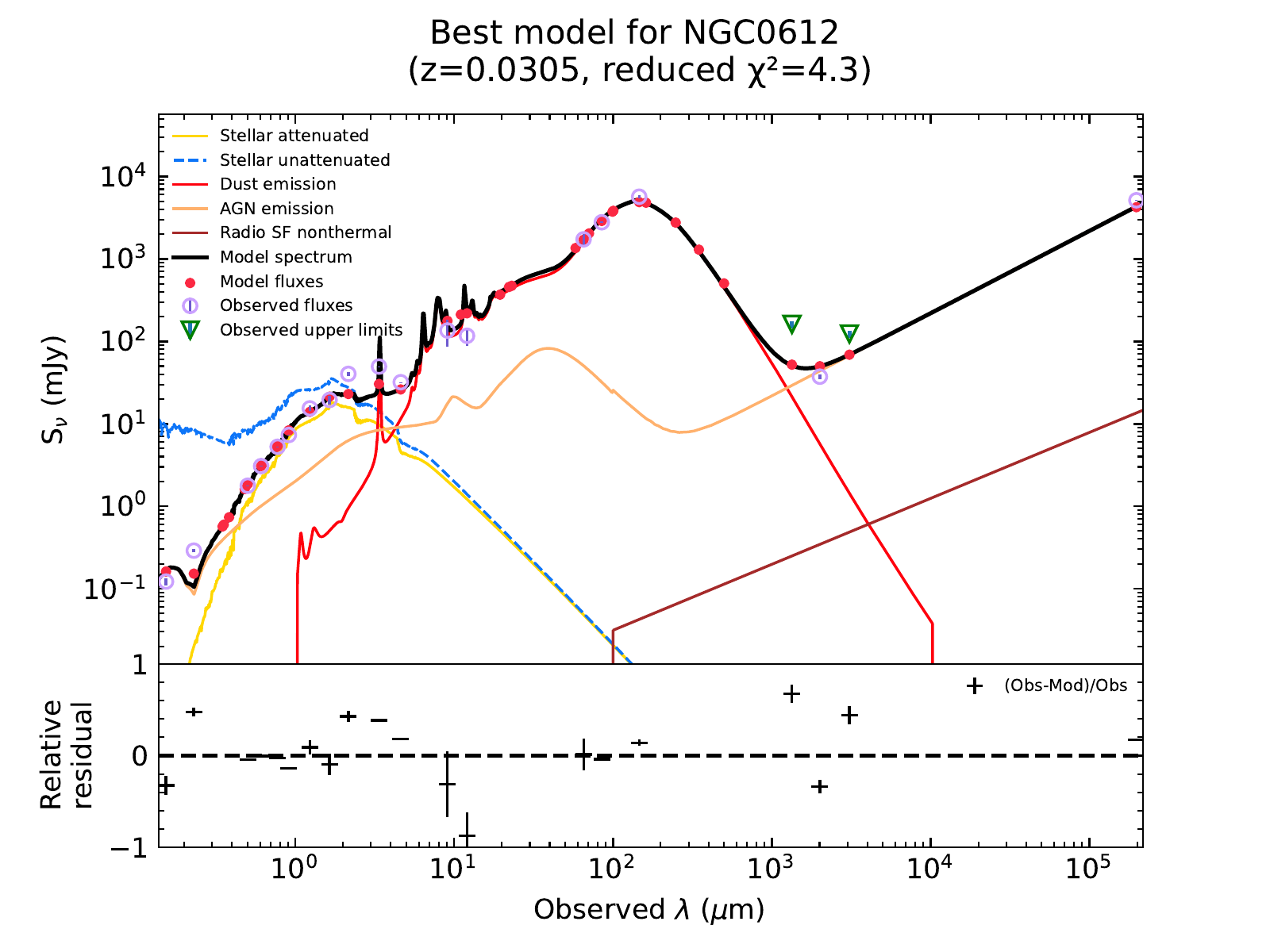}&
\includegraphics[scale=0.29]{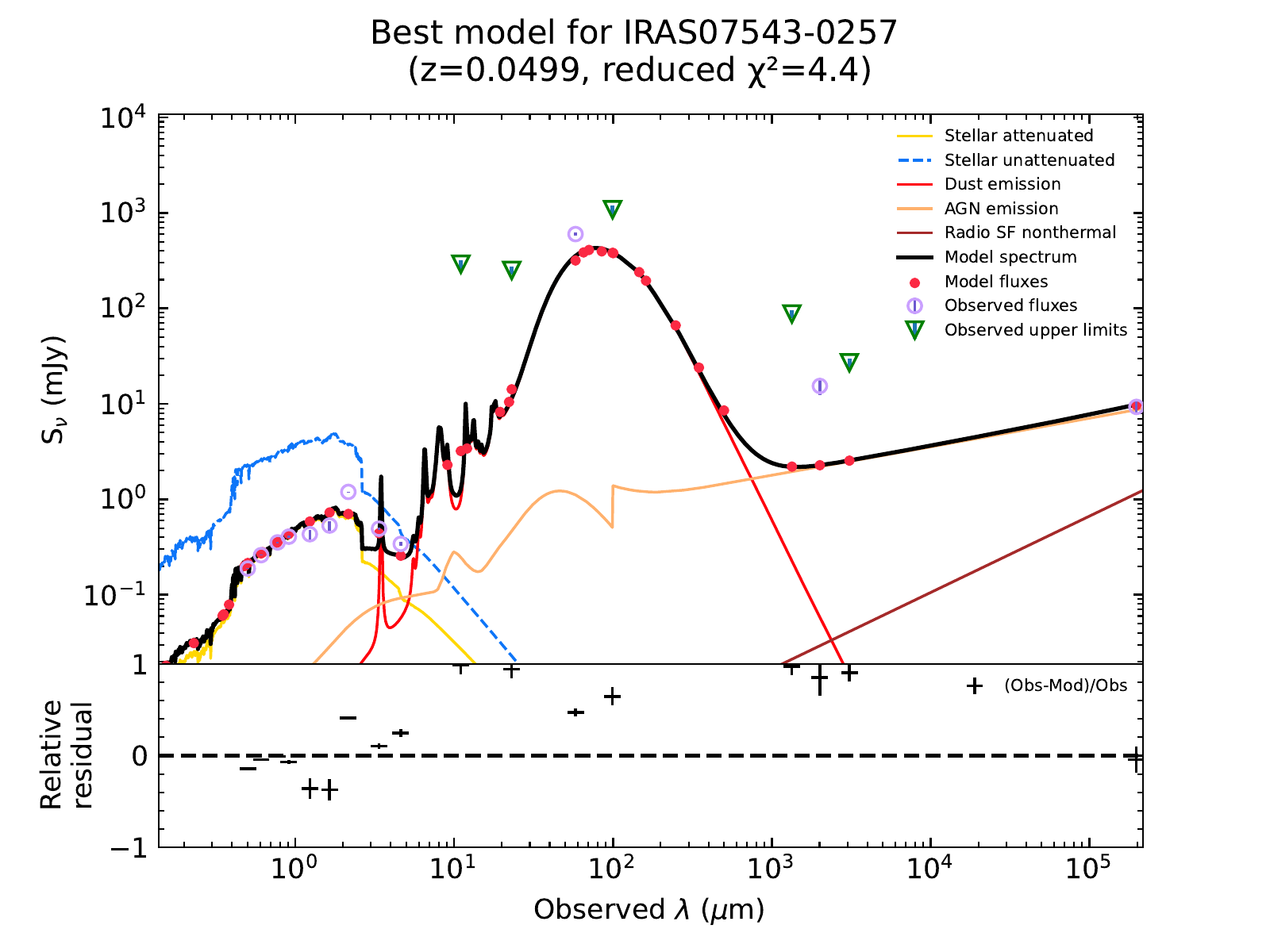}&
\includegraphics[scale=0.29]{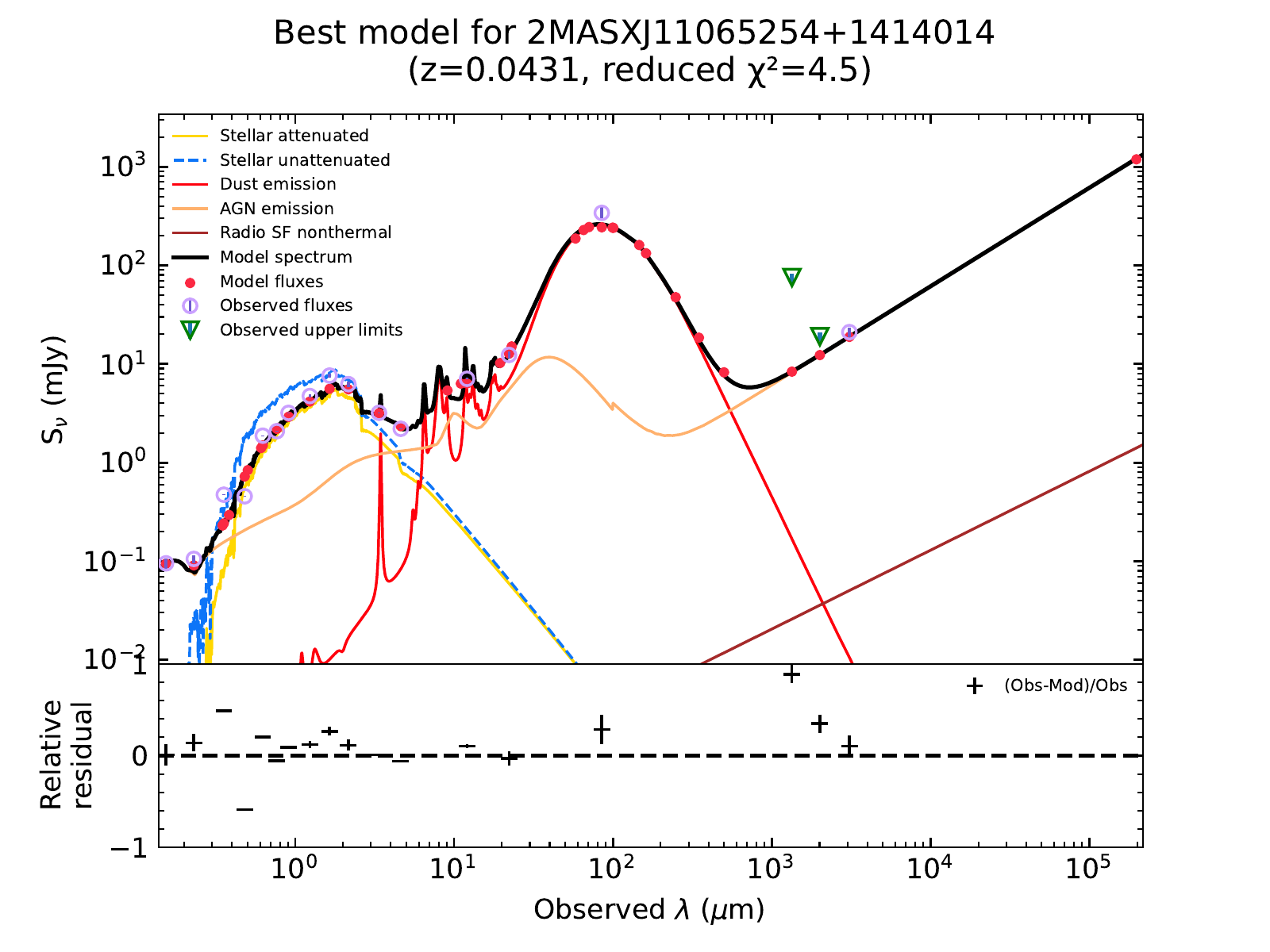}\\
\includegraphics[scale=0.29]{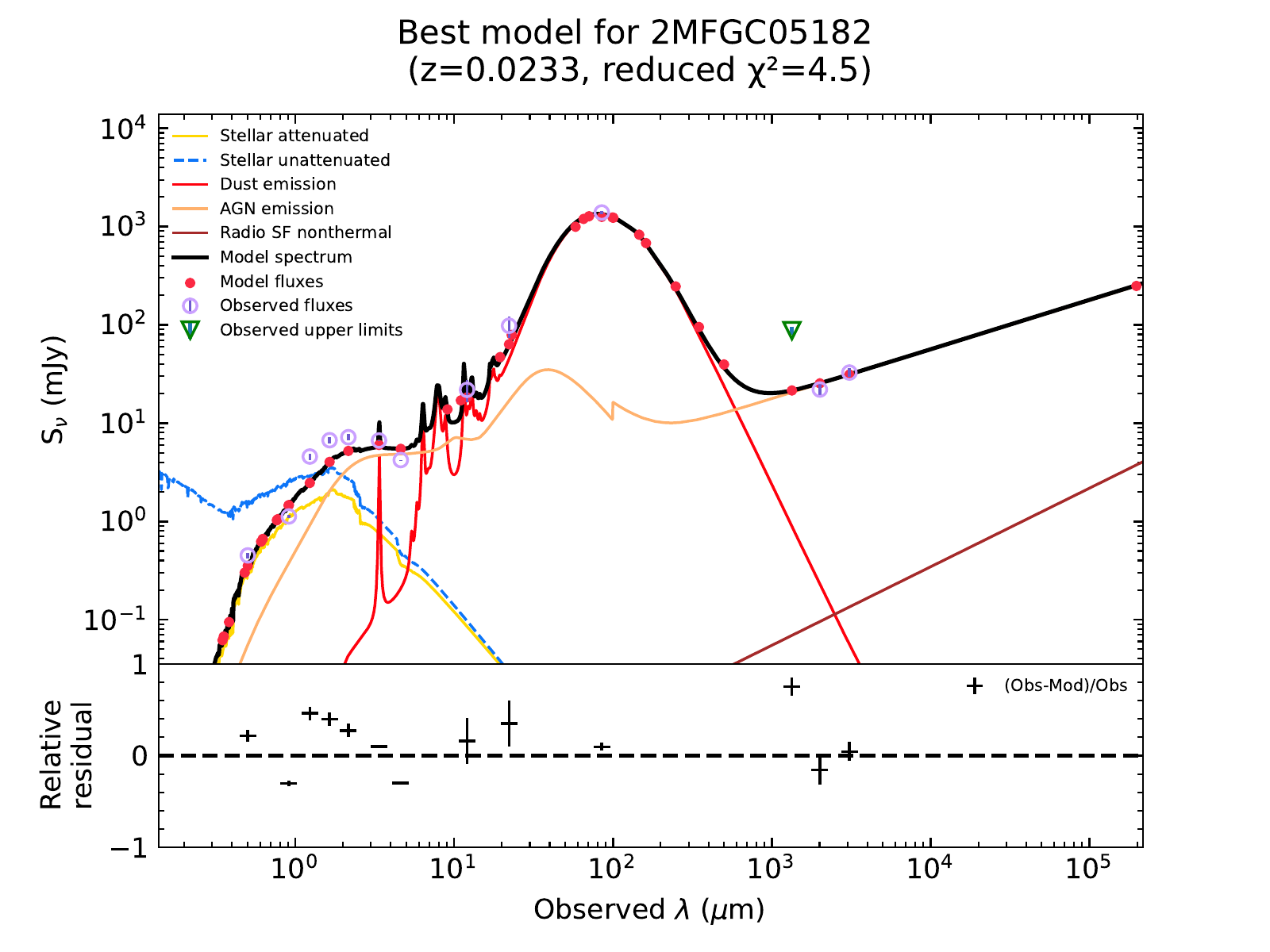}&
\includegraphics[scale=0.29]{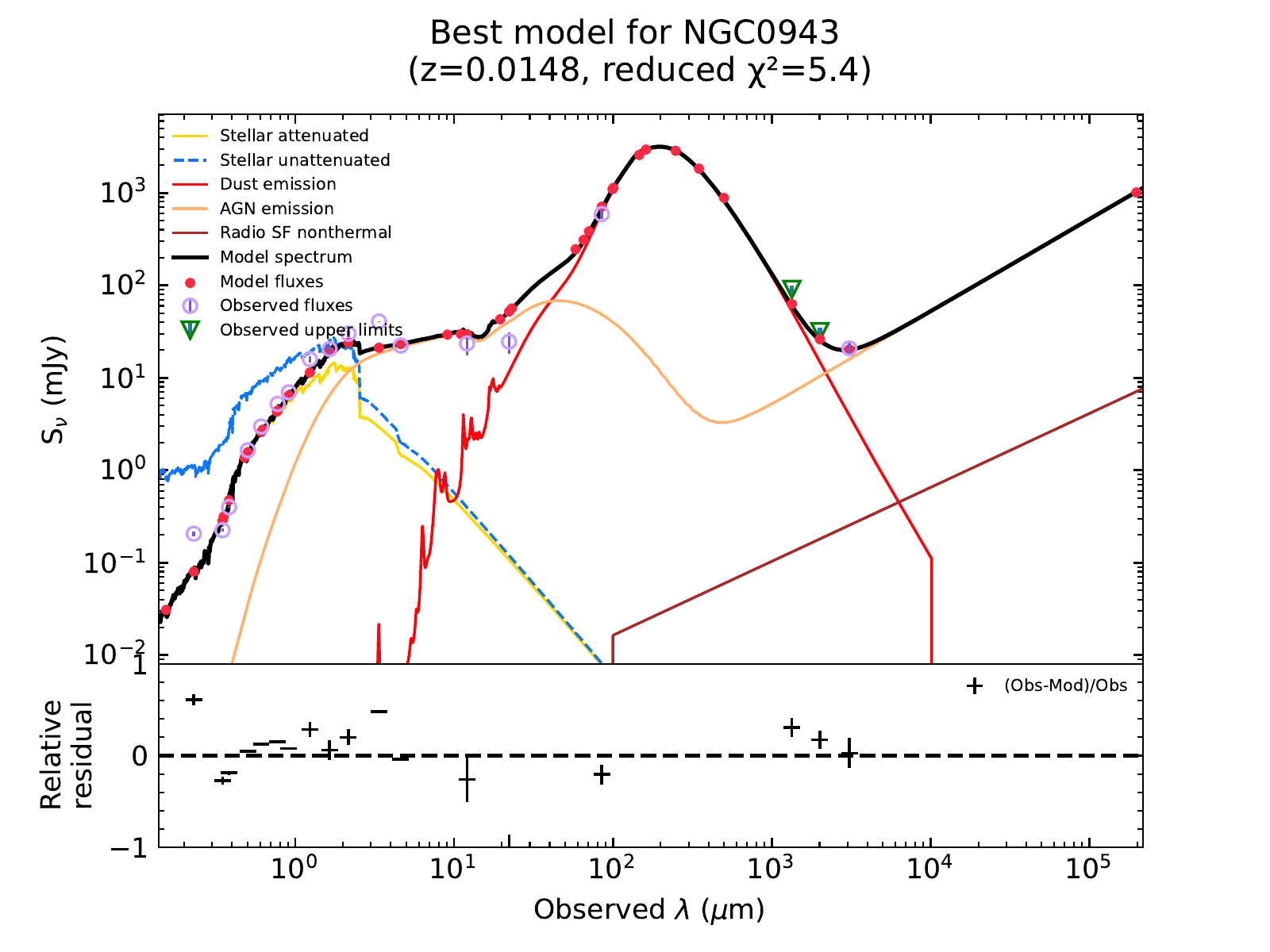}&
\includegraphics[scale=0.29]{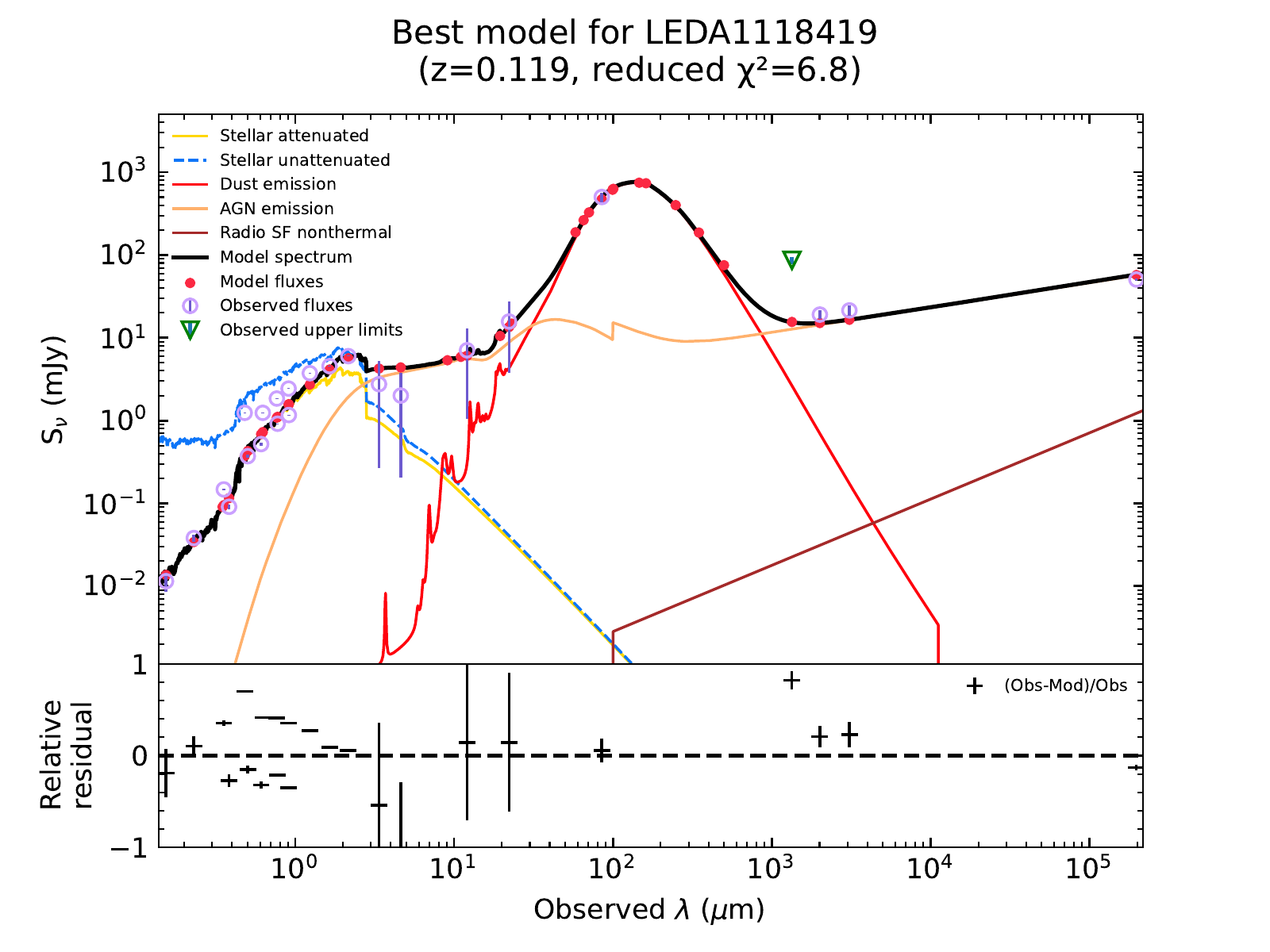}&
\includegraphics[scale=0.29]{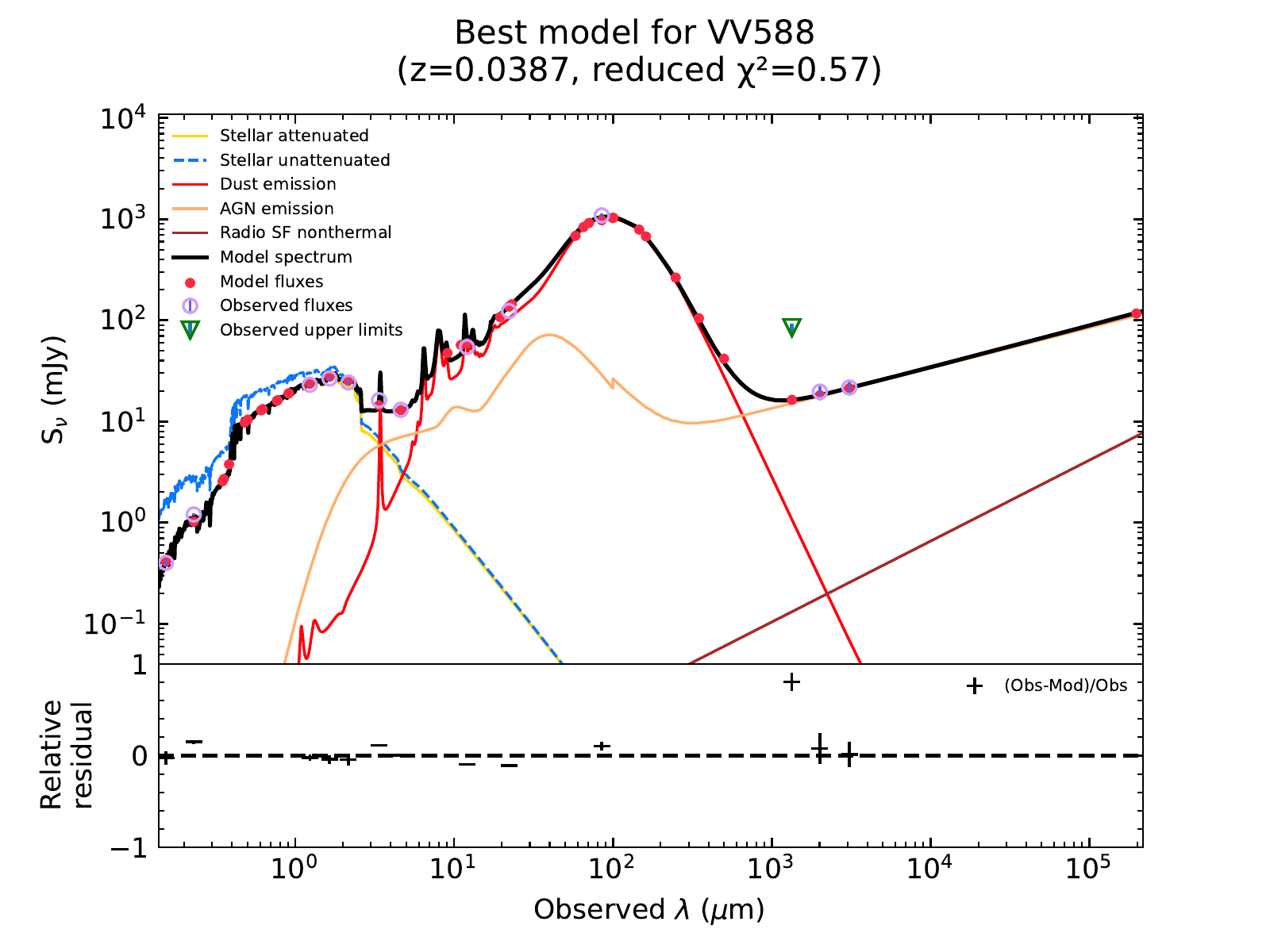}\\
\end{array}$
\end{center}
\caption{ Best-fitting models of radio-loud AGNs in our sample. The observed fluxes and best-fitting model fluxes are shown by the open violet circles and  the filled red circles, respectively. 
The unattenuated stellar emission, the dust emission, and the total AGN emission are shown by the dashed blue lines, the solid red lines, and the solid orange lines, respectively.}
\label{fig:fig5}
\end{figure*}

\begin{figure*}
\begin{center}$
\begin{array}{llll}
\includegraphics[scale=0.29]{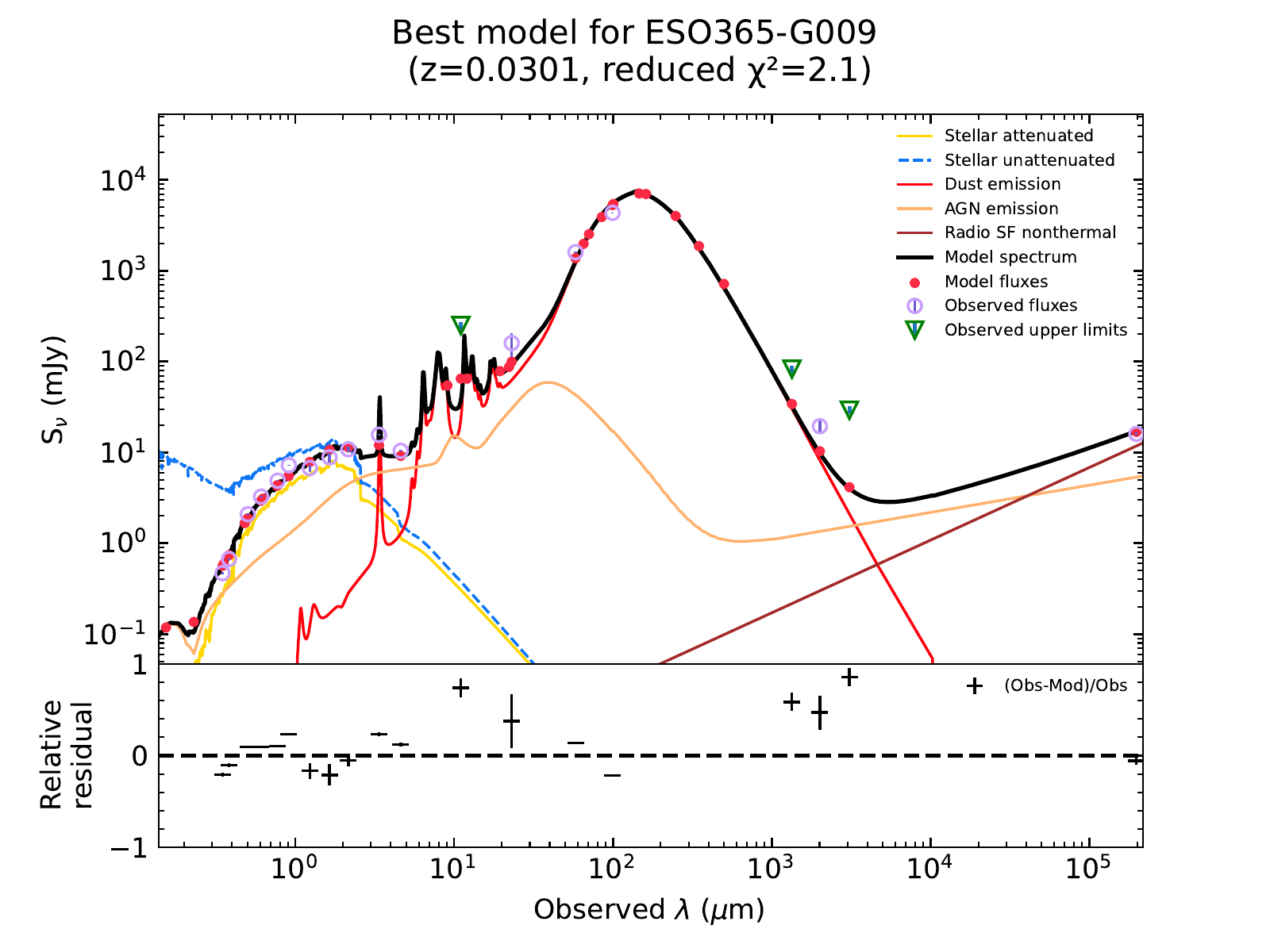}&
\includegraphics[scale=0.29]{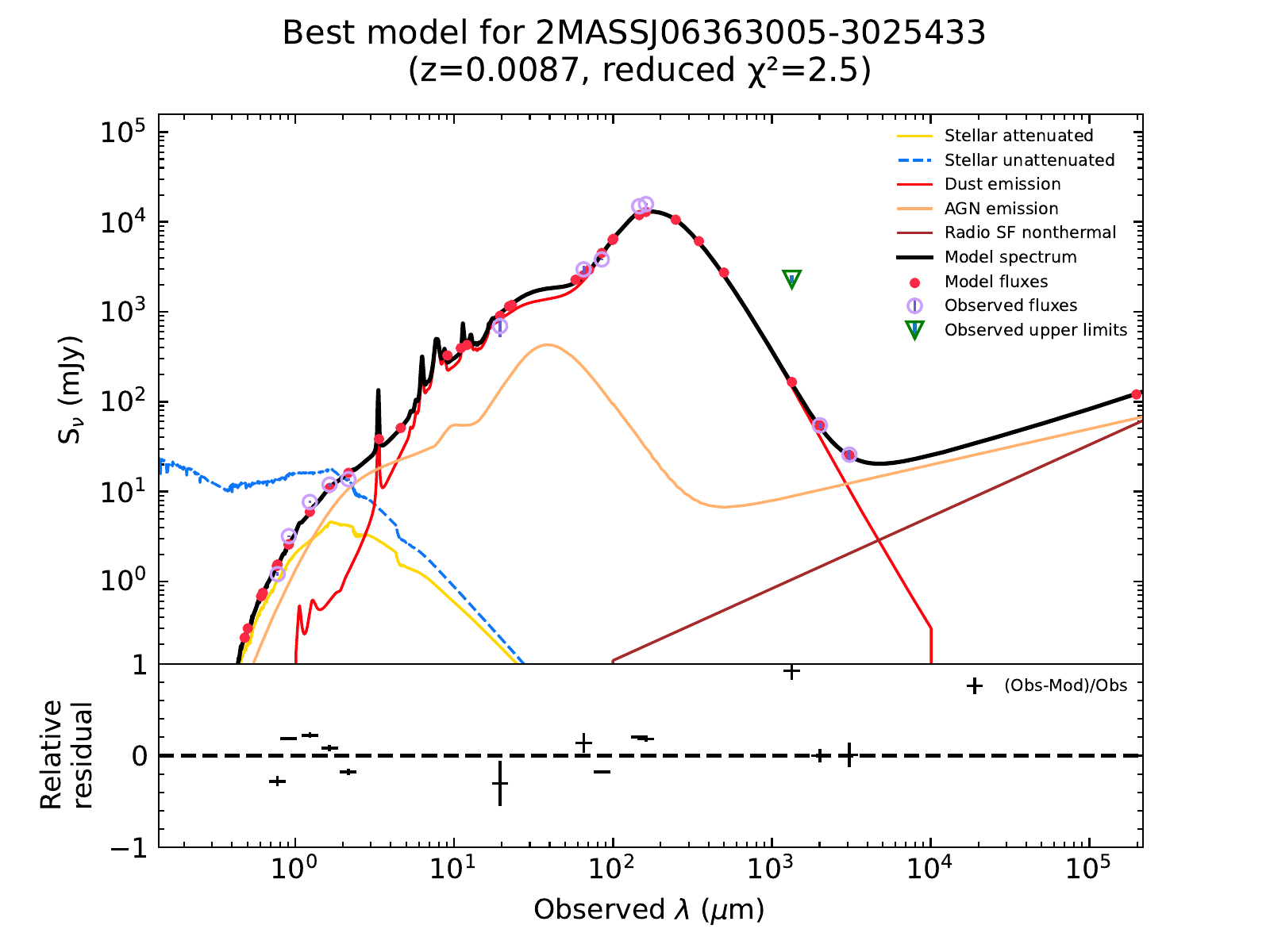}&
\includegraphics[scale=0.29]{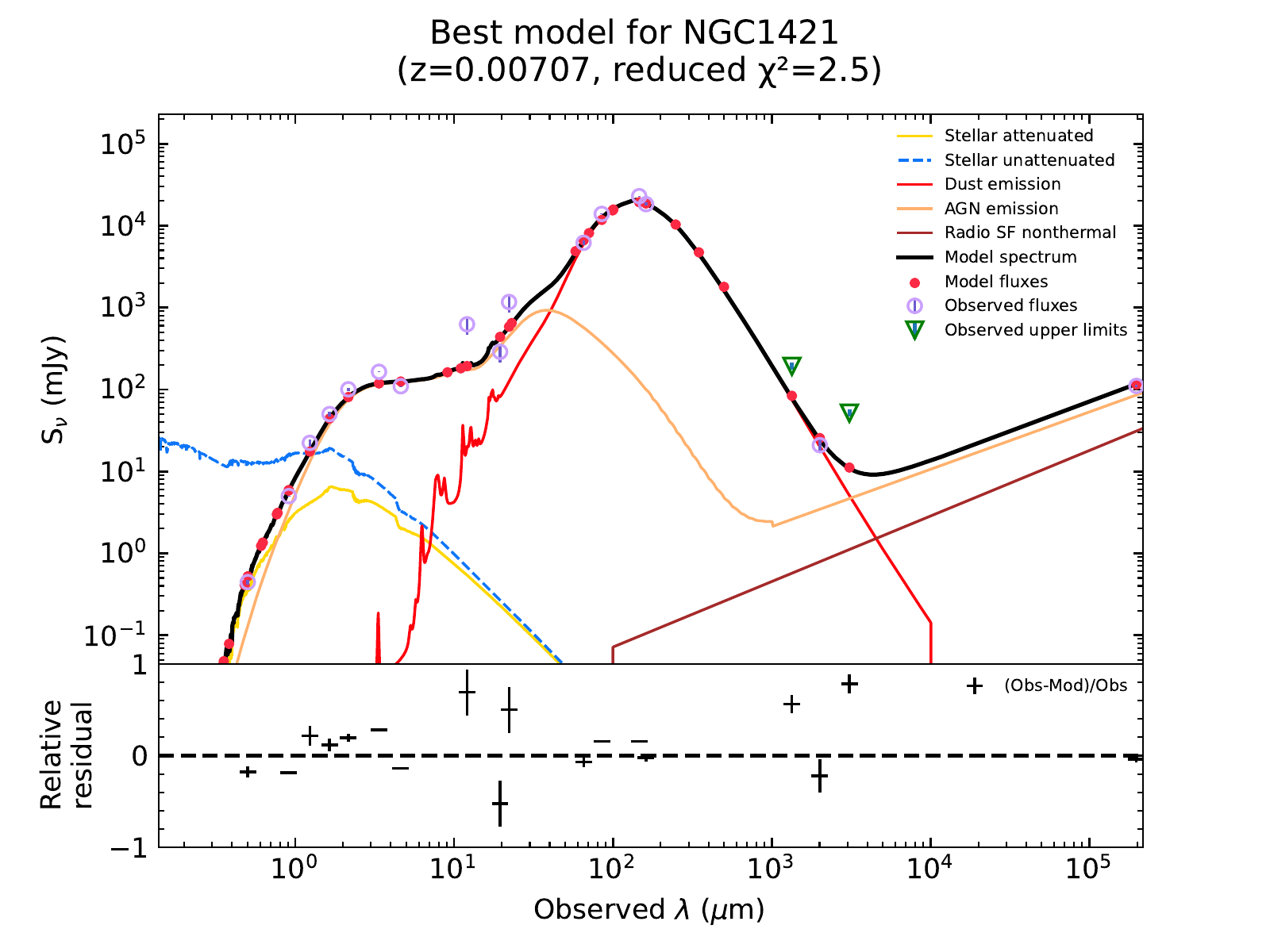}&
\includegraphics[scale=0.29]{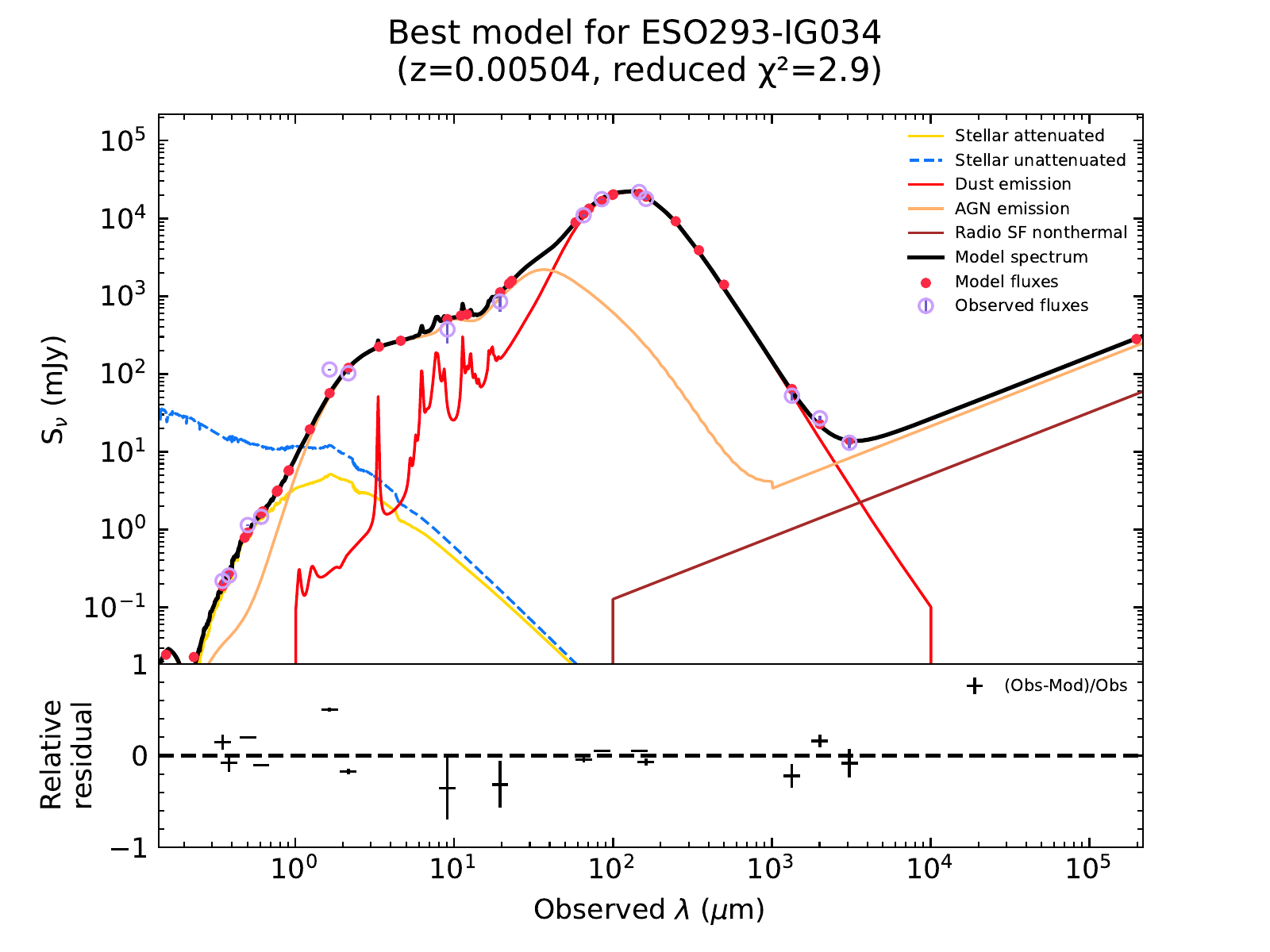}\\
\includegraphics[scale=0.29]{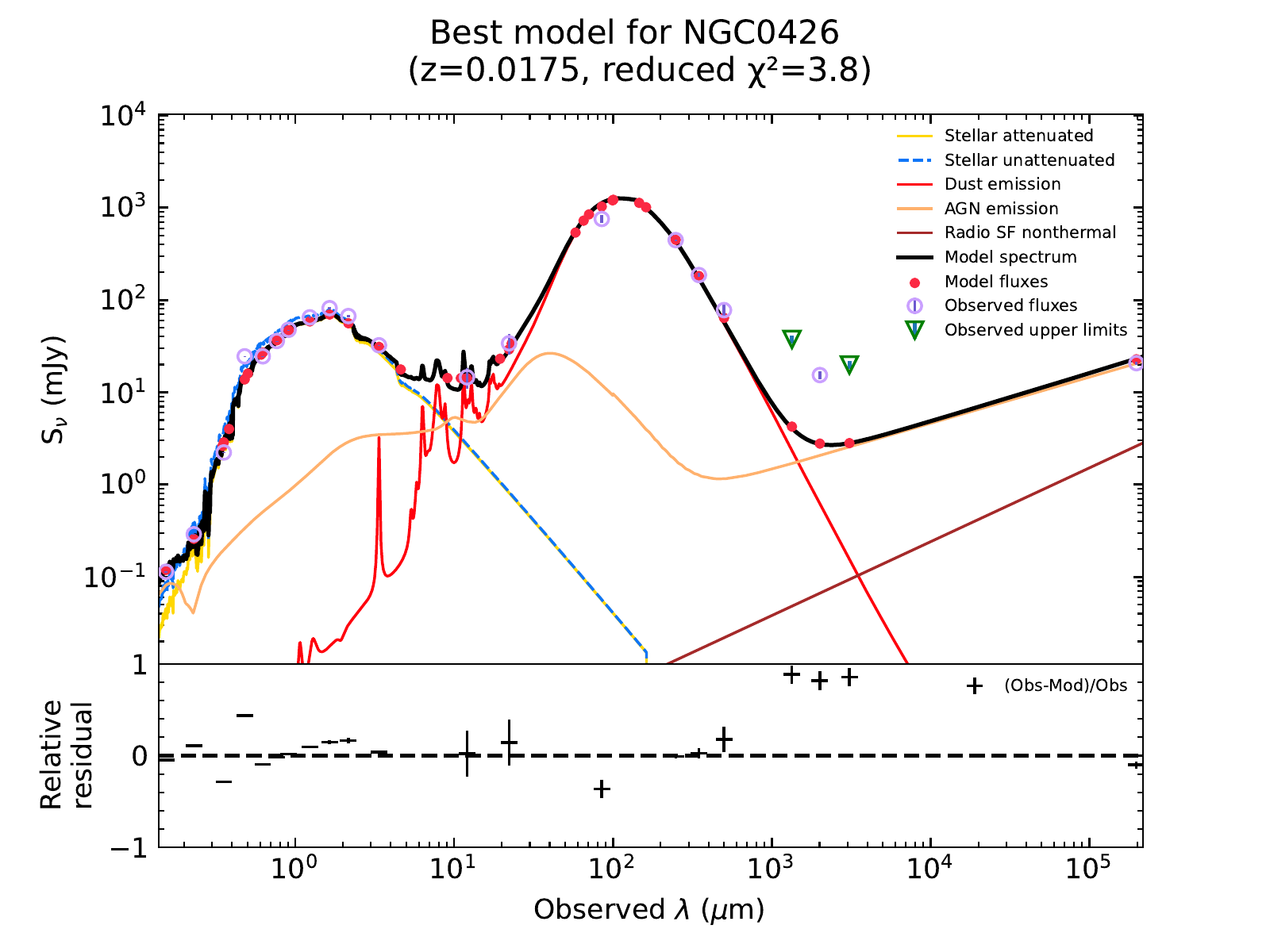}&
\includegraphics[scale=0.29]{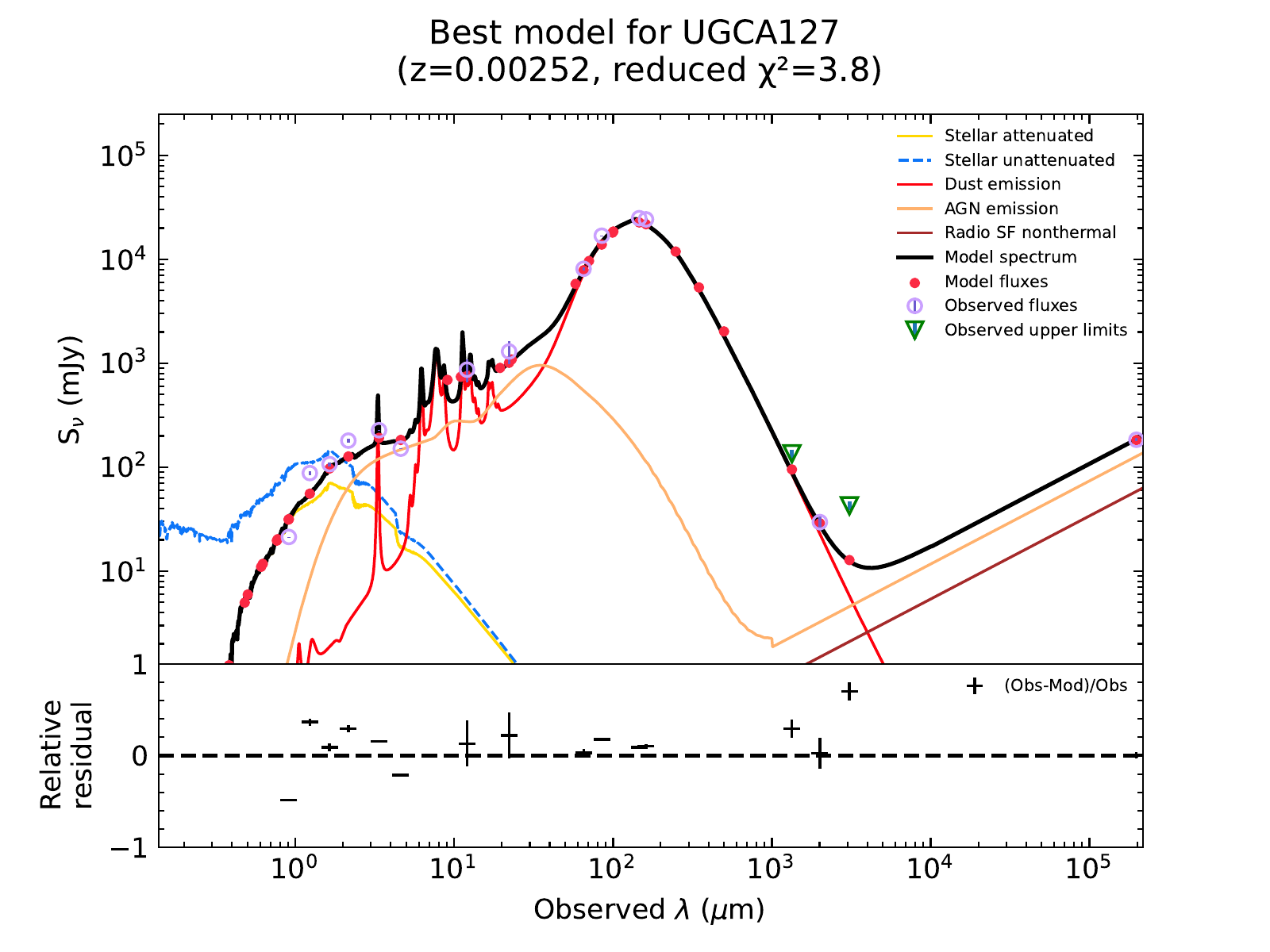}&
\includegraphics[scale=0.29]{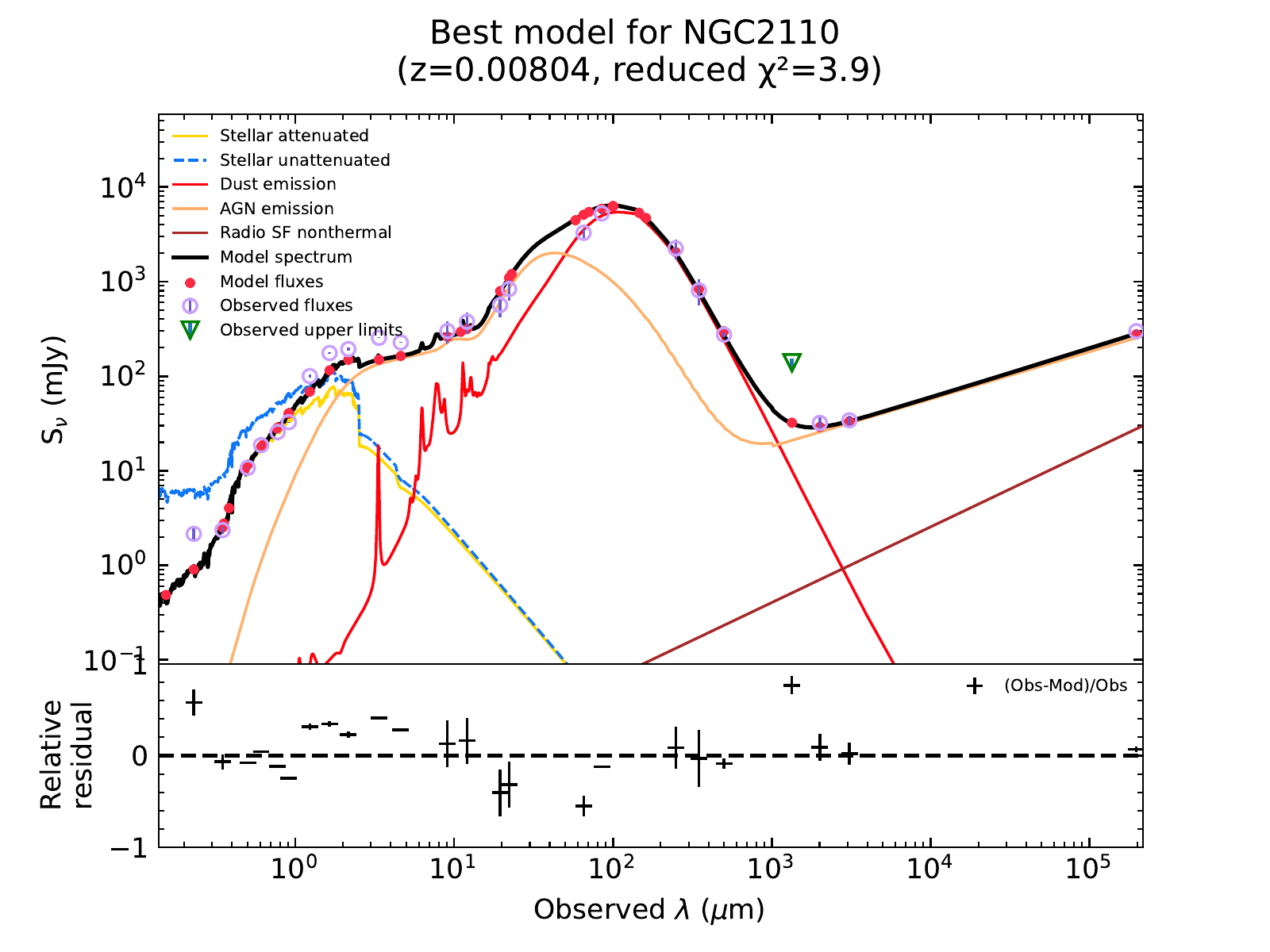}&
\includegraphics[scale=0.29]{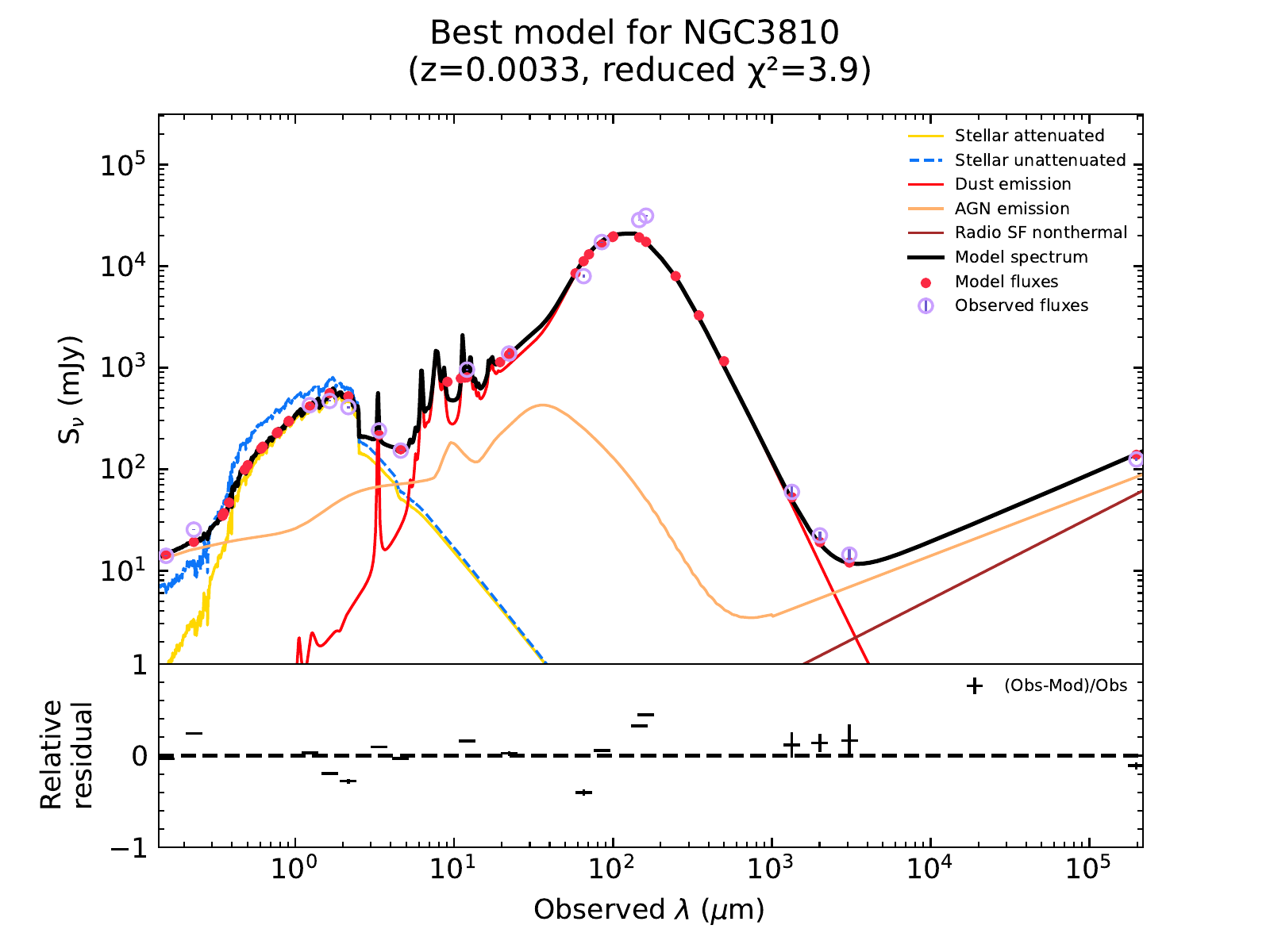}\\
\includegraphics[scale=0.29]{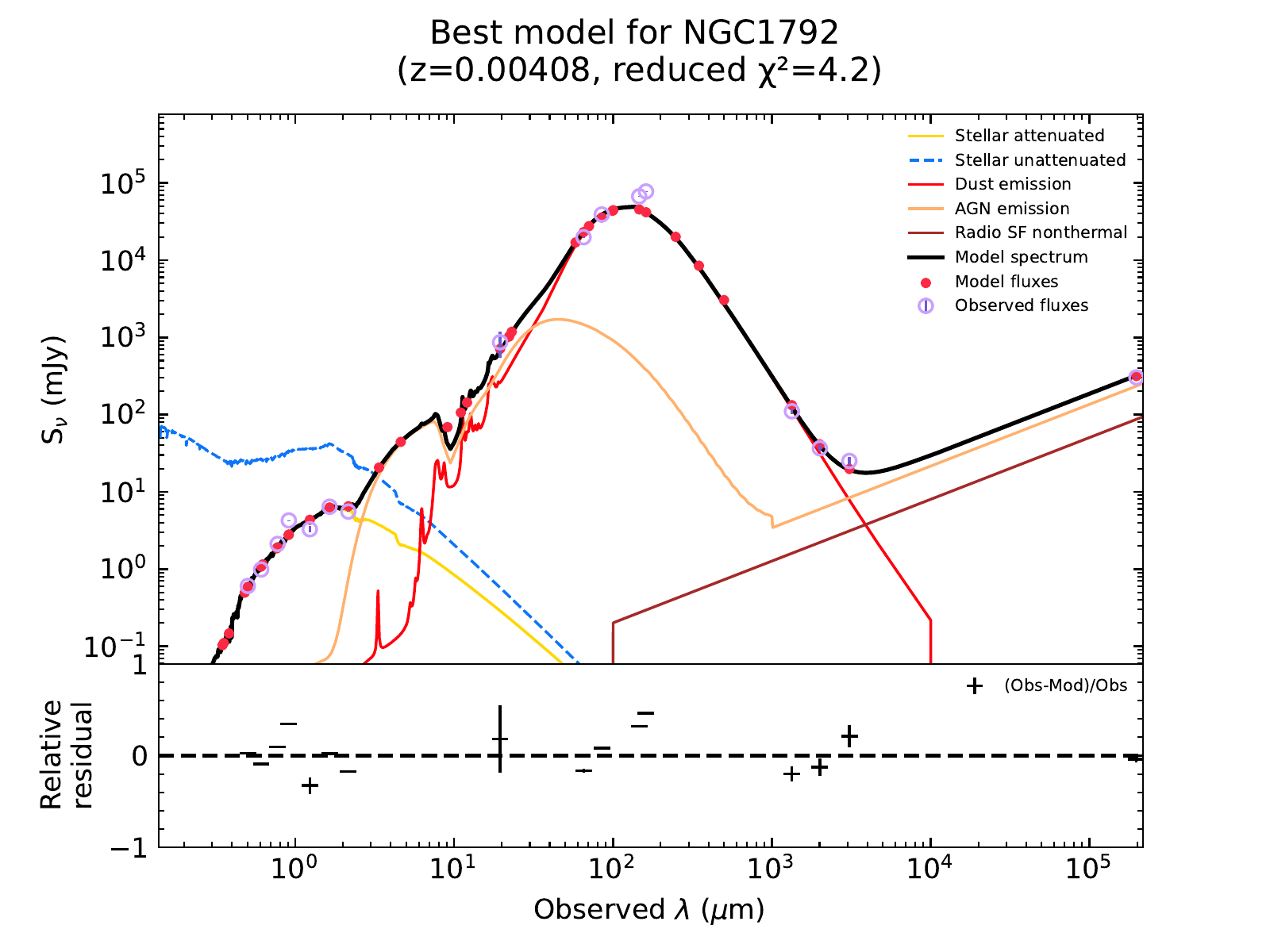}&
\includegraphics[scale=0.29]{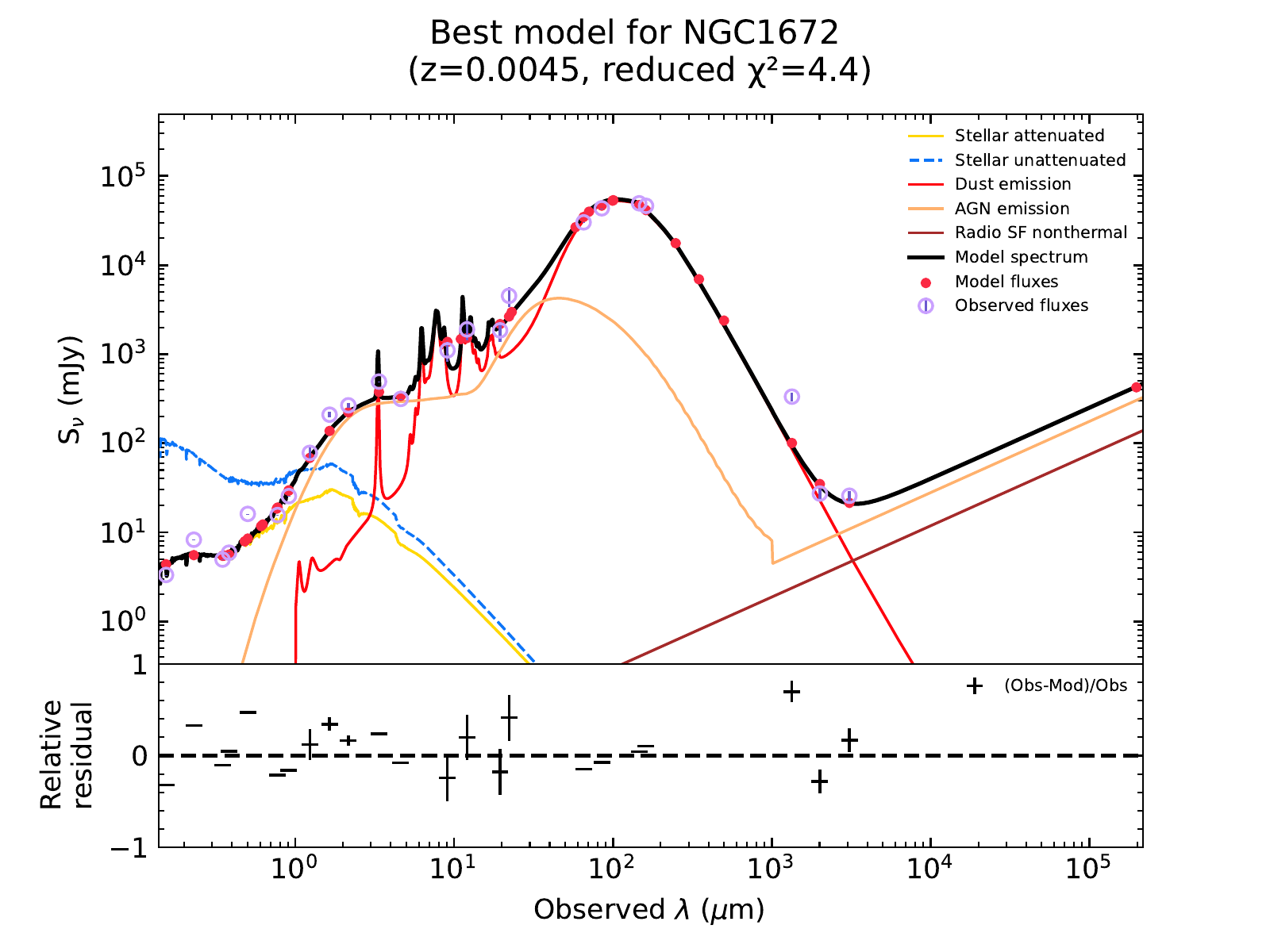}&
\includegraphics[scale=0.29]{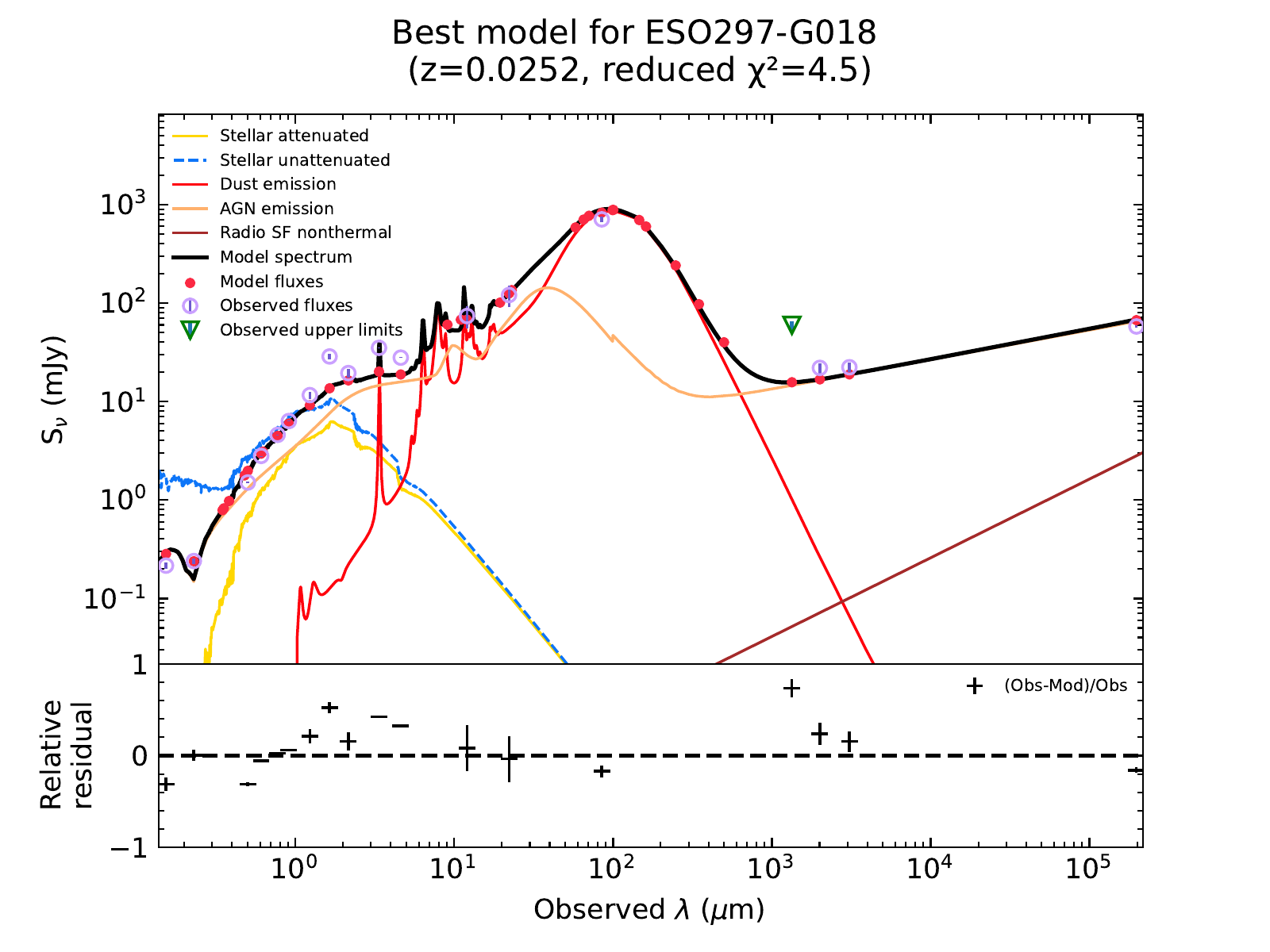}&
\includegraphics[scale=0.29]{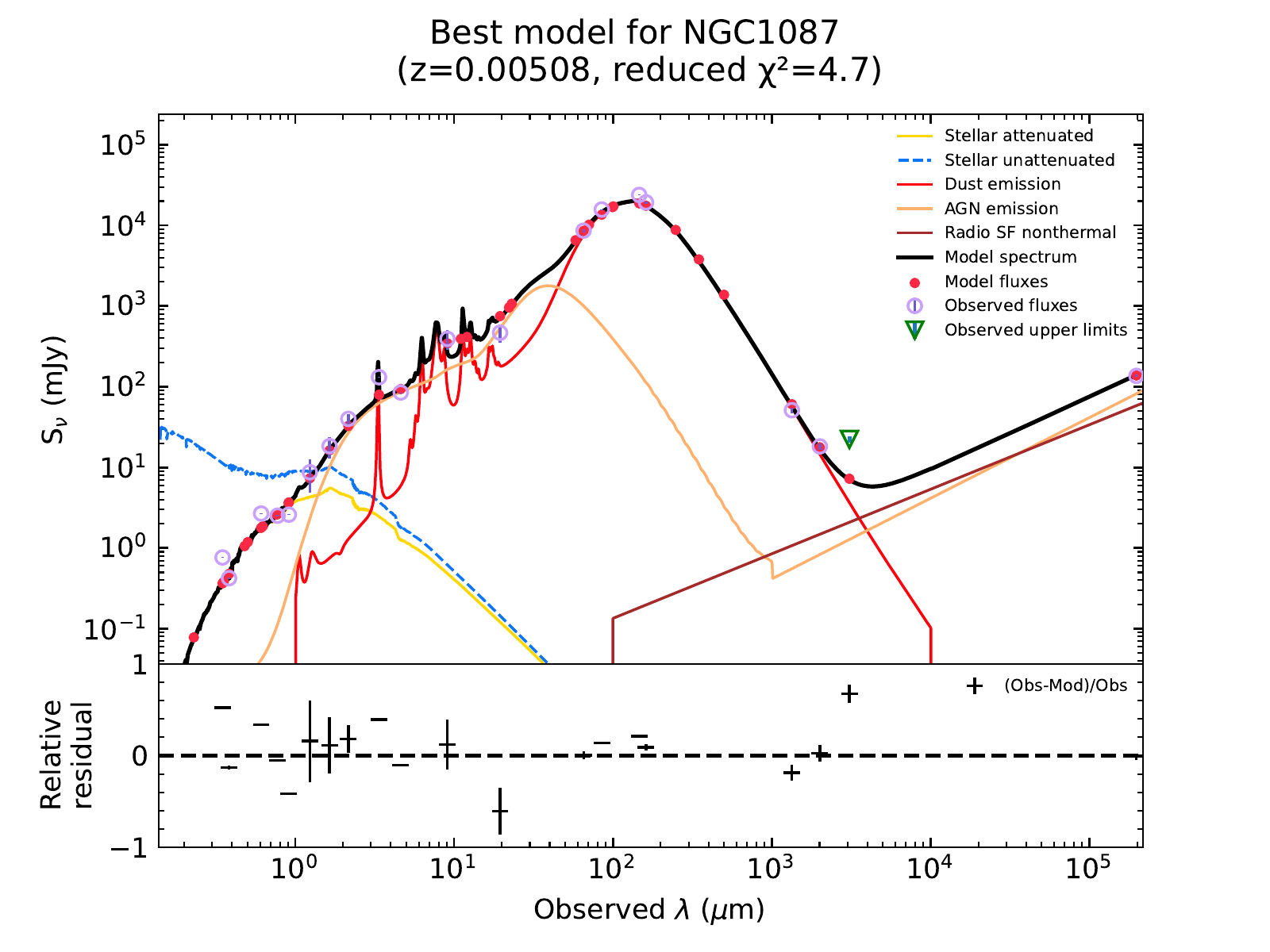}\\
\includegraphics[scale=0.29]{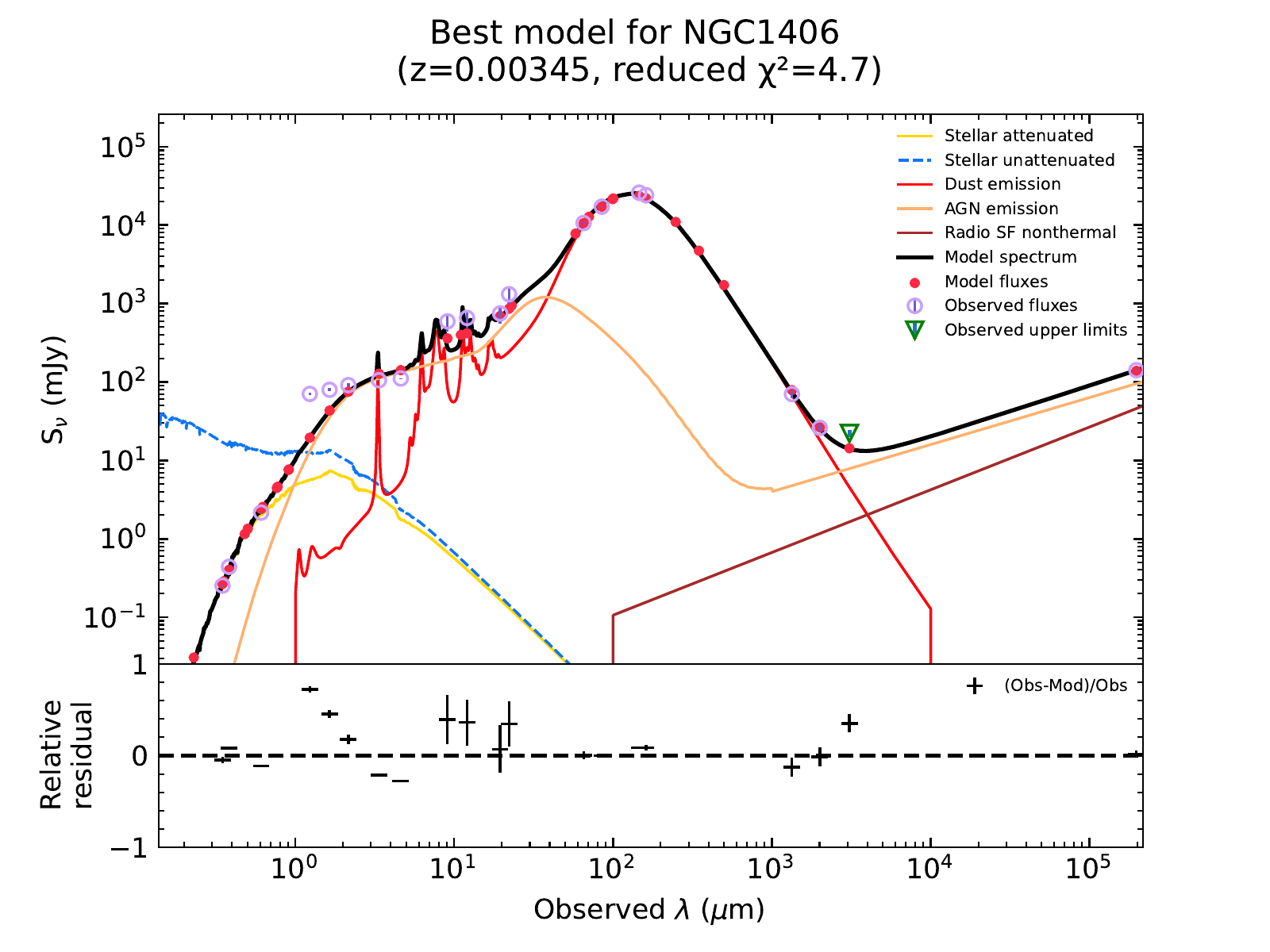}&
\includegraphics[scale=0.29]{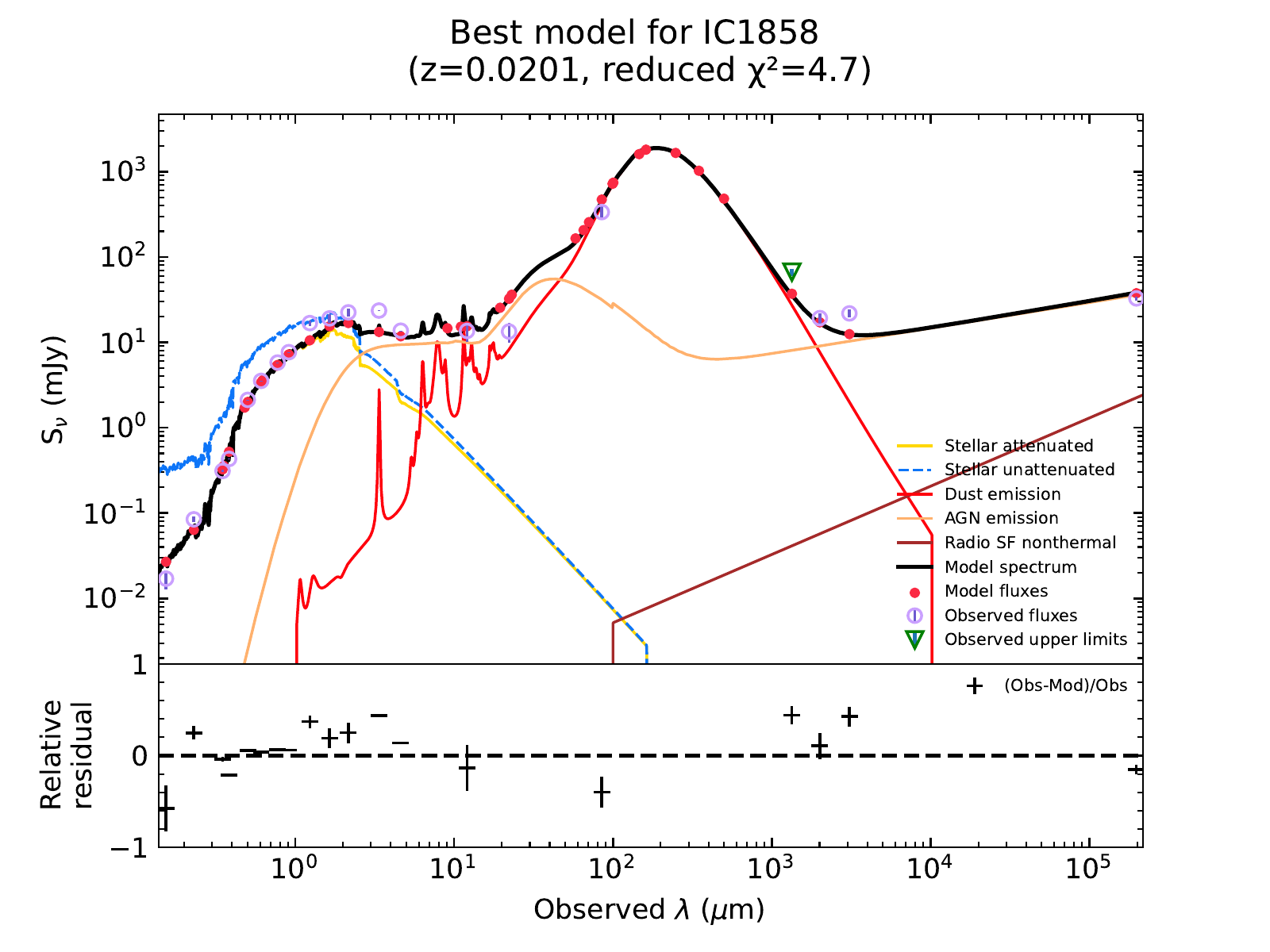}&
\includegraphics[scale=0.29]{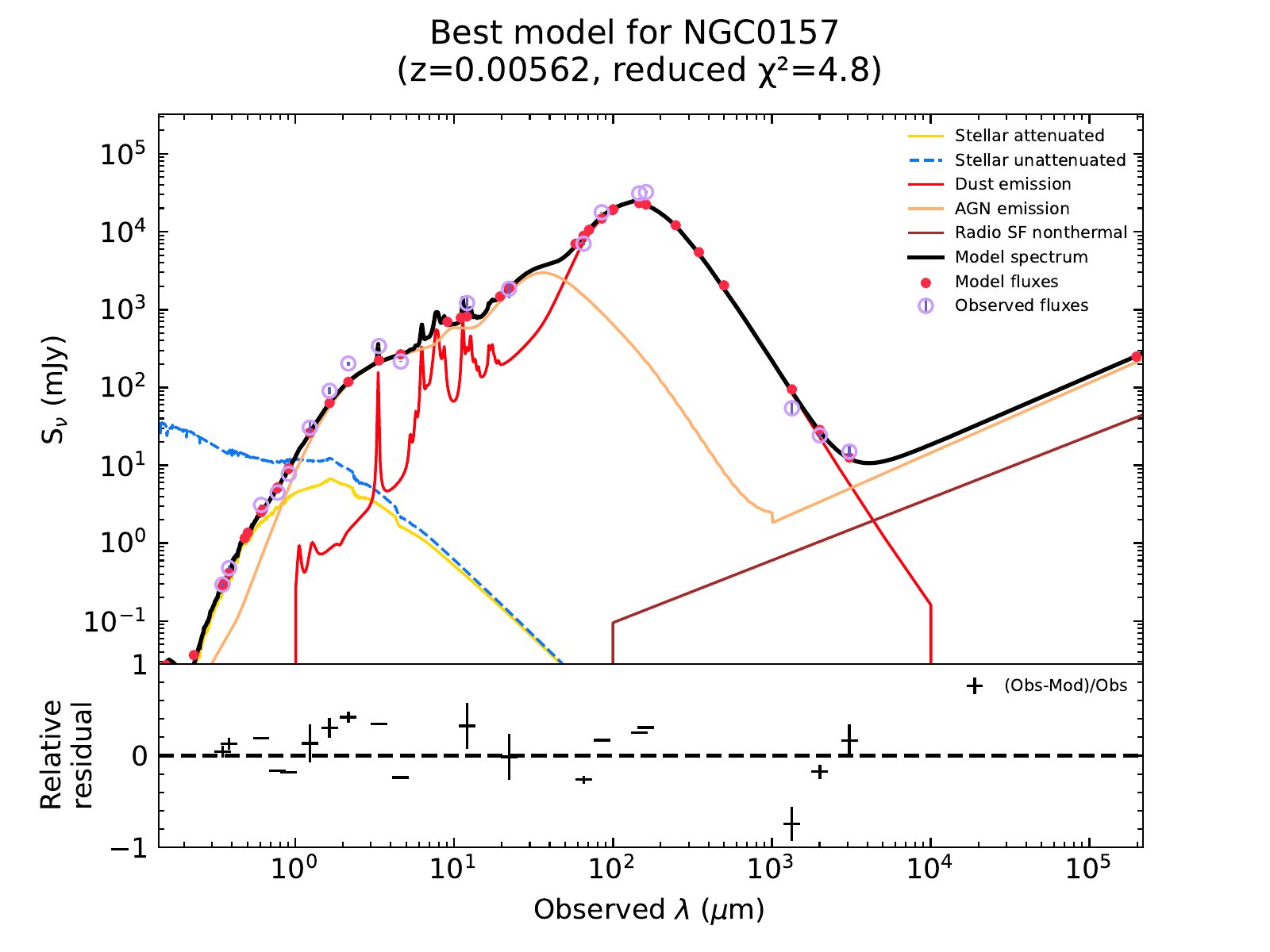}&
\includegraphics[scale=0.29]{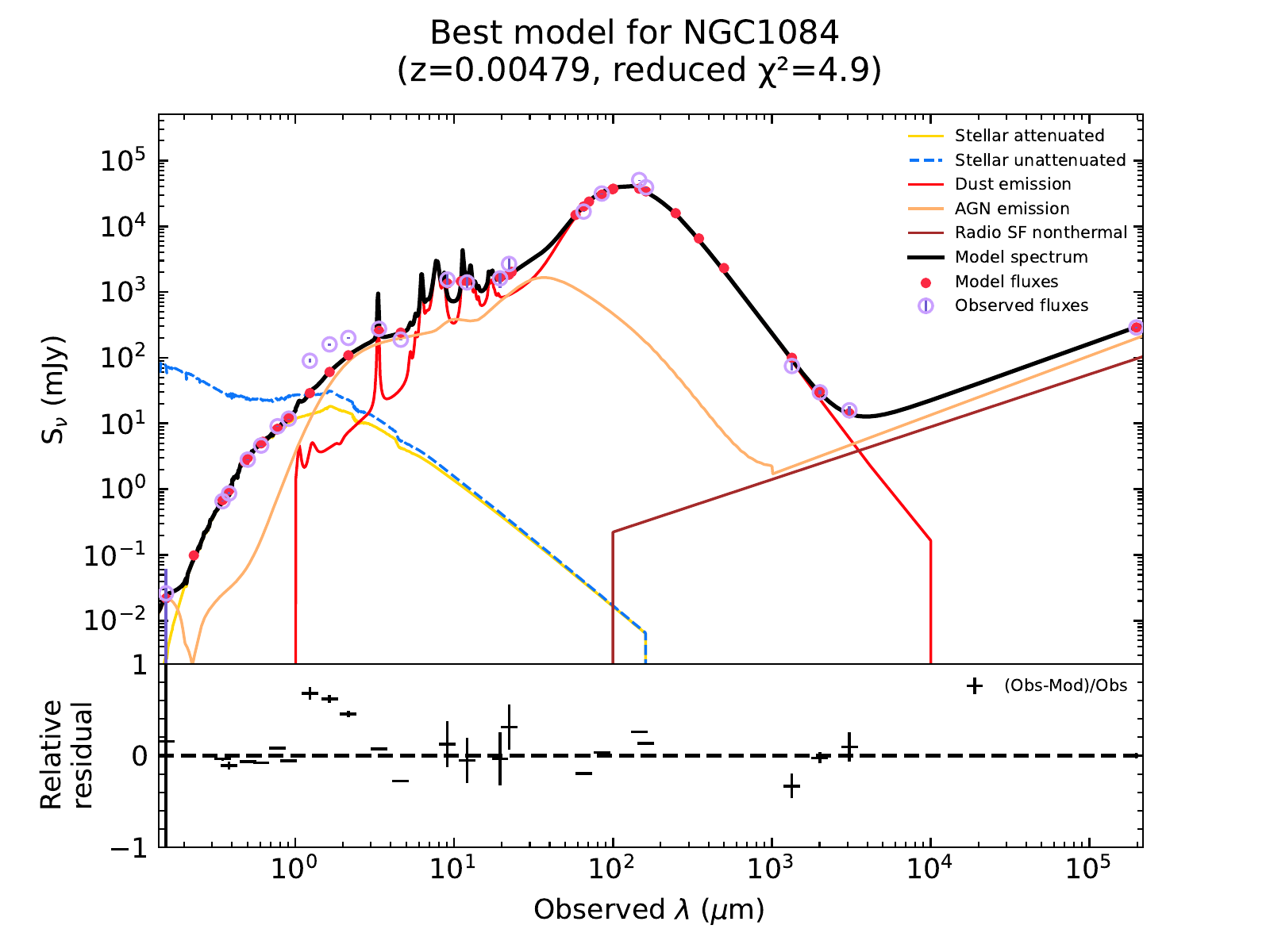}\\
\includegraphics[scale=0.29]{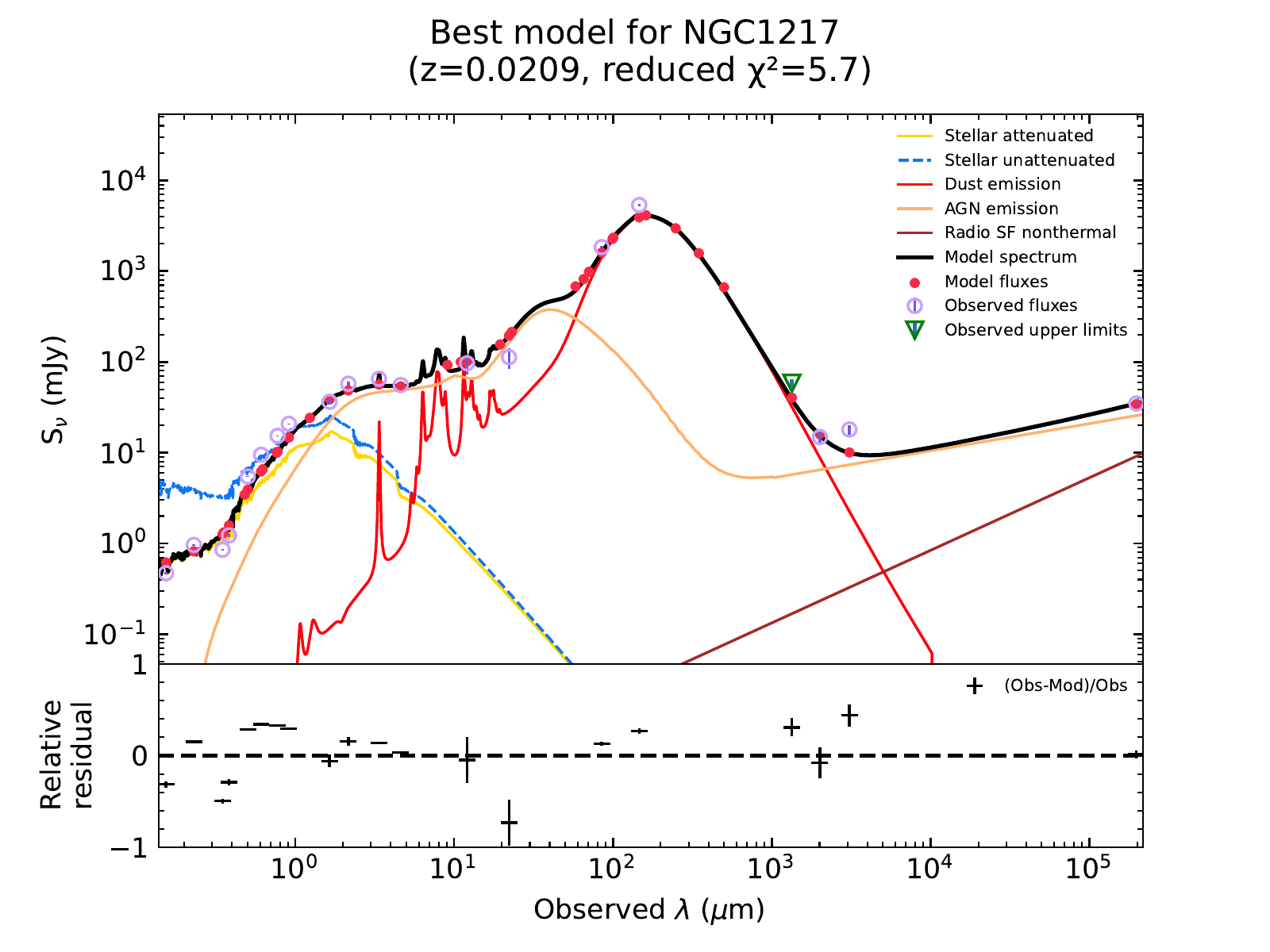}&
\includegraphics[scale=0.29]{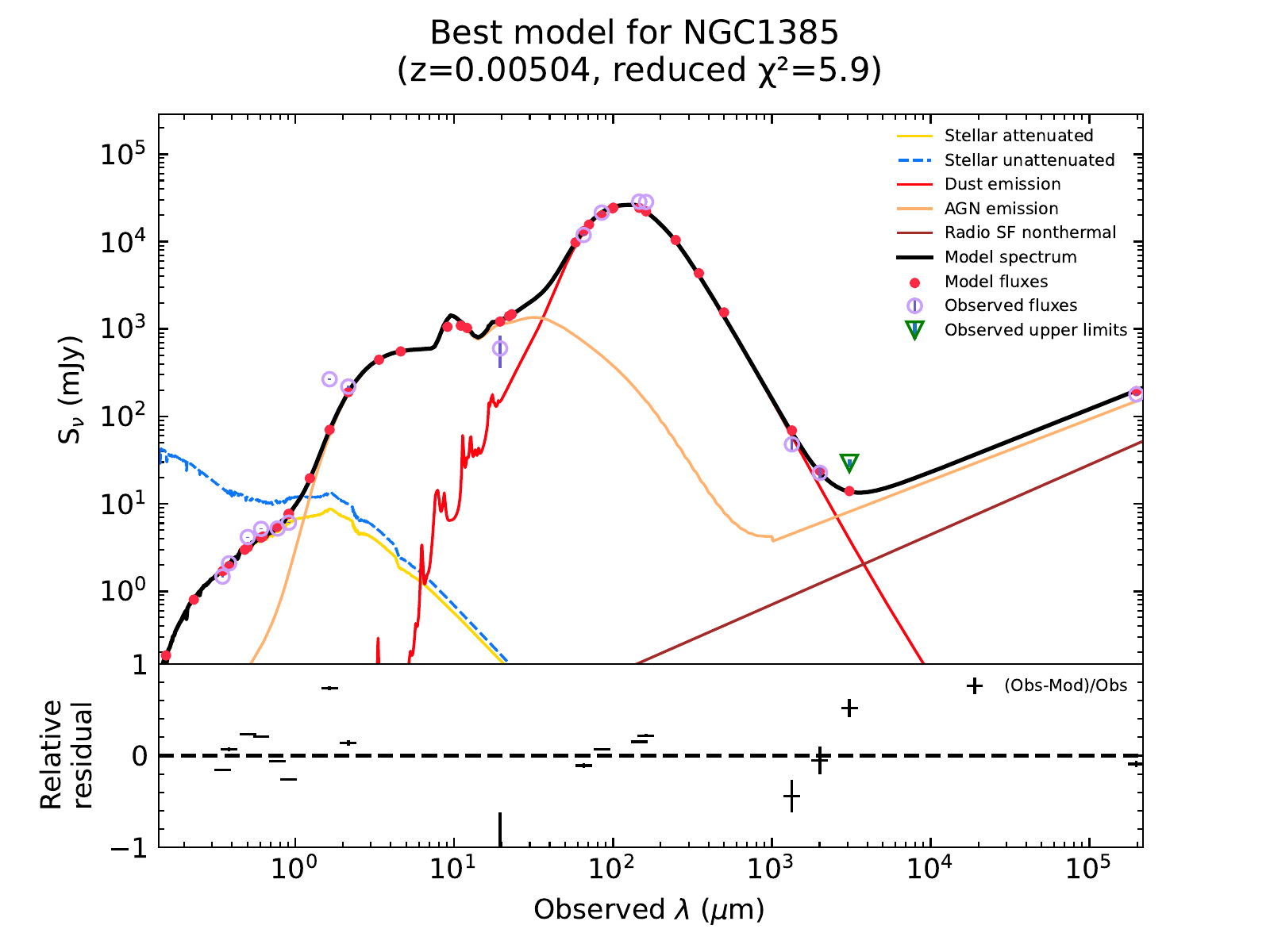}&
\includegraphics[scale=0.29]{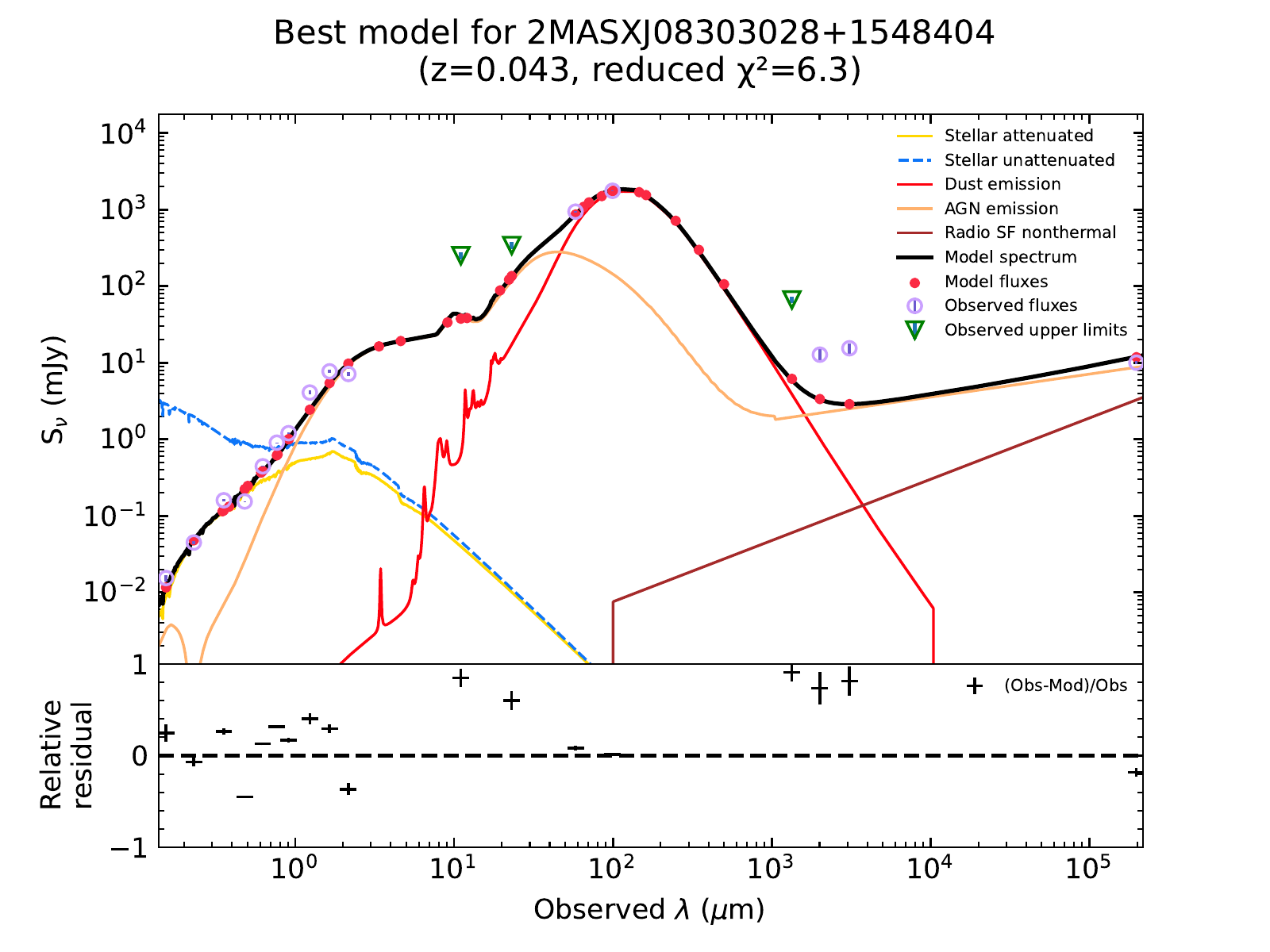}&
\includegraphics[scale=0.29]{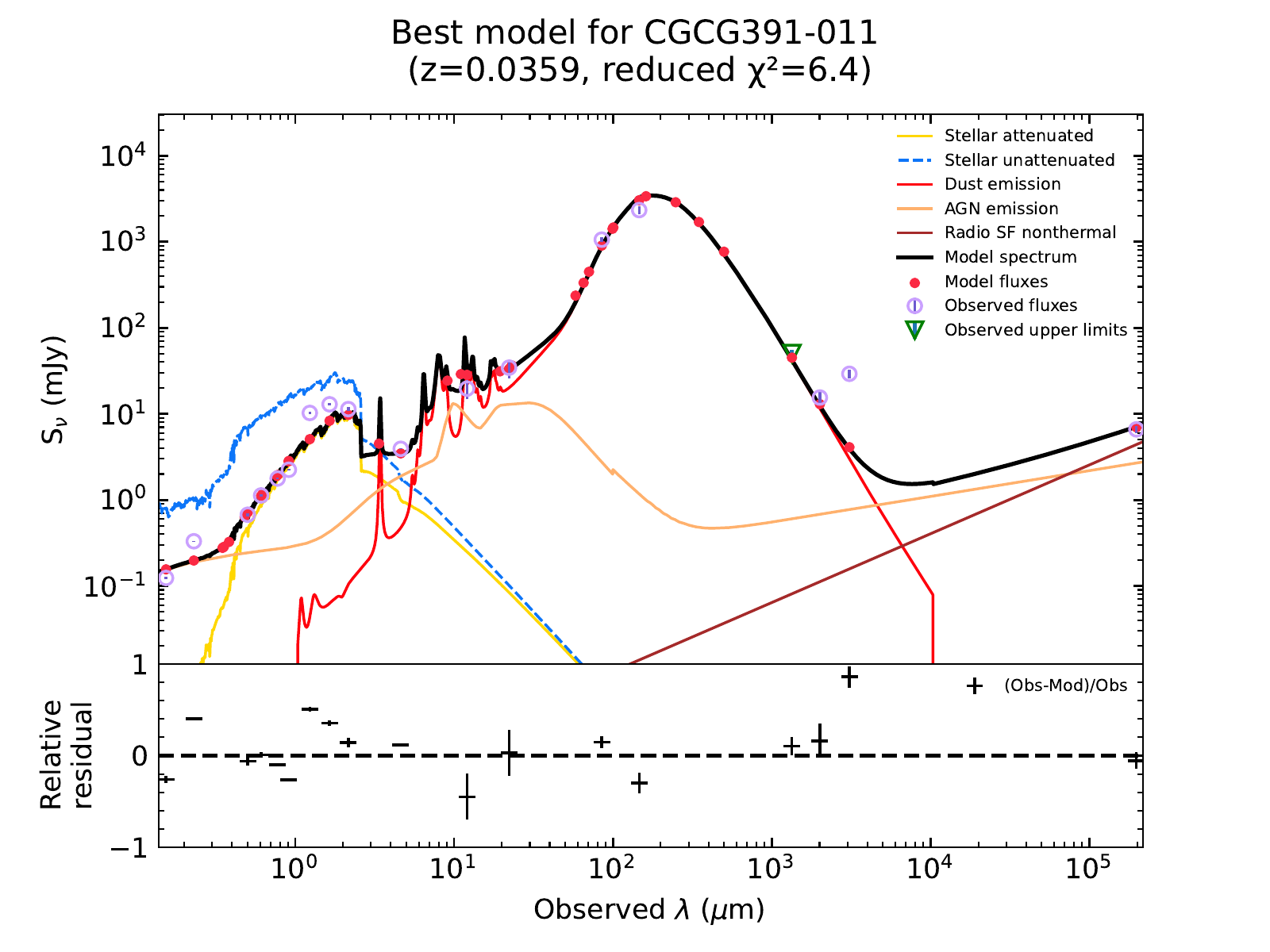}\\
\includegraphics[scale=0.29]{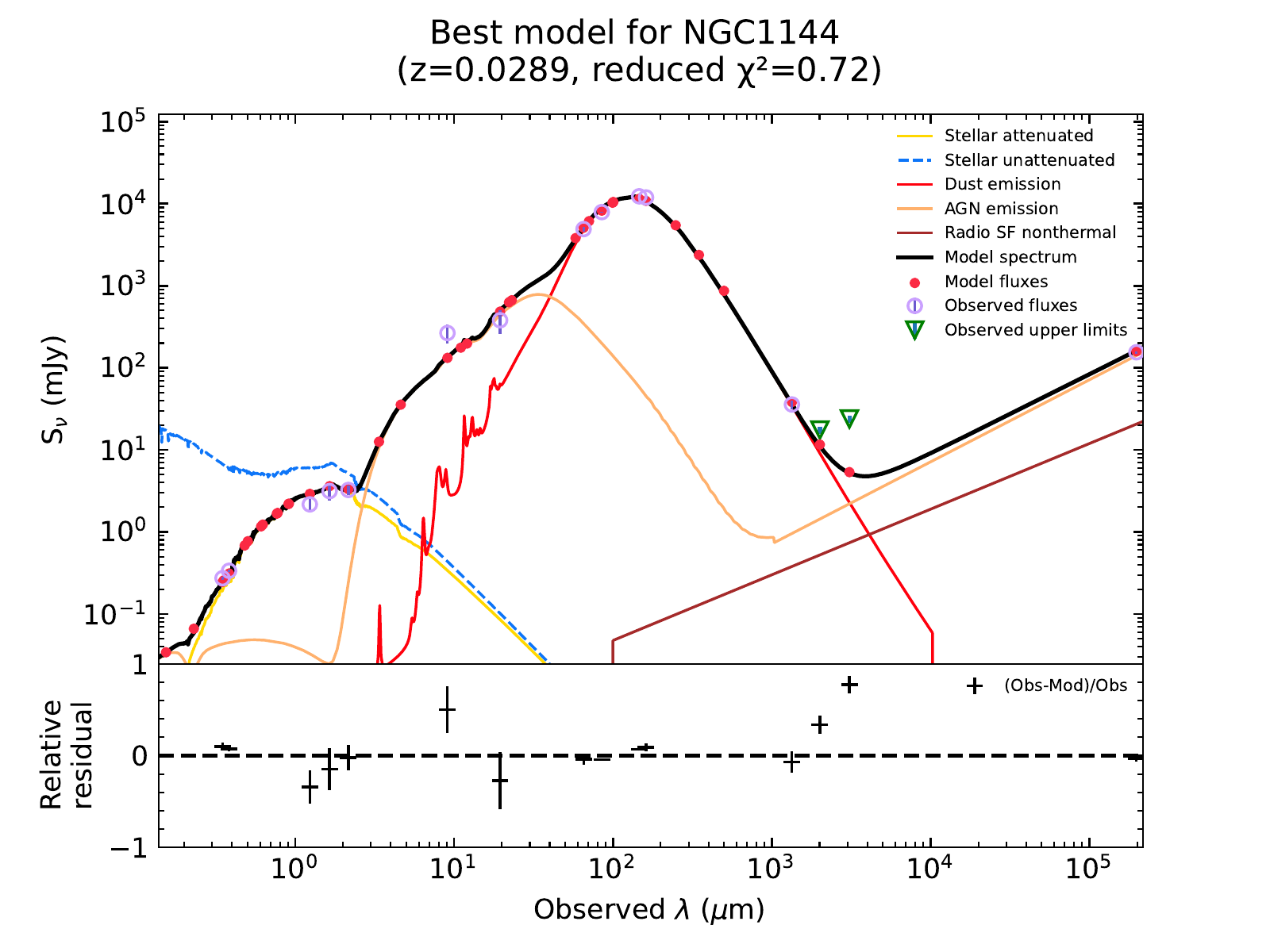} & & & \\
\end{array}$
\end{center}
\caption{ Best-fitting models of radio-quiet AGNs in our sample. The symbol styles for  the flux values and the color codes for the different model components are the same as in Figure \ref{fig:fig5}. }
\label{fig:fig6}
\end{figure*}

\begin{figure*}
\begin{center}$
\begin{array}{llll}
\includegraphics[scale=0.29]{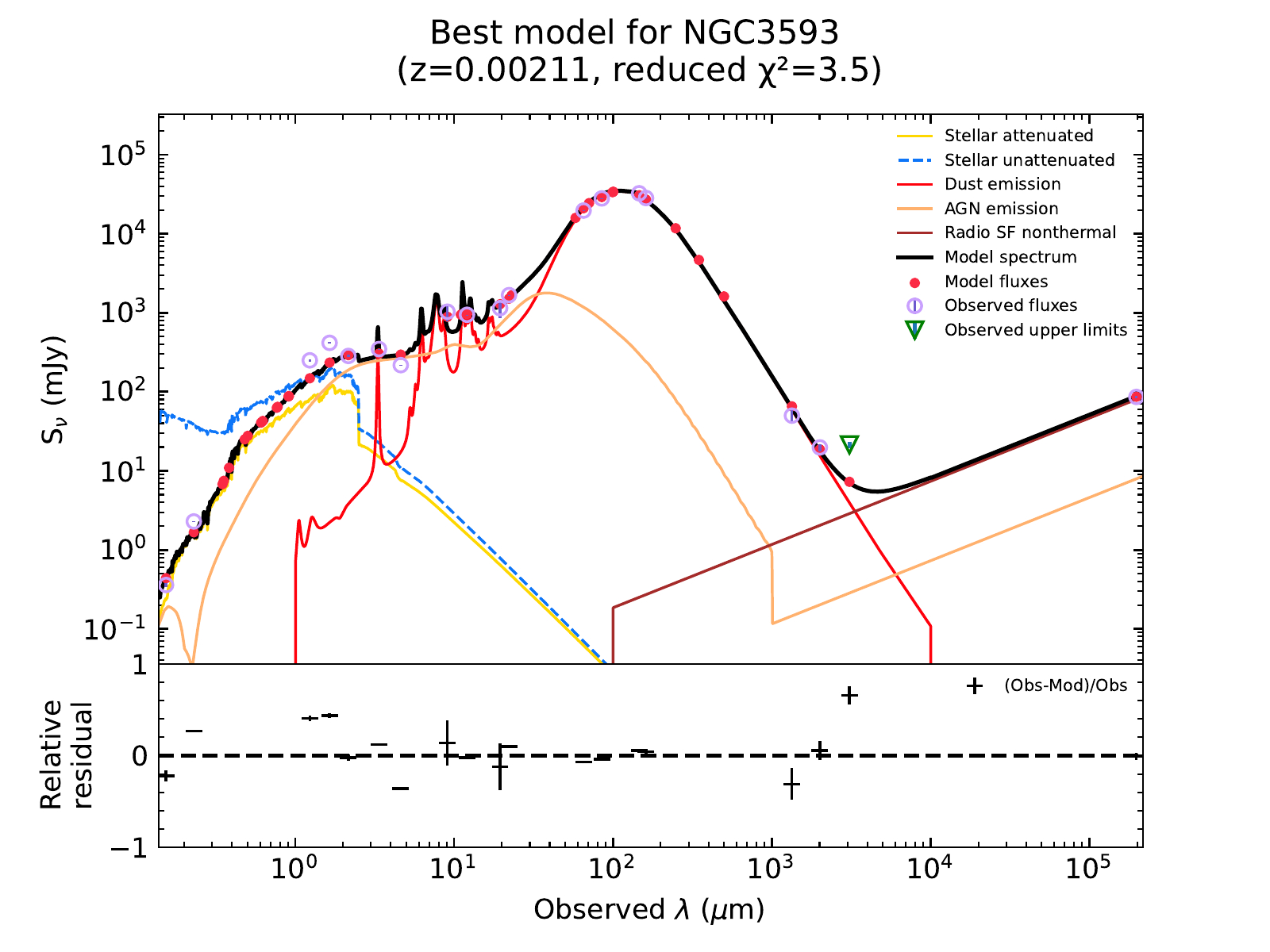}&
\includegraphics[scale=0.29]{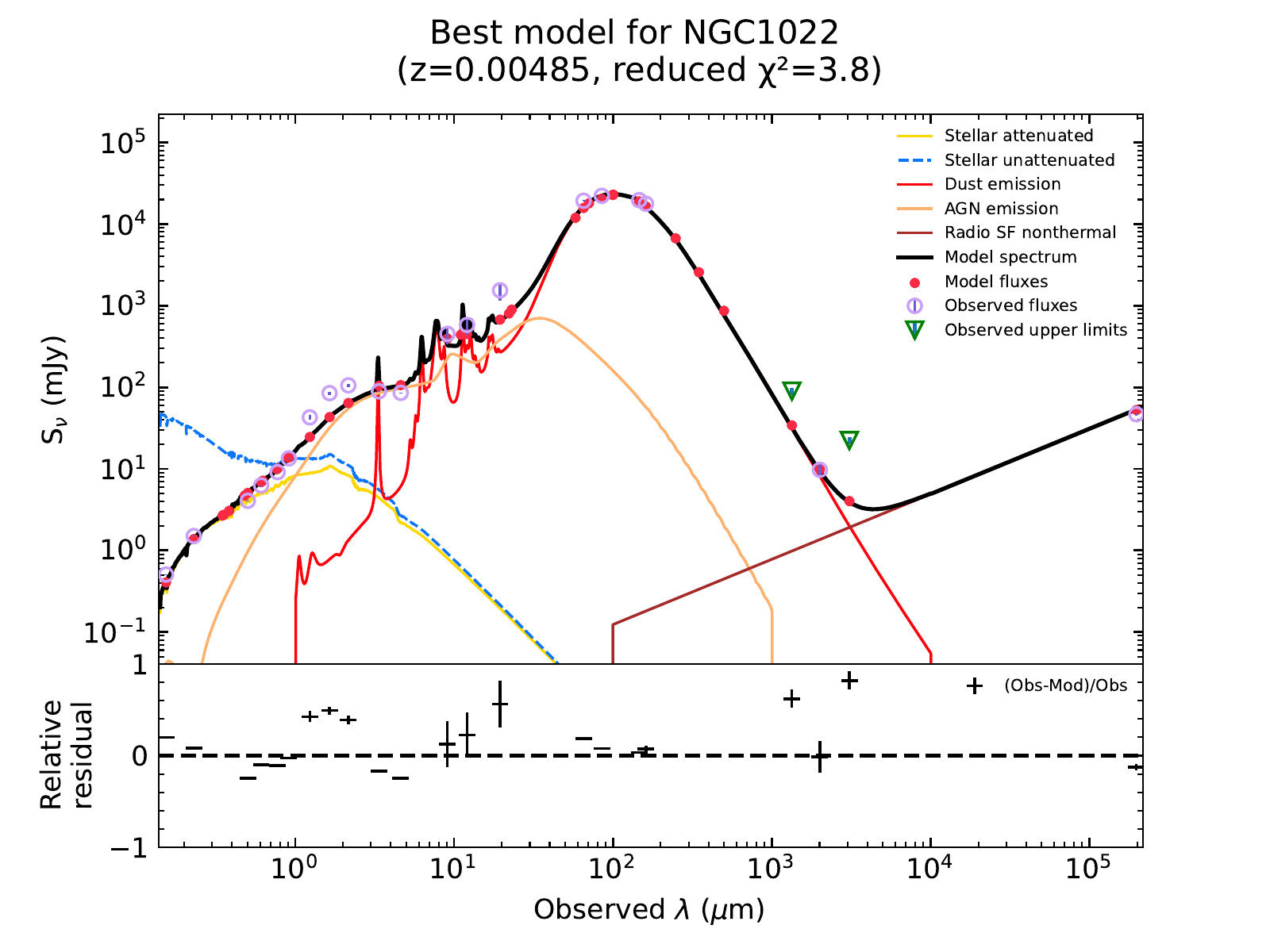}&
\includegraphics[scale=0.29]{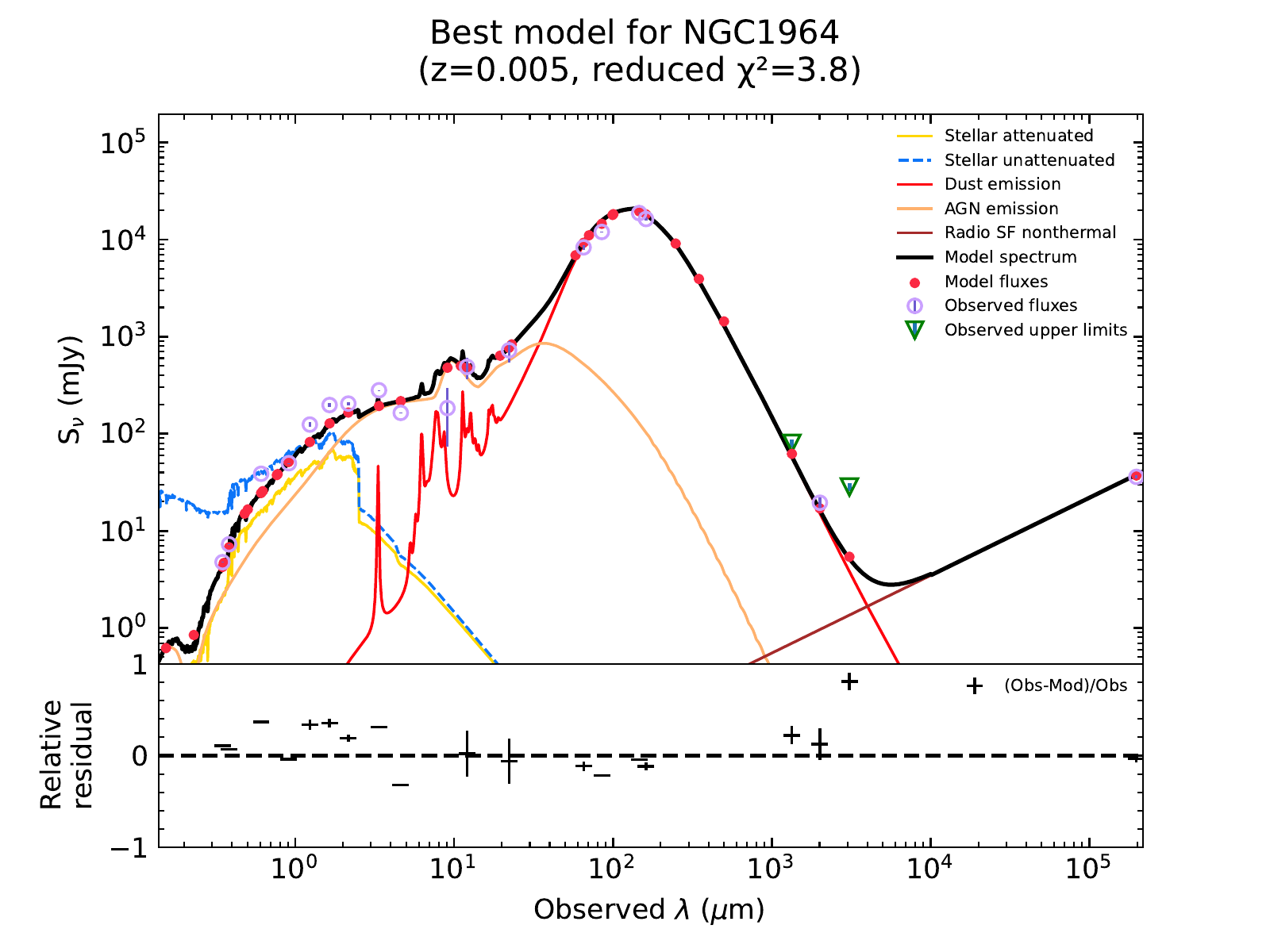}&
\includegraphics[scale=0.29]{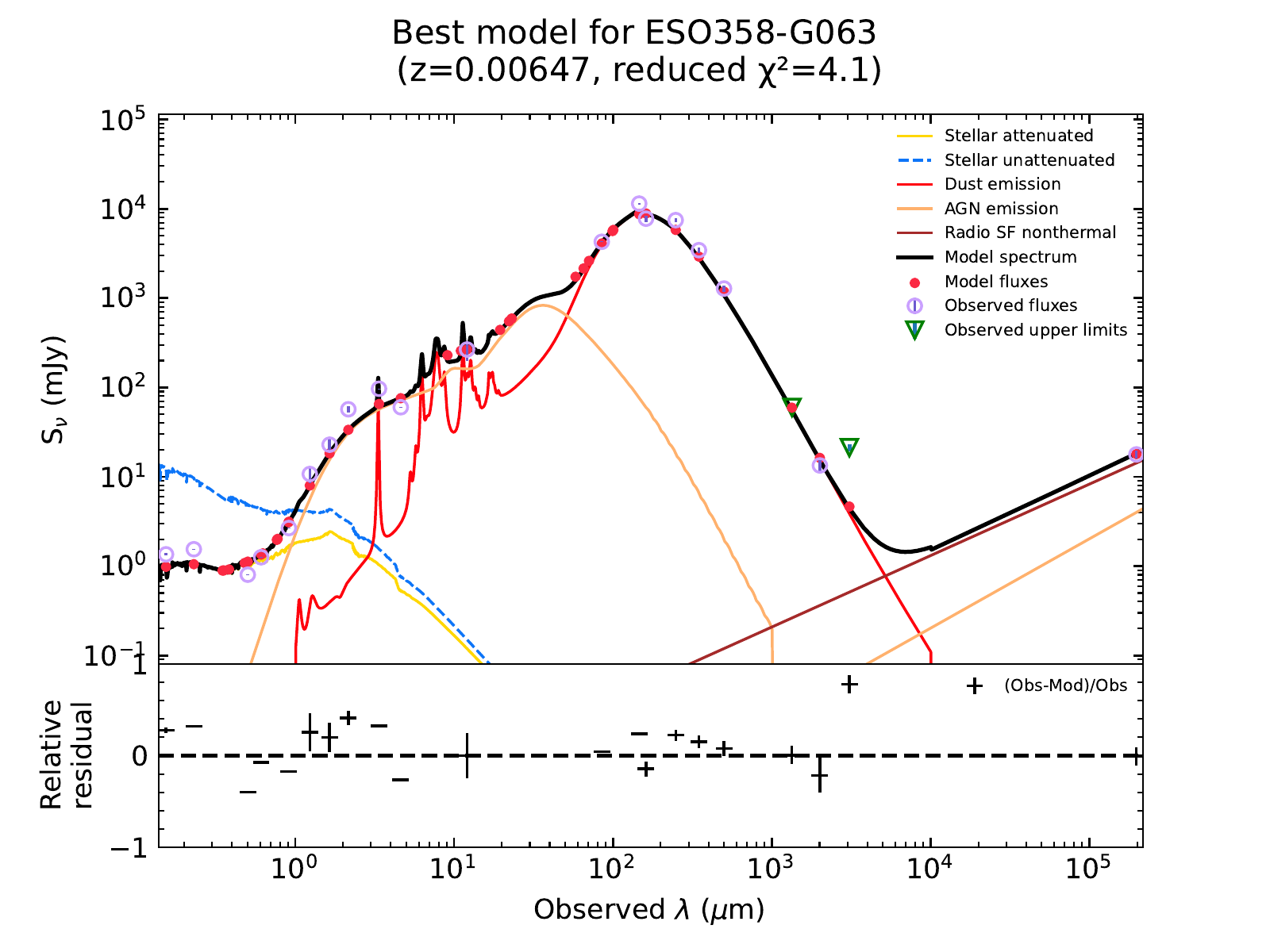}\\
\includegraphics[scale=0.29]{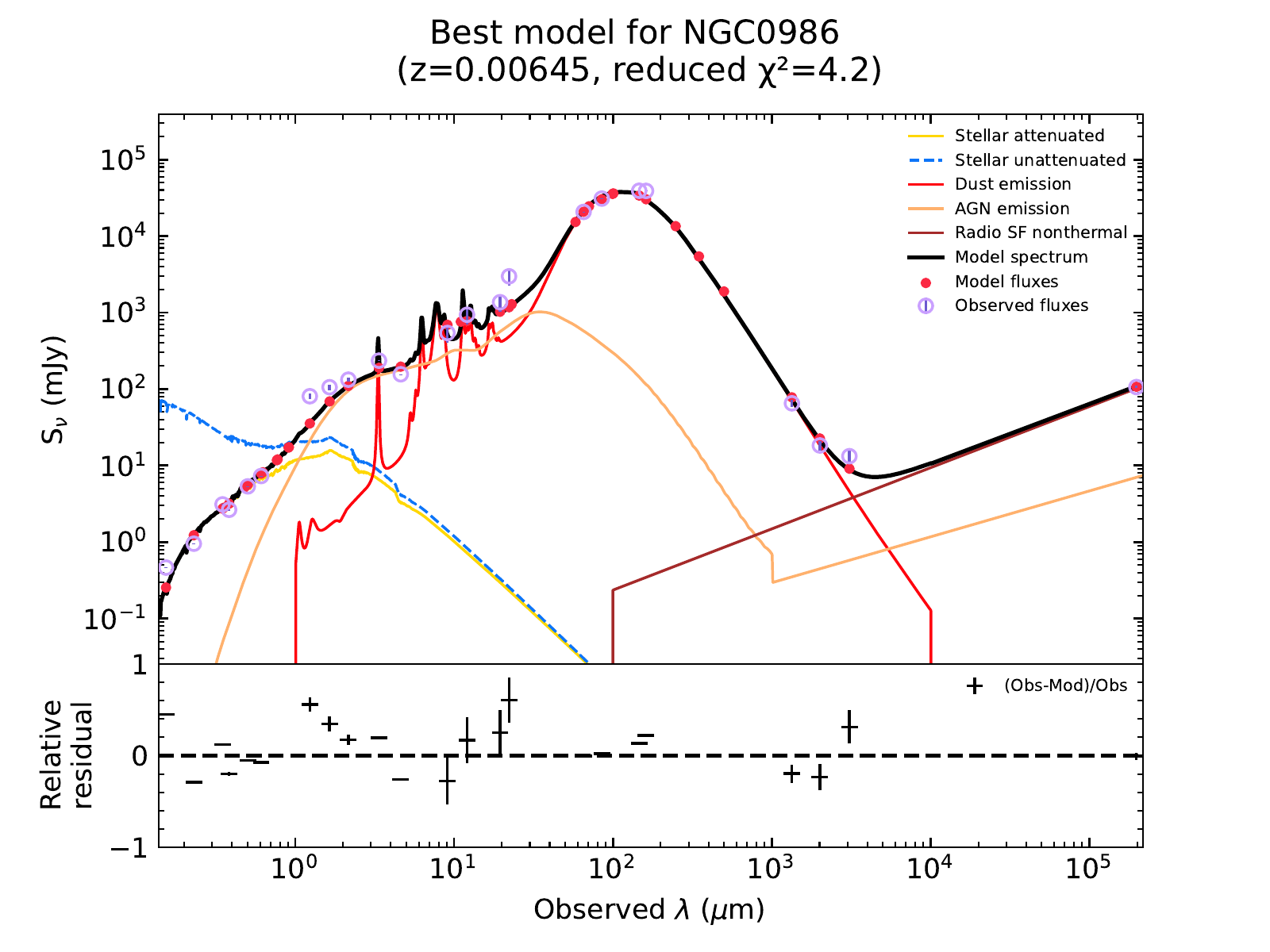}&
\includegraphics[scale=0.29]{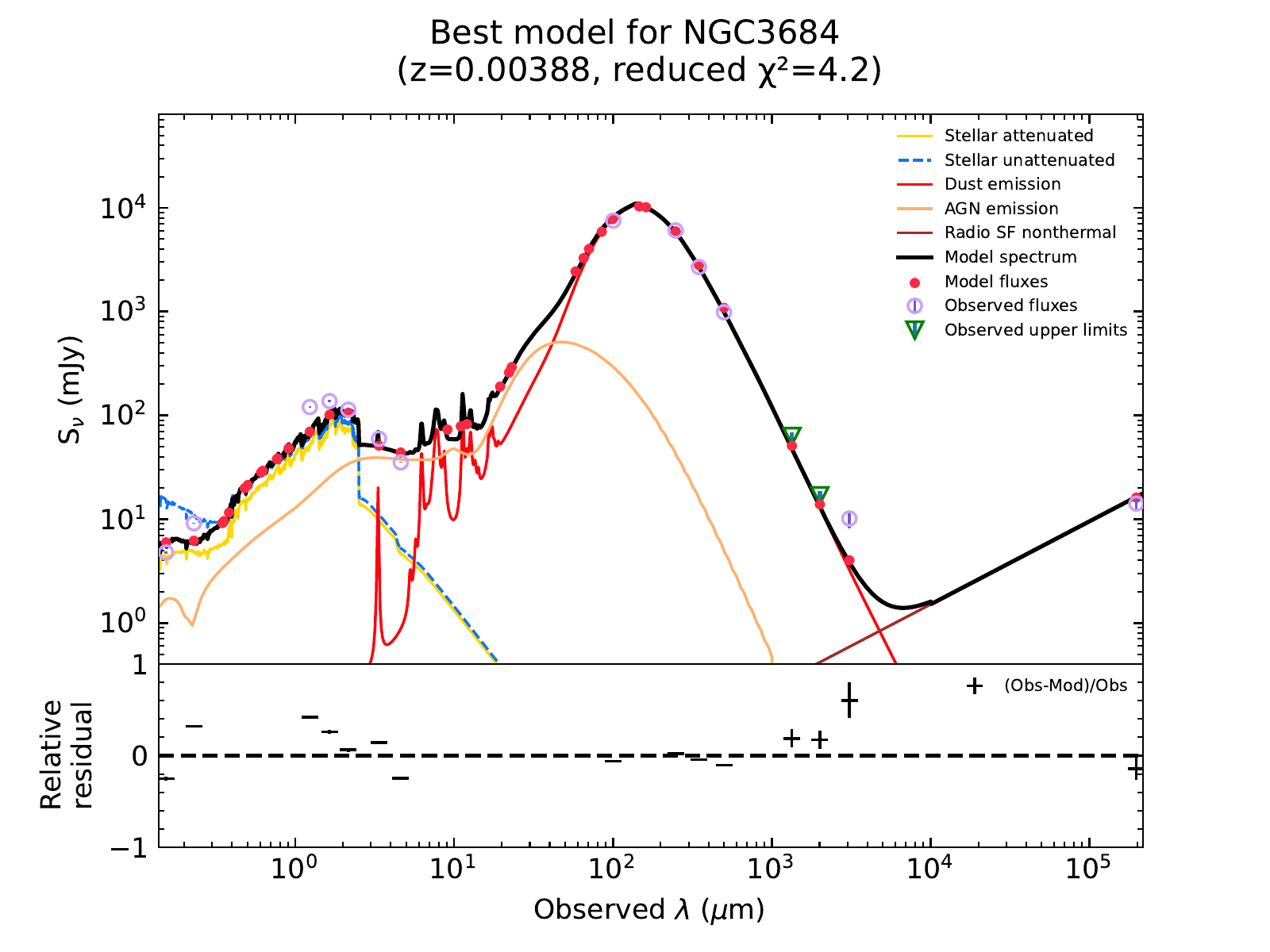}&
\includegraphics[scale=0.29]{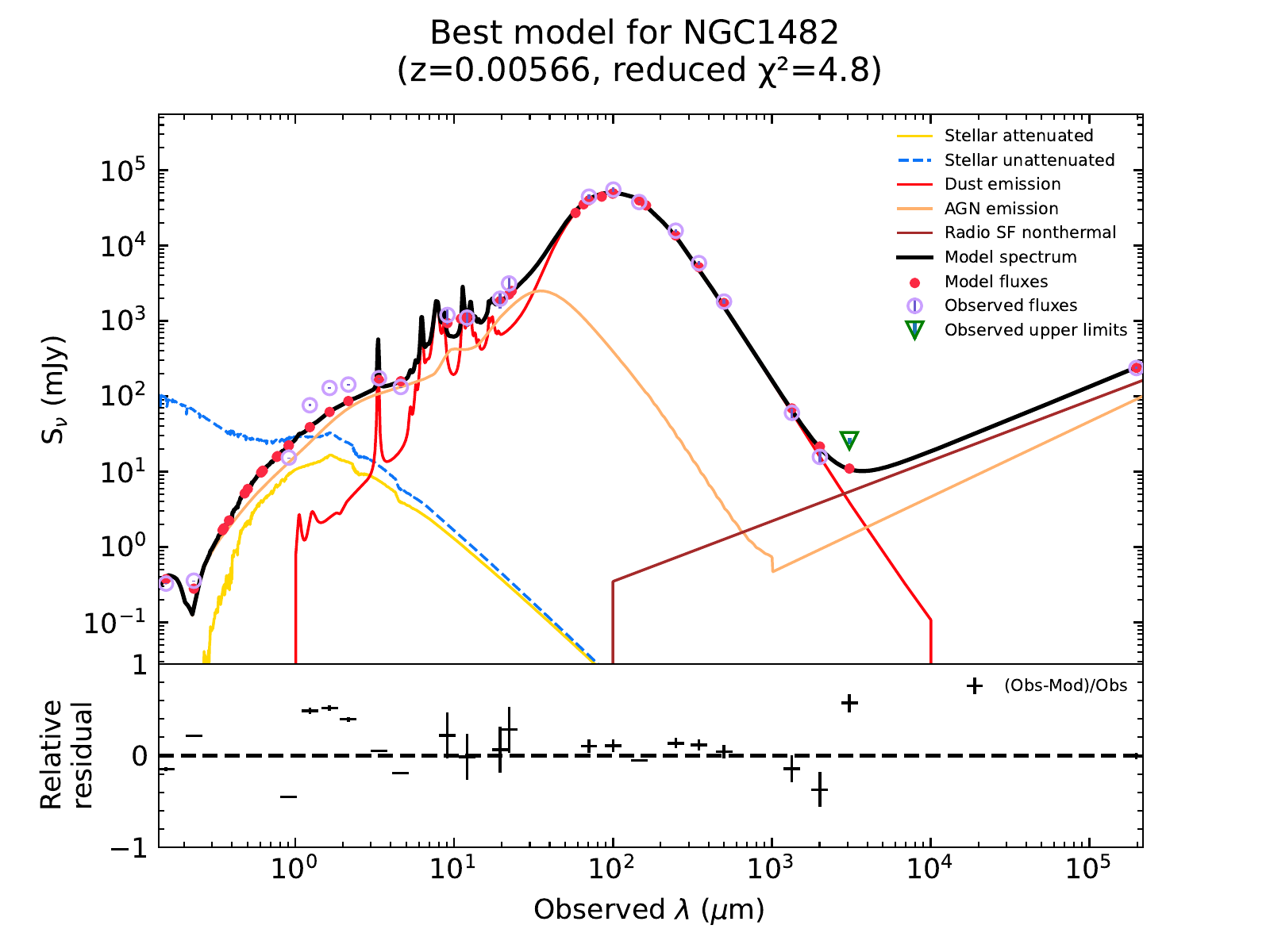}&
\includegraphics[scale=0.29]{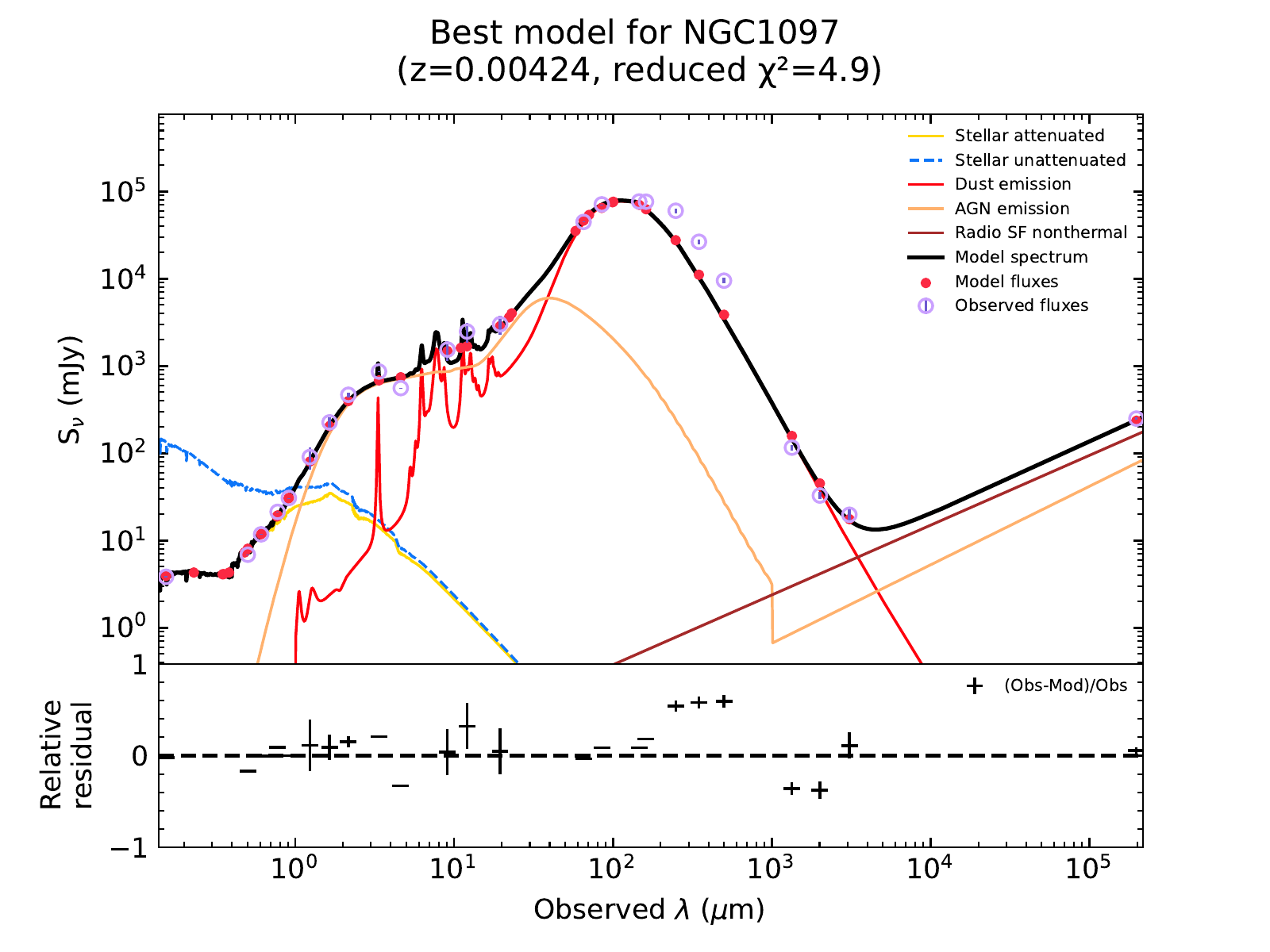}\\
\includegraphics[scale=0.29]{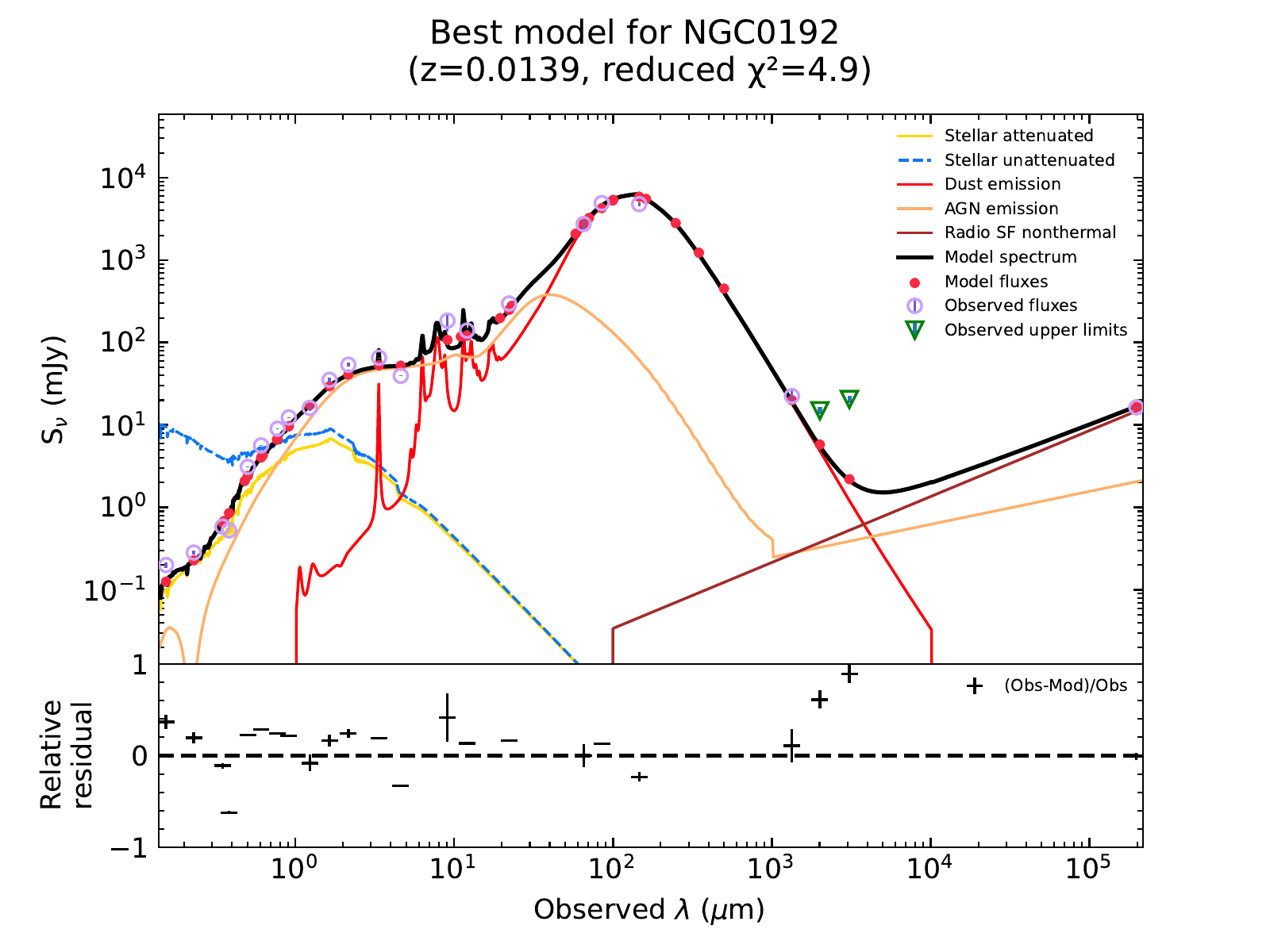}&
\includegraphics[scale=0.29]{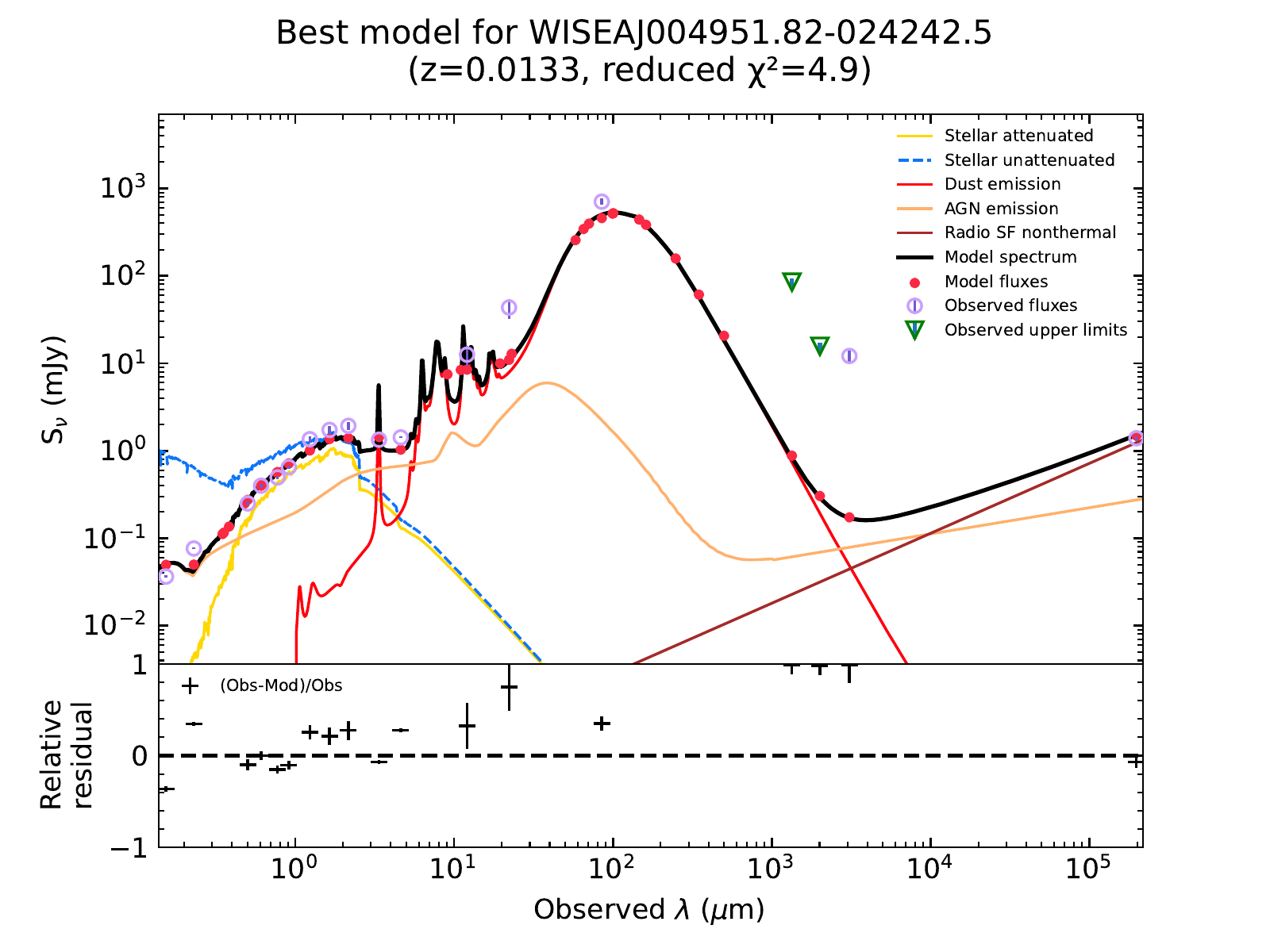}&
\includegraphics[scale=0.29]{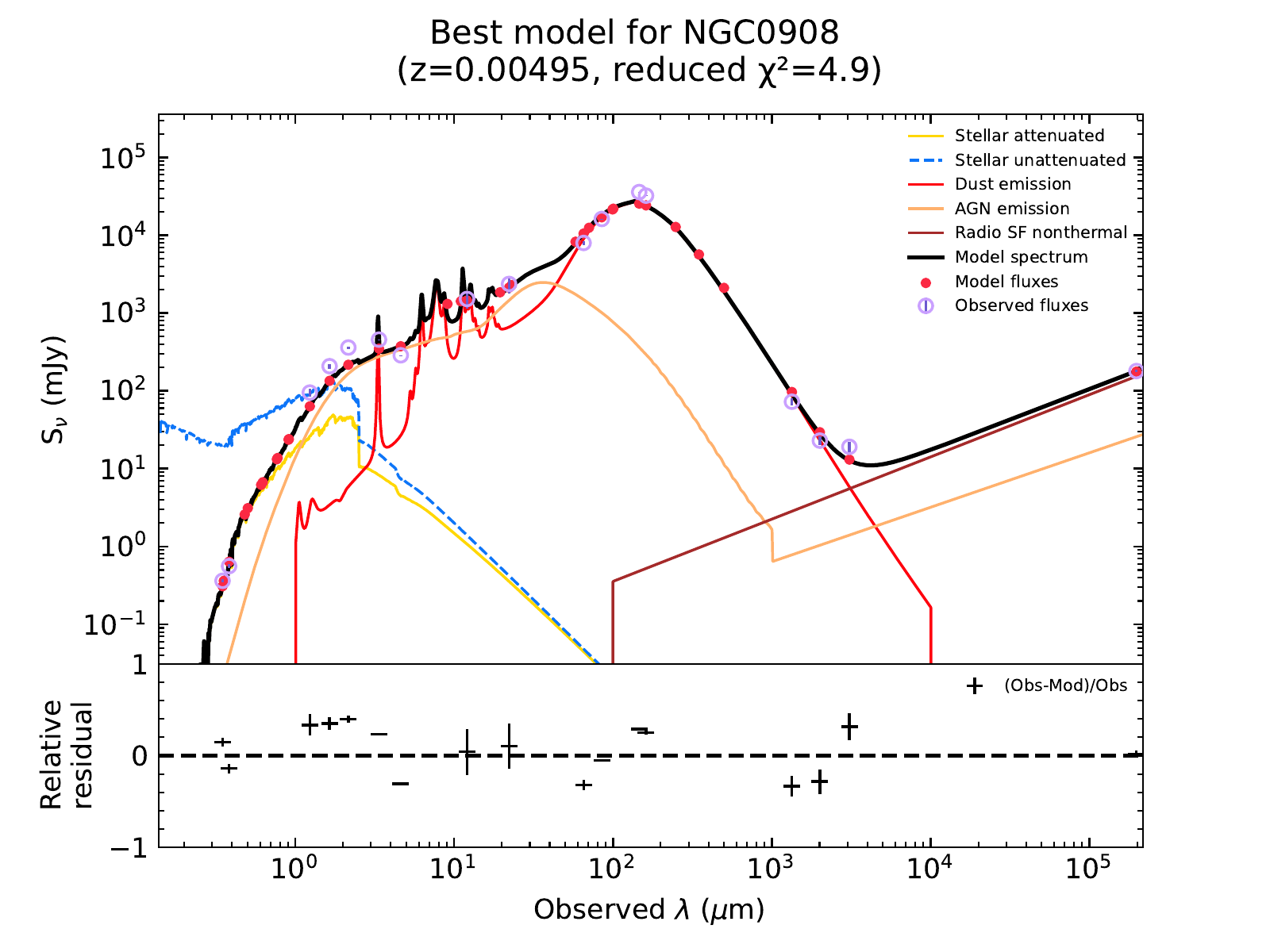}&
\includegraphics[scale=0.29]{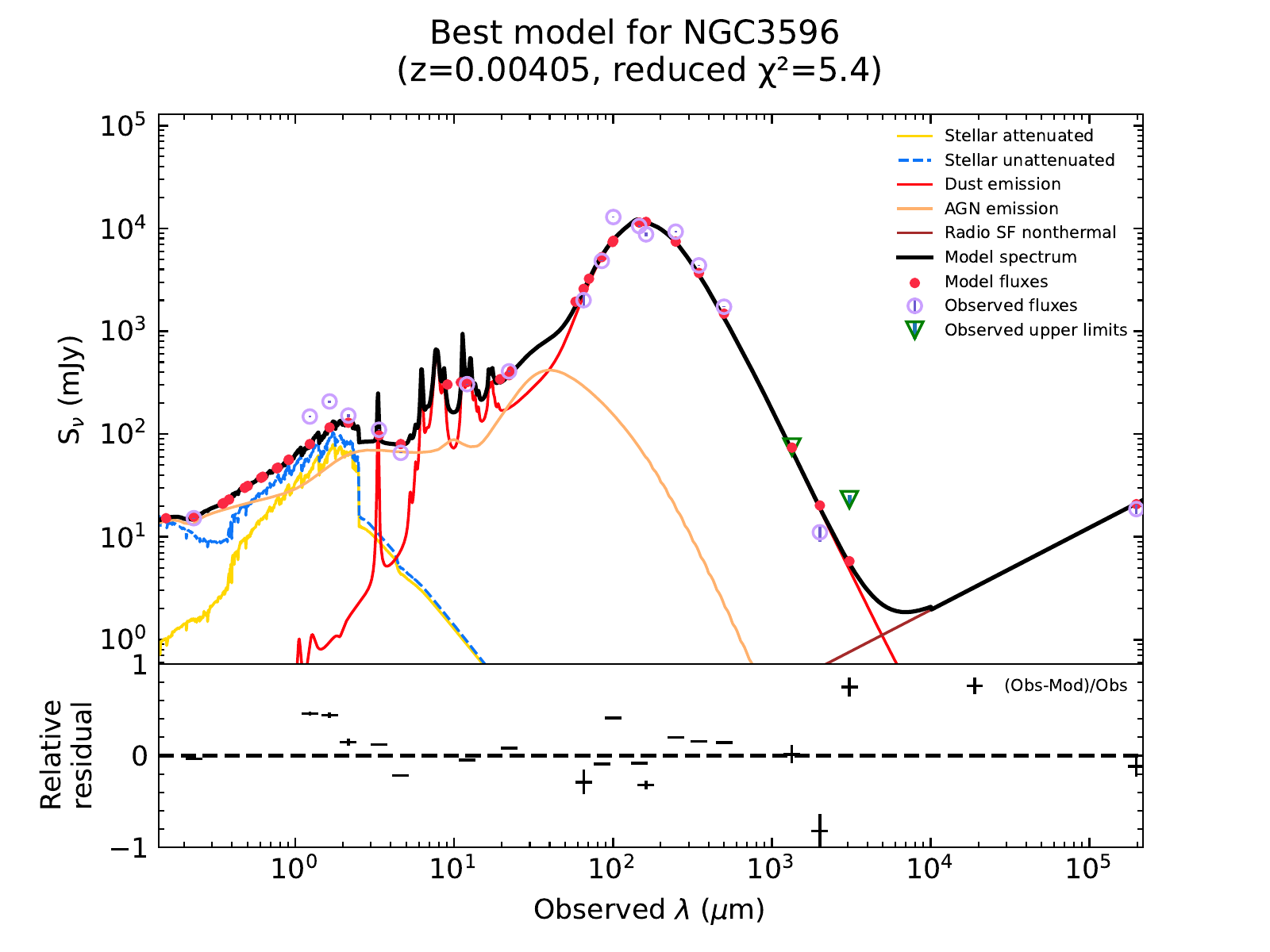}\\
\includegraphics[scale=0.29]{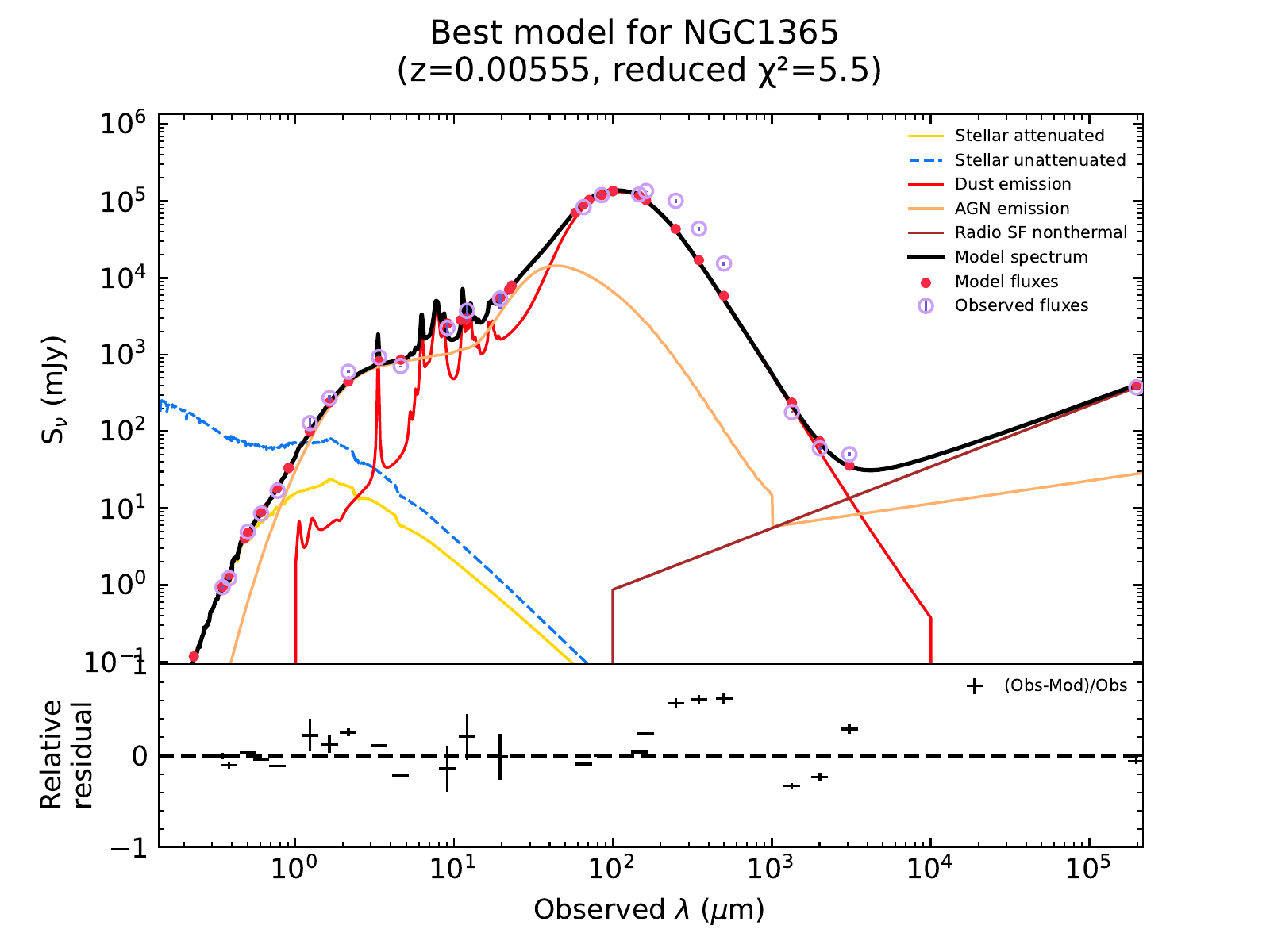}&
\includegraphics[scale=0.29]{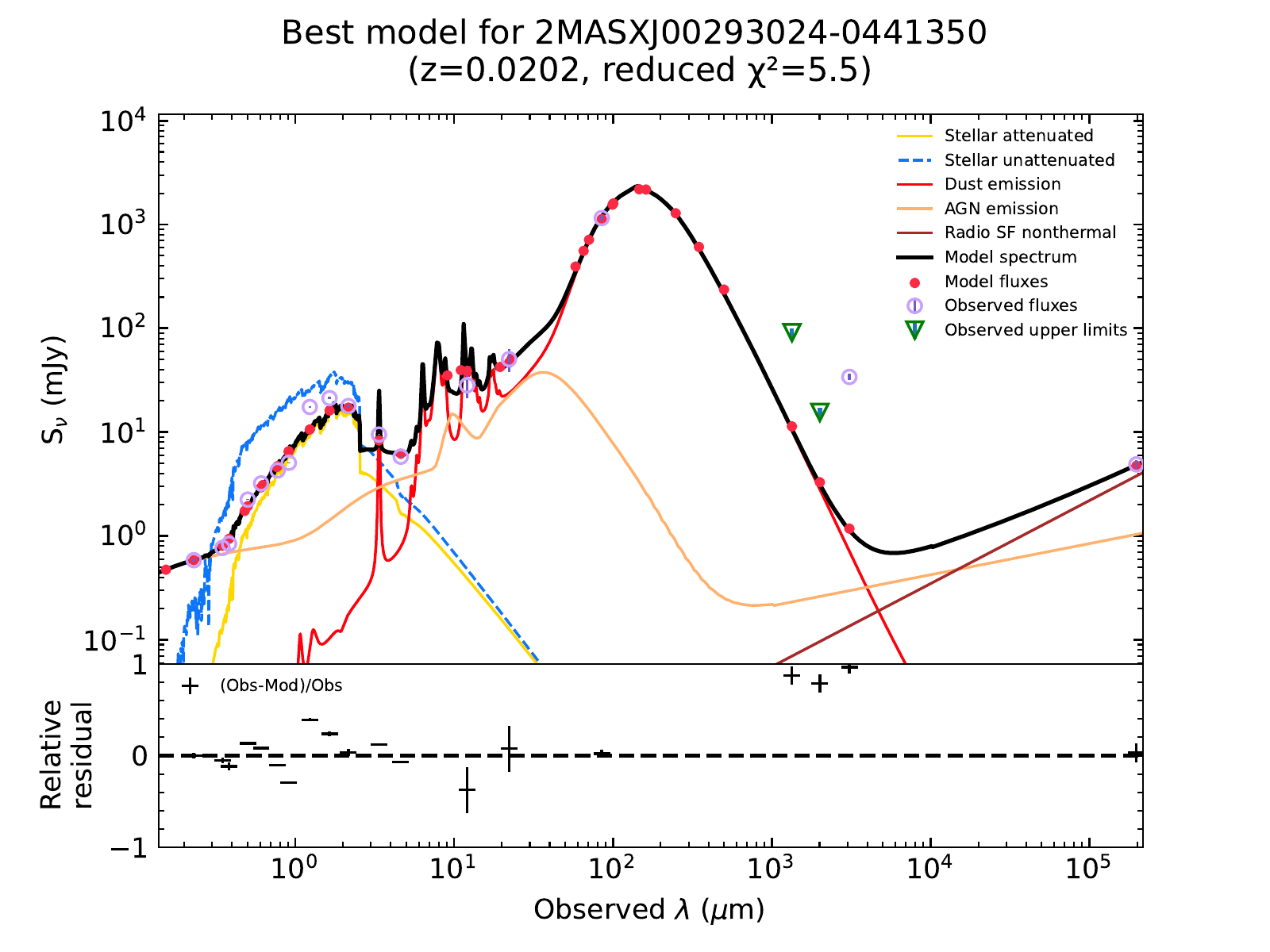}&
\includegraphics[scale=0.29]{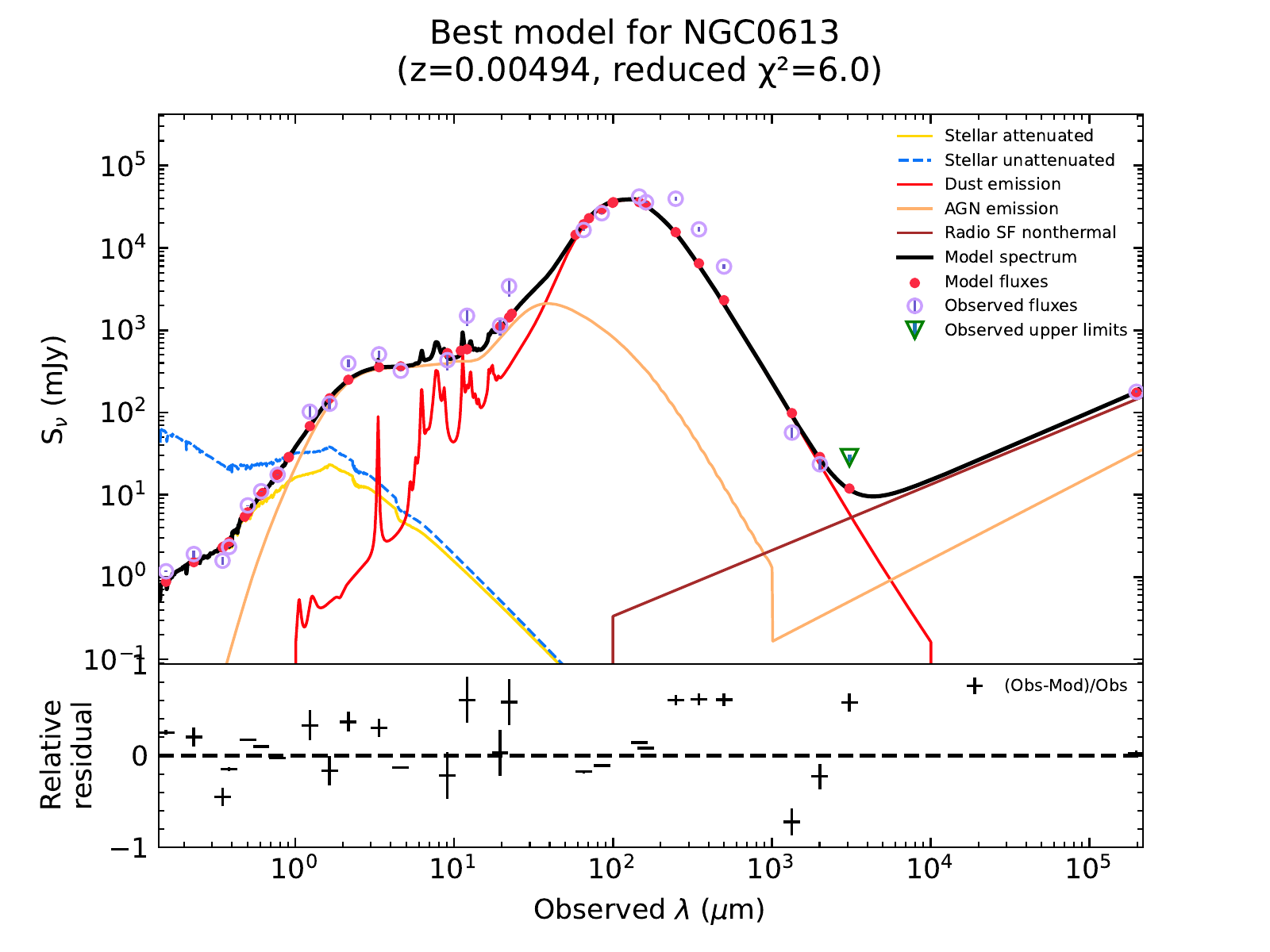}&
\includegraphics[scale=0.29]{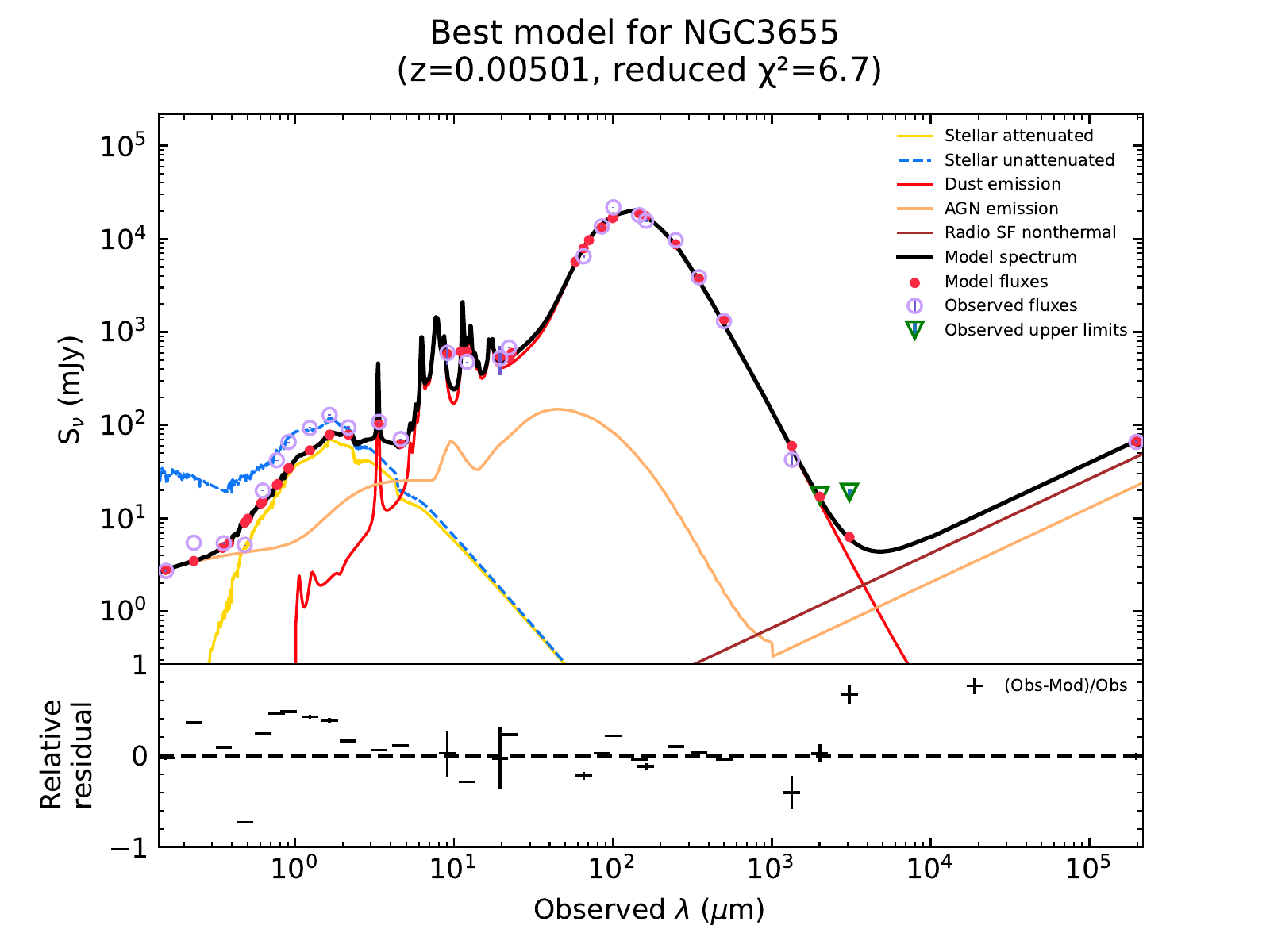}\\
\includegraphics[scale=0.29]{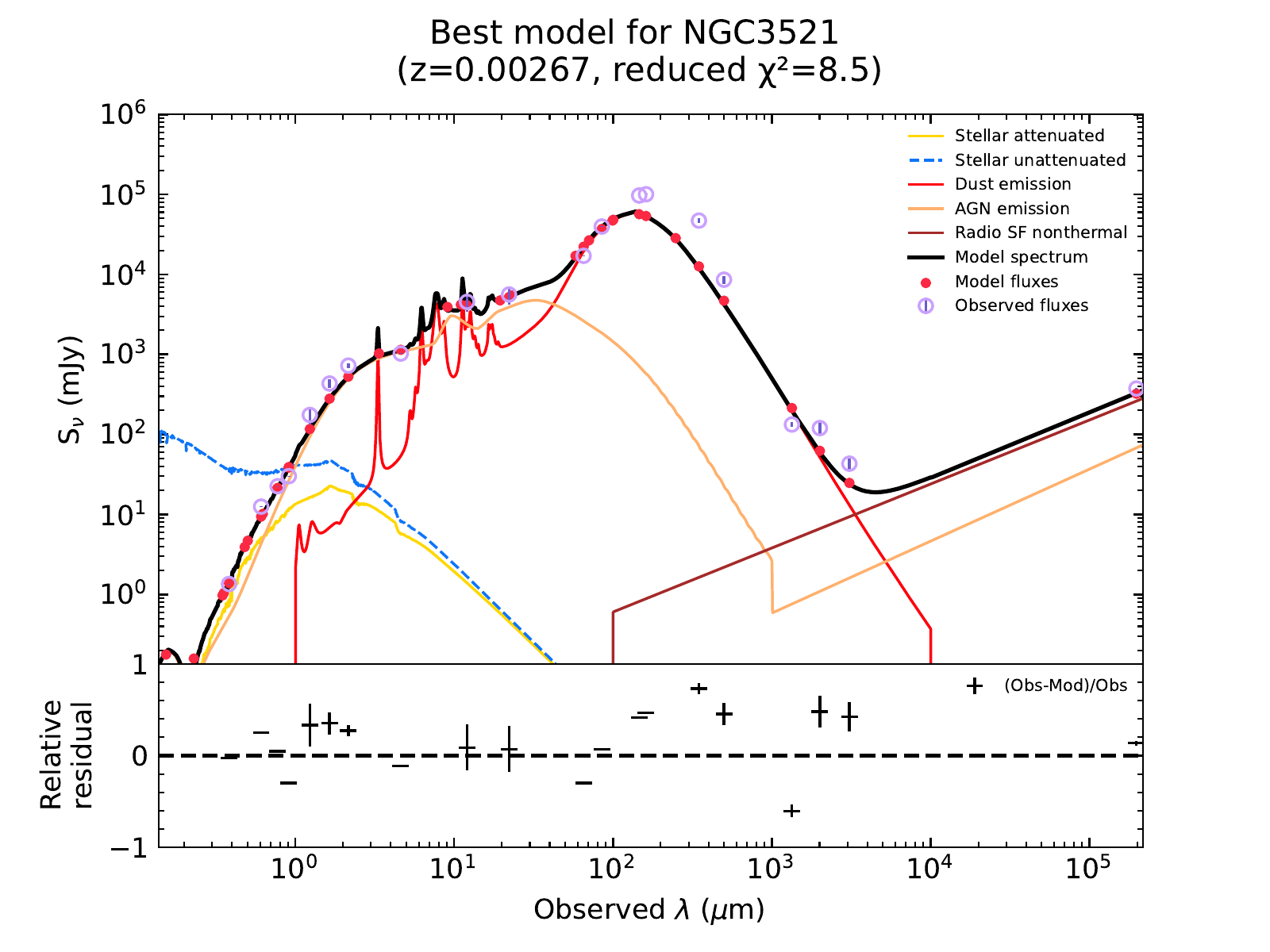}&
\includegraphics[scale=0.29]{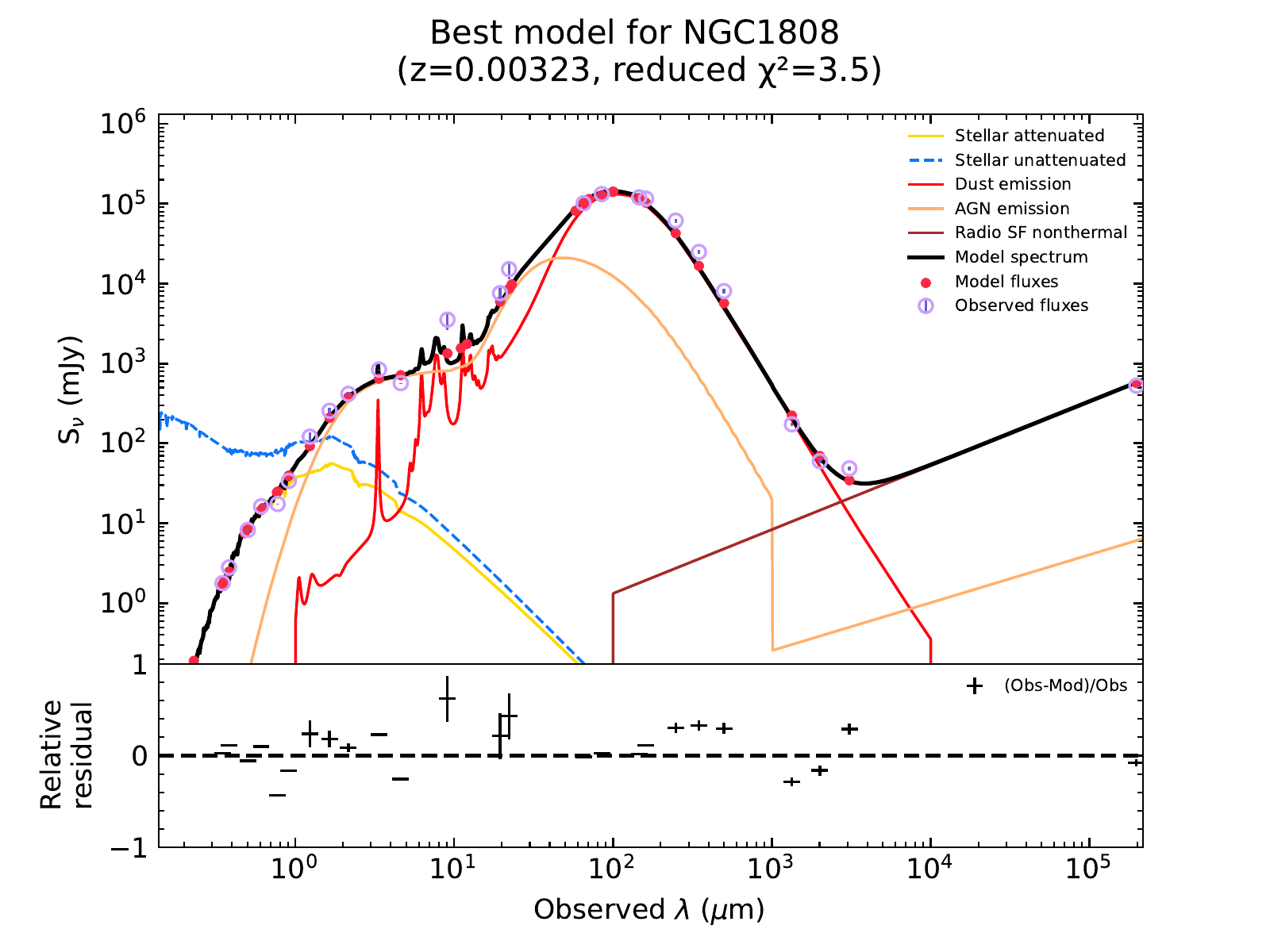}\\
\end{array}$
\end{center}
\caption{ Best-fitting models of radio star-forming galaxies in our sample. The symbol styles for the flux values and the color codes for the different model components are the same as in Figure \ref{fig:fig5}. }
\label{fig:fig7}
\end{figure*}

\begin{figure*}
\begin{center}$
\begin{array}{llll}
\includegraphics[scale=0.29]{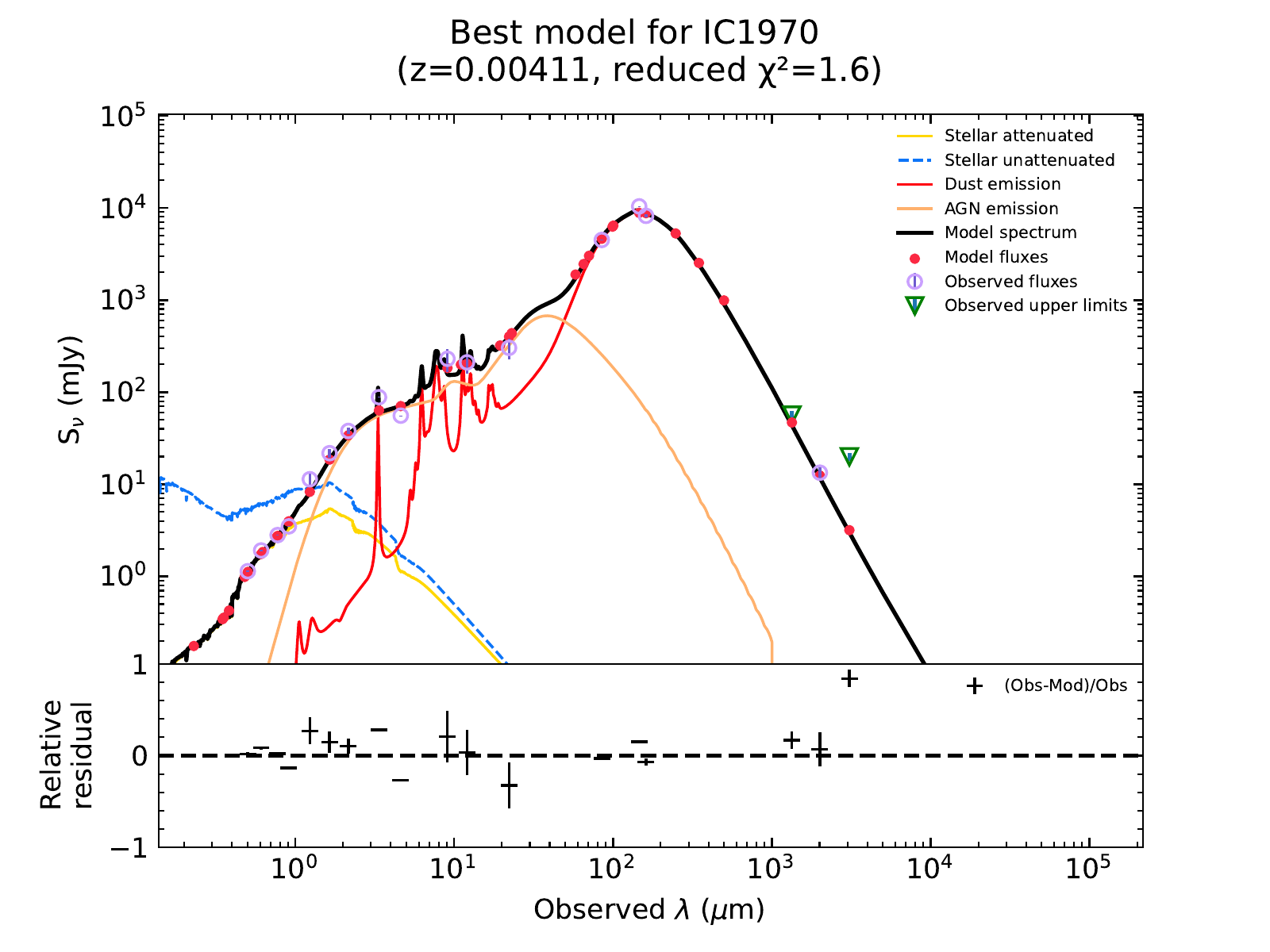}&
\includegraphics[scale=0.29]{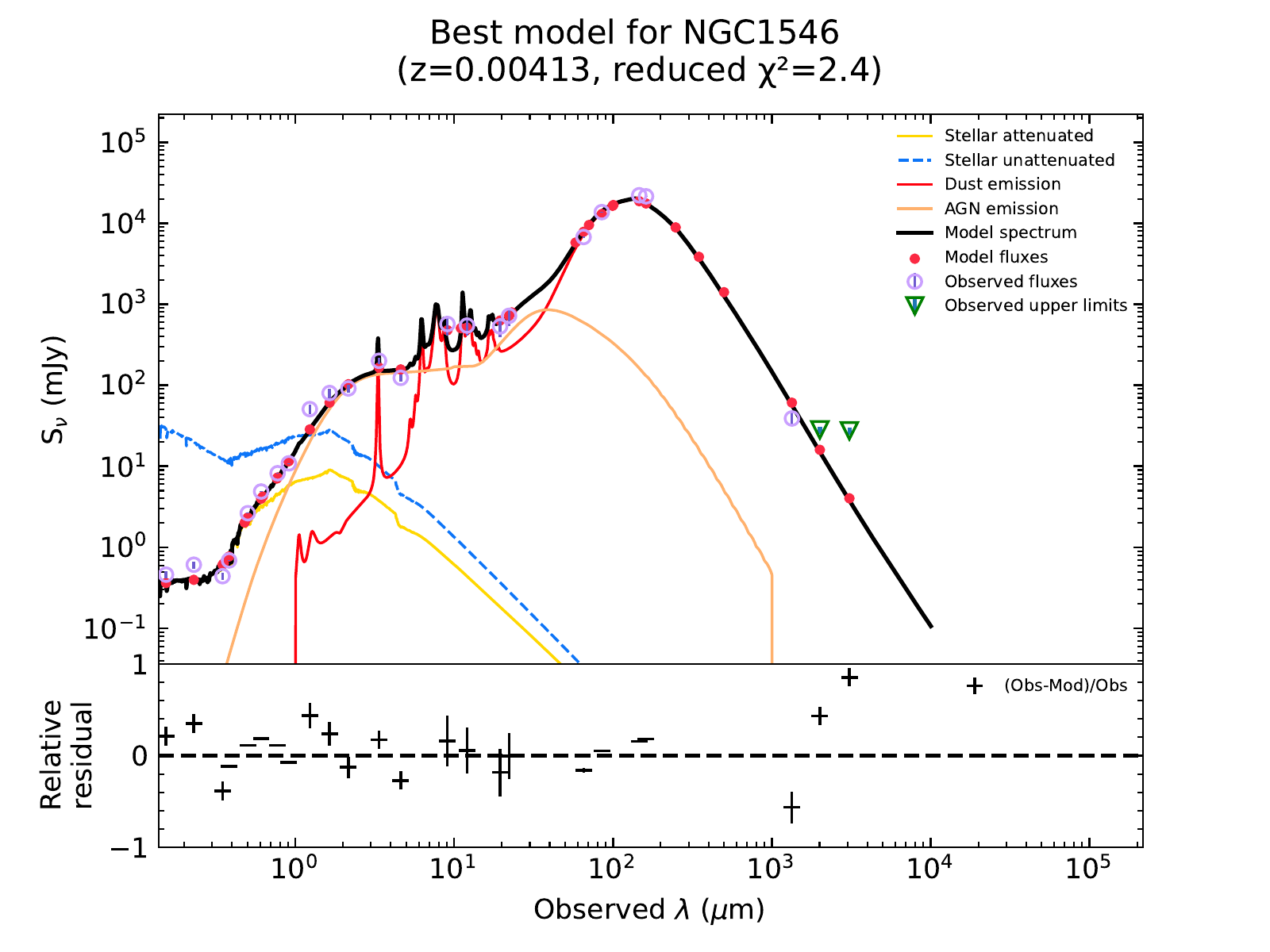}&
\includegraphics[scale=0.29]{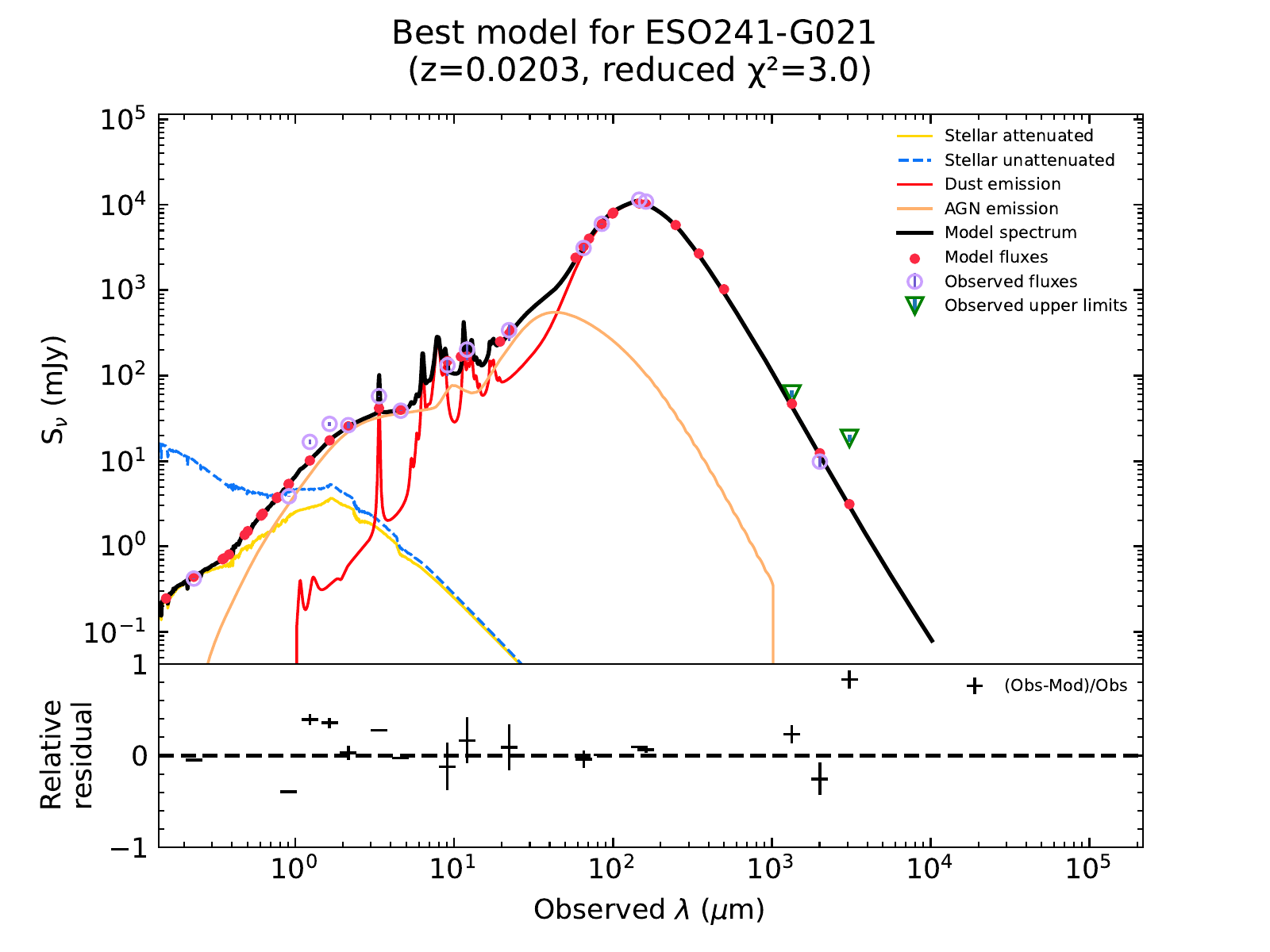}&
\includegraphics[scale=0.29]{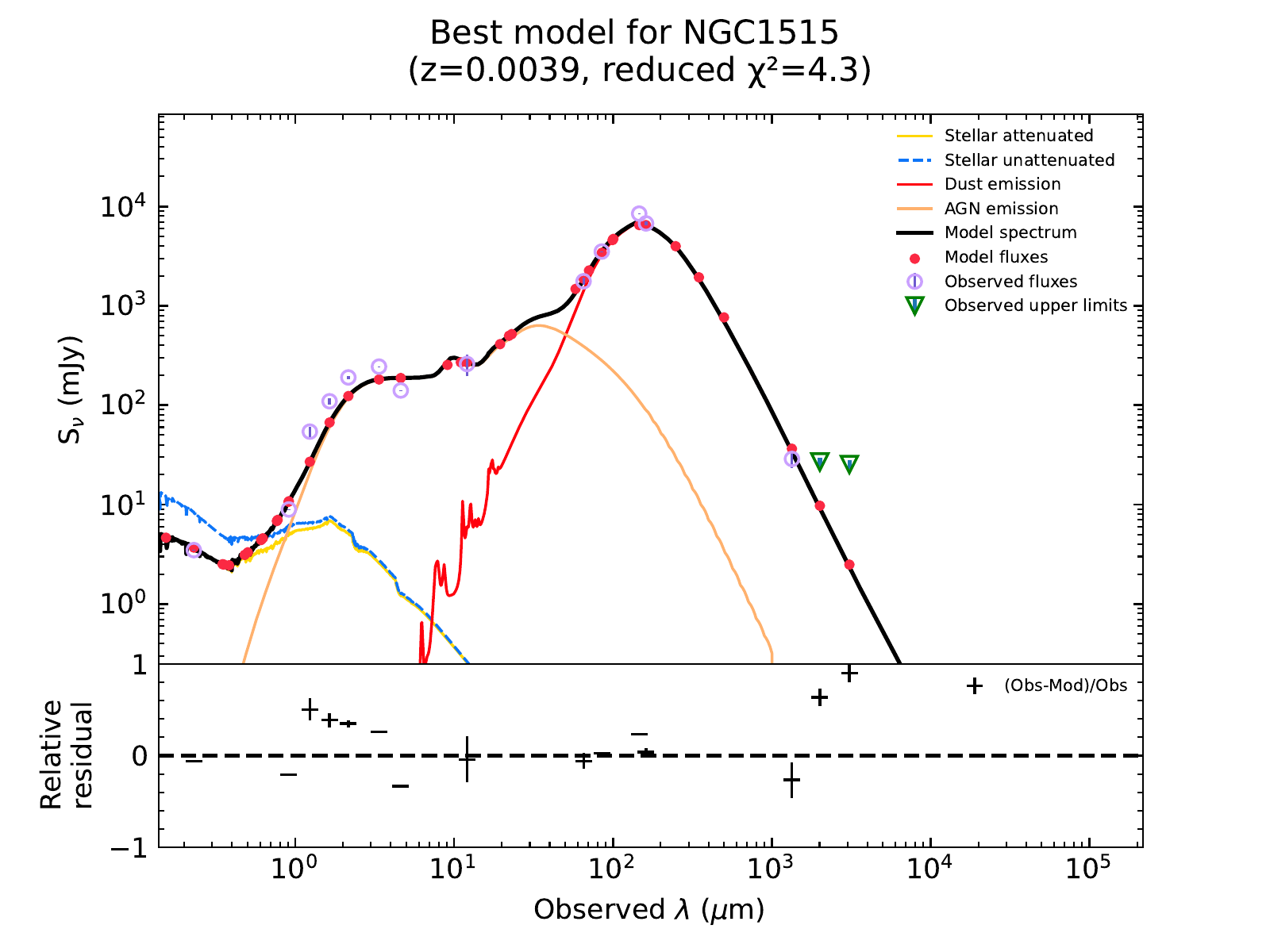}\\
\includegraphics[scale=0.29]{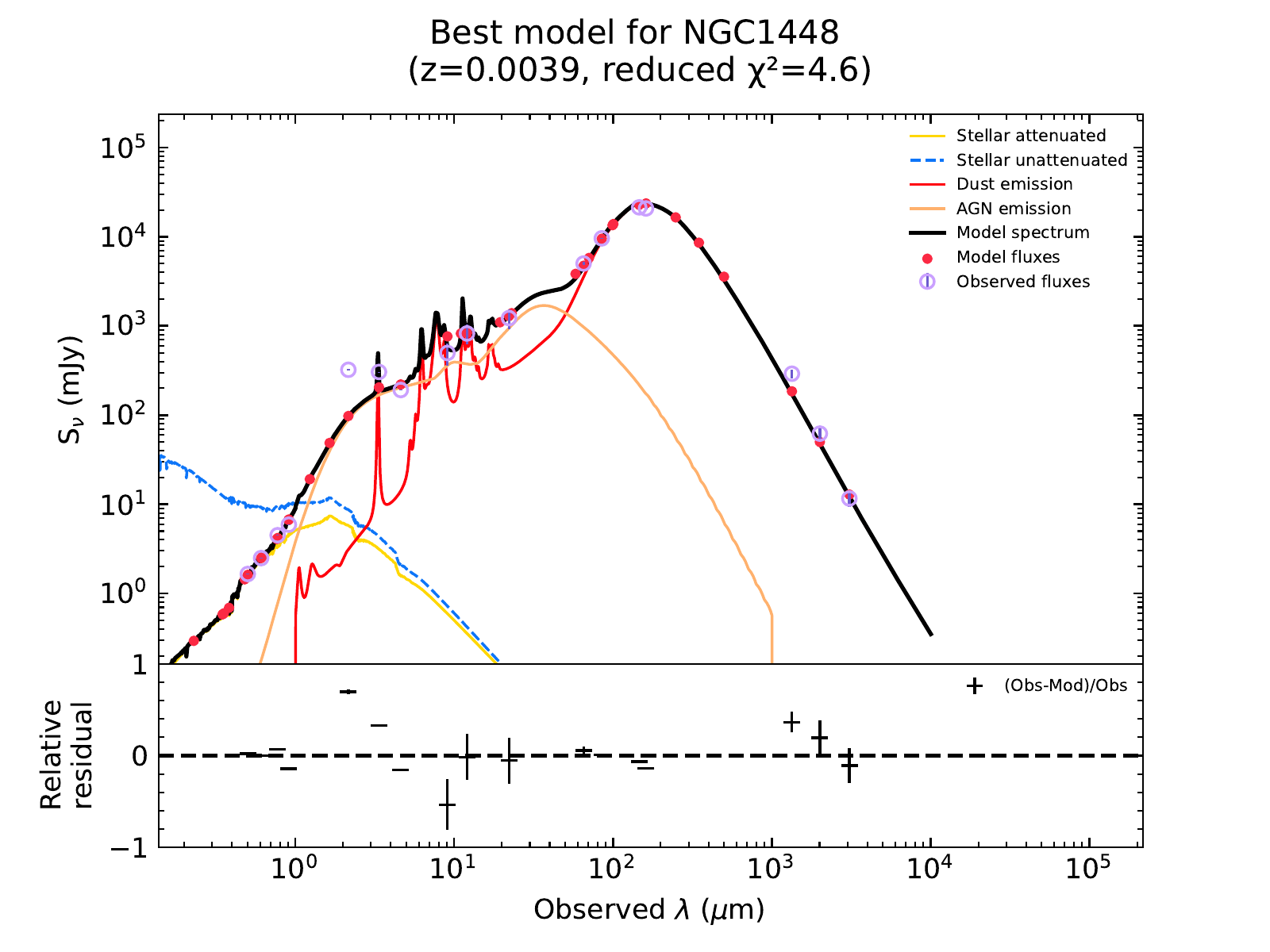}&
\includegraphics[scale=0.29]{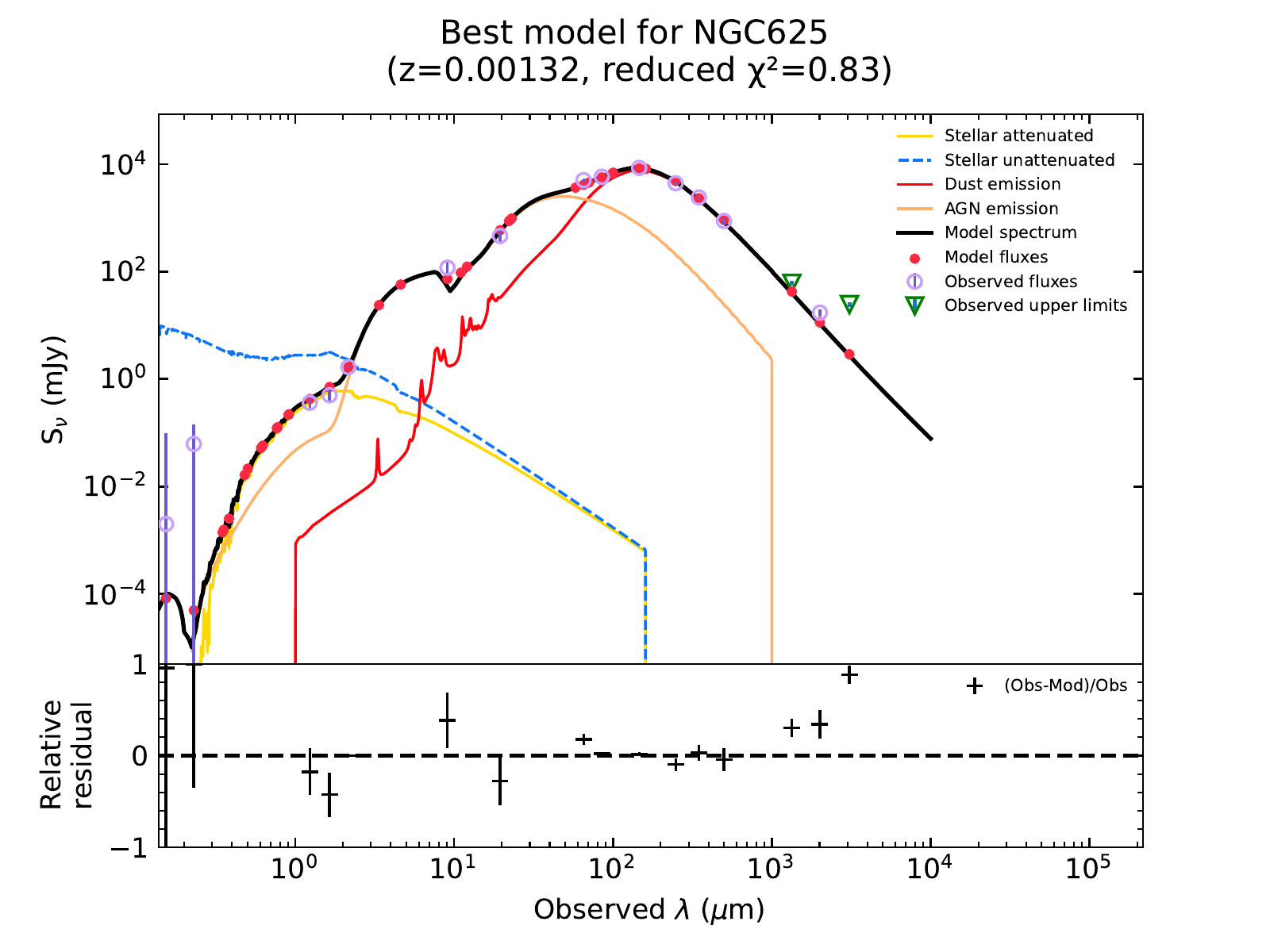}\\
\end{array}$
\end{center}
\caption{Best-fitting models of the galaxies for which the radio synchrotron emission component is not required in the SED, since the millimeter emission is consistent with the FIR dust emission and there is no radio detection at 1.4 GHz. The symbol styles for  flux the values and the color codes for the different model components are the same as in Figure \ref{fig:fig5}. }
\label{fig:fig8}
\end{figure*}

Our main goals are to  identify the origins of the millimeter emissions measured from the galaxies in our sample and to measure the physical parameters from the SEDs.  Since the radio flux at 1.4 GHz is crucial in order to obtain a reliable fit for the synchrotron radio component, it is included in the further SED analysis, when available.
As a result, we obtain statistically reliable (mostly $\chi_{reduced}^{2} \leq 5.0$, and only for a few cases $\chi_{reduced}^{2} \sim8$) fits for 69 galaxies. Among these statistically reliable SEDs, 63 have a radio component, either because it has a radio detection at 1.4 GHz or because its millimeter data require one (these being higher than the FIR dust component). We divide these 63 SEDs into two groups, based on the radio-loudness parameter of the AGN: (i) radio-loud $R_{\rm{AGN}} > 10$; and (ii) radio-quiet $R_{\rm{AGN}} \le 10$. In Figures \ref{fig:fig5} and \ref{fig:fig6}, example SEDs are shown for the radio-loud AGNs and the radio-quiet AGNs, respectively. There are radio SF galaxies among the radio-quiet AGNs whose AGN radio emissions are insignificant. We show example SEDs of these, as a third group, in Figure \ref{fig:fig7}. Only six SEDs do not have a radio component, since the millimeter emission is consistent with the FIR dust emission, and they do not have a radio detection at 1.4 GHz. Figure \ref{fig:fig8} shows aexample SEDs for the cases where a radio synchrotron emission component is not required. 

Our SED analysis shows that the millimeter data, which are usually lacking from galaxy SEDs, are highly valuable for accurately defining the SED shape, beyond the SED peak of around 100 $\mu$m.  For our sample, the millimeter emission mostly originates (for more than half of our sample) from the AGN synchrotron radio radiation (and in these cases, the radio emission from the SF still has a contribution). For a quarter of the sample at least, the origin is the SF synchrotron radio radiation. And for a minor group, the origin is the FIR dust emission. 

\begin{figure*}
\begin{center}$
\begin{array}{lcr}
\includegraphics[scale=0.62]{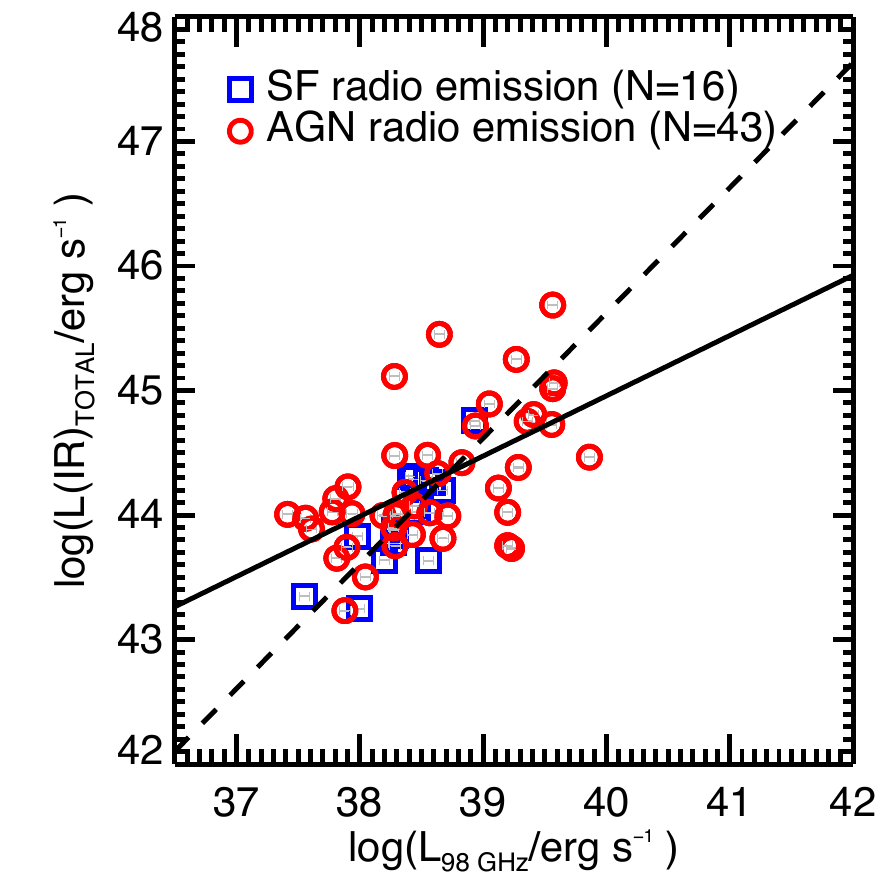}&
\includegraphics[scale=0.62]{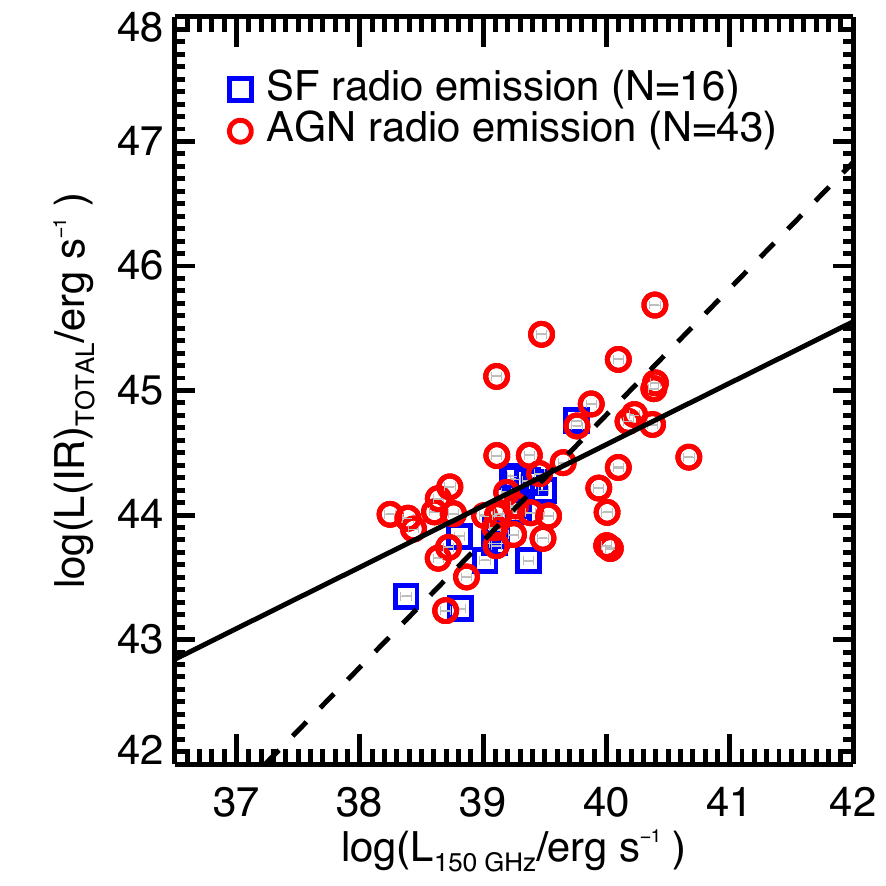}&
\includegraphics[scale=0.62]{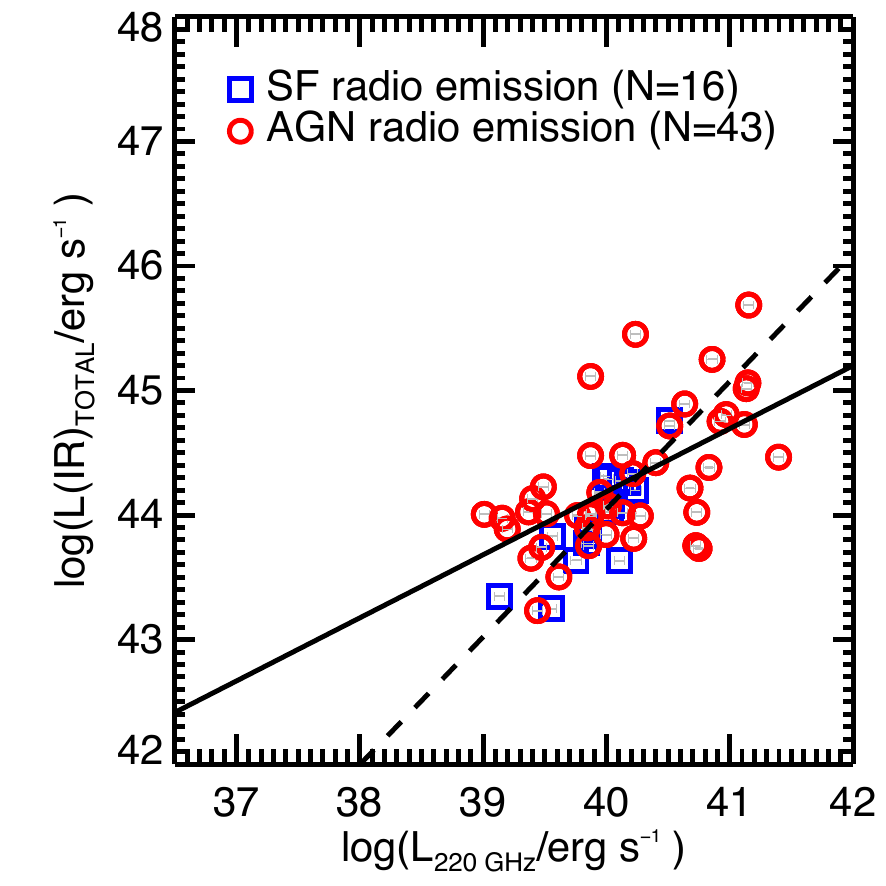}\\ 
\includegraphics[scale=0.62]{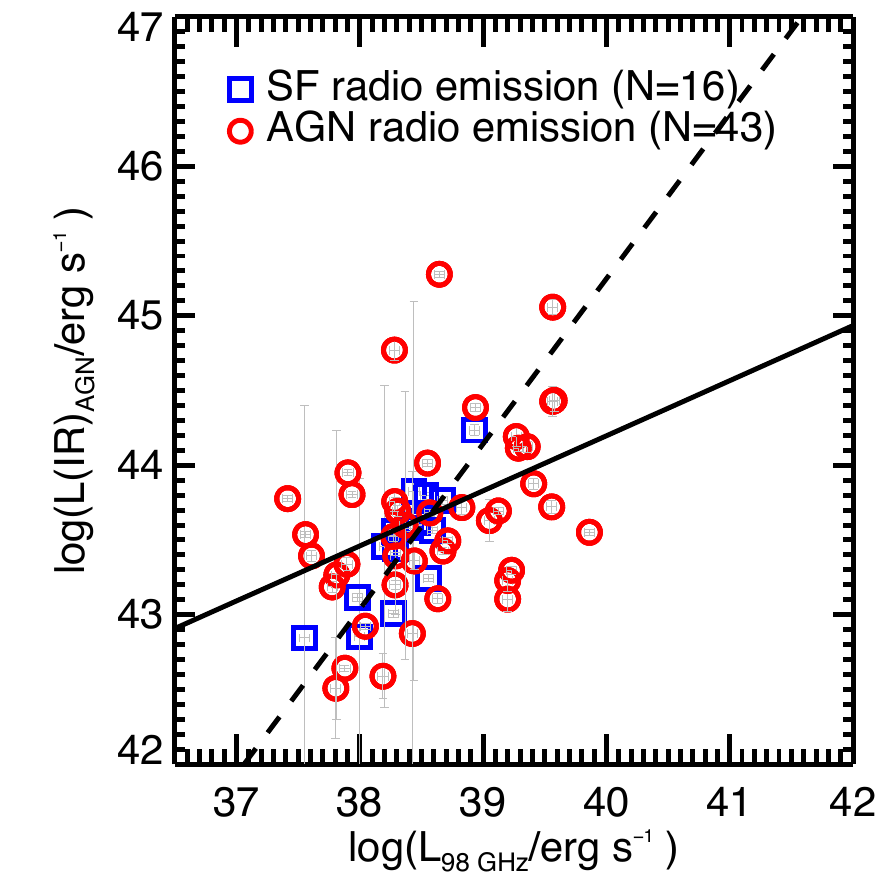}&
\includegraphics[scale=0.62]{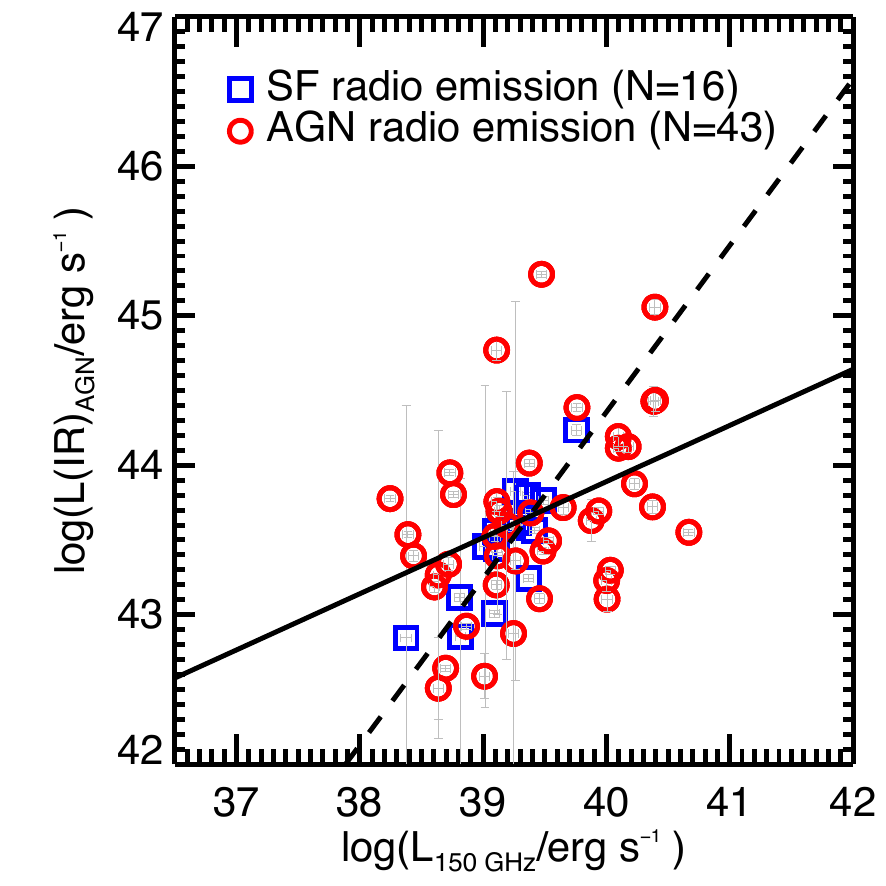}&
\includegraphics[scale=0.62]{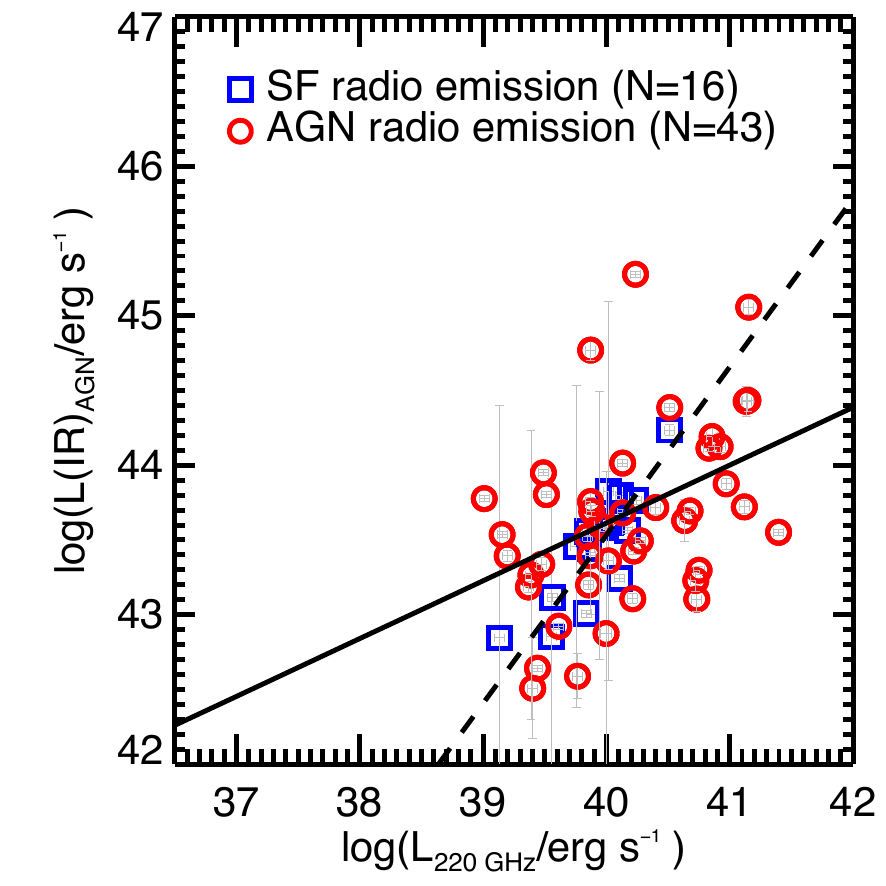}\\
\end{array}$
\end{center}
\caption{ The total IR luminosity (top) and the AGN luminosity (bottom) vs. the millimeter-band luminosities measured from the SED model at 98 GHz (left), 150 GHz (middle), and 220 GHz (right) for the 43 galaxies with AGN radio emission (red circles) and 
the 16 SF radio galaxies (blue squares).  These relationships can be used to estimate the AGN and total IR luminosities from the millimeter-band luminosities. The solid lines represent the relationships for the AGN radio emission, while the dashed lines show the relationships for the SF radio emission.}
\label{fig:fig9}
\end{figure*}

We measure the total IR luminosity $L\rm{(IR)}_{\rm{TOTAL}}$ as the sum of $L_{\rm{dust}}$ and  $L\rm{(IR)}_{\rm{AGN}}$ from best-fitted SED model.  We exclude four galaxies from this analysis, since the ACT millimeter data points do not fit well with the obtained best-fitting model. 
We separately quantify the relationships between the total IR luminosity, the AGN luminosity, and the millimeter-band luminosities at 98, 150, and 220 GHz for the 43 galaxies with AGN radio emission and 
the 16 SF radio galaxies, as shown in Figure \ref{fig:fig9}. The adopted form of the investigated luminosity relationships is as follows:  
\begin{equation}\label{Eq:2}
\log(L\rm{(IR)}_{\rm{TOTAL/AGN}})=B+\alpha_{L_\nu} \log(L_{\rm{\nu}}) +\epsilon_{L_\nu}.
\end{equation}
where the zeropoint is $B$, the slope is  $\alpha$, the estimated scatter is $\epsilon$ and $\nu$ is the frequency. The  Bayesian regression of \citep{Kelly2007} is used to obtain the best-fitting parameters. 
We list the obtained parameters in Table \ref{tab:table5}. 
As seen in Fig.  \ref{fig:fig9}, the 98 GHz, 150 GHz, and 220 GHz band luminosities measured from the SEDs are  in similar ranges for both the AGN and SF radio galaxies. Compared to the SF radio galaxies, the AGN radio emission galaxies have a larger scatter around the best-fitting relationship. Therefore, their slopes are different,  as listed in Table \ref{tab:table5}. 
We list the physical parameters measured from the SEDs in Table \ref{tab:table6}.  We check the reliability of these physical parameters with a mock analysis in CIGALE. In this procedure, CIGALE takes the data of each galaxy, based on the best-fitting SED, and adds a noise computed from a Gaussian distribution with the same standard deviation as the observed data. The CIGALE analysis is repeated for this mock catalog, based on the same set of user-defined parameters that were used for the original analysis. The obtained physical parameters from the mock analysis correlate well with the parameters obtained from the original analysis.

\begin{deluxetable*}{lccc}
 \label{tab:table5}
\tabletypesize{\scriptsize}
\tablewidth{0pt}
\tablecolumns{3}
\tablecaption{ Parameters of the Relationships between the Total IR Luminosity, AGN Luminosity, and the Millimeter-band Luminosities at 98, 150, and 220 GHz. 
	} 
\tablehead{       
\colhead{Luminosities} & \colhead{$\alpha_{L_\nu\,GHz}$} & \colhead{$B$} & \colhead{$\epsilon_{L_\nu\,GHz}$}} 
\startdata
          $ \log(L\rm{(IR)}_{\rm{TOTAL}}) - \log(L_{\rm{\ 98\,GHz}})$ AGN radio emission  & 0.48 $\pm$ 0.11 & 25.60 $\pm$ 4.35 & 0.46\\
          $ \log(L\rm{(IR)}_{\rm{TOTAL}}) - \log(L_{\rm{\ 98\,GHz}})$   SF radio emission & 1.01 $\pm$ 0.23 & 5.39 $\pm$ 8.68 & 0.26\\
           \hline
           $ \log(L\rm{(IR)}_{\rm{TOTAL}}) - \log(L_{\rm{\ 150\,GHz}})$ AGN radio emission  & 0.49 $\pm$ 0.11 & 24.85 $\pm$ 4.44 & 0.46\\
           $ \log(L\rm{(IR)}_{\rm{TOTAL}}) - \log(L_{\rm{\ 150\,GHz}})$  SF radio emission & 1.01 $\pm$ 0.22 & 4.24 $\pm$ 8.70 & 0.25\\
           \hline
           $ \log(L\rm{(IR)}_{\rm{TOTAL}}) - \log(L_{\rm{\ 220\,GHz}})$  AGN radio emission  & 0.51 $\pm$ 0.11 & 23.93 $\pm$ 4.53 & 0.45\\
           $ \log(L\rm{(IR)}_{\rm{TOTAL}}) - \log(L_{\rm{\ 220\,GHz}})$  SF radio emission & 1.03 $\pm$ 0.22 & 3.00 $\pm$ 8.60 & 0.24\\           
           \hline
           \hline
           $ \log(L\rm{(IR)}_{\rm{AGN}}) - \log(L_{\rm{\ 98\,GHz}})$ AGN radio emission  & 0.36 $\pm$ 0.14 & 29.67 $\pm$ 5.50 & 0.57\\
		  $ \log(L\rm{(IR)}_{\rm{AGN}}) - \log(L_{\rm{\ 98\,GHz}})$   SF radio emission& 1.10 $\pm$ 0.32 & 1.25 $\pm$ 12.30 & 0.23\\           
           \hline
           $ \log(L\rm{(IR)}_{\rm{AGN}}) - \log(L_{\rm{\ 150\,GHz}})$ AGN radio emission & 0.37 $\pm$ 0.14 & 29.09 $\pm$ 5.64 & 0.57\\
           $ \log(L\rm{(IR)}_{\rm{AGN}}) - \log(L_{\rm{\ 150\,GHz}})$  SF radio emission & 1.11 $\pm$ 0.32 & 0.01 $\pm$ 12.42 & 0.23\\           
           \hline
           $ \log(L\rm{(IR)}_{\rm{AGN}}) - \log(L_{\rm{\ 220\,GHz}})$  AGN radio emission  & 0.38 $\pm$ 0.14 & 28.37 $\pm$ 5.78 & 0.57\\
           $ \log(L\rm{(IR)}_{\rm{AGN}}) - \log(L_{\rm{\ 220\,GHz}})$  SF radio emission & 1.12 $\pm$ 0.31 & --1.28 $\pm$ 12.43 & 0.22\\
           \hline
\enddata
\end{deluxetable*}

\begin{deluxetable*}{llccccccccc}
\label{tab:table6}
\tablecolumns{11}
\tabletypesize{\scriptsize}
\setlength{\tabcolsep}{0.03in}
\tablewidth{0pt}
\tablecaption{ Physical parameters of 69 IR galaxies measured from the statistically reliable SED analysis with CIGALE.}
\tablehead{
\colhead{Optical Counterpart } &
\colhead{$\log (L\rm{(IR)}_{\rm{TOTAL}}$} &
\colhead{$\log(L_{\rm{dust}}$}&
\colhead{$\log(M_{\rm{dust}}$}&
\colhead{$\log(L\rm{(IR)}_{\rm{AGN}})$}&
\colhead{frac$_\mathrm{AGN}$} &
\colhead{$\psi_\mathrm{AGN}$} &
\colhead{Origin of} &
\colhead{$\alpha_{\rm{AGN}}$} &
\colhead{$R_{\rm{AGN}}$} \\
\colhead{Name} &
\colhead{  / erg s$^{-}$)} &
\colhead{  / erg s$^{-}$)}&
\colhead{/ kg)} & 
\colhead{  / erg s$^{-}$)} &
\colhead{} & 
\colhead{} & 
\colhead{Millimeter Emission} & 
\colhead{} & 
\colhead{} \\
\colhead{(1)} &
\colhead{(2)} &
\colhead{(3)} &
\colhead{(4)} &
\colhead{(5)} &
\colhead{(6)} &
\colhead{(7)} &
\colhead{(8)} &
\colhead{(9)} &
\colhead{(10)}}
\startdata
ESO293-IG034              &  43.90 $\pm$ 0.01  &  43.66 $\pm$ 0.01  &  37.65  &  43.52 $\pm$ 0.03  &  0.35  &  20.10  &             2  &   0.80  &        1  \\
SDSSJ000909.89-003705.3   &  44.48 $\pm$ 0.01  &  44.30 $\pm$ 0.01  &  37.83  &  44.01 $\pm$ 0.02  &  0.10  &  89.99  &             1  &   0.80  &     1000  \\
ESO241-G021               &  44.64 $\pm$ 0.01  &  44.50 $\pm$ 0.01  &  38.82  &  44.05 $\pm$ 0.02  &  0.20  &  30.10  &             4  &       .00  &              \\
2MASXJ00293024-0441350    &  43.98 $\pm$ 0.01  &  43.83 $\pm$ 0.01  &  38.20  &  43.46 $\pm$ 0.02  &  0.10  &  80.10  &             3  &   0.30  &        1  \\
NGC0157                   &  44.02 $\pm$ 0.01  &  43.75 $\pm$ 0.01  &  37.99  &  43.68 $\pm$ 0.02  &  0.40  &  30.10  &             2  &   0.90  &        1  \\
NGC0192                   &  44.20 $\pm$ 0.01  &  44.00 $\pm$ 0.01  &  38.07  &  43.77 $\pm$ 0.03  &  0.25  &  30.10  &             3  &   0.40  &        0  \\
WISEAJ004951.82-024242.5  &  43.05 $\pm$ 0.01  &  42.98 $\pm$ 0.01  &  36.52  &  42.21 $\pm$ 0.02  &  0.05  &  50.10  &             3  &   0.30  &        1  \\
NGC0349                   &  44.01 $\pm$ 0.01  &  43.62 $\pm$ 0.01  &  36.61  &  43.78 $\pm$ 0.02  &  0.45  &  30.10  &             1  &   0.60  &       11  \\
NGC0426                   &  43.66 $\pm$ 0.01  &  43.55 $\pm$ 0.01  &  37.29  &  43.00 $\pm$ 0.08  &  0.10  &  40.10  &             2  &   0.50  &        9  \\
6dFJ0113074-340055        &  44.00 $\pm$ 0.01  &  43.87 $\pm$ 0.01  &  37.63  &  43.39 $\pm$ 0.39  &  0.20  &  20.10  &             1  &   0.60  &      100  \\
\enddata
 \tablecomments{ The number codes for the origins of the millimeter emission are as follows: (1) Radio-loud AGN; (2) Radio-quiet AGN; (3) SF radio galaxy; (4) and dust emission. A machine-readable version of the full Table \ref{tab:table6} is available. Only a portion of this table is shown here for guidance regarding its form and content.}
\end{deluxetable*}

\subsection{FIR-to-Millimeter Wave and Millimeter-wave Color Distributions of Our Sample}\label{S:infraredACTcol}

\begin{figure*}
\begin{center}$
\begin{array}{lr}
\includegraphics[scale=0.8]{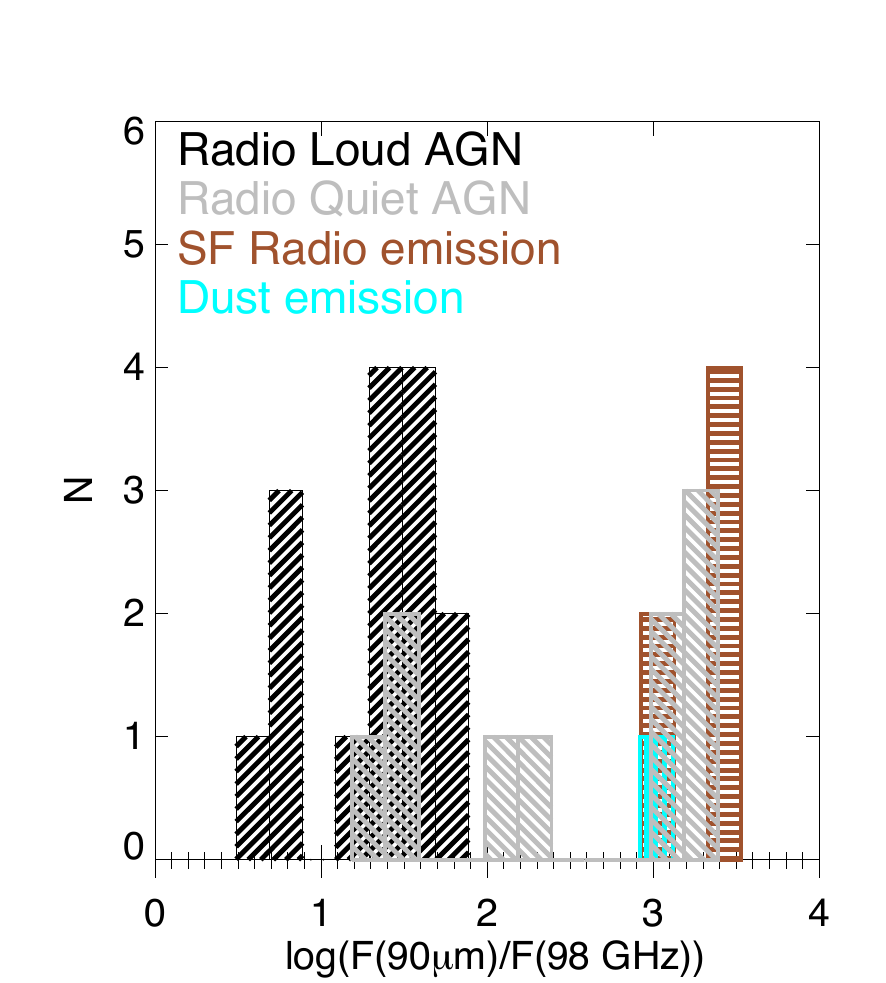}&
\includegraphics[scale=0.8]{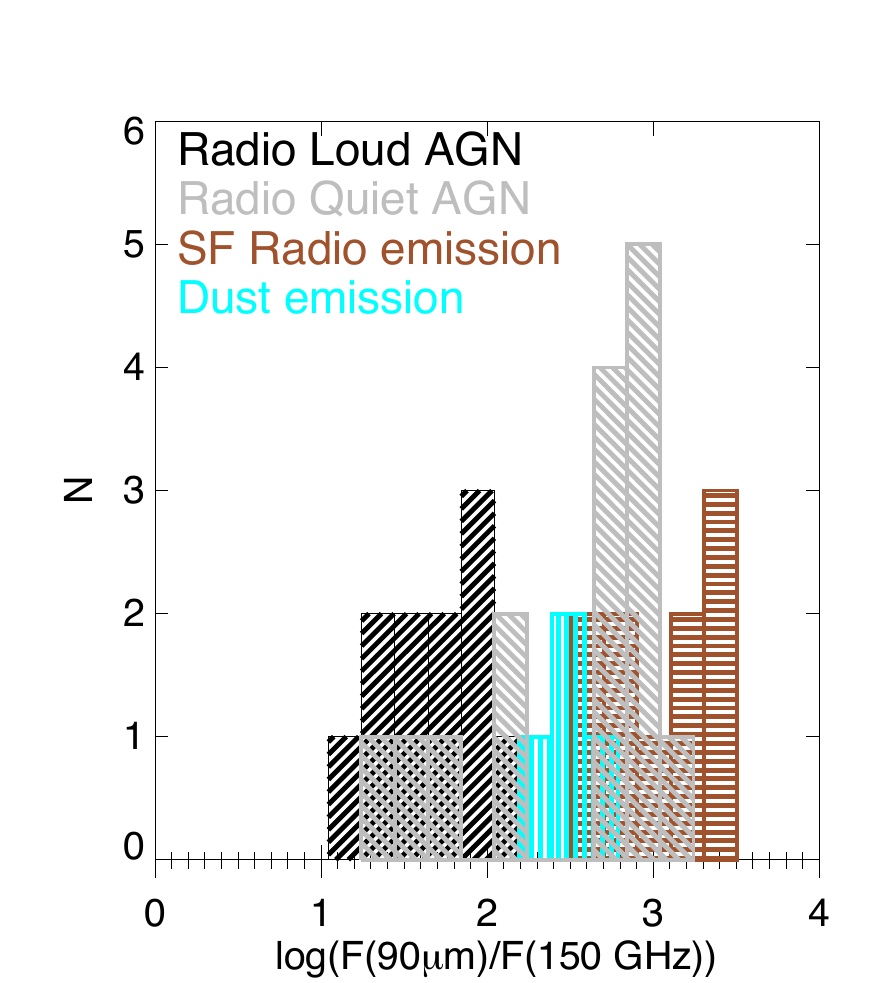}\\
\includegraphics[scale=0.8]{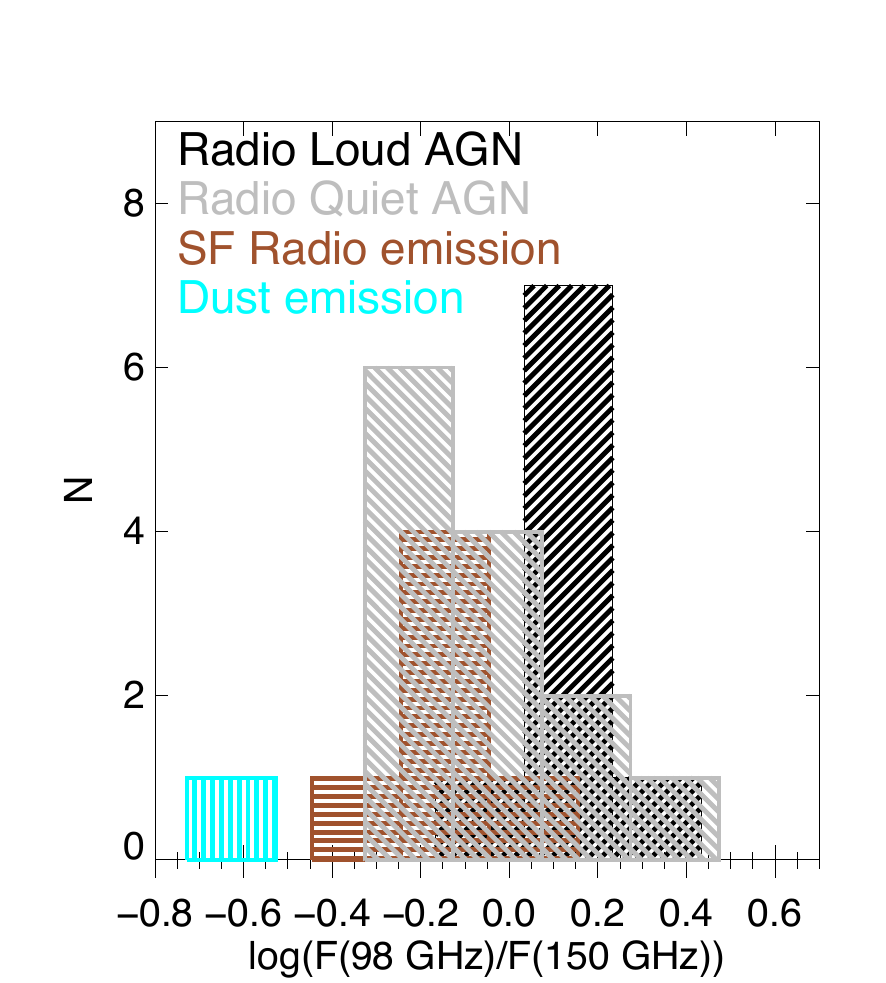}&
\includegraphics[scale=0.8]{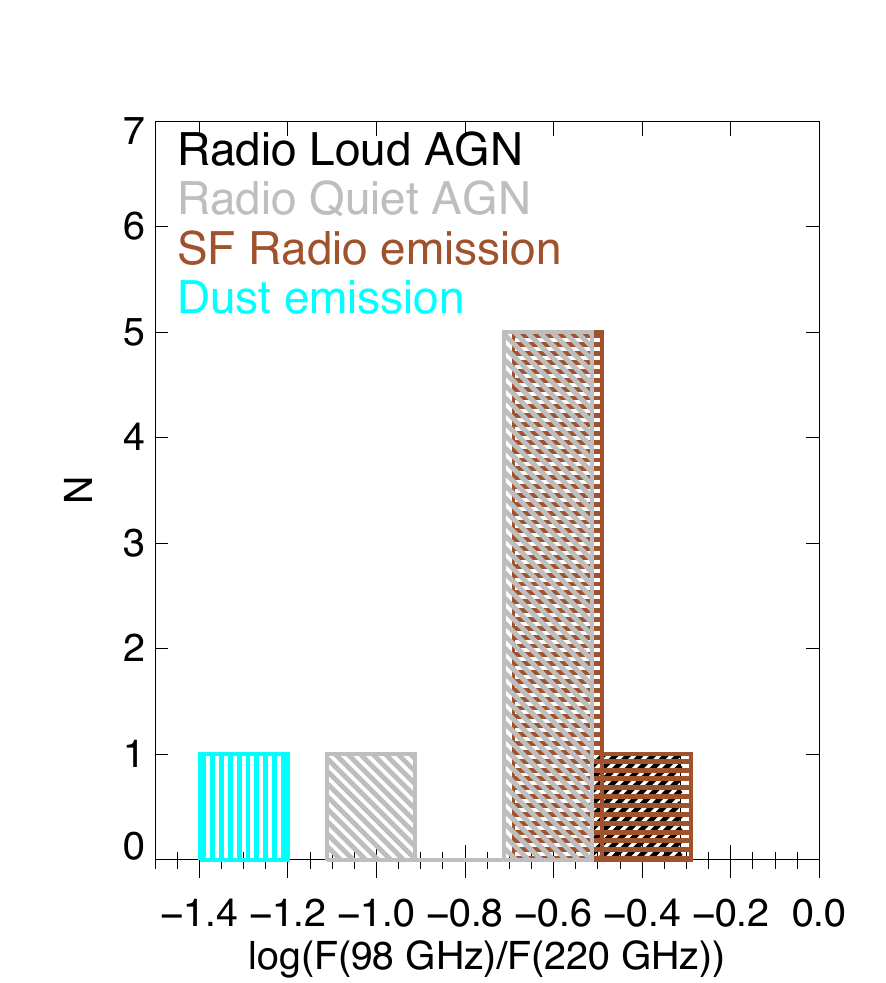}\\
\end{array}$
\end{center}
\caption{The log$(F_{\rm{90\,\mu m}}/F_{\rm{98\,GHz}})$ (top left), log$(F_{\rm{90\,\mu m}}/F_{\rm{150\,GHz}})$ (top right), log$(F_{\rm{98\,GHz}}/F_{\rm{150\,GHz}})$ (bottom left) and log$(F_{\rm{98\,GHz}}/F_{\rm{220\,GHz}})$ (bottom right) color distributions of our sample. The black distribution represents the radio-loud AGNs, the gray one represents the radio-quiet AGNs, the brown one represents the SF radio galaxies, and the cyan one represents the dust emission.  These color distributions are extremely useful for estimating the origins of the millimeter radiation, based only on the observed color, without any redshift requirement.} 
\label{fig:fig10}
\end{figure*}

We initially add the radio synchrotron emission component to the SEDs after visual inspection. This need  and the origin of the millimeter radiation can be quantified by the FIR--millimeter wave colors, so we therefore check the possible color combinations for our sample. The color distributions for different galaxy groups are shown in Fig. \ref{fig:fig10}. Here, the log$(F_{\rm{90\,\mu m}}/F_{\rm{98\,GHz}})$ color distribution in the top left panel and the log$(F_{\rm{90\,\mu m}}/F_{\rm{150\,GHz}})$ distribution in the top right panel similarly show that in the   log$(F_{\rm{90\,\mu m}}/F_{\rm{98\,GHz}}) < 2.0$ and log$(F_{\rm{90\,\mu m}}/F_{\rm{150\,GHz}}) < 2.0$ ranges, radio-loud and radio-quiet AGNs contribute significantly to the millimeter radiation, and that the synchrotron emission component is therefore needed in the SEDs. Since our sample includes only a few sources with log$(F_{\rm{90\,\mu m}}/F_{\rm{220\,GHz}})$ color, it is not large enough to determine a range for different distributions. 

The millimeter color distributions of our sample are shown in the bottom panels of Fig. \ref{fig:fig10}. Since our sample includes only a few SEDs with $F_{\rm{220\,GHz}}$ detections, the color combinations including this band do not provide a reliable separation range (as in the bottom right),  except for the clearly distinct region between $-1.4 < log(F_{\rm{98\,GHz}}/F_{\rm{220\,GHz}}) < -1.2$, for the dust emission. The log$(F_{\rm{98\,GHz}}/F_{\rm{150\,GHz}}) > 0.15$ region (bottom left) indicates a partial limit for radio-loud AGNs, while the log$(F_{\rm{98\,GHz}}/F_{\rm{150\,GHz}}) > -0.4$  region requires a radio component in the SEDs.

\subsection{Correlation Investigation of the FIR-to-Millimeter Wave and Millimeter-wave Colors}\label{S:correlation}

\begin{figure*}
\begin{center}$
\begin{array}{lr}
\includegraphics[scale=0.8]{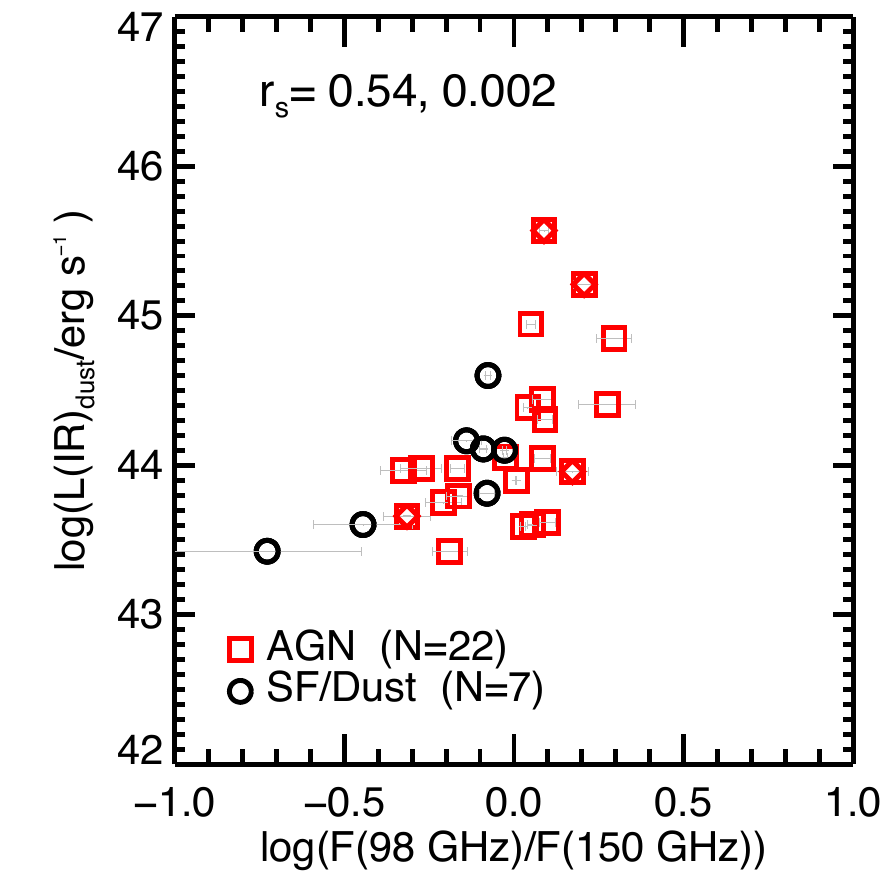}&
\includegraphics[scale=0.8]{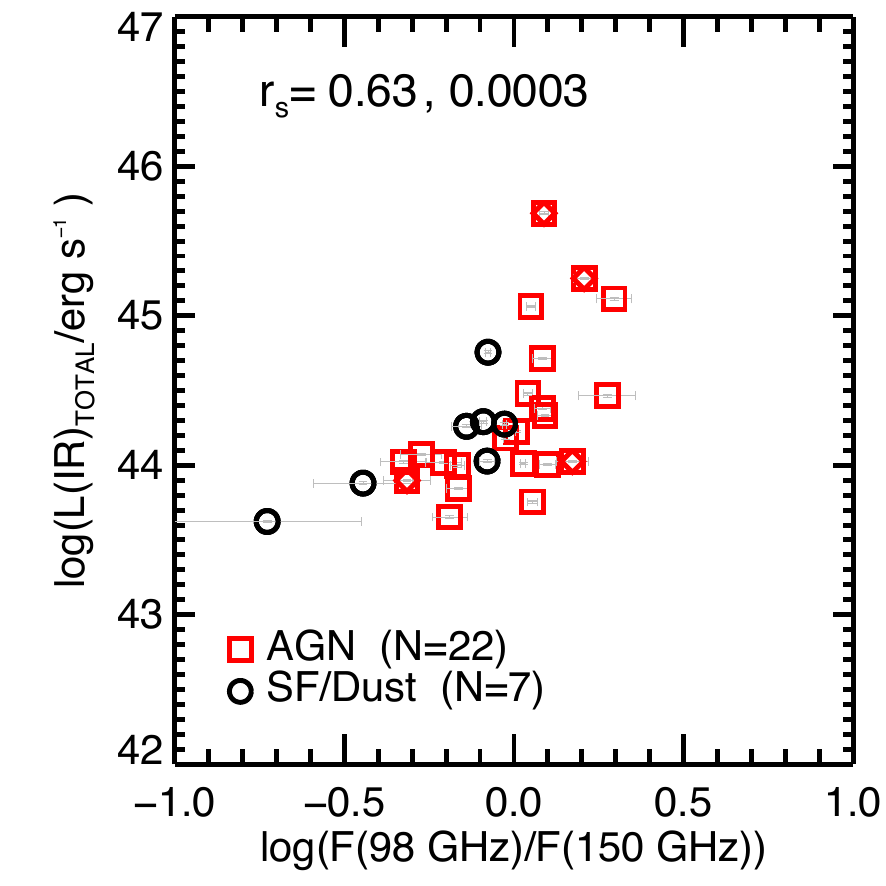}\\
\end{array}$
\end{center}
\caption{ The physical parameters $L_{\rm{dust}}$ and $L\rm{(IR)}_{\rm{TOTAL}}$ measured from the SEDs vs. the $F_{\rm{98\,GHz }}/F_{\rm{150\,GHz}}$ color. The red squares represent the AGNs and the black circles represent the SF galaxies. The Spearman rank correlation coefficient values $r_{s}$ are listed, with significances  of 0.002 (left) and 0.0003 (right). }
\label{fig:fig11}
\end{figure*}

Any possible correlations between the physical parameters $(L\rm{(IR)}_{\rm{TOTAL}}$, $L_{\rm{dust}}$, $L\rm{(IR)}_{\rm{AGN}}$,  frac$_\mathrm{AGN}$, and $\psi_\mathrm{AGN})$ measured directly from the SEDs and the FIR-to-millimeter wave and millimeter-wave colors are investigated. For this investigation, we include the AGN parameters of all galaxies, as measured from the SEDs, without applying any criteria for the AGN selection. 
The strengths of the correlations are measured using the Spearman rank correlation coefficient ($r_{s}$).  As a result, we only find moderate correlations between $\log(L\rm{(IR)}_{\rm{dust}})$ and the $\log(F_{\rm{98\,GHz}}/F_{\rm{150\,GHz}})$ color ($r_{s}$ = 0.54, with a significance\footnote[9]{The significance values show that there is less than a 0.1\% chance that the found $r_{s}$ occurred by chance, if there were no correlations between these parameters.} of 0.002) and
$\log(L\rm{(IR)}_{\rm{TOTAL}})$ and the $\log(F_{\rm{98\,GHz}}/F_{\rm{150\,GHz}})$ color ($r_{s}$ = 0.63, with a significance of 0.0003), as shown in Fig. \ref{fig:fig11}. These moderate correlations can be used as crude estimates for $L_{\rm{dust}}$ and $L\rm{(IR)}_{\rm{TOTAL}}$ from the millimeter-wave colors, in the absence of redshift information.

\section{Discussion}\label{S:dis}

\subsection{Comparison with Previous Work on Millimeter-selected Galaxies }

As reported by the previous studies of millimeter-selected galaxies \citep[e.g., ][]{Planck2011b,Gralla2020}, using ACT or Planck, a subset of these galaxies are DSFGs. Among those DSFGs, most sources are high-redshift lensed galaxies, while others are low-redshift star-forming IR galaxies. The main focuses of the earlier work on these low-redshift millimeter-selected IR galaxies have been detections in the millimeter maps, counterpart identifications, measuring the flux densities for the ACT bands, obtaining ACT spectral indices from the multifrequency data, and source classification (AGN/DSFG), based on the spectral indices \citep[e.g.,][]{Marriage2011a,Marsden2014,Datta2019,Gralla2020}. 

In \S \ref{S:spectralindices}, we computed millimeter-wave spectral indices and compared the sample mean values with the values that have been given by \citet{Marsden2014} and \citet{Gralla2020}. Although our sample size is much smaller than theirs (less than a quarter of their sample sizes), our  median value of $\alpha_{150-220} = 2.59 \pm 1.30$ (which is computed from the two overlapping ACT bands with their studies) is lower compared to their reported values (although it agrees within the sample standard deviation). 

The SEDs of ACT-detected millimeter DSFGs have also been investigated in previous studies, by including FIR data \citep[e.g., ][]{su2017,Gralla2020}. However, these studies only examined the SEDs of high-redshift lensed DSFGs. Therefore, we cannot make a direct comparison between the SEDs of low-redshift IR galaxies analyzed in this work and those studies.
 
\subsection{Implications of Including ACT Millimeter Bands in the SEDs}

When the traditional UV-to-FIR SEDs expand to millimeter wavelengths, depending on the FIR-to-millimete -wave and/or millimeter flux ratios, namely the colors, a radio synchrotron emission component is needed to obtain a valid SED model. Such color limits are determined in \S \ref{S:infraredACTcol}. As a following investigation, we check the color distributions of the sources with different origins. As shown in Figure \ref{fig:fig10}, the color distributions   for the dust, SF, and AGN emission show differences. Therefore, the color range that we obtained in \S \ref{S:infraredACTcol} can be used as a crude indication of the different origins of millimeter-band detections. However, it is valid to clearly separate the dust emission and  radio synchrotron component from the AGN and/or SF.

 We emphasize that the minor group of dust emission- originating millimeter emission galaxies constitute very important representatives of high-redshift star forming galaxies, which are our best tools for investigating these DSFGs beyond redshift $z= 2$. Additionally, our SF radio galaxy group shows that radio emission does not necessarily mean significant AGN contamination in the millimeter band, as such galaxies at higher redshifts should be considered as being SF-dominated. We recommend obtaining an SED decomposition for this type of galaxy.

\subsection{Future Prospects with JWST }
The James Webb Space Telescope (JWST) has unique capabilities in terms of observing the near-IR to mid-IR emission from these IR galaxies. For example, four nearby IR galaxies have already been selected for inclusion in the  Director's Discretionary Early Release Science Program \#1328. These IR galaxies are not in our sample, but their observations will show the full potential of JWST MIRI and NIRSpec observations, which will observe the IR SEDs in detail, by separating the IR dust continuum emission lines, like PAHs and other atomic and molecular emission lines. Most importantly, it will be possible to detect the AGN components in the mid-IR spectra obtained by JWST. Therefore, IR galaxies that have a radio component, as revealed by the millimeter data with small AGN fractions (obtained from our SED analysis), may reveal their possibly hidden AGNs by JWST. 

\section{Conclusions}\label{S:conc}

We search for the millimeter counterparts of  IR galaxies selected by AKARI and IRAS in the ACT DR5 maps at the 98 GHz, 150 GHz, and 220 GHz frequency bands.
In total, we report  167 millimeter counterparts, of which  134 are new identifications.
We list the flux density measurements in the three ACT frequency bands, when available. 
We analyze the UV-to-millimeter SEDs of 87 galaxies in our sample that fulfill the data requirements. We measure the total IR luminosity from the best-fitting SEDs and  investigate the origins of  the millimeter emission. 
The final conclusions of this work can be summarized as the follows: 
\begin{enumerate}

\item 
As a result of this work, the number of galaxies detected by ACT has been extended. In particular, the 98 GHz flux densities are new measurements from the recent DR5 maps for the reported  IR galaxies in our sample. 

\item
We report millimeter-wave spectral indices ($\alpha_{98-150}$, $\alpha_{98-220}$, and $\alpha_{150-220}$) for our IR galaxy sample, specifying the source type (AGN or DSFG), based on  $\alpha_{150-220}$. As expected (since our sample was selected as IR galaxies), the majority of the galaxies in our sample are consistent with the $\alpha_{150-220}$ values of DSFGs. The spectral indices computed from the 98 GHz band  $\alpha_{98-150}$ and  $\alpha_{98-220}$ are new measurements for ACT-detected galaxies.  

\item
 Using the latest version of CIGALE, V2022.1, we identify the origins of the millimeter emissions using SED fitting. We also show that CIGALE V2022.1 can model SEDs including millimeter bands, with its radio emission decomposition capability. Therefore,  we recommend it for studying higher-redshift IR galaxies selected by millimeter radiation. 

\item
We analyze the SEDs of IR galaxies selected by AKARI and IRAS and detected in the ACT DR 5 maps, for the first time, by including UV, optical, near-IR, mid-IR, FIR, and millimeter data in CIGALE v2022. As a result, we measure important physical parameters of these galaxies, including $L\rm{(IR)}_{\rm{TOTAL}}$, $L_{\rm{dust}}$, $L\rm{(IR)}_{\rm{AGN}}$,  frac$_\mathrm{AGN}$, and $\psi_\mathrm{AGN}$. We quantify the relationships between the $L\rm{(IR)}_{\rm{TOTAL}}$ and $L\rm{(IR)}_{\rm{AGN}}$ luminosities and the 98 GHz, 150 GHz, and 220 GHz band luminosities measured from the SED models.  These relations can be used as practical tools for inferring the AGN and total IR luminosities from the single millimeter-band luminosities, especially for galaxies with redshift measurements but without multi wavelength UV-to-radio data, to model their SEDs. 

\item
 Our analysis shows that the origins of the millimeter emission can be identified by the decomposition of the radio emission. We find that for more than half of our sample, the origin of the millimeter emission is from AGN and/or SF synchrotron emission, while in only six of 87 galaxies, the millimeter emission is produced by the dust emission. We also provide millimeter color ranges that indicate the different origins of the millimeter emission.

\end{enumerate}

\begin{acknowledgments}
 We thank the anonymous reviewer for the insightful and constructive comments. E.K. thanks O\u{g}uzhan Tekin for his technical support. 
T.G. acknowledges the support from the National Science and Technology Council of Taiwan, through grants 108-2628-M-007-004-MY3 and 110-2112-M-005-013-MY3. 
T.H. acknowledges support from the National Science and Technology Council of Taiwan, through grants 110-2112-M-005-013-MY3, 110-2112-M-007-034-, and 111-2123-M-001-008-.

This research is based on observations made with AKARI, a JAXA project, with the participation of ESA. 
This research has made use of the NASA/IPAC Extragalactic Database (NED), which is funded by the National Aeronautics and Space Administration and operated by the California Institute of Technology. 
Funding for the Sloan Digital Sky Survey IV has been provided by the Alfred P. Sloan Foundation, the U.S. Department of Energy Office of Science, and the Participating Institutions. 
SDSS-IV acknowledges support and resources from the Center for High Performance Computing  at the University of Utah. The SDSS website is www.sdss.org.

SDSS-IV is managed by the 
Astrophysical Research Consortium 
for the Participating Institutions 
of the SDSS Collaboration, including 
the Brazilian Participation Group, 
the Carnegie Institution for Science, 
Carnegie Mellon University, Center for 
Astrophysics | Harvard \& 
Smithsonian, the Chilean Participation 
Group, the French Participation Group, 
Instituto de Astrof\'isica de 
Canarias, The Johns Hopkins 
University, Kavli Institute for the 
Physics and Mathematics of the 
Universe (IPMU) / University of 
Tokyo, the Korean Participation Group, 
Lawrence Berkeley National Laboratory, 
Leibniz Institut f\"ur Astrophysik 
Potsdam (AIP),  Max-Planck-Institut 
f\"ur Astronomie (MPIA Heidelberg), 
Max-Planck-Institut f\"ur 
Astrophysik (MPA Garching), 
Max-Planck-Institut f\"ur 
Extraterrestrische Physik (MPE), 
National Astronomical Observatories of 
China, New Mexico State University, 
New York University, University of 
Notre Dame, Observat\'ario 
Nacional / MCTI, The Ohio State 
University, Pennsylvania State 
University, Shanghai 
Astronomical Observatory, United 
Kingdom Participation Group, 
Universidad Nacional Aut\'onoma 
de M\'exico, University of Arizona, 
University of Colorado Boulder, 
University of Oxford, University of 
Portsmouth, University of Utah, 
University of Virginia, University 
of Washington, University of 
Wisconsin, Vanderbilt University, 
and Yale University.

This publication makes use of data products from the Two Micron All Sky Survey, which is a joint project of the
University of Massachusetts and the Infrared Processing and Analysis Center/California Institute of Technology,
funded by the National Aeronautics and Space Administration and the National Science Foundation.
\end{acknowledgments}

%

\vspace{5mm}
\facilities{ACT, AKARI, WISE, Herschel, GALEX}

\software{CIGALE \citep{Boquien2019}, 
          Source-Extractor \citep{Bertin1996}
          }

\appendix
In the following we explain the adopted optimized apertures for the spatially extended sources
using the growth curves.

\section{Growth curves of spatially extended sources}\label{S:appendix}
 
\begin{figure*}[h]
\begin{center}$
\begin{array}{lll}
\includegraphics[width=0.32\columnwidth]{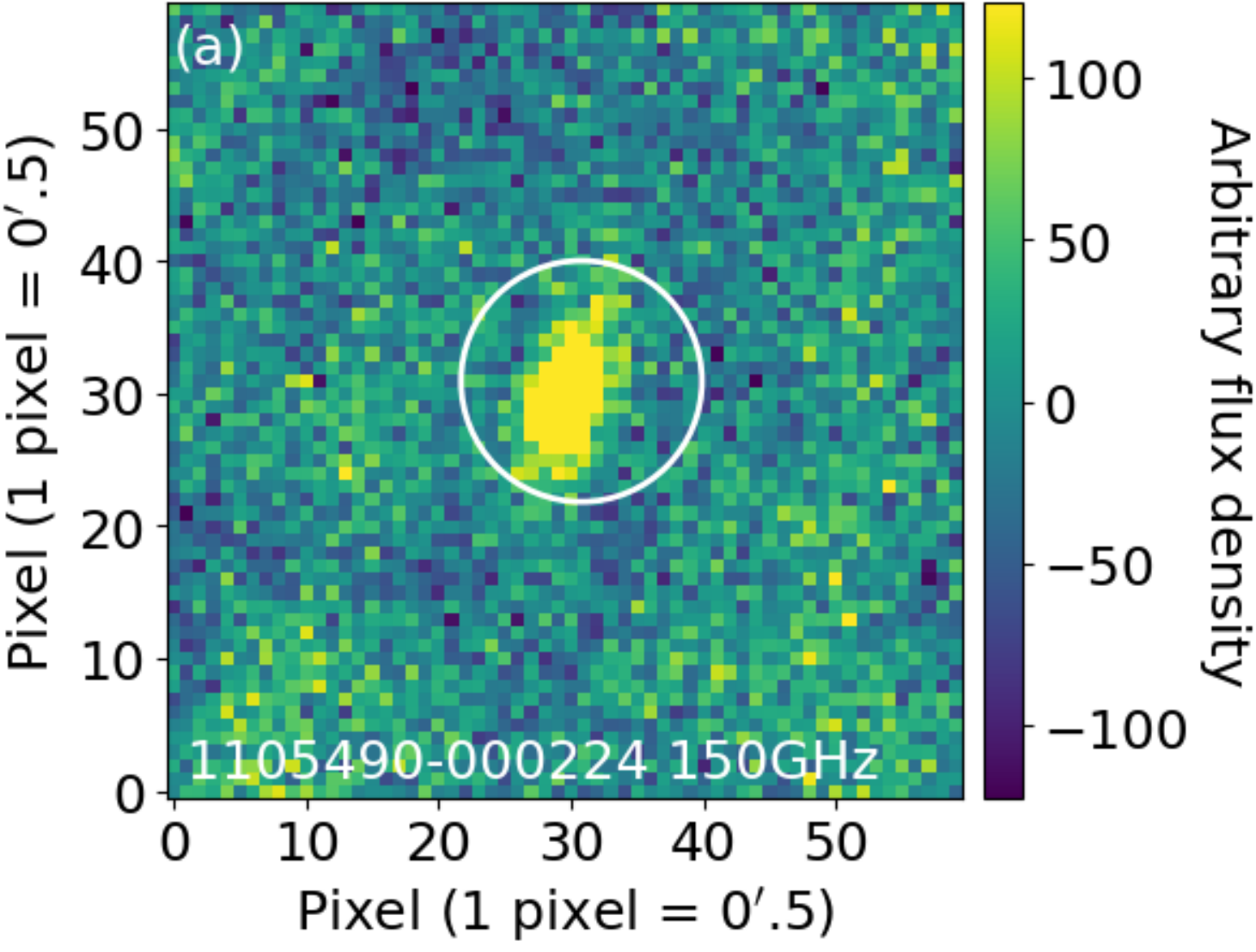} &
\includegraphics[width=0.32\columnwidth]{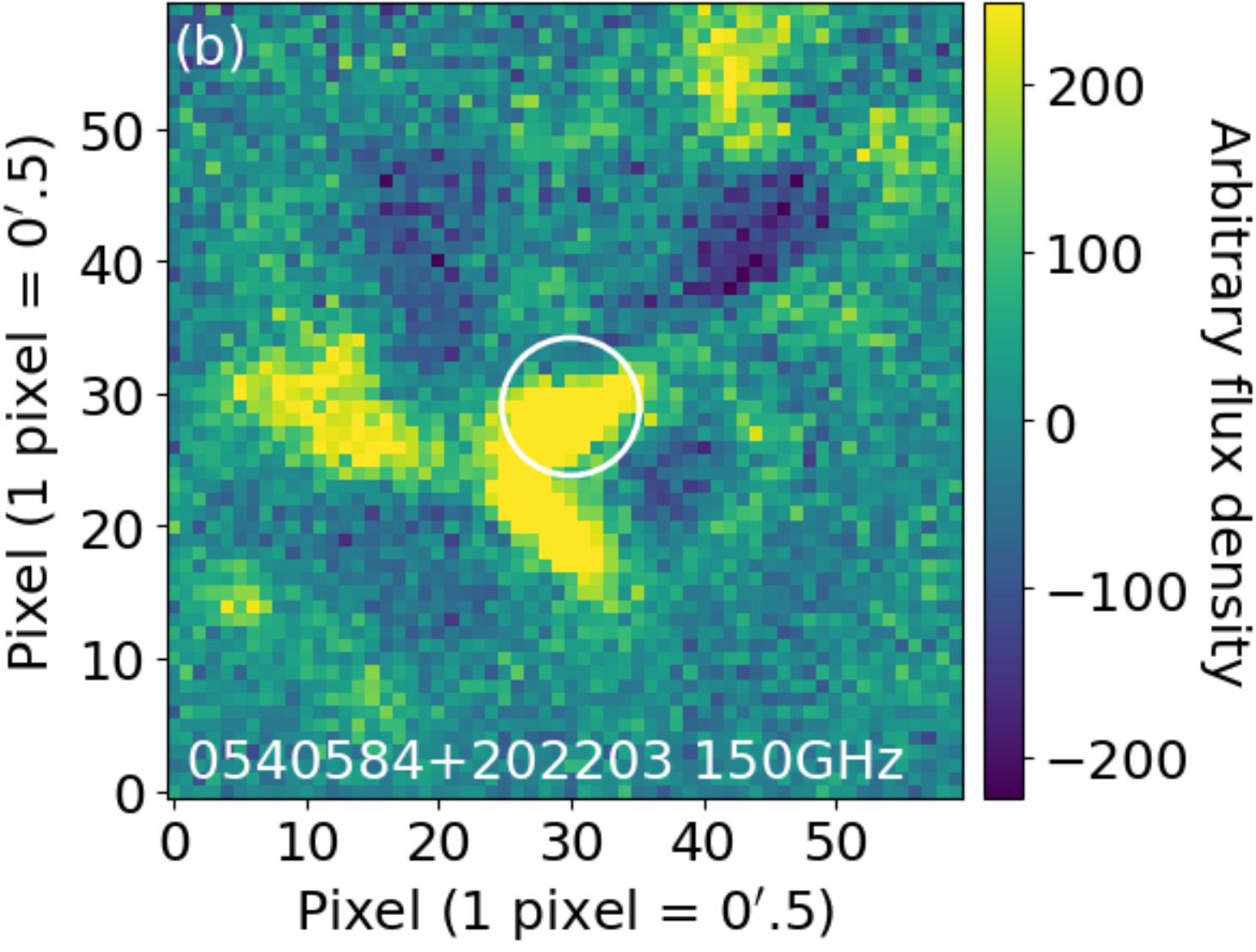} &
\includegraphics[width=0.32\columnwidth]{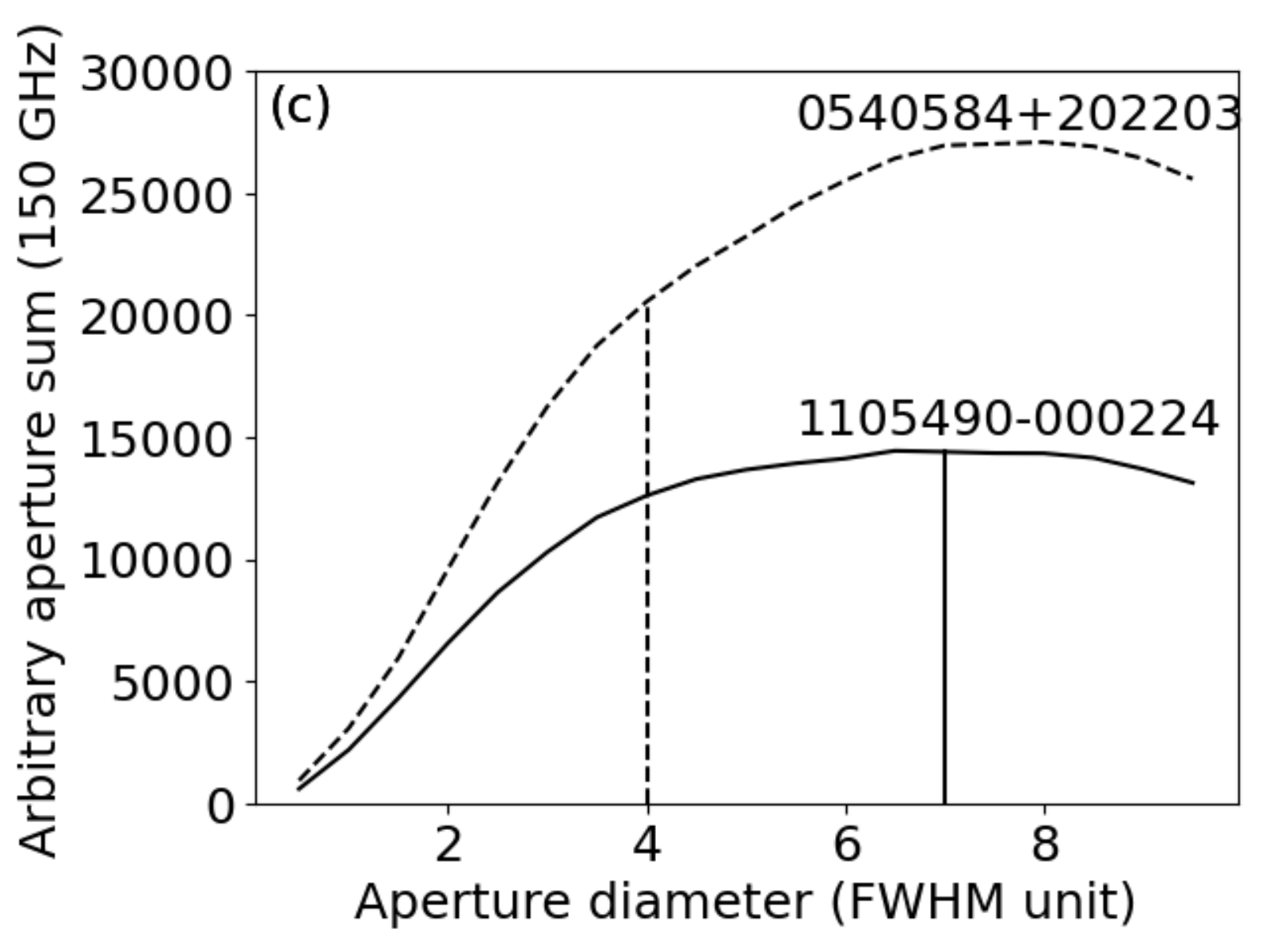}\\
\end{array}$
\end{center}
 \caption{
 Examples of spatially extended sources at 150 GHz.
Left:image of the isolated source matched with AKARI ID 1105490-000224.
 Middle: image of the merging system matched with AKARI ID 0540584+202203.
 The white circles indicate the aperture sizes adopted in our analysis.
 Right: the total flux density within the aperture as a function of the aperture size for sources matched with AKARI IDs 1105490-000224 and 0540584+202203.
 The solid and dashed lines correspond to 1105490-000224 and 0540584+202203, respectively.
 The adopted aperture size for 1105490-000224 (0540584+202203) is shown by the solid (dashed) vertical line.}
\label{fig:figA1}
\end{figure*}

The spatially extended sources are selected as described in Section \ref{S:sourceext}, and they include isolated sources and one merging system.
Panels (a) and (b) of Fig. \ref{fig:figA1} show examples of the spatially extended sources at 150 GHz: one of the isolated galaxies and one of the merging system.
Different aperture sizes are tested for the extended sources, to investigate ``growth curves''--i.e., the total flux density within the aperture as a function of the aperture size.
Panel (c) of Fig. \ref{fig:figA1} indicates the growth curves of the sources.
To estimate noises with different aperture sizes, the random aperture photometries are conducted with different aperture sizes, following the same manner as described in Section \ref{S:sourcedet}.
For the isolated extended sources, we adopt the aperture size at which the total flux densities converge (the solid vertical line in Fig. \ref{fig:figA1}c), unless the S/N is lower than 5.
The S/N decreases with increasing aperture size for the isolated extended samples in this work.
Therefore, some isolated extended sources show S/N $ < 5$ at the converged aperture sizes. 
In such cases, we use the largest aperture sizes, as long as the S/Ns are higher than 5.
The merging system presented in Fig. \ref{fig:figA1}b has a close companion. 
In this case, We adopt a flexible aperture size to avoid the contaminated flux density from the companion galaxy.
Using these optimized aperture sizes for the spatially extended sources, their flux densities are measured, instead of the point-source aperture correction.

\bibliography{actpaper}{}

\begin{thebibliography}{}
\expandafter\ifx\csname natexlab\endcsname\relax\def\natexlab#1{#1}\fi
\providecommand{\url}[1]{\href{#1}{#1}}
\providecommand{\dodoi}[1]{doi:~\href{http://doi.org/#1}{\nolinkurl{#1}}}
\providecommand{\doeprint}[1]{\href{http://ascl.net/#1}{\nolinkurl{http://ascl.net/#1}}}
\providecommand{\doarXiv}[1]{\href{https://arxiv.org/abs/#1}{\nolinkurl{https://arxiv.org/abs/#1}}}

\bibitem[{{Abdurro'uf} {et~al.}(2022){Abdurro'uf}, {Accetta}, {Aerts}, {Silva
  Aguirre}, {et~al.}}]{Abdurrouf2022}
{Abdurro'uf}, {Accetta}, K., {Aerts}, C., {Silva Aguirre}, V., {et~al.} 2022,
  \apjs, 259, 35, \dodoi{10.3847/1538-4365/ac4414}

\bibitem[{{AKARI Team}(2020)}]{AKARIIRC}
{AKARI Team}. 2020, AKARI/IRC Point Source Catalogue,  IPAC,
  \dodoi{10.26131/IRSA181}

\bibitem[{{Becker} {et~al.}(1995){Becker}, {White}, \& {Helfand}}]{Becker1995}
{Becker}, R.~H., {White}, R.~L., \& {Helfand}, D.~J. 1995, \apj, 450, 559,
  \dodoi{10.1086/176166}

\bibitem[{{Bertin} \& {Arnouts}(1996)}]{Bertin1996}
{Bertin}, E., \& {Arnouts}, S. 1996, \aaps, 117, 393,
  \dodoi{10.1051/aas:1996164}

\bibitem[{{B{\"o}hringer} {et~al.}(2004)}]{Bohringer2004}
{B{\"o}hringer}, H., {et~al.} 2004, \aap, 425, 367,
  \dodoi{10.1051/0004-6361:20034484}

\bibitem[{{Boquien} {et~al.}(2019)}]{Boquien2019}
{Boquien}, M., {et~al.} 2019, \aap, 622, A103,
  \dodoi{10.1051/0004-6361/201834156}

\bibitem[{{Cardelli} {et~al.}(1989){Cardelli}, {Clayton}, \&
  {Mathis}}]{Cardelli1989}
{Cardelli}, J.~A., {Clayton}, G.~C., \& {Mathis}, J.~S. 1989, \apj, 345, 245,
  \dodoi{10.1086/167900}

\bibitem[{{Carlstrom} {et~al.}(2011)}]{Carlstrom2011}
{Carlstrom}, J.~E., {et~al.} 2011, \pasp, 123, 568, \dodoi{10.1086/659879}

\bibitem[{{Casagrande} {et~al.}(2019){Casagrande}, {Wolf}, {Mackey},
  {Nordlander}, {Yong}, \& {Bessell}}]{Casagrande2019}
{Casagrande}, L., {Wolf}, C., {Mackey}, A.~D., {et~al.} 2019, \mnras, 482,
  2770, \dodoi{10.1093/mnras/sty2878}

\bibitem[{{Casey} {et~al.}(2021)}]{Casey2021}
{Casey}, C.~M., {et~al.} 2021, \apj, 923, 215, \dodoi{10.3847/1538-4357/ac2eb4}

\bibitem[{{Charlot} \& {Fall}(2000)}]{Charlot&Fall2000}
{Charlot}, S., \& {Fall}, S.~M. 2000, \apj, 539, 718, \dodoi{10.1086/309250}

\bibitem[{{Chary} \& {Elbaz}(2001)}]{Chary2001}
{Chary}, R., \& {Elbaz}, D. 2001, \apj, 556, 562, \dodoi{10.1086/321609}

\bibitem[{{Ciesla} {et~al.}(2015)}]{Ciesla2015}
{Ciesla}, L., {et~al.} 2015, \aap, 576, A10,
  \dodoi{10.1051/0004-6361/201425252}

\bibitem[{{Ciesla} {et~al.}(2016)}]{Ciesla2016}
---. 2016, \aap, 585, A43, \dodoi{10.1051/0004-6361/201527107}

\bibitem[{{Clark} {et~al.}(2018)}]{Clark2018}
{Clark}, C.~J.~R., {et~al.} 2018, \aap, 609, A37,
  \dodoi{10.1051/0004-6361/201731419}

\bibitem[{{Clements} {et~al.}(2019){Clements}, {Rowan-Robinson}, {Pearson},
  {et~al.}}]{Clements2019}
{Clements}, D.~L., {Rowan-Robinson}, M., {Pearson}, C., {et~al.} 2019, \pasj,
  71, 7, \dodoi{10.1093/pasj/psy099}

\bibitem[{{Condon} {et~al.}(1998)}]{Condon1998}
{Condon}, J.~J., {et~al.} 1998, \aj, 115, 1693, \dodoi{10.1086/300337}

\bibitem[{{Cutri} {et~al.}(2003){Cutri}, {Skrutskie}, {van Dyk},
  {et~al.}}]{Cutri2003}
{Cutri}, R.~M., {Skrutskie}, M.~F., {van Dyk}, S., {et~al.} 2003, VizieR Online
  Data Catalog, II/246

\bibitem[{{Cutri} {et~al.}(2021){Cutri}, {Wright}, {Conrow},
  {et~al.}}]{Cutri2013}
{Cutri}, R.~M., {Wright}, E.~L., {Conrow}, T., {et~al.} 2021, VizieR Online
  Data Catalog, II/328

\bibitem[{{da Cunha} {et~al.}(2008){da Cunha}, {Charlot}, \&
  {Elbaz}}]{daCunha2008}
{da Cunha}, E., {Charlot}, S., \& {Elbaz}, D. 2008, \mnras, 388, 1595,
  \dodoi{10.1111/j.1365-2966.2008.13535.x}

\bibitem[{{Dale} \& {Helou}(2002)}]{Dale2002}
{Dale}, D.~A., \& {Helou}, G. 2002, \apj, 576, 159, \dodoi{10.1086/341632}

\bibitem[{{Dale} {et~al.}(2014){Dale}, {Helou}, {Magdis}, {Armus},
  {D{\'\i}az-Santos}, \& {Shi}}]{Dale2014}
{Dale}, D.~A., {Helou}, G., {Magdis}, G.~E., {et~al.} 2014, \apj, 784, 83,
  \dodoi{10.1088/0004-637X/784/1/83}

\bibitem[{{Das} {et~al.}(2011)}]{Das2011}
{Das}, S., {et~al.} 2011, \apj, 729, 62, \dodoi{10.1088/0004-637X/729/1/62}

\bibitem[{{Datta} {et~al.}(2019)}]{Datta2019}
{Datta}, R., {et~al.} 2019, \mnras, 486, 5239, \dodoi{10.1093/mnras/sty2934}

\bibitem[{{Delvecchio} {et~al.}(2021)}]{Delvecchio2021}
{Delvecchio}, I., {et~al.} 2021, \aap, 647, A123,
  \dodoi{10.1051/0004-6361/202039647}

\bibitem[{{Draine} {et~al.}(2014)}]{Draine2014}
{Draine}, B.~T., {et~al.} 2014, \apj, 780, 172,
  \dodoi{10.1088/0004-637X/780/2/172}

\bibitem[{{Efstathiou} \& {Rowan-Robinson}(2003)}]{Efstathiou2003}
{Efstathiou}, A., \& {Rowan-Robinson}, M. 2003, \mnras, 343, 322,
  \dodoi{10.1046/j.1365-8711.2003.06679.x}

\bibitem[{{Everett} {et~al.}(2020)}]{Everett2020}
{Everett}, W.~B., {et~al.} 2020, \apj, 900, 55,
  \dodoi{10.3847/1538-4357/ab9df7}

\bibitem[{{Fixsen}(2009)}]{Fixsen2009}
{Fixsen}, D.~J. 2009, \apj, 707, 916, \dodoi{10.1088/0004-637X/707/2/916}

\bibitem[{{Fowler} {et~al.}(2007)}]{Fowler2007}
{Fowler}, J.~W., {et~al.} 2007, \ao, 46, 3444, \dodoi{10.1364/AO.46.003444}

\bibitem[{{Fritz} {et~al.}(2006){Fritz}, {Franceschini}, \&
  {Hatziminaoglou}}]{Fritz2006}
{Fritz}, J., {Franceschini}, A., \& {Hatziminaoglou}, E. 2006, \mnras, 366,
  767, \dodoi{10.1111/j.1365-2966.2006.09866.x}

\bibitem[{{Goto} {et~al.}(2010)}]{Goto2010}
{Goto}, T., {et~al.} 2010, \aap, 514, A6, \dodoi{10.1051/0004-6361/200913182}

\bibitem[{{Gralla} {et~al.}(2020)}]{Gralla2020}
{Gralla}, M.~B., {et~al.} 2020, \apj, 893, 104,
  \dodoi{10.3847/1538-4357/ab7915}

\bibitem[{{Griffin} {et~al.}(2010)}]{Griffin2010}
{Griffin}, M.~J., {et~al.} 2010, \aap, 518, L3,
  \dodoi{10.1051/0004-6361/201014519}

\bibitem[{{Hall} {et~al.}(2010)}]{Hall2010}
{Hall}, N.~R., {et~al.} 2010, \apj, 718, 632,
  \dodoi{10.1088/0004-637X/718/2/632}

\bibitem[{{Helfand} {et~al.}(2015){Helfand}, {White}, \&
  {Becker}}]{Helfand2015}
{Helfand}, D.~J., {White}, R.~L., \& {Becker}, R.~H. 2015, \apj, 801, 26,
  \dodoi{10.1088/0004-637X/801/1/26}

\bibitem[{{Helou} {et~al.}(1985){Helou}, {Soifer}, \&
  {Rowan-Robinson}}]{Helou1985}
{Helou}, G., {Soifer}, B.~T., \& {Rowan-Robinson}, M. 1985, \apjl, 298, L7,
  \dodoi{10.1086/184556}

\bibitem[{{Helou} \& {Walker}(1988)}]{HelouWalker1988}
{Helou}, G., \& {Walker}, D.~W. 1988, NASA RP-1190, 7, 0

\bibitem[{{Henderson} {et~al.}(2016)}]{Henderson2016}
{Henderson}, S.~W., {et~al.} 2016, Journal of Low Temperature Physics, 184,
  772, \dodoi{10.1007/s10909-016-1575-z}

\bibitem[{{Hilton} {et~al.}(2021)}]{Hilton2021}
{Hilton}, M., {et~al.} 2021, \apjs, 253, 3, \dodoi{10.3847/1538-4365/abd023}

\bibitem[{{Huchra} {et~al.}(2005){Huchra}, {Jarrett}, {Skrutskie},
  {et~al.}}]{Huchra2005}
{Huchra}, J., {Jarrett}, T., {Skrutskie}, M., {et~al.} 2005, in Astronomical
  Society of the Pacific Conference Series, Vol. 329, Nearby Large-Scale
  Structures and the Zone of Avoidance, ed. A.~P. {Fairall} \& P.~A. {Woudt},
  135

\bibitem[{{Huchra} {et~al.}(2012)}]{Huchra2012}
{Huchra}, J.~P., {et~al.} 2012, \apjs, 199, 26,
  \dodoi{10.1088/0067-0049/199/2/26}

\bibitem[{{Ishihara} {et~al.}(2010)}]{Ishihara2010}
{Ishihara}, D., {et~al.} 2010, \aap, 514, A1,
  \dodoi{10.1051/0004-6361/200913811}

\bibitem[{{Jarrett} {et~al.}(2000){Jarrett}, {Chester}, {Cutri}, {Schneider},
  {Skrutskie}, \& {Huchra}}]{Jarrett2000}
{Jarrett}, T.~H., {Chester}, T., {Cutri}, R., {et~al.} 2000, \aj, 119, 2498,
  \dodoi{10.1086/301330}

\bibitem[{{Jarvis} {et~al.}(2020)}]{Jarvis2020}
{Jarvis}, M.~E., {et~al.} 2020, \mnras, 498, 1560,
  \dodoi{10.1093/mnras/staa2196}

\bibitem[{{Jones} {et~al.}(2005){Jones}, {Saunders}, {Read}, \&
  {Colless}}]{Jones2005}
{Jones}, D.~H., {Saunders}, W., {Read}, M., \& {Colless}, M. 2005, \pasa, 22,
  277, \dodoi{10.1071/AS05018}

\bibitem[{{Jones} {et~al.}(2004)}]{Jones2004}
{Jones}, D.~H., {et~al.} 2004, \mnras, 355, 747,
  \dodoi{10.1111/j.1365-2966.2004.08353.x}

\bibitem[{{Jones} {et~al.}(2009)}]{Jones2009}
---. 2009, \mnras, 399, 683, \dodoi{10.1111/j.1365-2966.2009.15338.x}

\bibitem[{{Kelly}(2007)}]{Kelly2007}
{Kelly}, B.~C. 2007, \apj, 665, 1489, \dodoi{10.1086/519947}

\bibitem[{{Kilerci Eser} \& {Goto}(2018)}]{KilerciEser2018}
{Kilerci Eser}, E., \& {Goto}, T. 2018, \mnras, 474, 5363,
  \dodoi{10.1093/mnras/stx3110}

\bibitem[{{Ku{\'z}micz} {et~al.}(2018){Ku{\'z}micz}, {Jamrozy}, {Bronarska},
  {Janda-Boczar}, \& {Saikia}}]{Kuzmicz2018}
{Ku{\'z}micz}, A., {Jamrozy}, M., {Bronarska}, K., {Janda-Boczar}, K., \&
  {Saikia}, D.~J. 2018, \apjs, 238, 9, \dodoi{10.3847/1538-4365/aad9ff}

\bibitem[{{Le Floc'h} {et~al.}(2005)}]{LeFloch2005}
{Le Floc'h}, E., {et~al.} 2005, \apj, 632, 169, \dodoi{10.1086/432789}

\bibitem[{{Lianou} {et~al.}(2019){Lianou}, {Barmby}, {Mosenkov}, {Lehnert}, \&
  {Karczewski}}]{Lianou2019}
{Lianou}, S., {Barmby}, P., {Mosenkov}, A.~A., {Lehnert}, M., \& {Karczewski},
  O. 2019, \aap, 631, A38, \dodoi{10.1051/0004-6361/201834553}

\bibitem[{{Lonsdale} {et~al.}(2006){Lonsdale}, {Farrah}, \&
  {Smith}}]{Lonsdale2006}
{Lonsdale}, C.~J., {Farrah}, D., \& {Smith}, H.~E. 2006, {Ultraluminous
  Infrared Galaxies}, ed. J.~W. {Mason}, 285, \dodoi{10.1007/3-540-30313-8_9}

\bibitem[{{Ma{\l}ek} {et~al.}(2017)}]{Malek2017}
{Ma{\l}ek}, K., {et~al.} 2017, \aap, 598, A1,
  \dodoi{10.1051/0004-6361/201527969}

\bibitem[{{Maraston}(2005)}]{Maraston2005}
{Maraston}, C. 2005, \mnras, 362, 799, \dodoi{10.1111/j.1365-2966.2005.09270.x}

\bibitem[{{Maraston} {et~al.}(2009){Maraston}, {Str{\"o}mb{\"a}ck}, {Thomas},
  {Wake}, \& {Nichol}}]{Maraston2009}
{Maraston}, C., {Str{\"o}mb{\"a}ck}, G., {Thomas}, D., {Wake}, D.~A., \&
  {Nichol}, R.~C. 2009, \mnras, 394, L107,
  \dodoi{10.1111/j.1745-3933.2009.00621.x}

\bibitem[{{Marriage} {et~al.}(2011)}]{Marriage2011a}
{Marriage}, T.~A., {et~al.} 2011, \apj, 731, 100,
  \dodoi{10.1088/0004-637X/731/2/100}

\bibitem[{{Marsden} {et~al.}(2014)}]{Marsden2014}
{Marsden}, D., {et~al.} 2014, \mnras, 439, 1556, \dodoi{10.1093/mnras/stu001}

\bibitem[{{Martin} {et~al.}(2005)}]{Martin2005}
{Martin}, D.~C., {et~al.} 2005, \apjl, 619, L1, \dodoi{10.1086/426387}

\bibitem[{{Mauch} {et~al.}(2003)}]{Mauch2003}
{Mauch}, T., {et~al.} 2003, \mnras, 342, 1117,
  \dodoi{10.1046/j.1365-8711.2003.06605.x}

\bibitem[{{Mocanu} {et~al.}(2013)}]{Mocanu2013}
{Mocanu}, L.~M., {et~al.} 2013, \apj, 779, 61,
  \dodoi{10.1088/0004-637X/779/1/61}

\bibitem[{{Murakami} {et~al.}(2007)}]{Murakami2007}
{Murakami}, H., {et~al.} 2007, \pasj, 59, 369.
\newblock \doarXiv{0708.1796}

\bibitem[{{Murphy} {et~al.}(2011)}]{Murphy2011}
{Murphy}, E.~J., {et~al.} 2011, \apj, 737, 67,
  \dodoi{10.1088/0004-637X/737/2/67}

\bibitem[{{Murphy} {et~al.}(2012)}]{Murphy2012}
---. 2012, \apj, 761, 97, \dodoi{10.1088/0004-637X/761/2/97}

\bibitem[{{Murphy} {et~al.}(2010)}]{Murphy2010}
{Murphy}, T., {et~al.} 2010, \mnras, 402, 2403,
  \dodoi{10.1111/j.1365-2966.2009.15961.x}

\bibitem[{{Naess} {et~al.}(2020)}]{Naess2020}
{Naess}, S., {et~al.} 2020, \jcap, 2020, 046,
  \dodoi{10.1088/1475-7516/2020/12/046}

\bibitem[{{Noll} {et~al.}(2009){Noll}, {Burgarella}, {Giovannoli}, {Buat},
  {Marcillac}, \& {Mu{\~n}oz-Mateos}}]{Noll2009}
{Noll}, S., {Burgarella}, D., {Giovannoli}, E., {et~al.} 2009, \aap, 507, 1793,
  \dodoi{10.1051/0004-6361/200912497}

\bibitem[{{Onken} {et~al.}(2019)}]{Onken2019}
{Onken}, C.~A., {et~al.} 2019, \pasa, 36, e033, \dodoi{10.1017/pasa.2019.27}

\bibitem[{{Pilbratt} {et~al.}(2010)}]{Pilbratt2010}
{Pilbratt}, G.~L., {et~al.} 2010, \aap, 518, L1,
  \dodoi{10.1051/0004-6361/201014759}

\bibitem[{{Planck Collaboration} {et~al.}(2011){Planck Collaboration}, {Ade},
  {Aghanim}, {et~al.}}]{Planck2011}
{Planck Collaboration}, {Ade}, P.~A.~R., {Aghanim}, N., {et~al.} 2011, \aap,
  536, A1, \dodoi{10.1051/0004-6361/201116464}

\bibitem[{{Planck Collaboration} {et~al.}(2014{\natexlab{a}}){Planck
  Collaboration}, {Ade}, {Aghanim}, {et~al.}}]{Planck2014a}
---. 2014{\natexlab{a}}, \aap, 571, A1, \dodoi{10.1051/0004-6361/201321529}

\bibitem[{{Planck Collaboration} {et~al.}(2014{\natexlab{b}}){Planck
  Collaboration}, {Ade}, {Aghanim}, {et~al.}}]{Planck2014b}
---. 2014{\natexlab{b}}, \aap, 571, A15, \dodoi{10.1051/0004-6361/201321573}

\bibitem[{{Planck Collaboration} {et~al.}(2016){Planck Collaboration}, {Ade},
  {Aghanim}, {et~al.}}]{PlanckCollaboration2016b}
---. 2016, \aap, 594, A26, \dodoi{10.1051/0004-6361/201526914}

\bibitem[{{Planck Collaboration} {et~al.}(2020){Planck Collaboration},
  {Aghanim}, {Akrami, Y.}, {et~al.}}]{Planck2020}
{Planck Collaboration}, {Aghanim}, N., {Akrami, Y.}, {et~al.} 2020, \aap, 641,
  A1, \dodoi{10.1051/0004-6361/201833880}

\bibitem[{{{Planck Collaboration} and {Aatrokoski}, J. and {Ade}, P.~A.~R. and}
  {et~al.}(2011)}]{Planck2011b}
{{Planck Collaboration} and {Aatrokoski}, J. and {Ade}, P.~A.~R. and}, {et~al.}
  2011, \aap, 536, A15, \dodoi{10.1051/0004-6361/201116466}

\bibitem[{{Poglitsch} {et~al.}(2010)}]{Poglitsch2010}
{Poglitsch}, A., {et~al.} 2010, \aap, 518, L2,
  \dodoi{10.1051/0004-6361/201014535}

\bibitem[{{Randall} {et~al.}(2012)}]{Randall2012}
{Randall}, K.~E., {et~al.} 2012, \mnras, 421, 1644,
  \dodoi{10.1111/j.1365-2966.2012.20422.x}

\bibitem[{{Reuter} {et~al.}(2020)}]{Reuter2020}
{Reuter}, C., {et~al.} 2020, \apj, 902, 78, \dodoi{10.3847/1538-4357/abb599}

\bibitem[{{Rieke} {et~al.}(2009)}]{Rieke2009}
{Rieke}, G.~H., {et~al.} 2009, \apj, 692, 556,
  \dodoi{10.1088/0004-637X/692/1/556}

\bibitem[{{Rodighiero} {et~al.}(2010)}]{Rodighiero2010}
{Rodighiero}, G., {et~al.} 2010, \aap, 515, A8,
  \dodoi{10.1051/0004-6361/200912058}

\bibitem[{{Rowan-Robinson}(2001)}]{Rowan-Robinson2001}
{Rowan-Robinson}, M. 2001, \apj, 549, 745, \dodoi{10.1086/319450}

\bibitem[{{Salpeter}(1955)}]{Salpeter1955}
{Salpeter}, E.~E. 1955, \apj, 121, 161, \dodoi{10.1086/145971}

\bibitem[{{Sanders} \& {Mirabel}(1996)}]{Sanders1996}
{Sanders}, D.~B., \& {Mirabel}, I.~F. 1996, \araa, 34, 749,
  \dodoi{10.1146/annurev.astro.34.1.749}

\bibitem[{{Saunders} {et~al.}(2000)}]{Saunders2000}
{Saunders}, W., {et~al.} 2000, \mnras, 317, 55,
  \dodoi{10.1046/j.1365-8711.2000.03528.x}

\bibitem[{{Schlegel} {et~al.}(1998){Schlegel}, {Finkbeiner}, \&
  {Davis}}]{Schlegel1998}
{Schlegel}, D.~J., {Finkbeiner}, D.~P., \& {Davis}, M. 1998, \apj, 500, 525,
  \dodoi{10.1086/305772}

\bibitem[{{Serra} {et~al.}(2011)}]{Serra2011}
{Serra}, P., {et~al.} 2011, \apj, 740, 22, \dodoi{10.1088/0004-637X/740/1/22}

\bibitem[{{Skrutskie} {et~al.}(2003){Skrutskie}, {Cutri}, {Stiening},
  {et~al.}}]{2massallskypoint}
{Skrutskie}, M.~F., {Cutri}, R.~M., {Stiening}, R., {et~al.} 2003, 2MASS
  All-Sky Point Source Catalog,  IPAC, \dodoi{10.26131/IRSA2}

\bibitem[{{Spergel} {et~al.}(2003)}]{Spergel2003}
{Spergel}, D.~N., {et~al.} 2003, \apjs, 148, 175, \dodoi{10.1086/377226}

\bibitem[{{Story} {et~al.}(2013)}]{Story2013}
{Story}, K.~T., {et~al.} 2013, \apj, 779, 86,
  \dodoi{10.1088/0004-637X/779/1/86}

\bibitem[{{Su} {et~al.}(2017)}]{su2017}
{Su}, T., {et~al.} 2017, \mnras, 464, 968, \dodoi{10.1093/mnras/stw2334}

\bibitem[{{Sunyaev} \& {Zeldovich}(1972)}]{SunyaevZeldovich1972}
{Sunyaev}, R.~A., \& {Zeldovich}, Y.~B. 1972, Comments on Astrophysics and
  Space Physics, 4, 173

\bibitem[{{Swetz} {et~al.}(2011)}]{Swetz2011}
{Swetz}, D.~S., {et~al.} 2011, \apjs, 194, 41,
  \dodoi{10.1088/0067-0049/194/2/41}

\bibitem[{{Thornton} {et~al.}(2016)}]{Thornton2016}
{Thornton}, R.~J., {et~al.} 2016, \apjs, 227, 21,
  \dodoi{10.3847/1538-4365/227/2/21}

\bibitem[{{Toba} {et~al.}(2020)}]{Toba2020}
{Toba}, Y., {et~al.} 2020, \apj, 899, 35, \dodoi{10.3847/1538-4357/ab9cb7}

\bibitem[{{Two Micron All Sky Survey Science Team}(2020)}]{2massextended}
{Two Micron All Sky Survey Science Team}. 2020, 2MASS All-Sky Extended Source
  Catalog,  IPAC, \dodoi{10.26131/IRSA97}

\bibitem[{{Ulrich} {et~al.}(1997){Ulrich}, {Maraschi}, \& {Urry}}]{Ulrich1997}
{Ulrich}, M.-H., {Maraschi}, L., \& {Urry}, C.~M. 1997, \araa, 35, 445,
  \dodoi{10.1146/annurev.astro.35.1.445}

\bibitem[{{Urry} \& {Padovani}(1995)}]{Urry_Padovani1995}
{Urry}, C.~M., \& {Padovani}, P. 1995, \pasp, 107, 803, \dodoi{10.1086/133630}

\bibitem[{{van Velzen} {et~al.}(2012){van Velzen}, {Falcke}, {Schellart},
  {Nierstenh{\"o}fer}, \& {Kampert}}]{vanvelzen2012}
{van Velzen}, S., {Falcke}, H., {Schellart}, P., {Nierstenh{\"o}fer}, N., \&
  {Kampert}, K.-H. 2012, \aap, 544, A18, \dodoi{10.1051/0004-6361/201219389}

\bibitem[{{Vieira} {et~al.}(2010)}]{Vieira2010}
{Vieira}, J.~D., {et~al.} 2010, \apj, 719, 763,
  \dodoi{10.1088/0004-637X/719/1/763}

\bibitem[{{Viero} {et~al.}(2014)}]{Viero2014}
{Viero}, M.~P., {et~al.} 2014, \apjs, 210, 22,
  \dodoi{10.1088/0067-0049/210/2/22}

\bibitem[{{Vika} {et~al.}(2017){Vika}, {Ciesla}, {Charmandaris}, {Xilouris}, \&
  {Lebouteiller}}]{Vika2017}
{Vika}, M., {Ciesla}, L., {Charmandaris}, V., {Xilouris}, E.~M., \&
  {Lebouteiller}, V. 2017, \aap, 597, A51, \dodoi{10.1051/0004-6361/201629031}

\bibitem[{{Werner} {et~al.}(2004)}]{Werner2004}
{Werner}, M.~W., {et~al.} 2004, \apjs, 154, 1, \dodoi{10.1086/422992}

\bibitem[{{Wolf} {et~al.}(2018)}]{Wolf2018}
{Wolf}, C., {et~al.} 2018, \pasa, 35, e010, \dodoi{10.1017/pasa.2018.5}

\bibitem[{{Wright} {et~al.}(1994){Wright}, {Griffith}, {Burke}, \&
  {Ekers}}]{Wright1994}
{Wright}, A.~E., {Griffith}, M.~R., {Burke}, B.~F., \& {Ekers}, R.~D. 1994,
  \apjs, 91, 111, \dodoi{10.1086/191939}

\bibitem[{{Wright} {et~al.}(2010){Wright}, {Eisenhardt}, {Mainzer},
  {et~al.}}]{Wright2010}
{Wright}, E.~L., {Eisenhardt}, P.~R.~M., {Mainzer}, A.~K., {et~al.} 2010, \aj,
  140, 1868, \dodoi{10.1088/0004-6256/140/6/1868}

\bibitem[{{Wright, Edward L. et al.}(2019)}]{allwise}
{Wright, Edward L. et al.} 2019, AllWISE Source Catalog,  IPAC,
  \dodoi{10.26131/IRSA1}

\bibitem[{{Yamamura} {et~al.}(2018){Yamamura}, {Makiuchi}, {Koga}, \& {AKARI
  Team}}]{Yamamura2018}
{Yamamura}, I., {Makiuchi}, S., {Koga}, T., \& {AKARI Team}. 2018, in The
  Cosmic Wheel and the Legacy of the AKARI Archive: From Galaxies and Stars to
  Planets and Life, ed. T.~{Ootsubo}, I.~{Yamamura}, K.~{Murata}, \&
  T.~{Onaka}, 227--230

\bibitem[{{Yang} {et~al.}(2022)}]{Yang2022}
{Yang}, G., {et~al.} 2022, \apj, 927, 192, \dodoi{10.3847/1538-4357/ac4971}

\bibitem[{{York} {et~al.}(2000)}]{York2000}
{York}, D.~G., {et~al.} 2000, \aj, 120, 1579, \dodoi{10.1086/301513}

\bibitem[{{Zavala} {et~al.}(2021)}]{Zavala2021}
{Zavala}, J.~A., {et~al.} 2021, \apj, 909, 165,
  \dodoi{10.3847/1538-4357/abdb27}

\end{thebibliography}
\bibliographystyle{aasjournal}

\end{document}